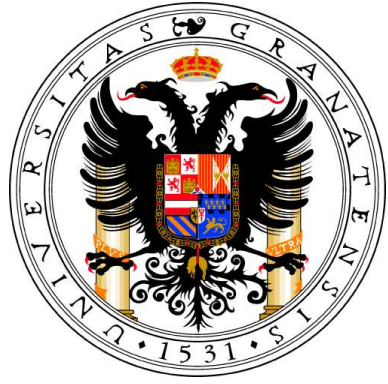

# *Information and Entanglement Measures in Quantum Systems with Applications to Atomic Physics*

TESIS DOCTORAL

por

# Daniel Manzano Diosdado

Departamento de Física Atómica, Molecular y Nuclear

Universidad de Granada

Marzo, 2010

A Raquel

**Tesis doctoral dirigida por:**

Dr. Jesús Sánchez-Dehesa

Dr. Ángel Ricardo Plastino

D. Jesús Sánchez-Dehesa Moreno-Cid, Doctor en Física, Doctor en Matemáticas y Catedrático del Departamento de Física Atómica, Molecular y Nuclear de la Facultad de Ciencias de la Universidad de Granada, y

D. Ángel Ricardo Plastino, Doctor en Astronomía e Investigador de Reconocida Valía de la Junta de Andalucía del Departamento de Física Atómica, Molecular y Nuclear de la Facultad de Ciencias de la Universidad de Granada.

MANIFIESTAN:

Que la presente Memoria titulada "Information and Entanglement Measures in Quantum Systems with Applications to Atomic Physics", presentada por Daniel Manzano Diosdado para optar al Grado de Doctor en Física, ha sido realizada bajo nuestra dirección en el Departamento de Física Atómica, Molecular y Nuclear de la Universidad de Granada y el Instituto Carlos I de Física Teórica y Computacional de la Universidad de Granada.

Granada, 3 de Marzo de 2010

Fdo.: Jesús Sánchez-Dehesa Moreno-Cid.

Fdo.: Ángel Ricardo Plastino.

Memoria presentada por Daniel Manzano Diosdado para optar al Grado de Doctor en Física por la Universidad de Granada.

Fdo.: Daniel Manzano Diosdado.

# Título de Doctor con Mención Europea

Con el fin de obtener la Mención Europea en el Título de Doctor, se han cumplido, en lo que atañe a esta Tesis Doctoral y a su Defensa, los siguientes requisitos:

1. La tesis está redactada en inglés con una introducción en español.
2. Dos de los miembros del tribunal provienen de una universidad europea no española.
3. Una parte de la defensa se ha realizado en inglés.
4. Una parte de esta Tesis Doctoral se ha realizado en Austria, en el Instituto de Óptica e Información Cuántica de Viena.

# Agradecimientos

# Contents

# Foreword

This thesis is a multidisciplinary contribution to the information theory of single-particle Coulomb systems (see Chapters 1, 2 and 3), to the theory of special functions of mathematical physics (Chapter 4), to quantum computation (Chapter 5) and to quantum information and atomic physics (Chapters 6 and 7). The notions of information, complexity and entanglement play a central role.

The Chapters are self-contained with their own Introduction and Conclusions. They may be read in arbitrary order and correspond to one (Chapters 1, 2 and 5) or two (Chapters 3, 4 and 7) scientific publications.

In Chapter 1 we explore both analytically and numerically the internal disorder of the hydrogenic atom which is associated to the non-uniformity of the quantum-mechanical probability density of its physical states, and which gives rise to the great diversity of three-dimensional geometries of its configuration orbitals. This is done for the ground and excited stationary states not only with the variance and the disequilibrium, but also by means of the Fisher-Shannon, Cramer-Rao and LMC shape complexities in position and momentum spaces. The dependence of these composite information-theoretic measures on the nuclear charge $Z$ and the three quantum numbers $(n, l, m)$ of the orbitals is carefully examined. Briefly, it is found that the three complexity measures do not depend on $Z$ and that the Fisher-Shannon measure quadratically depends on $n$. Moreover, the explicit expression of the shape complexity is obtained and sharp bounds to the Fisher-Shannon measure are given.

Chapter 2 generalizes to $D$ dimensions some of the themes from the preceding chapter. In it we provide the mathematical description of a formalism to calculate the LMC shape complexity of arbitrary stationary states of the $D$-dimensional hydrogenic system in terms of certain entropic functionals of the Laguerre and Gegenbauer or ultrashperical polynomials. We emphasize the ground and circular states, where the shape complexity is explicitly calculated and discussed in terms of the dimensionality and the quantum numbers. Then, the dimensional and Rydberg energy limits as well as their associated uncertainty products are explicitly given.

In Chapter 3 we extend the work done in the two previous chapters to include the relativistic effects at the Klein-Gordon level. Precisely, we investigate the relativistic charge compression of spinless Coulomb particles by means of various single (variance, Shannon entropy and Fisher information) and composite (Fisher-Shannon and LMC shape complexity) information-theoretic measures both qualitative and quantitatively. The three single charge spreading measures show that the relativistic effects are bigger (i.e. the charge compresses more towards the origin) for the low-lying energetic states and when the nuclear charge increases. Moreover, the relativistic effects enhance the variance and the Shannon information when $n$ ($l$) decreases for fixed $l$ ($n$); the Fisher information has the opposite behaviour. Moreover, the Fisher-Shannon complexity increases with the magnetic quantum number for fixed ($n$, $l$). It is observed, once the Lorentz invariance is appropriately taken into account, that the Shannon-Fisher and LMC complexities increases with the nuclear charge, contrary to the non-relativistic case.

In Chapter 4 we introduce the direct spreading measures of a family of special functions of the mathematical physics, the orthogonal hypergeometric polynomials, which quantify the spread of their associated Rakhmanov probability density all over their orthogonality interval in various complementary ways. Then, we emphasize the Hermite and Laguerre polynomials where we calculate not only the ordinary moments and standard deviation, but also the information-theoretic lengths of Renyi, Shannon and Fisher types. This is done for the Renyi measure by the use of a general methodology which uses the multivariable Bell polynomials so useful in Combinatorics and, in the Laguerre case, the linearization technique of Srivastava and Niukkanen. For the Shannon length, which cannot be analytically calculated because of its logarithmic functional form, its asymptotics and some upper bounds are obtained. The Fisher length is explicitly given. Later on, all the direct spreading measures of these polynomials are mutually compared and computationally analyzed.

Chapter 5 is a contribution in the field of learning processes for quantum computers that realizes classical operations. We propose a new model for making a quantum automaton for the processing of classical information. It is based in a machine that can make any unitary operation in one qubit. The operation used for testing the learning procedure is the $k^{th}$ root of NOT (logical negation) of a bit, that is a well defined classical operation. It is proved that the learning time for the quantum machine is independent of the root of the operation, but for any classical machine it will scale quadratically with $k$ for $k = 2^m$, being $m$ a natural number. Finally the speed of both a classical learning model and the quantum learning model is compared.

In Chapter 6 the state of the art in the field of entanglement and its applications to fermionic systems is reviewed. Special emphasis is done on separability criteria and entanglement measures for systems with distinguishable and indistinguishable subsystems and in the definition of entanglement for identical fermionic systems.

In Chapter 7, attention turns to the study of the entanglement properties of multi-fermionic systems. First, based on the linear and von Neumann entropies of the single particle reduced density matrix, we found separability criteria for pure states of $N$ identical fermions which are much simpler than the criteria recently proposed in the literature. Then we derive some inequalities for these entropies which allow us to propose natural entanglement measures for $N$-fermion pure states. Moreover, they connection with classical Hartree-Fock results are pointed out. After that, the new measures are used to study the entanglement properties of both ground and excited states of two exactly solvable systems of two charged fermions and compare them with those of the helium-like atoms by use of high-quality Kinoshita-like eigenfunctions. The dependence of the entanglement on the strength of the confining potential has been studied. Briefly, it is found in all cases that the amount of entanglement tends to grow when the energy is increasing. The dependence of the entanglement on the parameter of the models as well as on the nuclear charge is also investigated. Finally, a new criterion proposed by Walborn et al [1] in 2009 for the detection of entanglement in quantum systems with continuous variables is numerically analyzed.

# Introducción

Esta tesis es una contribución multidisciplinar a la teoría de la información de sistemas monoparticulares Coulombianos (ver Capítulos 1, 2 y 3), a la teoría de información de las funciones especiales de la física matemática (Capítulo 4), a la computación cuántica (Capítulo 5) y a la información cuántica de sistemas atómicos (Capítulos 6 y 7). Los conceptos de información, complejidad y entrelazamiento juegan un papel principal.

Los Capítulos son autocontenidos, con su propia introducción y conclusiones. Cada uno corresponde a una (Capítulos 1, 2 y 5) o dos (Capítulos 3, 4 y 7) publicaciones científicas.

En el Capítulo 1 exploramos analítica y numéricamente el desorden interno del átomo de hidrógeno, que está asociado a la no uniformidad de la densidad de probabilidad mecano-cuántica de sus estados, la cual está relacionada con la gran diversidad de geometrías tridimensionales de los orbitales atómicos. Este estudio se realiza para el estado fundamental y estados excitados no sólo mediante la determinación de la varianza y el desequilibrio, sino también por medio de las complejidades Fisher-Shannon, Cramer-Rao y LMC, tanto en el espacio de momentos como en el de posiciones. Se examina cuidadosamente la dependencia de estas tres medidas con la carga nuclear $Z$ y los números cuánticos $(n, l, m)$. En resumen, se encuentra que estas tres medidas de complejidad no dependen de $Z$, y que la medida de Fisher-Shannon depende cuadráticamente de $n$. Además se dan expresiones explícitas de la medida LMC así como cotas precisas a la medida de Fisher-Shannon.

El Capítulo 2 generaliza a sistemas $D$ dimensionales algunos de los resultados del capítulo anterior. En él se describe un formalismo fisico-matemático de cálculo de la complejidad LMC de estados estacionarios arbitrarios de un sistema hidrogenoide $D$-dimensional en términos de ciertos funcionales entrópicos de los polinomios de Laguerre y Gegenbauer o ultraesféricos. Se hace hincapié en el estado fundamental y los estados circulares, donde la complejidad LMC se calcula explícitamente y es analizada en función de la dimensionalidad y de los números cuánticos.

El Capítulo 3 extiende el trabajo hecho en los dos capítulos anteriores para incluir los efectos relativistas de tipo Klein-Gordon. Concretamente, investigamos la compresión de la carga de partículas Columbianas sin espín mediante varias medidas teórico-informacionales de tipo simple (varianza, entropía de Shannon e información de Fisher) y compuesto (complejidades Fisher-Shannon y LMC). Las tres medidas simples muestran que los efectos relativistas son más importantes (i.e. la carga se comprime más hacia el origen) para los estados de baja energía y cuando la carga nuclear aumenta. Además los efectos relativistas aumentan las medidas de dispersión simple (varianza y entropía de Shannon) cuando $n(l)$ disminuye para un $l(n)$ fijo, mientras la información de Fisher tiene un comportamiento opuesto. La medida de Fisher-Shannon también aumenta con el número cuántico magnético para un $(n, l)$ fijo. Se observa también que las complejidades de Fisher-Shannon y LMC aumentan con el número atómico $Z$, al contrario que en el caso no relativista.

En el Capítulo 4 introducimos las medidas de esparcimiento directas de una familia de funciones especiales de la física-matemática, los polinomios ortogonales hipergeométricos, que cuantifican de varias maneras la distribución de sus densidades de probabilidad de Rakhmanov en todo su intervalo de ortogonalidad. Hacemos hincapié en el caso de los polinomios de Hermite y Laguerre, donde calculamos no sólo los momentos ordinarios y la desviación estándar, sino también las longitudes de las medidas teórico-informacionales de Renyi, Shannon y Fisher. Esto se realiza para la medida de Renyi mediante el uso de una metodología general que usa los polinomios de Bell multivariables y, en el caso de los polinomios de Laguerre, mediante la fórmula de linealización de Srivastava y Niukkanen. Para la longitud de Shannon, que no puede ser calculada analíticamente debido a que es un funcional logarítmico, se determinan su asintótica y cotas superiores. La longitud de Fisher se obtiene explícitamente. Finalmente, todas estas medidas son comparadas entre si y analizadas computacionalmente.

El Capítulo 5 es una contribución al campo de los procesos de aprendizaje para computadores cuánticos que realizan operaciones clásicas. Proponemos un nuevo modelo para realizar un autómata cuántico para el procesado de información clásica. Este se basa en una máquina que puede realizar una operación arbitraria en un qubit. La operación usada para testear el proceso de aprendizaje es la raíz k-ésima de la operación NOT (negación lógica) de un bit, que es una operación clásica. Se puede probar que el tiempo de aprendizaje para la máquina cuántica es independiente de la raíz de la operación $k$; por otro lado la máquina clásica escala cuadráticamente con $k$ si $k = 2^m$, siendo $m$ un número natural. Finalmente se compara la velocidad de aprendizaje de un modelo clásico con la del modelo cuántico propuesto.

En el Capítulo 6 se revisa brevemente el concepto de entrelazamiento cuántico, haciendo énfasis en las diferencias existentes entre el concepto de entrelazamiento en sistemas constituidos por subsistemas distinguibles y el correspondiente concepto en sistemas de fermiones idénticos.

En el Capítulo 7 se investiga el entrelazamiento de sistemas multifermiónicos y de variables continuas. Primero, basándonos en las entropías lineal y de von Neumann de la matriz densidad reducida, encontramos criterios de separabilidad para estados puros de $N$ fermiones idénticos mucho más simples que otros criterios recientemente propuestos en la literatura. Derivamos unas desigualdades para estas entropías que nos permiten proponer medidas de entrelazamiento para sistemas puros de $N$ fermiones. Además se analizan las conexiones existentes entre estos resultados y ciertos resultados clásicos de la teoría de Hartree-Fock. Estas nuevas medidas se aplican al estudio de las propiedades de entrelazamiento, tanto para el estado fundamental como para estados excitados, de dos modelos resolubles de dos fermiones idénticos interactuantes, y se comparan los resultados con un modelo del helio basado en funciones de onda de tipo Kinoshita altamente precisas. Se explora la dependencia del entrelazamiento con la intensidad del potencial de confinamiento. Brevemente, se encuentra que en todos estos casos el entrelazamiento crece al aumentar la energía. También se estudia la dependencia del entrelazamiento con los parámetros de los modelos, así como con la carga nuclear. Finalmente se exploran numéricamente diversos aspectos un criterio de separabilidad recientemente propuesto por Walborn et al [1] para sistemas cuánticos de variables continuas, analizándose su eficiencia en función de diferentes parámetros.

# Author's publications

A.R. Plastino, D. Manzano and J.S. Dehesa. *Separability Criteria and Entanglement Measures for Pure States of N Identical Fermions.* EPL (Europhysics Letters), **86** 20005 (2009).

S. López-Rosa, D. Manzano and J.S. Dehesa. *Complexity of D-dimensional hydrogenic systems in position and momentum spaces.* Physica A, **388** 3273 (2009).

J.S. Dehesa, S. López-Rosa and D. Manzano. *Configuration Complexity of Hydrogenic Atoms.* European Physical Journal D, **55** 539 (2009).

S. López-Rosa, D. Manzano and J.S. Dehesa. *Multidimensional hydrogenic complexities.* Mathematical Physics and Field Theory: Julio Abad, in memoriam. Editors M. Asorey et al. Zaragoza: Prensas Universitarias de Zaragoza, (2009). See also arXiv:0907.0570.

D. Manzano, M. Pawłowski and Č. Brukner. *The speed of quantum and classical learning for performing the k-th root of NOT.* New Journal of Physics, **11** 113018 (2009).

P. Sánchez-Moreno, J.S. Dehesa, D. Manzano and R.J. Yáñez. *Spreading lengths of Hermite polynomials.* Journal of Computational and Applied Mathematics, **233** 2136 (2010).

D. Manzano, J.S. Dehesa and R.J. Yáñez. *Relativistic Klein-Gordon charge effects by information-theoretic measures.* New Journal of Physics, **12** 023014 (2010).

J.S. Dehesa, S. López-Rosa and D. Manzano. *Entropy and complexity analyses of D-dimensional quantum systems.* In special issue: "Statistical Complexities: Application to Electronic Structure", edited by K.D. Sen. Springer, Berlin, (2010).

D. Manzano, A.R. Plastino, J.S. Dehesa and T. Koga. *Quantum entanglement in two-electron atomic models.* J. Phys. A (2010).

D. Manzano, S. López-Rosa and J.S. Dehesa. *Complexity analysis of Klein-Gordon single particle systems.* Submitted for publication.

P. Sánchez-Moreno, D. Manzano and J.S. Dehesa. *Direct spreading measures of Laguerre polynomials.* Submitted for publication.

D. Manzano, A.R. Plastino and J.S. Dehesa. *Some features of two entropic separability criteria for continuous quantum systems.* Preprint.

# Chapter 1

# Configuration complexities of hydrogenic atoms

A basic problem in information theory of natural systems is the identification of the proper quantifier(s) of their complexity or internal disorder at their physical states. Presently this remains open not only for a complicated system, like e.g. a nucleic acid (either DNA or its single-strand lackey, RNA) in its natural (decidedly non-crystalline) state, but also for the simplest quantum-mechanical realistic systems, including the hydrogenic atom. Indeed there does not yet exist any quantity to properly measure the rich variety of three-dimensional geometries of the hydrogenic orbitals, which are described by means of three integer numbers: the principal, orbital and magnetic or azimuthal quantum numbers usually denoted by $n$, $l$ and $m$, respectively.

The root-mean-square or standard deviation does not measure the extent to which the electronic distribution is in fact concentrated, but rather the separation of the region(s) of concentration from a particular point of the distribution (the centroid or mean value), so that it is only useful for the nodeless ground state. In general, for excited states (whose probability densities are strongly oscillating) it is a misleading (and, at times, undefined) uncertainty measure. To take care of these defects, some information-theoretic quantities have been proposed: the Shannon entropic power [2, 3] defined by

$$H[\rho] = \exp\{S[\rho]\}; \qquad \text{with} \qquad S[\rho] = -\int \rho(\vec{r}) \log \rho(\vec{r}) \, d\vec{r}, \qquad (1.1)$$

the averaging density or disequilibrium [4–9] defined by

$$\langle \rho \rangle = \int [\rho(\vec{r})]^2 \, d\vec{r}, \qquad (1.2)$$

and the Fisher information [10] defined by

$$I[\rho] = \int \rho(\vec{r}) \left[\vec{\nabla} \log \rho(\vec{r})\right]^2 d\vec{r}. \tag{1.3}$$

The two former quantities measure differently the total extent or spreading of the electronic distribution. Moreover, they have a global character because they are quadratic and logarithmic functionals of the associated probability density $\rho(\vec{r})$. On the contrary, the Fisher information has a locality property because it is a gradient functional of the density, so that it measures the pointwise concentration of the electronic cloud and quantifies its gradient content, providing a quantitative estimation of the oscillatory character of the density. Moreover, the Fisher information measures the bias to particular points of the space, i.e. it gives a measure of the local disorder.

These three information-theoretic elements, often used as uncertainty measures, have shown (i) to be closely connected to various fundamental and/or experimentally measurable quantities (e.g., kinetic energy, ionization potential,..) (see e.g. [11, 12]) and (ii) to exhibit the periodicity of the atomic shell structure (see e.g. [11, 13, 14]). More recently, various composite information-theoretic measures have been introduced which have shown not only these properties but also other manifestations of the complexity of the atomic systems. Let us just mention the Fisher-Shannon measure defined by

$$C_{FS}[\rho] = I[\rho] \times J[\rho], \qquad \text{with} \qquad J[\rho] = \frac{1}{2\pi e} \exp(2S[\rho]/3), \tag{1.4}$$

the Cramer-Rao or Fisher-Heisenberg measure (see e.g. [11, 15, 16]) defined by

$$C_{CR}[\rho] = I[\rho] \times V[\rho], \qquad \text{with} \qquad V[\rho] = \langle r^2 \rangle - \langle r \rangle^2, \tag{1.5}$$

and the LMC shape complexity [6, 17] defined by

$$C_{SC}[\rho] = \langle \rho \rangle \times H[\rho]. \tag{1.6}$$

They quantify different facets of the internal disorder of the system which are manifest in the diverse and complex three-dimensional geometries of its orbitals. The Fisher-Shannon measure grasps the oscillatory nature of the electronic probability cloud together with its total extent in the configuration space. The Cramer-Rao quantity takes also into account the gradient content but jointly with the electronic spreading around the centroid. The shape complexity measures the combined effect of the average height and the total spreading of the probability density; so, being insensitive to the electronic oscillations. This measure exhibits the important property of scale invariance, which the original LMC measure [6] lacks, as it was first pointed out by Anteneodo and Plastino [4].

However, it has not yet been proved its usefulness to disentangle among the rich three-dimensional atomic geometries of any physical system, not even for the hydrogenic atom although some properties have been recently found [18]. In this Section we will investigate this issue by means of the three composite information-theoretic measures just mentioned for general hydrogenic orbitals in position and momentum spaces. Briefly, let us advance that here we find that the Fisher-Shannon measure turns out to be the most appropriate measure to describe the (intuitive) complexity of the three dimensional geometry of hydrogenic orbitals.

Nevertheless we should immediately say that these three measures are complementary in the sense that, according to its composition, they grasp different facets of the internal disorder of the system which are manifest in the great diversity and complexity of configuration shapes of the probability density $\rho(\vec{r})$ corresponding to its orbitals $(n, l, m)$. The Fisher-Shannon and Cramer-Rao measures have an ingredient of local character (namely, the Fisher information) and another one of global character (the modified Shannon entropic power in the Fisher-Shannon case and the variance in the Cramer-Rao case). The shape complexity is composed by two global ingredients: the disequilibrium and the Shannon entropic power; so, this quantity is not well prepared to grasp the oscillating nature of the hydrogenic orbitals but it takes into account the average height and the total extent of the electron distribution. The Fisher-Shannon measure appropriately describes the oscillating nature together with the total extent of the probability cloud of the orbital. The Cramer-Rao measure takes into account the gradient content jointly with the spreading of the probability density around its centroid.

The structure of the Chapter is the following. First, in Section 1.1, the hydrogenic problem is briefly reviewed to fix notations and to gather the known results about the information-theoretic measures of the hydrogenic orbitals. In Section 1.2, the three composite measures mentioned above are discussed both numerically and analytically for the ground and excited hydrogenic states. In Section 1.3, various sharp upper bounds for these composite measures are provided in terms of the three quantum numbers of the orbital. Finally, some conclusions are given.

## 1.1 The hydrogenic problem: Information-theoretic measures

In this Section we first describe the hydrogenic orbitals in the configuration space to fix notations; then we gather some known results for various spreading measures (variance, Fisher information and Shannon entropy) of the system in terms of the quantum numbers $(n, l, m)$ of the orbital.

The position hydrogenic orbitals (i.e., the solutions of the non-relativistic, time-independent Schrödinger equation describing the quantum mechanics for the motion of an electron in the Coulomb field of a nucleus with charge $+Ze$) corresponding to stationary states of the hydrogenic system in the configuration space are characterized within the infinite-nuclear-mass approximation by the energetic eigenvalues

$$E = -\frac{Z^2}{2n^2}, \quad\quad\quad n = 1, 2, 3, ..., \quad\quad\quad (1.7)$$

and the spatial eigenfunctions

$$\Psi_{n,l,m}(\vec{r}) = R_{n,l}(r)Y_{l,m}(\Omega), \quad\quad\quad (1.8)$$

where $n = 1, 2, ...$, $l = 0, 1, ..., n-1$ and $m = -l, -l+1, ..., l-1, l$, and $r = |\vec{r}|$ and the solid angle $\Omega$ is defined by the angular coordinates $(\theta, \varphi)$. The radial eigenfunction, duly normalized to unity, is given by

$$R_{n,l}(r) = \frac{2Z^{3/2}}{n^2}\left[\frac{\omega_{2l+1}(\tilde{r})}{\tilde{r}}\right]^{1/2}\tilde{L}_{n-l-1}^{(2l+1)}(\tilde{r}), \quad\quad\quad (1.9)$$

with $\tilde{r} = \frac{2Zr}{n}$, and $\{\tilde{L}_k^{(\alpha)}(x)\}$ denote the Laguerre polynomials orthonormal with respect to the weight function $\omega_\alpha(x) = x^\alpha e^{-x}$ on the interval $[0, \infty)$; that is, they satisfy the orthogonality relation

$$\int_0^\infty dx\, \omega_\alpha(x)\tilde{L}_n^{(\alpha)}(x)\tilde{L}_m^{(\alpha)}(x) = \delta_{nm}. \quad\quad\quad (1.10)$$

The angular eigenfunction $Y_{l,m}(\theta, \varphi)$ are the renowned spherical harmonics which describe the bulky shape of the system and are given by

$$Y_{l,m}(\theta, \varphi) = \frac{1}{\sqrt{2\pi}}e^{im\varphi}\tilde{C}_{l-m}^{(m+1/2)}(\cos\theta)(\sin\theta)^m, \quad\quad\quad (1.11)$$

where $\{\tilde{C}_k^{(\lambda)}(x)\}$ denotes the Gegenbauer or ultraspherical polynomials, which are orthonormal with respect to the weight function $(1-x^2)^{\lambda-1/2}$ on the interval $[-1, +1]$. Then, the probability to find the electron between $\vec{r}$ and $\vec{r} + d\vec{r}$ is

$$\rho(\vec{r})\,d\vec{r} = |\Psi_{n,l,m}(\vec{r})|^2\,d\vec{r} = D_{n,l}(r)dr \times \Theta_{l,m}(\theta)d\theta d\varphi,$$

where

$$D_{n,l}(r) = R_{n,l}^2(r)r^2, \quad\quad \text{and} \quad\quad \Theta_{l,m}(\theta) = |Y_{l,n}(\theta, \varphi)|^2 \sin\theta, \quad\quad (1.12)$$

are the known radial and angular probability densities, respectively. So, the total probability density of the hydrogenic atom is given by

$$\rho(\vec{r}) = \frac{4Z^3}{n^4}\frac{\omega_{2l+1}(\tilde{r})}{\tilde{r}}\tilde{L}_{n-l-1}^{(2l+1)}(\tilde{r})\,|Y_{l,m}(\theta, \varphi)|^2. \quad\quad\quad (1.13)$$

Let us now gather the known results for the following spreading measures of our system: the variance and the Fisher information. They have the values

$$V[\rho] = \frac{n^2(n^2+2) - l^2(l+1)^2}{4Z^2}, \qquad (1.14)$$

for the variance [19, 20], and

$$I[\rho] = \frac{4Z^2}{n^3}[n - |m|], \qquad (1.15)$$

for the Fisher information [21, 22].

The Shannon information of the hydrogenic atom $S[\rho]$ is composed by the radial part given by

$$S(R_{n,l}) = A_1(n,l) + \frac{1}{2n}E_1\left(\tilde{L}_{n-l-1}^{(2l+1)}\right) - 3\log Z, \qquad (1.16)$$

with

$$A_1(n,l) = \log\left(\frac{n^4}{4}\right) + \frac{3n^2 - l(l+1)}{n} - 2l\left[\frac{2n-2l-1}{2n} + \Psi(n+l+1)\right],$$

where $\psi(x) = \Gamma'(x)/\Gamma(x)$ is the digamma function, and the angular part given by

$$S(Y_{l,m}) = A_2(l,m) + E_0\left(\tilde{C}_{l-|m|}^{(|m|+1/2)}\right), \qquad (1.17)$$

$$A_2(l,m) = \log\left(2^{2|m|+1}\pi\right) - 2|m|\left[\psi(l+m+1) - \psi(l+1/2) - \frac{1}{2l+1}\right].$$

Then, from Eqs. (1.16)-(1.17), one has the value for the Shannon information of the state $(n,l,m)$:

$$\begin{aligned}S[\rho] &= S(R_{n,l}) + S(Y_{l,m}) \\ &= A(n,l,m) + \frac{1}{2n}E_1\left(\tilde{L}_{n-l-1}^{(2l+1)}\right) + E_0\left(\tilde{C}_{l-|m|}^{(|m|+1/2)}\right) - 3\log Z, \quad (1.18)\end{aligned}$$

with

$$\begin{aligned}A(n,l,m) &= A_1(n,l) + A_2(l,m) \\ &= \log\left(2^{2|m|-1}\pi n^4\right) + \frac{3n^2 - l(l+1)}{n} \\ &\quad - 2l\left[\frac{2n-2l-1}{2n} + \psi(n+l+1)\right] \\ &\quad - 2|m|\left[\psi(l+m+1) - \psi(l+1/2) - \frac{1}{2l+1}\right]. \quad (1.19)\end{aligned}$$

The symbols $E_i(\tilde{y}_n)$, $i = 0$ and 1, denote the following entropic integrals of the polynomials $\{\tilde{y}_n\}$ orthonormal with respect to the respective weight function $\omega(x)$ on $x \in [a, b]$

$$E_i(\tilde{y}_n) = \int_a^b x^i \omega(x) \tilde{y}_n^2(x) \log \tilde{y}_n^2(x) dx, \quad (1.20)$$

whose calculation is a difficult, not-yet-accomplished analytical task for polynomials of generic degree in spite of numerous efforts [23–26]. As a particular case, let us mention that for the ground state ($n = 1, l = m = 0$), Eqs. (1.14), (1.15) and (1.18)-(1.20) yield the following values

$$V[\rho_{g.s.}] = \frac{3}{4Z^2}, \quad I[\rho_{g.s.}] = 4Z^2 \quad \text{and} \quad S[\rho_{g.s.}] = 3 + \log \pi - 3 \log Z,$$

for the variance, Fisher information and Shannon entropy, respectively.

Let us now calculate the disequilibrium or averaging density $\langle \rho \rangle$ of the hydrogenic orbital $(n, l, m)$. From Eqs. (1.2) and (1.8) one has

$$\begin{aligned}\langle \rho \rangle &= \int_{\Re^3} \rho^2(\vec{r}) d^3r = \int_0^\infty r^2 |R_{nl}(r)|^4 dr \times \int_\Omega |Y_{lm}(\Omega)|^4 d\Omega \\ &\equiv \langle \rho \rangle_R \times \langle \rho \rangle_Y.\end{aligned} \quad (1.21)$$

Let us begin with the calculation of the radial part $\langle \rho \rangle_R$. For purely mathematical convenience we use the notation $n_r = n - l - 1$ and the change of variable $\tilde{r} = \frac{2Z}{n}r$. Then one has

$$\langle \rho \rangle_R = \left(\frac{n}{2Z}\right)^3 \int_0^\infty |R_{nl}(\tilde{r})|^4 \tilde{r}^2 d\tilde{r} = \frac{2Z^3}{n^5} \left(\frac{n_r!}{(n+l)!}\right)^2 K(n_r, l), \quad (1.22)$$

where $K(n_r, l)$ denotes the integral

$$\begin{aligned}K(n_r, l) &= \int_0^\infty e^{-2\tilde{r}} \tilde{r}^{4l+2} \left[L_{n_r}^{(2l+1)}(\tilde{r})\right]^4 d\tilde{r} \\ &= 2^{-4l-3} \left[\frac{\Gamma(2l+n_r+2)}{2^{2n_r} n_r!}\right]^2 \sum_{k=0}^{n_r} \binom{2n_r - 2k}{n_r - k}^2 \frac{(2k)! \Gamma(4l+2k+3)}{(k!)^2 \Gamma^2(2l+k+2)} \quad (1.23)\end{aligned}$$

For the second equation, see Appendix A. Then, the substitution of Eq. (1.23) into Eq. (1.22) yields the following value

$$\langle \rho \rangle_R = \frac{Z^3 2^{2-4n}}{n^5} \sum_{k=0}^{n_r} \binom{2n_r - 2k}{n_r - k}^2 \frac{(k+1)_k}{k!} \frac{\Gamma(4l+2k+3)}{\Gamma^2(2l+k+2)}, \quad (1.24)$$

for the radial part of the disequilibrium. Remark that we have used the Pochhammer symbol $(x)_k = \Gamma(x+k)/\Gamma(x)$.

The angular contribution to the disequilibrium is

$$\begin{aligned}\langle\rho\rangle_Y &= \int_0^{2\pi} d\phi \int_0^{\pi} \sin\theta d\theta \, |Y_{lm}(\theta,\phi)|^4 \\ &= \sum_{l'=0}^{2l} \left(\frac{\hat{l}^2 \hat{l'}}{\sqrt{4\pi}}\right)^2 \begin{pmatrix} l & l & l' \\ 0 & 0 & 0 \end{pmatrix}^2 \begin{pmatrix} l & l & l' \\ m & m & -2m \end{pmatrix}^2,\end{aligned} \quad (1.25)$$

for the angular part of the disequilibrium. This expression is considerably much more transparent and simpler than its equivalent $_3F_2(1)$-form recently obtained [27] by other means. See Appendix A for further details.

Finally, the combination of Eqs. (1.21), (1.24) and (1.25) yields the value

$$\langle\rho\rangle = Z^3 D(n,l,m), \quad (1.26)$$

for the total disequilibrium of the hydrogenic orbital $(n,l,m)$, where $D(n,l,m)$ is given by

$$\begin{aligned}D(n,l,m) &= \frac{(2l+1)^2}{2^{4n}\pi n^5} \sum_{k=0}^{n_r} \binom{2n_r - 2k}{n_r - k}^2 \frac{(k+1)_k}{k!} \frac{\Gamma(4l+2k+3)}{\Gamma^2(2l+k+2)} \\ &\quad \times \sum_{l'=0}^{2l} (2l'+1) \begin{pmatrix} l & l & l' \\ 0 & 0 & 0 \end{pmatrix}^2 \begin{pmatrix} l & l & l' \\ m & m & -2m \end{pmatrix}^2,\end{aligned} \quad (1.27)$$

where $n_r = n - l - 1$. Note that for the ground state, the disequilibrium is $\langle\rho_{g.s.}\rangle = \frac{Z^3}{8\pi}$.

## 1.2 Composite information-theoretic measures of hydrogenic orbitals

Let us here discuss both analytical and numerically the three following composite information-theoretic measures of a general hydrogenic orbital with quantum numbers $(n,l,m)$: the Cramer-Rao or Fisher-Heisenberg and Fisher-Shannon measures and the shape complexity. Briefly, let us highlight in particular that these three quantities do not depend on the nuclear charge $Z$. Moreover, (a) the Cramer-Rao measure is given explicitly, (b) the Fisher-Shannon measure is shown to quadratically depend on the principal quantum number $n$, and (c) the shape complexity, which is a modified version of the LMC complexity [6], is carefully analyzed in terms of the quantum numbers. In this way we considerably extend the recent finding of Sañudo and López-Ruíz [18] relative to the fact that the Fisher-Shannon and shape complexities have their minimum values for the orbitals with the highest orbital momentum.

The Cramer-Rao measure is obtained in a straightforward manner from Eqs. (1.5), (1.14) and (1.15), having the value

$$C_{CR}[\rho] = \frac{n-|m|}{n^3}\left[n^2(n^2+2) - l^2(l+1)^2\right]. \tag{1.28}$$

Now, from Eqs. (1.4), (1.15) and (1.16)-(1.19) one has that the Fisher-Shannon measure has the value

$$C_{FS}[\rho] = \frac{4(n-|m|)}{n^3}\frac{1}{2\pi e}e^{\frac{2}{3}B(n,l,m)}, \tag{1.29}$$

where

$$B(n,l,m) = A(n,l,m) + \frac{1}{2n}E_1\left(\tilde{L}_{n-l-1}^{(2l+1)}\right) + E_0\left(\tilde{C}_{l-|m|}^{(|m|+1/2)}\right). \tag{1.30}$$

The symbols $E_i(\tilde{y}_n)$ denote the entropic integrals given by Eq. (1.20). Similarly, taking into account Eqs. (1.1), (1.6), (1.18)-(1.19) and (1.26) one has the value

$$C_{SC}[\rho] = D(n,l,m)e^{B(n,l,m)}, \tag{1.31}$$

where the explicit expression of $D(n,l,m)$ is given by Eq. (1.27). In particular, for the ground state we have the values

$$C_{CR}[\rho_{g.s.}] = 3, \qquad C_{FS}[\rho_{g.s.}] = \frac{2e}{\pi^{1/3}}, \qquad \text{and} \qquad C_{SC}[\rho_{g.s.}] = \frac{e^3}{8},$$

for the three composite information-theoretic measures mentioned above. Let us highlight from Eqs. (1.29)-(1.31) that the three composite information-theoretic measures do not depend on the nuclear charge. Moreover, it is known that $C_{FS}[\rho] \geq 3$ for all three-dimensional densities [3, 15] but also $C_{CR}[\rho] \geq 3$ for any hydrogenic orbital as one can easily show from Eq. (1.28).

Let us now discuss numerically the Fisher-Shannon, Cramer-Rao and shape complexity measures of hydrogenic atoms for various specific orbitals in terms of their corresponding quantum numbers $(n,l,m)$. To make possible the mutual comparison among these measures and to avoid problems with physical dimensions, we study the dependence of the ratio between the measures $C[\rho_{n,l,m}] \equiv C(n,l,m)$ of the orbital we are interested in and the corresponding measure $C[\rho_{1,0,0}] \equiv C(1,0,0) \equiv C(g.s.)$ of the ground state, that is:

$$\zeta(n,l,m) := \frac{C(n,l,m)}{C(1,0,0)},$$

on the three quantum numbers. The results are shown in Figures 1.1, 1.4 and 1.5, where the relative values of the three composite information-theoretic measures are plotted in terms of $n$, $m$ and $l$, respectively. More specifically, in Figure 1.1, we have given the three measures for various ns-states (i.e., with $l = m = 0$). Therein, we observe that (a) the Fisher-Shannon and Cramer-Rao measures have an increasing parabolic behaviour when

$n$ is increasing while the shape complexity is relatively constant, and (b) the following inequalities

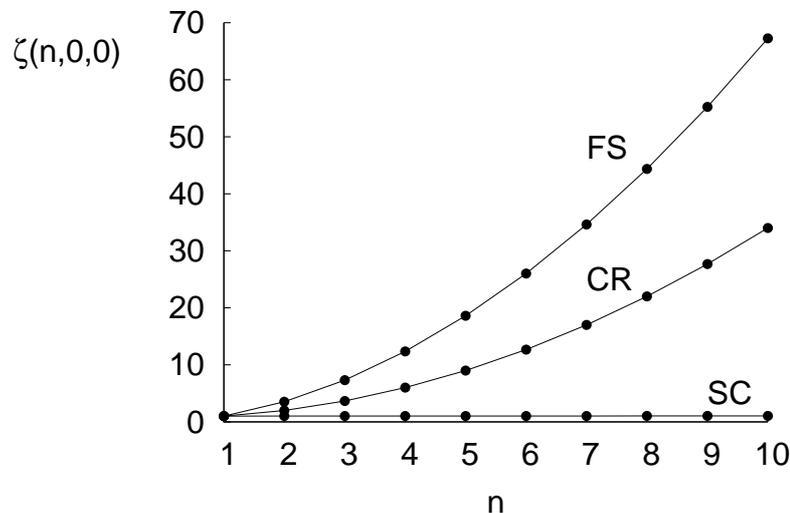

FIGURE 1.1: Relative Fisher-Shannon measure $\zeta_{FS}(n,0,0)$, Cramer-Rao measure $\zeta_{CR}(n,0,0)$ and shape complexity $\zeta_{SC}(n,0,0)$ of the ten lowest hydrogenic states s as a function of $n$. See text.

$$\zeta_{FS}(n,0,0) > \zeta_{CR}(n,0,0) > \zeta_{SC}(n,0,0),$$

are fulfilled for fixed $n$. Similar characteristics are shown by states $(n,l,m)$ other than $(n,0,0)$. Both to understand this behaviour and to gain a deeper insight into the internal complexity of the hydrogenic atom which is manifest in the three-dimensional geometry of its configuration orbitals (and so, in the spatial charge distribution density of the atom at different energies), we have drawn the radial $D_{n,l} = R_{n,l}^2(r)r^2$ and angular $\Theta_{l,m}(\theta) = |Y_{l,m}(\theta,\varphi)|^2 \sin\theta$ densities (see Eq. (1.12)) in Figures 1.2 and 1.3, respectively, for the three lowest energetic levels of hydrogen.

From Figure 1.2 we realize that when $n$ is increasing and $l$ is fixed, both the oscillatory character (so, the gradient content and its associated Fisher information) and the spreading (so, the Shannon entropic power) of the radial density certainly grow while its variance hardly does so and the average height (which controls $\langle \rho \rangle$) clearly decreases. Taking into account these radial observations and the graph of $\Theta_{0,0}(\theta)$ at the top line of Figure 1.3, we can understand the parabolic growth of the Fisher-Shannon and Cramer-Rao measures as well as the lower value and relative constancy of the shape complexity for ns-states shown in Figure 1.1 when $n$ is increasing. In fact, the gradient content (mainly because of its radial contribution) and the spreading of the radial density of these states contribute constructively to the Fisher-Shannon measure of hydrogen, while the spreading and the average height almost cancel one to another, making the shape complexity to have a very small, almost constant value; we should say, for completeness, that $\zeta_{SC}(n,0,0)$ increases from 1 to 1.04 when $n$ varies from 1 to 10. In the Cramer-Rao case, the parabolic growth is practically only due to the increasing behaviour of the gradient content , so to its Fisher information ingredient.

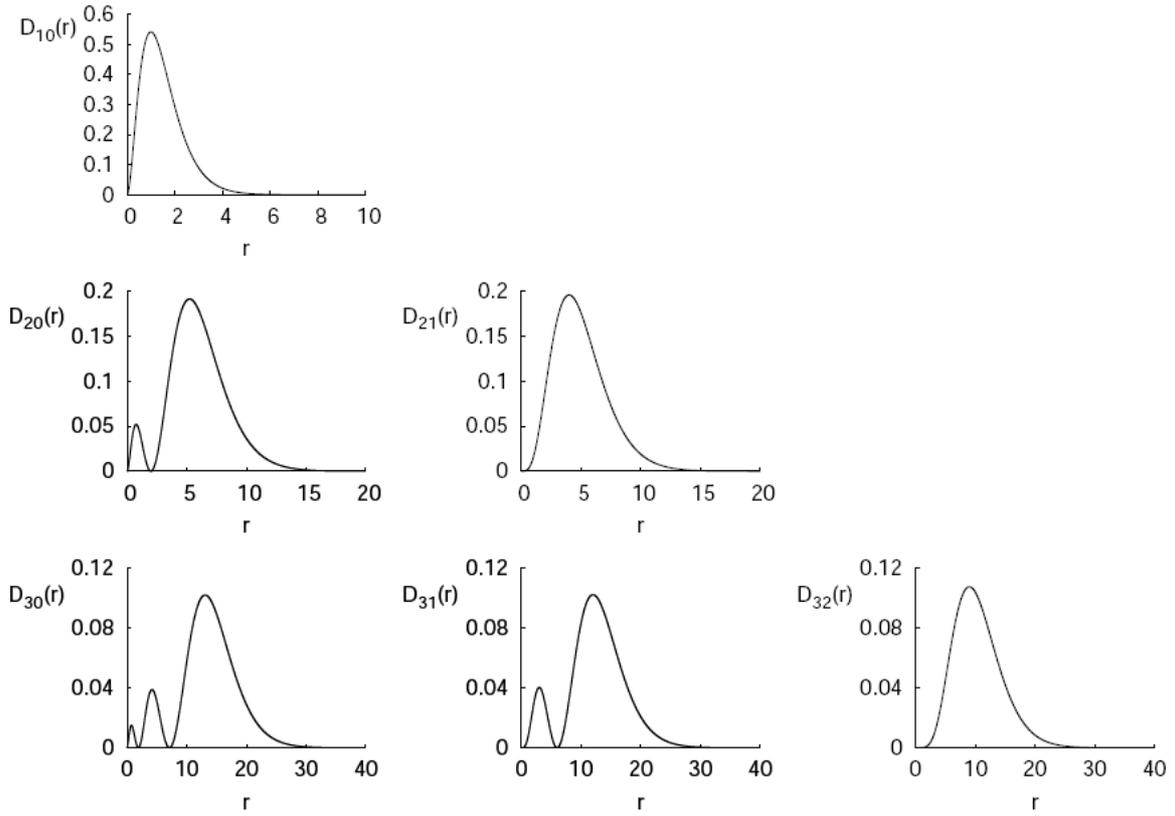

FIGURE 1.2: Radial distribution $D_{n,l}(r) = R_{n,l}^2(r)r^2$ of all the electronic orbitals corresponding to the three lowest energy levels of hydrogen. Atomic units have been used.

Let us now explain and understand the linear decreasing behaviour of the Fisher-Shannon and Cramer-Rao measures as well as the practical constancy of the shape complexity for the hydrogen orbital ($n = 20, l = 17, m$) when $|m|$ is increasing, as shown in Figure 1.4. These phenomena purely depend on the angular contribution due to the analytical form of the angular density $\Theta_{17,m}(\theta)$ since the radial contribution (i.e. that due to the radial density $R_{n,l}(r)$) is constant when $m$ varies. A straightforward extrapolation of the graphs corresponding to the angular densities $\Theta_{l,m}(\theta)$ contained in Figure 1.3, shows that when $l$ is fixed and $|m|$ is increasing, both the gradient content and spreading of this density decrease while the average height and the probability concentration around its centroid are apparently constant. Therefore, the Fisher-Shannon and Cramer-Rao have a similar decreasing behaviour as shown in Figure 1.4 although with a stronger rate in the former case, because its two ingredients (Fisher information and Shannon entropic power) contribute constructively while in the Cramer-Rao case, one of the ingredients (namely, the variance) does not contribute at all. Keep in mind, by the way, that the relations (1.14) and (1.15) show that the total variance does not depend on $m$ and the Fisher information linearly decreases when $|m|$ is increasing, respectively. On the other hand, Figure 1.3 shows that the angular average height increases while the spreading decreases so that the overall combined contribution of these two ingredients to the shape complexity is relatively constant and very small when $|m|$ varies; in fact, $\zeta_{SC}(20, 17, m)$ parabolically decreases from 1 to 0.6 when $|m|$ varies from 0 to 17.

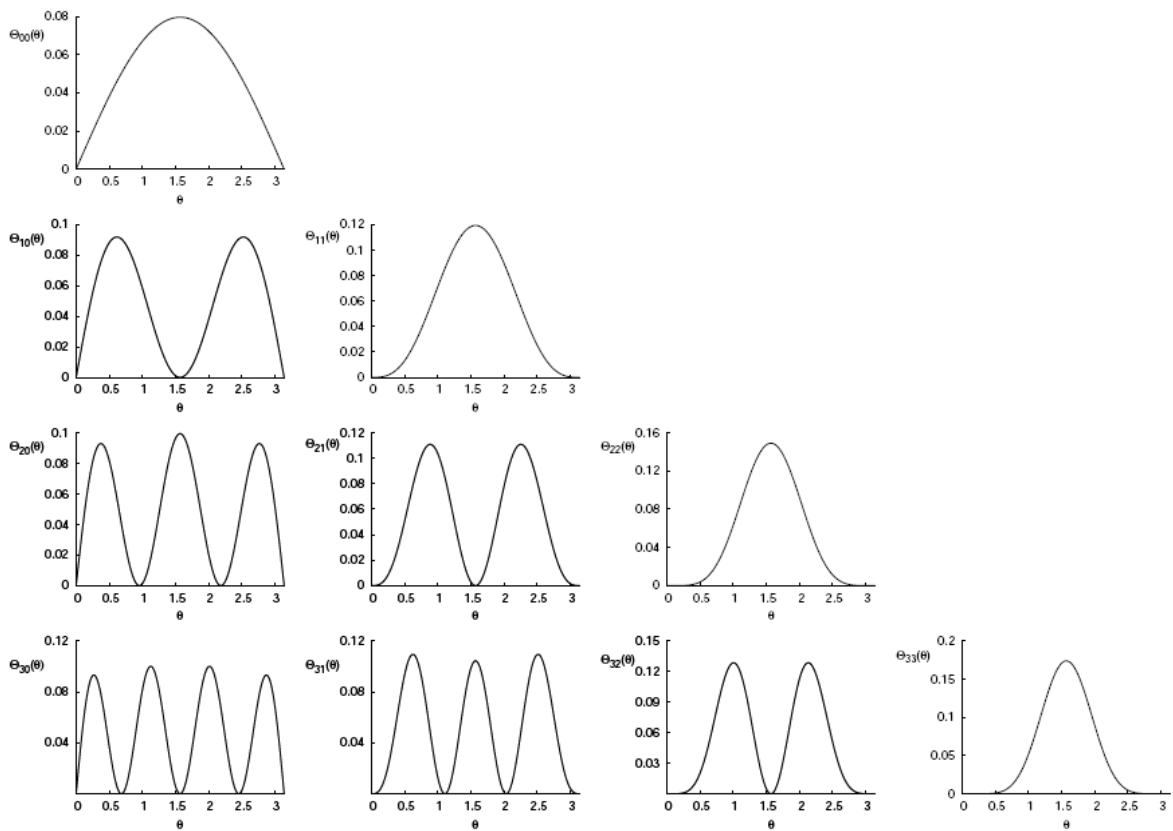

FIGURE 1.3: Angular distribution $\Theta_{l,m}(\theta) = |Y_{l,m}(\theta, \varphi)|^2 \sin\theta$ of all electronic orbitals corresponding to the four lowest lying energy levels of hydrogen. Atomic units have been used.

In Figure 1.5 it is shown that the Fisher-Shannon and Cramer-Rao measures have a concave decreasing form and the shape complexity turns out to be comparatively constant for the orbital $(n = 20, l, m = 1)$ when the orbital quantum number $l$ varies. We can understand these phenomena by taking into account the graphs, duly extrapolated, of the lines of Figure 1.2 and the columns of Figure 1.3 where the radial density for fixed $n$ and the angular density for fixed $m$ are shown. Herein we realize that when $l$ is increasing, (a) the radial gradient content decreases while the corresponding angular quantity increases, so that the gradient content of the total density $\rho(\vec{r})$ does not depend on $l$ in accordance to its Fisher information as given by Eq. (1.15); (b) the radial and angular spreadings have decreasing and constant behaviours, respectively, so that the overall effect is that the Shannon entropic power of the total density $\rho(\vec{r})$ increases,

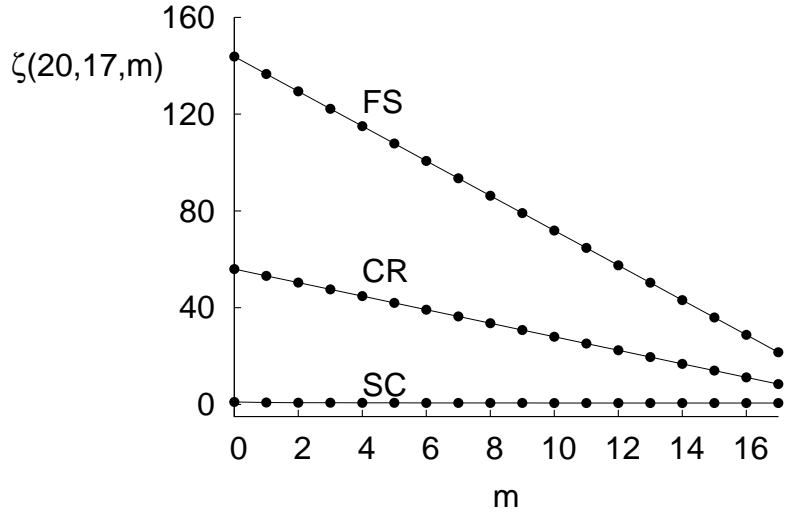

FIGURE 1.4: Relative Fisher-Shannon measure $\zeta_{FS}(20,17,m)$, Cramer-Rao measure $\zeta_{CR}(20,17,m)$ and shape complexity $\zeta_{SC}(20,17,m)$ of the manifold of hydrogenic levels with $n = 20$ and $l = 17$ as a function of the magnetic quantum number $m$. See text.

|  | $a$ | $b$ | $c$ | $R$ |
|---|---|---|---|---|
| $C_{FS}(n,0,0)$ | 0.565 | 1.202 | -1.270 | 0.999996 |
| $C_{FS}(n,3,1)$ | 0.451 | 0.459 | -4.672 | 0.999998 |

TABLE 1.1: Fisher-Shannon measure of the hydrogenic orbitals $(n,l,m) = (n,0,0)$ and $(n,3,1)$

(c) both the radial and the angular average height increase, so that the total averaging density $\rho(\vec{r})$ increases, and a similar phenomenon occurs with the concentration of the radial and angular probability clouds around their respective mean value, so that the total variance $V[\rho]$ decreases very fast (as Eq. (1.14) analytically shows). Taking into account these observations into the relations (1.4), (1.5) and (1.6) which define the three composite information-theoretic measures under consideration, we can immediately explain the decreasing dependence of the Fisher-Shannon and Cramer-Rao measures on the orbital quantum number as well as the relative constancy of the shape complexity, as illustrated in Figure 1.5; in fact, $\zeta_{SC}(20,l,1)$ also decreases but within the small interval $(1, 0.76)$ when $l$ goes from 0 to 19.

Finally, for completeness, we have numerically studied the dependence of the Fisher-Shannon measure on the principal quantum number $n$. We have found the fit

$$C_{FS}(n,l,m) = a_{lm}n^2 + b_{lm}n + c_{lm},$$

where the parameters a,b,c are given in Table 1.1 for two particular states with the corresponding correlation coefficient $R$ of the fit. It would be extremely interesting to show this result from Eqs. (1.29)-(1.30) in a rigorous mathematical way.

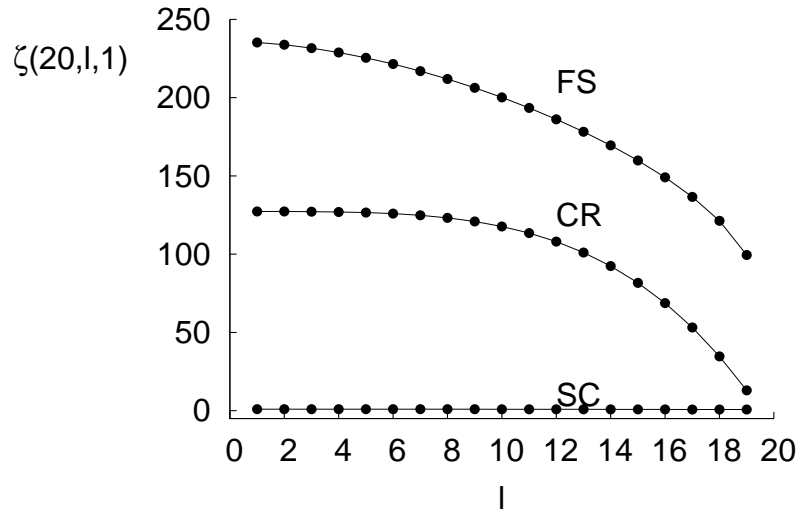

FIGURE 1.5: Relative Fisher-Shannon measure $\zeta_{FS}(20,l,1)$, Cramer-Rao measure $\zeta_{CR}(20,l,1)$ and shape complexity $\zeta_{SC}(20,l,1)$ of the hydrogenic states with $n=20$ and $m=1$ as a function of the orbital quantum number $l$. See text.

## 1.3 Upper bounds to the Fisher-Shannon measure and shape complexity

We have seen previously that, contrary to the Cramer-Rao measure whose expression can be calculated explicitly in terms of the quantum numbers $(n,l,m)$, the Fisher-Shannon measure and the shape complexity have not yet been explicitly found. This is basically because one of their two ingredients (namely, the Shannon entropic power) has not yet been computed directly in terms of the quantum numbers. Here we will calculate rigorous upper bounds to these two composite information-theoretic measures by means of the three quantum numbers of a generic hydrogenic orbital. Let us first gather the expressions

$$C_{FS}[\rho] = \frac{4Z^2}{n^3}(n-|m|)\frac{1}{2\pi e}e^{\frac{2}{3}S[\rho]}, \quad (1.32)$$

for the Fisher-Shannon measure and

$$C_{SC}[\rho] = Z^3 D(n,l,m) e^{S[\rho]}, \quad (1.33)$$

for the shape complexity of the hydrogenic orbital $(n,l,m)$, where $S[\rho]$ denotes the Shannon information entropy given by Eq. (1.1) and $D(n,l,m)$ has the exact value given by Eq. (1.27). To write down these two expressions, we have taken into account Eqs. (1.4) and (1.15) and Eqs. (1.1), (1.6) and (1.27), respectively. The exact calculation of the Shannon entropy $S[\rho]$ is a formidable open task, not yet accomplished in spite of numerous efforts [24–26]. Nevertheless, variational bounds to this information-theoretic quantity have been found [28–30] by means of one and two radial expectation values.

$$S[\rho] \leqslant \log\left[8\pi\left(\frac{e\langle r\rangle}{3}\right)^3\right]. \tag{1.34}$$

Then, taking into account that the expectation value $\langle r\rangle$ of the hydrogenic orbital $(n,l,m)$ is given [19, 31] by

$$\langle r\rangle = \frac{1}{2Z}\left[3n^2 - l(l+1)\right], \tag{1.35}$$

so that

$$S[\rho] \leqslant \log\left\{\frac{\pi e^3}{27Z^3}\left[3n^2 - l(l+1)\right]^3\right\}. \tag{1.36}$$

Now, from Eqs. (1.32), (1.33) and (1.36), we finally obtain the upper bounds

$$C_{FS}[\rho] \leqslant B_{FS} = \frac{2e}{9\pi^{1/3}} \frac{n-|m|}{n^3}\left[3n^2 - l(l+1)\right]^2, \tag{1.37}$$

to the Fisher-Shannon measure, and

$$C_{SC}[\rho] \leqslant B_{SC} = \frac{\pi e^3}{27}\left[3n^2 - l(l+1)\right]^3 \times D(n,l,m), \tag{1.38}$$

to the shape complexity. It is worth noting that these two inequalities saturate at the ground state, having the values $\frac{2e}{\pi^{1/3}}$ and $\frac{e^3}{8}$ for the Fisher-Shannon and shape complexity cases, respectively, when $n=1$, $l=0$, and $m=0$.

For the sake of completeness we plot in Figure 1.6 and Figure 1.7 the values of the ratios

$$\xi_{FS}(n,l,m) = \frac{B_{FS} - C_{FS}[\rho]}{C_{FS}[\rho]},$$

and

$$\xi_{CS}(n,l,m) = \frac{B_{SC} - C_{SC}[\rho]}{C_{SC}[\rho]},$$

for the Fisher-Shannon and the shape complexity measures, respectively, in the case $(n,l,m)$ for $n = 1, 2, 3, 4, 5$ and $6$, and all allowed values of $l$. Various observations are apparent. First, the two ratios vanish when $n = 1$ indicating the saturation of the inequalities (1.33) and (1.38) just mentioned. Second, for a manifold with fixed $n$ the greatest accuracy occurs for the states $s$. Moreover, the accuracy of the bounds decreases when $l$ is increasing up to the centroid of the manifold and then it decreases. Finally, the Fisher-Shannon bound is always more accurate than the Cramer-Rao bound for the same hydrogenic orbital.

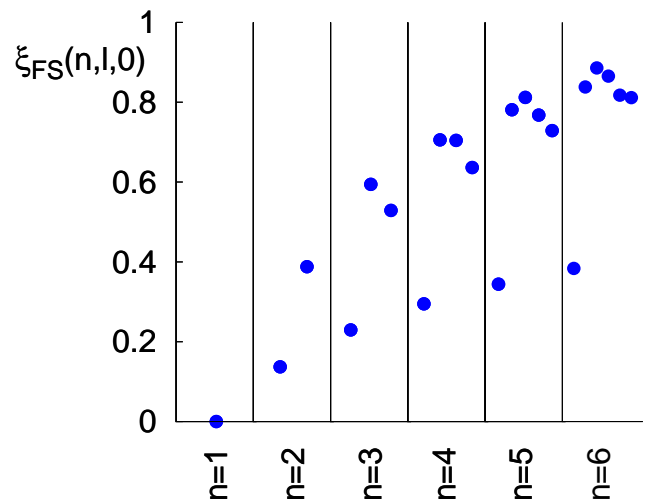

FIGURE 1.6: Dependence of the Fisher-Shannon ratio, $\xi_{FS}(n,l,0)$, on the quantum numbers $n$ and $l$.

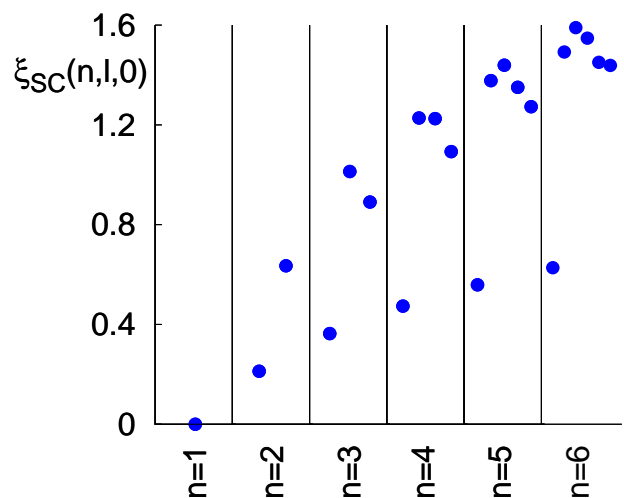

FIGURE 1.7: Dependence of the shape-complexity ratio, $\xi_{SC}(n,l,0)$, on the quantum numbers $n$ and $l$.

## 1.4 Conclusions

In this Chapter we have investigated both analytically and numerically the internal disorder of a hydrogenic atom which gives rise to the great diversity and complexity of three-dimensional geometries for its configuration orbitals $(n,l,m)$. This is done by means of the following composite information-theoretic quantities: the Fisher-Shannon and the Cramer-Rao measures, and the LMC shape complexity. The two former ones have a common ingredient of local character (the Fisher information) and a measure of global character, namely the Shannon entropic power in the Fisher-Shannon case and the variance in the Cramer-Rao case. The LMC shape complexity is composed by two quantities of global character: the disequilibrium (whose explicit expression is here calculated for the first time in terms of the quantum numbers $n$, $l$ and $m$ of the orbital) and the Shannon entropic power.

We have studied the dependence of these three composite quantities in terms of the quantum numbers $n$, $l$ and $m$. It is found that: (i) when $(l, m)$ are fixed, all of them have an increasing behaviour as a function of the principal quantum number $n$, with a rate of growth which is bigger in the Cramer-Rao and (even more emphatic) Fisher-Shannon cases; this is mainly because of the increasingly strong radial oscillating nature (when $n$ gets bigger), what is appropriately grasp by the Fisher ingredient of these two composite quantities; (ii) all of them decrease when the magnetic quantum number $|m|$ is increasing, and the decreasing rate is much faster in the Cramer-Rao case and more emphatically in the Fisher-Shannon case; this is basically because of the increasingly weak angular oscillating nature when $|m|$ decreases, what provokes the lowering of the Fisher ingredient of these two quantities; and (iii) all of them decrease when the orbital quantum number $l$ is increasing, and again, the decreasing rate is much faster in the Cramer-Rao and Fisher-Shannon cases; here, however, the physical interpretation is much more involved as it is duly explained in Section 1.3.

Finally, for completeness, we have used some variational bounds to the Shannon entropy to find sharp, saturating upper bounds to the Fisher-Shannon measure and to the LMC shape complexity .

# Chapter 2

# Complexity of $D$-dimensional hydrogenic systems in position and momentum spaces

The hydrogenic system (i.e., a negatively-charged particle moving around a positively-charged core which electromagnetically binds it in its orbit) with dimensionality $D \geq 1$, plays a central role in $D$-dimensional quantum physics and chemistry [32, 33]. It includes not only a large variety of three-dimensional physical systems (e.g., hydrogenic atoms and ions, exotic atoms, antimatter atoms, Rydberg atoms) but also a number of nanoobjects so much useful in semiconductor nanostructures (e.g., quantum wells, wires and dots) [34, 35] and quantum computation (e.g., qubits) [36]. Moreover it has a particular relevance for the dimensional scaling approach in atomic and molecular physics [33] as well as in quantum cosmology [38] and quantum field theory [39, 40]. Let us also say that the existence of hydrogenic systems with non standard dimensionalities has been shown for $D < 3$ [35] and suggested for $D > 3$ [41]. We should also highlight the use of $D$-dimensional hydrogenic wavefunctions as complete orthonormal sets for many-body problems [42, 43] in both position and momentum spaces, explicitly for three-body Coulomb systems (e.g. the hydrogen molecular ion and the helium atom); generalizations are indeed possible in momentum-space orbitals as well as in their role as Sturmians in configuration spaces.

The internal disorder of this system, which is manifest in the non-uniformity quantum-mechanical density and in the so distinctive hierarchy of its physical states, is being increasingly investigated beyond the root-mean-square or standard deviation (also called Heisenberg measure) by various information-theoretic elements; first, by means of the Shannon entropy [24, 25, 44] and then, by other individual information and/or spreading measures as the Fisher information and the power and logarithmic moments [45], as it is described in Ref. [31] where the information theory of $D$-dimensional hydrogenic systems is reviewed in detail. Just recently, further complementary insights have been shown to be obtained in the three-dimensional hydrogen atom by means of composite information-theoretic measures, such as the Fisher-Shannon and the LMC shape complexity [18, 46]. In particular, Sañudo and Lopez-Ruiz [18] have found some numerical evidence that, contrary to the energy, both the Fisher-Shannon measure and the LMC shape complexity in the position space do not present any accidental degeneracy (i.e. they do depend on the orbital quantum number $l$); moreover, they take on their minimal values at the circular states (i.e., those with the highest $l$). In fact, the position Shannon entropy by itself has also these two characteristics as it has been numerically pointed out long ago [25], where the dependence on the magnetic quantum number is additionally studied for various physical states.

The LMC shape complexity [17] occupies a very special position not only among the composite information-theoretic measures in general, but also within the class of measures of complexity. This is because of the following properties: (i) invariance under replication, translation and rescaling transformations, (ii) minimal value for the simplest probability densities (namely, e.g. uniform and Dirac's delta in one-dimensional case), and (iii) simple mathematical structure: it is given as the product of the disequilibrium or averaging density and the Shannon entropy power of the system.

In this Chapter we provide the analytical methodology to calculate the LMC shape complexity of the stationary states of the $D$-dimensional hydrogenic system in the two reciprocal position and momentum spaces and later we apply it to a special class of physical states which includes the ground state and the circular states (i.e. states with the highest hyperangular momenta allowed within a given electronic manifold). First, in Section 2.1, we briefly describe the known expressions of the quantum-mechanical density of the system in both spaces. In Section 2.2 we show that the computation of the two shape complexities for arbitrary $D$-dimensional hydrogenic stationary states boils down to the evaluation of some entropic functionals of Laguerre and Gegenbauer polynomials. To have the final expressions of these complexity measures in terms of the dimensionality $D$ and the quantum numbers characterizing the physical state under consideration, we need to compute the values of these polynomial entropic functionals what is, in general, a formidable open task. However, in Section 2.3, we succeed to do it for the important cases of ground and circular states. It seems that for the latter ones the shape complexity has the minimal values, at least in the three-dimensional case as indicated above. It is also shown that our results always fulfill the uncertainty relation satisfied by the position and momentum shape complexities [46]. In Section 2.4, the shape complexities are numerically studied and their dimensionality dependence is discussed. Finally, some conclusions are given.

## 2.1 The $D$-dimensional hydrogenic quantum-mechanical densities

Let us consider an electron moving in the $D$-dimensional Coulomb ($D \geqslant 2$) potential $V(\vec{r}) = -\frac{Z}{r}$, where $\vec{r} = (r, \theta_1, \theta_2, ..., \theta_{D-1})$ denotes the electronic vector position in polar coordinates. The stationary states of this hydrogenic system are described by the wavefunctions

$$\Psi_{n,l,\{\mu\}}(\vec{r}, t) = \psi_{n,l,\{\mu\}}(\vec{r}) \exp(-iE_n t),$$

where $(E_n, \Psi_{n,l,\{\mu\}})$ denote the physical solutions of the Schrödinger equation of the system [31–33]. The energies are given by

$$E = -\frac{Z^2}{2\eta^2}, \quad \text{with} \quad \eta = n + \frac{D-3}{2}; \quad n = 1, 2, 3, ..., \tag{2.1}$$

and the eigenfunctions can be expressed as

$$\Psi_{n,l,\{\mu\}}(\vec{r}) = R_{n,l}(r)\mathcal{Y}_{l,\{\mu\}}(\Omega_{D-1}), \tag{2.2}$$

where $(l, \{\mu\}) \equiv (l \equiv \mu_1, \mu_2, ..., \mu_{D-1}) \equiv (l, \{\mu\})$ denote the hyperquantum numbers associated to the angular variables $\Omega_{D-1} \equiv (\theta_1, \theta_2, ..., \theta_{D-1} \equiv \varphi)$, which may have all values consistent with the inequalities $l \equiv \mu_1 \geq \mu_2 \geq ... \geq |\mu_{D-1}| \equiv |m| \geq 0$. The radial function is given by

$$R_{n,l}(r) = \left(\frac{\lambda^{-D}}{2\eta}\right)^{1/2} \left[\frac{\omega_{2\mathcal{L}+1}(\hat{r})}{\hat{r}^{D-2}}\right]^{1/2} \tilde{L}_{\eta-\mathcal{L}-1}^{(2\mathcal{L}+1)}(\hat{r}), \tag{2.3}$$

where $\tilde{L}_k^{(\alpha)}(x)$ denotes the Laguerre polynomials of degree $k$ and parameter $\alpha$, orthonormal with respect to the weight function $\omega_\alpha(x) = x^\alpha e^{-x}$, and the grand orbital angular momentum hyperquantum number $\mathcal{L}$ and the adimensional parameter $\hat{r}$ are

$$\mathcal{L} = l + \frac{D-3}{2}, \quad l = 0, 1, 2, ... \quad \text{and} \quad \hat{r} = \frac{r}{\lambda}, \quad \text{with} \quad \lambda = \frac{\eta}{2Z}. \tag{2.4}$$

The angular part $\mathcal{Y}_{l,\{\mu\}}(\Omega_{D-1})$ is given by the hyperspherical harmonics [32, 47]

$$\mathcal{Y}_{l,\{\mu\}}(\Omega_{D-1}) = \frac{1}{\sqrt{2\pi}} e^{im\varphi} \prod_{j=1}^{D-2} \tilde{C}_{\mu_j - \mu_{j+1}}^{(\alpha_j + \mu_{j+1})}(\cos\theta_j)(\sin\theta_j)^{\mu_{j+1}}, \tag{2.5}$$

with $\alpha_j = \frac{1}{2}(D - j - 1)$ and $\tilde{C}_k^{(\lambda)}(x)$ denotes the orthonormal Gegenbauer polynomials of degree $k$ and parameter $\lambda$.

Then, the quantum-mechanical probability density of the system in position space is

$$\rho_{n,l,\{\mu\}}(\vec{r}) = \left|\Psi_{n,l,\{\mu\}}(\vec{r})\right|^2 = R_{n,l}^2(r)\left|\mathcal{Y}_{l,\{\mu\}}(\Omega_{D-1})\right|^2, \tag{2.6}$$

In momentum space the eigenfunction of the system is [31, 32, 48, 49]

$$\tilde{\Psi}_{nl\{\mu\}}(\vec{p}) = \mathcal{M}_{n,l}(p)\mathcal{Y}_{l\{\mu\}}(\Omega_{D-1}), \tag{2.7}$$

where the radial part is

$$\mathcal{M}_{n,l}(p) = 2^{\mathcal{L}+2}\left(\frac{\eta}{Z}\right)^{D/2}\frac{(\eta\tilde{p})^l}{(1+\eta^2\tilde{p}^2)^{\mathcal{L}+2}}\tilde{C}^{(\mathcal{L}+1)}_{\eta-\mathcal{L}-1}\left(\frac{1-\eta^2\tilde{p}^2}{1+\eta^2\tilde{p}^2}\right) \tag{2.8}$$

$$= \left(\frac{\eta}{Z}\right)^{D/2}(1+y)^{3/2}\left(\frac{1+y}{1-y}\right)^{\frac{D-2}{4}}\omega^{*1/2}_{\mathcal{L}+1}(y)\tilde{C}^{(\mathcal{L}+1)}_{\eta-\mathcal{L}-1}(y),$$

with $y=\frac{1-\eta^2\tilde{p}^2}{1+\eta^2\tilde{p}^2}$ and $\tilde{p}=\frac{p}{Z}$ (here the electron momentum $p$ is assumed to be expressed in units of $p_\mu$, where $p_{\mu_r}=\frac{\mu_r}{m_e}p_0=\mu_r$ m.a.u, since $m_e=1$ and the momentum atomic unit is $p_0=\frac{\hbar}{a_0}=\frac{m_e e^2}{\hbar}$; $\mu_r$ is the reduced mass of the system). The symbol $\tilde{C}^{(\alpha)}_m(x)$ denotes the Gegenbauer polynomial of order $k$ and parameter $\alpha$ orthonormal with respect to the weight function $\omega^*_\alpha(x)=(1-x^2)^{\alpha-\frac{1}{2}}$ on the interval $[-1,+1]$. The angular part is again an hyperspherical harmonic as in the position case, but with the angular variables of the vector $\vec{p}$. Then, one has the following expression

$$\gamma(\vec{p}) = \left|\tilde{\Psi}_{n,l,\{\mu\}}(\vec{p})\right|^2 = \mathcal{M}^2_{n,l}(p)\left[\mathcal{Y}_{l\{\mu\}}(\Omega_{D-1})\right]^2, \tag{2.9}$$

for the quantum-mechanical probability density of the system in momentum space.

## 2.2 The LMC shape complexity of the D-dimensional hydrogenic system

Here we describe the methodology to compute the position and momentum LMC shape complexity of our system in an arbitrary physical state characterized by the hyperquantum numbers $(\eta,\mu_1,...,\mu_{D-1})$. We show that the calculation of the position and momentum hydrogenic shape complexities ultimately reduce to the evaluation of some entropic functionals of Laguerre and Gegenbauer polynomials.

### 2.2.1 Position space

The LMC shape complexity $C[\rho]$ of the position probability density $\rho(\vec{r})$ is defined [17] as

$$C[\rho] = \langle\rho\rangle\exp(S[\rho]), \tag{2.10}$$

where

$$\langle\rho\rangle = \int[\rho(\vec{r})]^2\,d\vec{r}, \tag{2.11}$$

and
$$S[\rho] = -\int \rho(\vec{r}) \log \rho(\vec{r}) d\vec{r}, \tag{2.12}$$

denote the first-order frequency moment (also called averaging density or disequilibrium, among other names) and the Shannon entropy of $\rho(\vec{r})$, respectively. Then, this composite information-theoretic quantity measures the complexity of the system by means of a combined balance of the average height (as given by $\langle\rho\rangle$) and the total bulk extent (as given by $S[\rho]$) of the corresponding quantum-mechanical probability density $\rho(\vec{r})$.

Let us first calculate $\langle\rho\rangle$. From (2.2) and (2.11) one obtains that

$$\langle\rho\rangle = \frac{2^{D-2}}{\eta^{D+2}} Z^D K_1(D, \eta, \mathcal{L}) K_2(l, \{\mu\}), \tag{2.13}$$

where
$$K_1(D, \eta, \mathcal{L}) = \int_0^\infty x^{-D-5} \left\{ \omega_{2\mathcal{L}+1}(x) \left[ \tilde{L}_{\eta-\mathcal{L}-1}^{2\mathcal{L}+1}(x) \right]^2 \right\}^2 dx, \tag{2.14}$$

and
$$K_2(l, \{\mu\}) = \int_\Omega \left| \mathcal{Y}_{l\{\mu\}}(\Omega_{D-1}) \right|^4 d\Omega_{D-1}. \tag{2.15}$$

The Shannon entropy of $\rho(\vec{r})$ has been recently shown [50] to have the following expression
$$S[\rho] = S[R_{nl}] + S\left[\mathcal{Y}_{l\{\mu\}}\right], \tag{2.16}$$

with the radial part
$$S[R_{n,l}] = -\int_0^\infty r^{D-1} R_{n,l}^2(r) \log R_{n,l}^2 dr$$
$$= A(n,l,D) + \frac{1}{2\eta} E_1\left[\tilde{L}_{\eta-\mathcal{L}-1}^{(2\mathcal{L}+1)}\right] - D\log Z, \tag{2.17}$$

and the angular part
$$S\left[\mathcal{Y}_{l,\{\mu\}}\right] = -\int_{S_{D-1}} \left|\mathcal{Y}_{l,\{\mu\}}(\Omega_{D-1})\right|^2 \log \left|\mathcal{Y}_{l,\{\mu\}}(\Omega_{D-1})\right|^2 d\Omega_{D-1}$$
$$= B(l, \{\mu\}, D) + \sum_{j=1}^{D-2} E_0\left[\tilde{C}_{\mu_j-\mu_{j+1}}^{(\alpha_j+\mu_{j+1})}\right], \tag{2.18}$$

where $A(n,l,D)$ and $B(l, \{\mu\}, D)$ have the following values

$$A(n,l,D) = -2l\left[\frac{2\eta - 2\mathcal{L} - 1}{2\eta} + \psi(\eta + \mathcal{L} + 1)\right] + \frac{3\eta^2 - \mathcal{L}(\mathcal{L}+1)}{\eta} - \log\left[\frac{2^{D-1}}{\eta^{D+1}}\right],$$

and

$$B(l, \{\mu\}, D) = \log 2\pi - 2 \sum_{j=1}^{D-2} \mu_{j+1}$$
$$\times \left[\psi(2\alpha_j + \mu_j + \mu_{j+1}) - \psi(\alpha_j + \mu_j) - \log 2 - \frac{1}{2(\alpha_j + \mu_j)}\right],$$

with $\psi(x) = \frac{\Gamma'(x)}{\Gamma(x)}$ is the digamma function. The entropic functionals $E_i[\tilde{y}_n]$, $i = 1$ and 2, of the polynomials $\{\tilde{y}_n\}$, orthonormal with respect to the weight function $\omega(x)$, are defined [44, 51] by

$$E_1[\tilde{y}_n] = -\int_0^\infty x\omega(x)\tilde{y}_n^2(x) \log \tilde{y}_n^2(x) dx, \tag{2.19}$$

and

$$E_0[\tilde{y}_n] = -\int_{-1}^{+1} \omega(x)\tilde{y}_n^2(x) \log \tilde{y}_n^2(x) dx, \tag{2.20}$$

respectively.

Finally, from Eqs. (2.10), (2.13) and (2.16)-(2.18), we obtain the following value for the position shape complexity of our system:

$$C[\rho] = \frac{2^{D-2}}{\eta^{D+2}} K_1(D, \eta, \mathcal{L}) K_2(\mathcal{L}, \{\mu\}) \tag{2.21}$$
$$\times \exp\left[A(n, l, D) + \frac{1}{2\eta} E_1\left[\tilde{L}_{\eta-\mathcal{L}-1}^{(2\mathcal{L}+1)}\right] + S\left[\mathcal{Y}_{l,\{\mu\}}\right]\right],$$

where the entropy of the hyperspherical harmonics $S[\mathcal{Y}_{l,\{\mu\}}]$, given by Eq. (2.18), is controlled by the entropy of Gegenbauer polynomials $E_0\left[\tilde{C}_k^{(\alpha)}\right]$ defined by Eq. (2.20). It is important to remark that the position complexity $C[\rho]$ does not depend on the strength of the Coulomb potential, that is, on the nuclear charge $Z$.

### 2.2.2 Momentum space

The shape complexity $C[\gamma]$ of the momentum probability density $\gamma(\vec{p})$ is given by

$$C[\gamma] = \langle\gamma\rangle \exp(S[\gamma]), \tag{2.22}$$

where the momentum averaging density $\langle\gamma\rangle$ can be obtained from Eq. (2.9) as follows:

$$\langle\gamma\rangle = \int \gamma^2(\vec{p}) d\vec{p} = \frac{2^{4\mathcal{L}+8}\eta^D}{Z^D} K_3(D, \eta, \mathcal{L}) K_2(l, \{\mu\}), \tag{2.23}$$

with $K_2$ is given by Eq. (2.15), and $K_3$ can be expressed as

$$K_3(D, \eta, \mathcal{L}) = \int_0^\infty \frac{y^{4l+D-1}}{(1+y^2)^{4\mathcal{L}+8}} \left[\tilde{C}_{\eta-\mathcal{L}-1}^{(\mathcal{L}+1)}\left(\frac{1-y^2}{1+y^2}\right)\right]^4 dy. \tag{2.24}$$

On the other hand, the momentum Shannon entropy $S[\gamma]$ can be calculated in a similar way as in the position case. We have obtained that

$$\begin{aligned} S[\gamma] &= -\int \gamma(\vec{p}) \log \gamma(\vec{p})\, d\vec{p} = S[\mathcal{M}_{nl}] + S[\mathcal{Y}_{l,\{\mu\}}] \\ &= F(n,l,D) + E_0\left[\tilde{C}^{(\mathcal{L}+1)}_{\eta-\mathcal{L}-1}\right] + D\log Z + S[\mathcal{Y}_{l,\{\mu\}}], \end{aligned} \qquad (2.25)$$

where $F(n,l,D)$ has been found to have the value

$$\begin{aligned} F(n,l,D) &= -\log \frac{\eta^D}{2^{2\mathcal{L}+4}} - (2\mathcal{L}+4)\left[\psi(\eta+\mathcal{L}+1) - \psi(\eta)\right] \\ &+ \frac{\mathcal{L}+2}{\eta} - (D+1)\left[1 - \frac{2\eta(2\mathcal{L}+1)}{4\eta^2 - 1}\right]. \end{aligned} \qquad (2.26)$$

Then, from Eqs. (2.22), (2.23) and (2.25) we finally have the following value for the momentum shape complexity

$$\begin{aligned} C[\gamma] &= 2^{4\mathcal{L}+8}\eta^D K_3(D,\eta,\mathcal{L})\, K_2(\mathcal{L},\{\mu\}) \\ &\quad \times \exp\left\{F(n,l,D) + E_0\left[\tilde{C}^{(\mathcal{L}+1)}_{\eta-\mathcal{L}-1}\right] + S[\mathcal{Y}_{l,\{\mu\}}]\right\}. \end{aligned} \qquad (2.27)$$

Notice that, here again, this momentum quantity does not depend on the nuclear charge $Z$. Moreover the momentum complexity $C[\rho]$ is essentially controlled by the entropy of the Gegenbauer polynomials $E_0\left[\tilde{C}^{(\alpha)}_k\right]$, since the entropy of hyperspherical harmonics $S[\mathcal{Y}_{l,\{\mu\}}]$ reduces to that of these polynomials according to Eq. (2.18).

## 2.3 LMC shape complexities of ground and circular states

Here we apply the general expressions (2.21) and (2.27) found for the position and momentum shapes complexities of an arbitrary physical state of the $D$-dimensional hydrogenic system, respectively, to the ground state ($n=1, \mu_i = 0, \forall i = 1...D-1$) and to the circular states. A circular state is a single-electron state with the highest hyperangular momenta allowed within a given electronic manifold, i.e. a state with hyperangular momentum quantum numbers $\mu_i = n-1$ for all $i = 1,...,D-1$.

### 2.3.1 Ground state

In this case $\eta - \mathcal{L} - 1 = 0$, so that the Laguerre polynomial involved in the radial wavefunction is a constant. Then, the probability density of the ground state in position space given by Eqs. (2.3), (2.5) and (2.6) reduces as follows:

$$\rho_{g.s.}(\vec{r}) = \left(\frac{2Z}{D-1}\right)^D \frac{1}{\pi^{\frac{D-1}{2}}\Gamma\left(\frac{D+1}{2}\right)} e^{-\frac{4Z}{D-1}r}, \qquad (2.28)$$

which has been also found by various authors (see e.g. [32, 33]).

The expressions (2.13)-(2.15), which provide the averaging density of arbitrary quantum-mechanical state, reduce to the value

$$\langle \rho_{g.s.} \rangle = \frac{Z^D}{(D-1)^D} \frac{1}{\pi^{\frac{D-1}{2}}\Gamma\left(\frac{D+1}{2}\right)}, \qquad (2.29)$$

for the ground-state averaging density. Moreover, the angular part of the entropy is

$$S\left[\mathcal{Y}_{0,\{0\}}\right] = \log \frac{2\pi^{D/2}}{\Gamma\left(\frac{D}{2}\right)}, \qquad (2.30)$$

so that it is equal to $\log 2\pi$ and $\log 4\pi$ for $D = 2$ and $3$, respectively. Then, the formulas (2.16)-(2.20) of the Shannon entropy of arbitrary physical state of our system simplify as

$$S[\rho_{g.s.}] = \log\left(\frac{(D-1)^D}{2^D}\pi^{\frac{D-1}{2}}\Gamma\left(\frac{D+1}{2}\right)\right) + D - D\log Z, \qquad (2.31)$$

for the ground-state Shannon entropy. Finally, from Eq. (2.21) or from its own definition together with (2.29)-(2.31) we obtain that the position shape complexity of $D$-dimensional hydrogenic ground state has the value

$$C[\rho_{g.s.}] = \left(\frac{e}{2}\right)^D. \qquad (2.32)$$

In momentum space we can operate in a similar manner. First we have seen that the ground-state probability density is

$$\gamma_{g.s.}(\vec{p}) = \frac{(D-1)^D \Gamma\left(\frac{D+1}{2}\right)}{Z^D \pi^{\frac{D+1}{2}} \Gamma\left(\frac{D}{2}\right)} \frac{1}{\left(1 + \frac{(D-1)^2}{4}\tilde{p}^2\right)^{D+1}}, \qquad (2.33)$$

which has been also given by Aquilanti et al [48], among others. Then, we have found the values

$$\langle \gamma_{g.s.} \rangle = \left(\frac{2D-2}{Z}\right)^D \frac{1}{\pi^{\frac{D+2}{2}}} \frac{\Gamma^2\left(\frac{D+1}{2}\right)\Gamma\left(2+\frac{3D}{2}\right)}{\Gamma(2D+2)}, \qquad (2.34)$$

for the momentum averaging density, and

$$S[\gamma_{g.s.}] = \log \frac{\pi^{\frac{D+1}{2}}}{(D-1)^D \Gamma\left(\frac{D+1}{2}\right)} + (D+1)\left[\psi(D+1) - \psi\left(\frac{D}{2}+1\right)\right] + D\log Z, \quad (2.35)$$

for the momentum Shannon entropy, directly from Eq. (2.29) or from Eqs. (2.23)-(2.24) and (2.25)-(2.26), respectively. Finally. from Eq. (2.27) or by means of Eqs. (2.34)-(2.35) we have the following value

$$C\left[\gamma_{g.s.}\right] = \frac{2^D \Gamma\left(\frac{D+1}{2}\right) \Gamma\left(2+\frac{3D}{2}\right)}{\pi^{1/2} \Gamma(2D+2)} \exp\left\{(D+1)\left[\psi(D+1) - \psi\left(\frac{D+2}{2}\right)\right]\right\}, \quad (2.36)$$

for the ground-state D-dimensional hydrogenic shape complexity in momentum space. In particular, this quantity has the values

$$C_2(\gamma_{g.s.}) = \frac{2e^{3/2}}{5} = 1.7926$$

$$C_3(\gamma_{g.s.}) = \frac{66}{e^{10/3}} = 2.3545$$

$$C_4(\gamma_{g.s.}) = \frac{e^{35/12}}{6} = 3.0799$$

for the hydrogenic system with dimensionalities $D = 2, 3$ and $4$, respectively. Let us here mention that the three-dimensional value agrees with that calculated in [17].

### 2.3.2 Circular states

Following a parallel process with circular states, we have obtained

$$\rho_{c.s.}(\vec{r}) = \frac{2^{D+2-2n} Z^D}{\pi^{\frac{D-1}{2}} (2n+D-3)^D \Gamma(n) \Gamma\left(n+\frac{D-1}{2}\right)} e^{-\frac{r}{\lambda}} \left(\frac{r}{\lambda}\right)^{2n-2} \prod_{j=1}^{D-2} (\sin\theta_j)^{2n-2},$$

for the position probability density, and

$$\gamma_{c.s.}(\vec{p}) = \frac{2^{2n-2}(2n+D-3)^D \Gamma\left(n+\frac{D-1}{2}\right)}{Z^D \pi^{\frac{D+1}{2}} \Gamma(n)} \frac{(\eta p/Z)^{2n-2}}{(1+\frac{\eta^2 p^2}{Z^2})^{2n+D-1}} \prod_{j=1}^{D-2} (\sin\theta_j)^{2n-2},$$

for the momentum probability density of a D-dimensional hydrogenic circular state with the principal quantum number $n$. Moreover, we have found the values

$$\langle \rho_{c.s.} \rangle = \frac{Z^D \Gamma\left(n-\frac{1}{2}\right) \Gamma\left(2n+\frac{D-3}{2}\right)}{2^{2n-2} \pi^{\frac{D}{2}} (2n+D-3)^D \Gamma(n) \Gamma^2\left(n+\frac{D-1}{2}\right)}, \quad (2.37)$$

and

$$\langle \gamma_{c.s.} \rangle = \frac{2^{4n+D-4}(2n+D-3)^D \Gamma^2\left(n+\frac{D-1}{2}\right) \Gamma(2n-1) \Gamma\left(2n+\frac{3D}{2}\right)}{Z^D \pi^{\frac{D+2}{2}} \Gamma^2(n) \Gamma(4n+2D-2)}, \quad (2.38)$$

for the position and momentum averaging densities of our system. On the other hand, we have also been able to express the position and momentum entropies as

$$S[\rho_{c.s.}] = 2n + D - 2 - (n-1)\left[\psi(n) + \psi\left(n + \frac{D-1}{2}\right)\right] - D\log 2 \qquad (2.39)$$
$$+ \log\left[(2n+D-3)^D \pi^{\frac{D-1}{2}} \Gamma(n)\Gamma\left(n + \frac{D-1}{2}\right)\right] - D\log Z,$$

and

$$S[\gamma_{c.s.}] = A(n,D) + \log\left[\frac{2^{D+1} Z^D \pi^{\frac{D+1}{2}} \Gamma(n)}{(2n+D-3)^D \Gamma\left(n + \frac{D-1}{2}\right)}\right], \qquad (2.40)$$

where the constant $A(n,D)$ is given by

$$A(n,D) = \frac{2n+D-1}{2n+D-3} - \frac{D+1}{2n+D-2} - (n-1)\psi(n)$$
$$- \left(\frac{D+1}{2}\right)\psi\left(n + \frac{D-2}{2}\right) + \left(n + \frac{D-1}{2}\right)\psi\left(n + \frac{D-3}{2}\right). \qquad (2.41)$$

Finally, from Eqs. (2.37)-(2.40) or from Eqs. (2.21) and (2.27) we have the values

$$C[\rho_{c.s.}] = \frac{\Gamma\left(n - \frac{1}{2}\right)\Gamma\left(2n + \frac{D-3}{2}\right)}{2^{2n+D-2} \pi^{1/2} \Gamma\left(n + \frac{D-1}{2}\right)}$$
$$\times \exp\left\{2n + D - 2 - (n-1)\left[\psi(n) + \psi\left(n + \frac{D-1}{2}\right)\right]\right\}, \qquad (2.42)$$

and

$$C[\gamma_{c.s.}] = \frac{2^{4n+2D-3} \Gamma\left(n + \frac{D-1}{2}\right) \Gamma(2n-1) \Gamma\left(2n + \frac{3D}{2}\right)}{\pi^{1/2} \Gamma(n) \Gamma(4n+2D-2)} \exp[A(n,D)], \qquad (2.43)$$

for the position and momentum shape complexity of a D-dimensional hydrogenic system in an arbitrary circular state. It is worthwhile remarking for checking purposes that Eqs. (2.42) and (2.43) reduce to Eqs. (2.32) and (2.36) in case that $n = 1$, respectively, as expected; in this sense we have to use the two following properties of the digamma function: $\psi(2z) = \frac{1}{2}\left[\psi(z) + \psi\left(z + \frac{1}{2}\right)\right] + \log 2$ and $\psi(z+1) = \psi(z) + \frac{1}{z}$.

## 2.4 Numerical study and physical discussion

Here we discuss the general complexity expressions obtained in the previous Subsection in terms of (a) the dimensionality for a given circular state (i.e., for fixed $n$), and (b) the principal quantum number $n$ for a given dimensionality.

Let us begin with the dimensional analysis of the position and momentum complexities, $C_D[\rho_{c.s.}]$ and $C_D[\gamma_{c.s.}]$, given by Eqs. (2.42) and (2.43), respectively. The resulting position complexity as a function of the dimensionality is drawn at Figure (2.1) for the ground state ($n=1$) and the circular states with $n = 2$ and 3. It shows a parabolic growth for all states when $D$ is increasing, being always greater than unity; the minimum value of $C[\rho]$ is $\left(\frac{e}{2}\right)^2 = 1.847$, what occurs for $D = 2$.

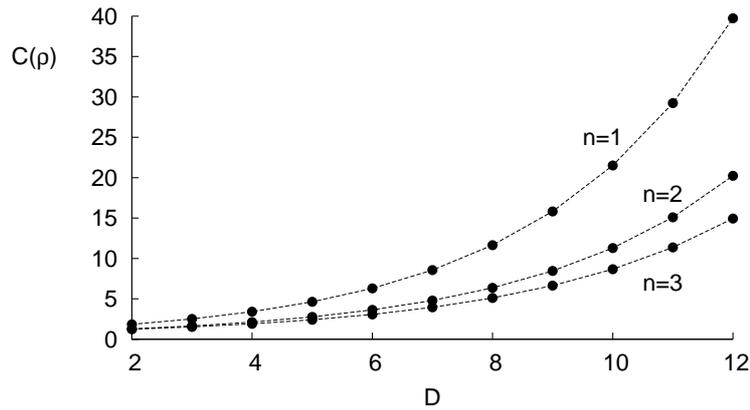

FIGURE 2.1: Variation of the shape complexity in position space with the dimension $D$ for three circular states. Atomic units are used.

The shape of the momentum complexity (whose minimum value $\frac{2}{5}e^{\frac{3}{2}} = 1.793$ corresponds to the case $n = 1$ and $D = 2$) appears to have a strong resemblance with the position one, mainly because the two ingredients of each complexity have opposite behaviours when $D$ varies. This is shown in Figure (2.2), where the Shannon entropies $S[\rho]$ and $S[\gamma]$ as well as the logarithmic values of the position and momentum values of the disequilibrium are plotted for the ground state in terms of $D$. Keep in mind that $C[\rho] = \exp(S[\rho] + \log\langle\rho\rangle)$ in position space and a similar form in momentum space. We observe that the Shannon entropies and the disequilibrium logarithmic measures have opposite behaviours in the two reciprocal spaces, so that the combined exponential effect which gives rise to the corresponding complexities is very similar qualitatively and almost quantitatively. Moreover, it happens that, for a given dimensionality, the relative contribution of the disequilibrium (entropic power) is smaller than that of the entropic power (disequilibrium) in position (momentum) space. This indicates that the relative contribution of the bulk extent of the position (momentum) probability density is more (less) powerful than its average height.

In addition, from Figure 2.2, we observe that the inequalities

$$C_D[\rho_{c.s.}; n = 3] < C_D[\rho_{c.s.}; n = 2] < C_D[\rho_{g.s.}],$$

are fulfilled in position space, and similarly in momentum space. This decreasing phenomenon of the complexity for the circular states when the quantum number $n$ is increasing, can be more clearly observed in the left graph of Figure (2.3) where the values of position complexity for the states with $n = 1$-15 are given at the dimensionalities $D = 2$, 5 and 15. Therein we remark that when the quantum number n is increasing, the radial density behaves so that its maximum height decreases and its spreading increases at different rates in such a way that it overall occurs the phenomenon pointed out by this chain of inequalities; namely, the larger $n$ is, the smaller is the shape complexity of the corresponding circular state.

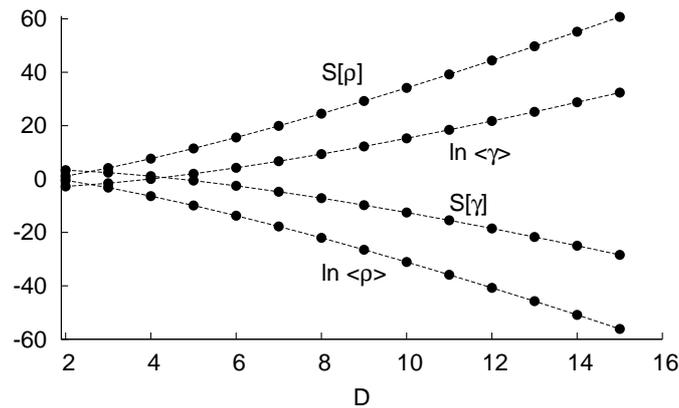

FIGURE 2.2: Ground state Shannon entropy ($S(\rho)$, $S(\gamma)$) and disequilibrium ($\langle\rho\rangle$, $\langle\gamma\rangle$) in position and momentum spaces as a function of the dimension $D$. Atomic units are used.

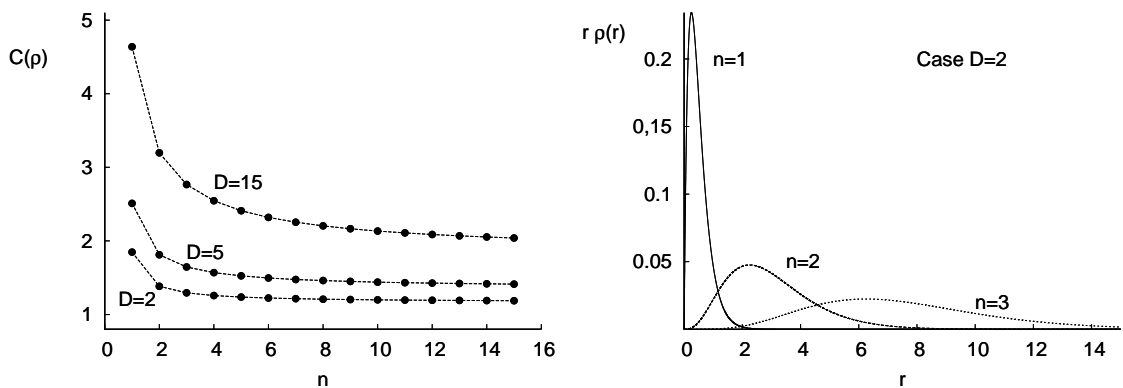

FIGURE 2.3: Variation of the position shape complexity of circular states with the principal quantum number $n$ for various dimensionalities. (Right) Radial probability density in position space for various two-dimensional circular states.

These dimensional and energetic (quantum number $n$) behaviours of the position complexity turn out to be a delicate overall balance of the average height and the bulk spreading of the system given by its two information-theoretic ingredients: the disequilibrium $\langle\rho\rangle$ and the Shannon entropic power, respectively.

Now we would like to find the dimensional (i.e., when $D \to \infty$) limit, and the high energy or Rydberg ($n \to \infty$) limit of the position and momentum complexities of our system. The former one plays a relevant role in the dimensional scaling methods in atomic and molecular physics [33], and the latter one for the Rydberg states which lie down at the region where the transition classical-quantum takes place. The large $D$ limit is closed to (but not the same) the conventional classical limit obtained by $\hbar \to 0$ for a fixed dimension [33].

For the ground state, whose energy is $E_{g.s.} = -2\left(\frac{Z}{D-1}\right)^2$, the position complexity is, according to Eq. (2.32), $C_D[\rho_{g.s.}] = \left(\frac{e}{2}\right)^D$. So that at the pseudoclassical limit, in which the electron is located at a fixed radial distance, the energy vanishes while the position complexity diverges. In momentum space, the shape complexity given by Eq. (2.36) has the following behaviour

$$C[\gamma_{g.s.}] \sim \frac{3^{\frac{3}{2}(D-1)}}{2^{2D-\frac{3}{2}}\sqrt{e}}, \qquad D \to \infty \qquad (2.44)$$

for the pseudoclassical limit.

A similar asymptotic analysis of Eq. (2.42) has allowed us to find the following values for the position shape complexity of a general circular state (characterized by the quantum number $n$)

$$C[\rho_{c.s.}] \sim \left(\frac{e}{2}\right)^{D+2n-2} e^{(1-n)\psi(n)} \frac{\Gamma\left(n-\frac{1}{2}\right)}{\sqrt{\pi}}, \qquad D \to \infty \qquad (2.45)$$

at the dimensional limit, and the value

$$C[\rho_{c.s.}] \sim \left(\frac{e}{2}\right)^{\frac{D-1}{2}}, \qquad n \gg 1 \qquad (2.46)$$

for the circular Rydberg states of a D-dimensional hydrogenic system.

Operating with Eq. (2.43) in a parallel way, we have obtained the values

$$C[\gamma_{c.s.}] \sim \left(\frac{3^{3/2}}{4}\right)^D \frac{3^{2n-\frac{1}{2}}\Gamma(2n-1)}{2^{4n-\frac{5}{2}}\Gamma(n)} e^{(1-n)\psi(n)-\frac{1}{2}}, \qquad D \to \infty \qquad (2.47)$$

for the momentum shape complexity of a circular state with quantum number $n$ at the pseudoclassical limit, and the value

$$C[\gamma_{c.s.}] \sim \left(\frac{e}{2}\right)^{\frac{D-1}{2}}, \qquad n \gg 1 \qquad (2.48)$$

for the momentum shape complexity of a circular Rydberg state.

Let us also make some comments about the uncertainty products of the position and momentum shape complexities $C[\rho]\,C[\gamma]$ for the ground and circular states. The general expressions are readily obtained from Eqs. (2.32) and (2.42) in position space, and from Eqs. (2.36) and (2.43) in momentum space. Moreover, this uncertainty product behaves

as
$$C\left[\rho_{c.s.}\right]C\left[\gamma_{c.s.}\right] \sim \left(\frac{e}{2}\right)^{D-1}, \qquad n \gg 1$$

at the Rydberg limit, and as

$$C\left[\rho_{c.s.}\right]C\left[\gamma_{c.s.}\right] \sim \left(\frac{3^{3/2}e}{2^3}\right)^D \frac{3^{2n-\frac{1}{2}}}{2^{4n-\frac{5}{2}}} \frac{\Gamma^2\left(n-\frac{1}{2}\right)}{\pi} e^{2n-\frac{5}{2}-2(n-1)\psi(n)}, \qquad D \to \infty$$

at the dimensional limit for circular states, where Eqs. (2.46) and (2.48), and (2.45) and (2.47) have been taken into account. The last expression yields

$$C\left[\rho_{g.s.}\right]C\left[\gamma_{g.s.}\right] \sim \left(\frac{3^{3/2}e}{2^3}\right)^D \left(\frac{3}{2e^{1/3}}\right)^{\frac{3}{2}}, \qquad D \to \infty$$

for the ground state uncertainty product. Finally, for completeness, let us remark that the complexity uncertainty product is always not less than $\frac{e}{2} = 1.359$.

## 2.5 Conclusions

The LMC shape complexity of the hydrogenic system in $D$-dimensional position and momentum spaces has been investigated. We have seen that the explicit computation of this complexity is a formidable open task, mainly because the analytical evaluation of the entropic functionals of the Laguerre and Gegenbauer polynomials, $E_1\left[\tilde{L}_k^{(\alpha)}\right]$ and $E_0\left[\tilde{C}_k^{(\alpha)}\right]$, involved in the calculation of the Shannon entropy, has not yet been accomplished.

The general methodology presented here is used to find explicit expressions for the position and momentum complexities of the ground and circular states in terms of the dimensionality and the principal quantum number. Then, these information-theoretic quantities are numerically discussed for various states and dimensionalities as well as for the dimensional and high-lying energy (Rydberg) limits. Briefly, we find that both position and momentum complexities increase (decrease) when the dimensionality (the quantum number of the state) is increasing. This phenomenon is the result of a delicate balance of the average height and the bulk spreading of the system given by their two information-theoretic ingredients, the disequilibrium and the entropic power, respectively. Finally, the uncertainty product of the position and momentum LMC shape complexities is examined.

# Chapter 3

# Entropy and complexity analyses of Klein-Gordon systems

The interplay of quantum mechanics, relativity theory and information theory is a most important topic in the present-day theoretical physics [52–58]. While the link between the two former theories is well known for everybody, it seems that the information theory has not yet percolated sufficiently in the scientific community as a whole. However, it is well established by now that the information-theoretic approach provides deeper insights for numerous other physical problems and poses new unsolved issues. Information is physical [59]. We refer to the excellent monographs of e.g., Peres and Terno [57] and Nalewajski [56], where the physical interest of this approach in relativistic quantum-mechanical phenomena and the quantum theory of electronic structure is explicitly shown and discussed in detail, respectively. Let us just mention, for illustration, the recent information-theoretic interpretations of the paradox of quantum black holes [60], the natural ultraviolet cutoff at the Planck scale [61, 62], and numerous physical phenomena (avoided crossings of atoms in external fields [63], periodicity and shell structure troughout the periodic table [11, 13], molecular similarities [64, 65],..). In particular, the information-theoretic treatment gives rise to the information-representation of the molecular states, which complements the conventional energy-representation of the density-functional and wave-function theories. Recently, the information-theoretic approach has allowed us (i) to predict the transition state structure and other stationary points so as to reveal the bond breaking/forming regions of chemical reactions [66, 67], (ii) to explain the growing behaviour of nanostructured molecules of polyamidoamine dendrimers, starting from monomers, dimers, trimers and tetramers up to generations of G0, G1, G2 and G3 with 84, 228, 516 and 1092 atoms respectively [68] and (iii) to study the entanglement properties of many-fermion systems [69].

Special relativity provokes both important restrictions on the transfer of information between distant systems [57] and severe changes on the integral structure of physical systems [70].This is mainly because the relativistic effects produce a spatial redistribution of the single-particle density $\rho(\vec{r})$ of the corresponding quantum-mechanical states, which substantially alter the spectroscopic and macroscopic properties of the systems. The quantitative study of the relativistic modification of the spatial extent of the charge density of atomic and molecular systems by information-theoretic means is a widely open field [55, 56]. The only works published up to now have calculated the ground-state relativistic effects on hydrogenic [55] and many-electron neutral atoms [18, 52] in different settings by use of the renowned standard deviation (or Heisenberg measure) as well as various information-theoretic measures.

Here we quantify the relativistic effects of the ground and excited states of the spinless single-particle charge spreading by the comparison of the Klein-Gordon and Schrödinger values for three qualitatively different measures: the Heisenberg measure $\sigma[\rho]$, the Shannon entropic power $N[\rho]$ [37] and the Fisher information $I[\rho]$ [10, 15]. While the Heisenberg quantity gives the spreading with respect to the centroid of the charge distribution, the Shannon and Fisher measures do not refer to any specific point.

The Shannon entropic power $N[\rho]$, which is essentially given by the exponential of the Shannon entropy $S[\rho] = -\langle \log \rho(\vec{r}) \rangle$, measures the total extent to which the distribution is in fact concentrated [15, 71]. This quantity has various relevant features. First, it avoids the dimensionality troubles of $S[\rho]$, highlighting its physical meaning. Second, it exists when $\sigma$ does not. Third, it is finite whenever $\sigma$ is. Thus, as a measure of uncertainty the use of the Shannon entropic power allows a wider quantitative range of applicability than the Heisenberg measure [72]. Contrary to the Shannon and Heisenberg measures, which are insensitive to electronic oscillations, the translationally invariant Fisher information [10] has a locality property because it is a gradient functional of the density, so that it measures the pointwise concentration of the electronic cloud and quantifies its gradient content, providing a quantitative estimation of the oscillatory character of the density. Moreover, the Fisher information measures the bias to particular points of the space, i.e. it gives a measure to the local disorder.

The structure of this Chapter is the following. In Section 3.1, the quantum-mechanical motion Klein-Gordon equation of a spinless relativistic particle with a negative electric charge in a Coulomb potential is described and its Lorentz-invariant charge density is given. In Section 3.2, we compute the ordinary moments of general excited states, with emphasis on the Heisenberg measure for circular and $S$-states. These quantities, which are well-known in the Dirac case [73, 74], are only known [75] for the non-Lorentz-invariant density in the Klein-Gordon case. Here we study them for the Lorentz-invariant Klein-Gordon density. Then, we compute the following single information-theoretic measures of the system: the Shannon entropy and the Fisher information. Finally, in Section 3.3, the relativistic effects are analysed by means of the Fisher-Shannon and LMC shape complexities.

## 3.1 Klein-Gordon equation for Coulomb systems: Basics

To calculate the measures of the charge spreading in a relativistic quantum-mechanical system we have to tackle the problem of the very concept of quantum probability consistent with Lorentz covariance. The general formulation and interpretation of this problem is still a currently discussed issue [76]. In this Chapter we avoid this problem following the relativistic quantum mechanics [70] by restricting ourselves to study the stationary states of a spinless relativistic particle with a negative electric charge in a spherically symmetric Coulomb potential $V(r) = -\frac{Ze^2}{r}$, which are the solutions of the relativistic scalar wave equation, usually called the Klein-Gordon equation [77–80],

$$[\epsilon - V(r)]\,\psi(\vec{r}) = (-\hbar^2 c^2 \nabla^2 + m_0^2 c^4)\psi(\vec{r}), \qquad (3.1)$$

appropriately normalized to the particle charge. The symbols $m_0$ and $\epsilon$ denote the mass and the relativistic energy eigenvalue, respectively. We will work in spherical coordinates, taking the ansatz $\psi(r,\theta,\phi) = r^{-1}u(r)Y_{lm}(\theta,\phi)$, where $Y_{lm}(\theta,\phi)$ denotes the spherical harmonics of order $(l,m)$. Then, to highlight the resemblance with the non-relativistic Schrödinger equation, we let [37, 75]

$$\beta \equiv \frac{2}{\hbar c}(m_0^2 c^4 - \epsilon^2)^{\frac{1}{2}} = \frac{2m_0 c^2}{\hbar c}\sqrt{1 - \left(\frac{\epsilon}{m_0 c^2}\right)^2}, \qquad (3.2)$$

$$\lambda \equiv \frac{2\epsilon Z e^2}{\hbar^2 c^2 \beta}, \qquad (3.3)$$

and substitute the radial variable $r$ by the dimensionless variable $s$ through the transformation

$$r \to s: \qquad s = \beta r. \qquad (3.4)$$

So, the radial Klein-Gordon equation satisfied by $u(s)$ can be written in the form

$$\frac{d^2 u(s)}{ds^2} - \left[\frac{l'(l'+1)}{s^2} - \frac{\lambda}{s} + \frac{1}{4}\right]u(s) = 0, \qquad (3.5)$$

where we have used the notation

$$l' = \sqrt{\left(l + \frac{1}{2}\right)^2 - \gamma^2} - \frac{1}{2}, \qquad \text{with} \quad \gamma \equiv Z\alpha, \qquad (3.6)$$

being $\alpha = \frac{e^2}{\hbar c}$ the fine structure constant. The physical solutions corresponding to the bound states (whose energy eigenvalues fulfil $|\epsilon| < m_0 c^2$) require that the radial eigenfunctions $u_{nl}(r)$ vanish both at the origin and at infinity, so that they have the

form [37]

$$u_{nl}(s) = \mathcal{N} s^{(l'+1)} e^{-\frac{s}{2}} \widetilde{L}_{n-l-1}^{(2l'+1)}(s), \tag{3.7}$$

where $\widetilde{L}_k^{(\alpha)}(s)$ denotes the orthonormal Laguerre polynomials of degree $k$ and parameter $\alpha$. The energy eigenvalues $\epsilon \equiv \epsilon_{nl}(Z)$ of the stationary bound states with wavefunctions $\Psi_{nlm}(\vec{r}, t) = \psi_{nlm}(\vec{r}) exp(-\frac{i}{\hbar}\epsilon t)$ are known to have the form [37]

$$\epsilon = \frac{m_0 c^2}{\sqrt{1 + \left(\frac{\gamma}{n-l+l'}\right)^2}}. \tag{3.8}$$

The constant $\mathcal{N}$ is determined not by the normalization of the wavefunction to unity as in the non-relativistic case, but by the charge conservation carried out by $\int_{\mathbb{R}^3} \rho(\vec{r}) d^3 r = e$ to preserve the Lorentz invariance [70], where the charge density of the negatively charged particle (e.g., a $\pi^-$-meson; $q = -e$) is given by

$$\rho_{nlm}(\vec{r}) = \frac{e}{m_0 c^2} \left[\epsilon - V(r)\right] |\psi_{nlm}(\vec{r})|^2. \tag{3.9}$$

Then, the charge normalization imposes the following restriction on the radial eigenfunctions

$$\begin{aligned} 1 &= \int_0^\infty \frac{\epsilon - V(r)}{m_0 c^2} u_{\epsilon l}^2(r) dr \\ &= \frac{1}{m_0 c^2} \int_0^\infty \left(\frac{\epsilon}{\beta} + \frac{\gamma \hbar c}{s}\right) u_{\epsilon l}^2(s) ds. \end{aligned} \tag{3.10}$$

The substitution of the expression (3.7) for $u_{\epsilon l}(s)$ into Eq. (3.10) provides the following normalization constant

$$\begin{aligned} \mathcal{N}^2 &= m_0 c^2 \left[\frac{2\epsilon}{\beta}(n + l' - l) + \gamma \hbar c\right]^{-1} \\ &= \frac{m_0 c^2 \gamma}{\hbar c} \frac{1}{(n + l' - l)^2 + \gamma^2}, \end{aligned} \tag{3.11}$$

where we have used for the second equality the relation

$$\frac{\epsilon}{\beta} = \frac{\hbar c}{2} \frac{n + l' - l}{\gamma}. \tag{3.12}$$

Let us emphasize that the resulting Lorentz-invariant charge density $\rho_{LI}(\vec{r})$ given by Eq. (3.9) is always (i.e. for any observer's velocity $v$) appropriately normalized while the density $\rho_{NLI}(\vec{r}) = |\psi_{nlm}(\vec{r})|^2$ (used in [75]) is not. This is numerically illustrated in Figure 3.1 for a pionic atom with nuclear charge $Z = 68$ in the infinite nuclear mass approximation ($\pi^-$ -meson mass=273.132054 a.u.).

For completeness we have plotted in Figure 3.2 the radial density of the charge distribution for two different states ($n = 1$, $l = 0$) and ($n = 4$, $l = 1$) of a pionic system with nuclear charge $Z = 68$ in the infinite nuclear mass approximation, respectively. Moreover, we have also made in these figures a comparison with the corresponding Schrödinger density functions. We observe that the relativistic effects other than spin (i) tend to compress the charge towards the origin, and (ii) they are most apparent for states $S$.

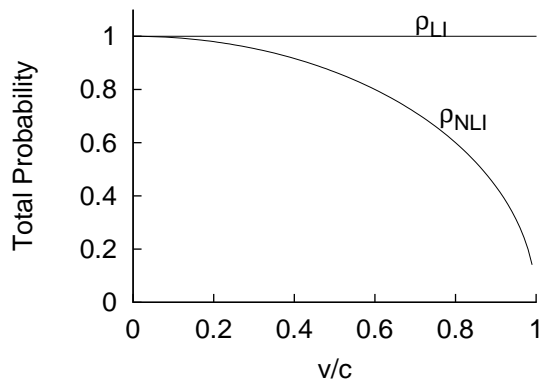

FIGURE 3.1: Normalization of the charge density for the Lorentz invariant (LI) and the non-Lorentz invariant (NLI) charge densities for different velocities of the observer.

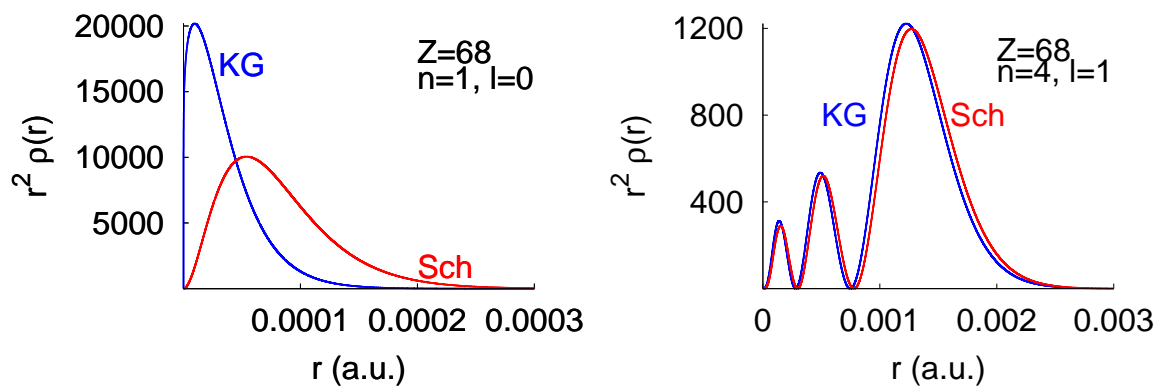

FIGURE 3.2: Comparison of the charge Klein-Gordon and Schrödinger radial density for the states $1S$ (left) and $4P$ (right) of the pionic system with $Z = 68$. Atomic units ($\hbar = m_e = e = 1$) are used.

## 3.2 Relativistic charge effects by information measures

In this Section we quantify this relativistic charge compression by three different means. First, in Subsection 3.2.1, we compute the ordinary moments or radial expectation values $\langle r^k \rangle$ for general $(n, l, m)$ states, making emphasis in the Heisenberg measure for circular $(l = n-1)$ and $S$-states ($l = 0$). These quantities, which are well-known in the Dirac case [73, 74], are strikingly only known [75] for the non-Lorentz-invariant density $\rho_{NLI}(\vec{r})$ in the Klein-Gordon case. Here we will study them for the Lorentz-invariant Klein-Gordon density $\rho_{LI}(\vec{r})$. Then in Subsection 3.2.2, we study numerically the most relevant charge information-theoretic measures of the system; namely the Shannon entropy and the Fisher information.

### 3.2.1   Radial expectation values and Heisenberg's measure.

The charge distribution of the Klein-Gordon particles in a Coulomb potential can be completely characterized by means of the ordinary radial expectation values $\langle r^k \rangle$, $k \in \mathbb{N}$, given by

$$
\begin{aligned}
\langle r^k \rangle &:= \int_{\mathbb{R}^3} r^k \rho_{nlm}(\vec{r}) d^3 r \\
&= \frac{1}{m_0 c^2} \int_0^\infty \left( \epsilon + \frac{Ze^2}{r} \right) r^k u_{nl}^2(r) dr \\
&= \frac{1}{m_0 c^2} \frac{1}{\beta^k} \int_0^\infty \left( \frac{\epsilon}{\beta} + \frac{\gamma \hbar c}{s} \right) s^k u_{nl}^2(s) ds \\
&= \frac{\mathcal{N}^2}{m_0 c^2} \frac{1}{\beta^k} \left[ \frac{\epsilon}{\beta} \mathscr{J}_{nl}(k) + \gamma \hbar c \mathscr{J}_{nl}(k-1) \right],
\end{aligned}
\qquad (3.13)
$$

where we have used Eqs. (3.7) and (3.9), and the symbol $\mathscr{J}_{nl}(k)$ denotes the integral [37]

$$
\begin{aligned}
\mathscr{J}_{nl}(k) &:= \int_0^\infty x^{2l'+k+2} e^{-x} \left[ \widetilde{L}_{n-l-1}^{(2l'+1)}(x) \right]^2 dx \\
&= \frac{(n-l-1)!}{\Gamma(n-l+2l'+1)} \\
&\quad \times \sum_{j=n-l-k-2}^{n-l-1} \binom{k+1}{n-l-j-1}^2 \frac{\Gamma(2l'+k+j+3)}{j!}.
\end{aligned}
\qquad (3.14)
$$

For the lowest values of $k$ we have

$$
\begin{aligned}
\mathscr{J}_{nl}(0) &= 2(n+l'-l) \\
\mathscr{J}_{nl}(1) &= 2\left[3(n-l)^2 + l'(6n+2l'-6l-1)\right] \\
\mathscr{J}_{nl}(2) &= 4(n+l'-l) \\
&\quad \times \left[1 + 5(n-l)^2 + l'(10n+2l'-10l-3)\right].
\end{aligned}
$$

Then, besides the normalization $\langle r^0 \rangle = 1$, we have the following value

$$\langle r \rangle = \frac{\mathcal{N}^2}{m_0 c^2} \frac{1}{\beta} \left[ \frac{\epsilon}{\beta} \mathscr{I}_{nl}(1) + \gamma \hbar c \, \mathscr{I}_{nl}(0) \right], \tag{3.15}$$

for the centroid of the charge density, and

$$\langle r^2 \rangle = \frac{\mathcal{N}^2}{m_0 c^2} \frac{1}{\beta^2} \left[ \frac{\epsilon}{\beta} \mathscr{I}_{nl}(2) + \gamma \hbar c \, \mathscr{I}_{nl}(1) \right], \tag{3.16}$$

for the second-order moment, so that the Heisenberg measure $\sigma_{nl}$ which quantifies the charge spreading around the centroid is given by

$$\begin{aligned} \sigma_{nl}^2 \equiv \sigma[\rho_{nlm}] &= \langle r^2 \rangle - \langle r \rangle^2 \\ &= \frac{\mathcal{N}^2}{m_0 c^2} \frac{1}{\beta^2} \left\{ \frac{\epsilon}{\beta} \mathscr{I}_{nl}(2) + \gamma \hbar c \, \mathscr{I}_{nl}(1) - \right. \\ &\left. - \frac{\mathcal{N}^2}{m_0 c^2} \left[ \frac{\epsilon}{\beta} \mathscr{I}_{nl}(1) + \gamma \hbar c \, \mathscr{I}_{nl}(0) \right]^2 \right\}. \end{aligned} \tag{3.17}$$

To gain insight into these general expressions we are going to discuss two particular classes of quantum-mechanical states, the circular (i.e., $l = n - 1$) states and the ns-states (i.e., $l = 0$).

For circular states we have that

$$\begin{aligned} l' &= \sqrt{\left(n - \tfrac{1}{2}\right)^2 - \gamma^2} - \tfrac{1}{2} \\ \epsilon &= \frac{m_0 c^2}{\sqrt{1 + \left(\frac{\gamma}{l'+1}\right)^2}}; \; \frac{\epsilon}{\beta} = \frac{\hbar c}{2\gamma}(l'+1), \end{aligned}$$

so that

$$\frac{\mathcal{N}^2}{m_0 c^2} = \frac{\gamma}{\hbar c} \frac{1}{(l'+1)^2 + \gamma^2},$$

and the integrals

$$\begin{aligned} \mathscr{I}_{nl}(0) &= 2l' + 2 \\ \mathscr{I}_{nl}(1) &= (2l'+2)(2l'+3) \\ \mathscr{I}_{nl}(2) &= (2l'+2)(2l'+3)(2l'+4). \end{aligned}$$

Then, the centroid of the charge distribution is, according to Eq. (3.15),

$$\langle r \rangle = \frac{\hbar c}{4 m_0 c^2} \frac{1}{\gamma \sqrt{1 + \left(\frac{\gamma}{l'+1}\right)^2}} \left[ (2l'+2)(2l'+3) + 4\gamma^2 \right], \tag{3.18}$$

and the second-order moment, according to equation (3.16), becomes

$$\langle r^2 \rangle = \left(\frac{\hbar c}{m_0 c^2}\right)^2 \frac{1}{2\gamma^2}(l'+1)(2l'+3)\left[(l'+1)(l'+2)+\gamma^2\right], \quad (3.19)$$

so that the Heisenberg measure for circular states $\sigma_n^2 \equiv \sigma_{n,n-1}^2$ has the following value

$$\begin{aligned}
\sigma_n^2 &= \left(\frac{\hbar c}{m_0 c^2}\right)^2 \left(\frac{l'+1}{4\gamma^2}\right) \\
&\times \frac{(l'+1)(2l'+3)\left[(l'+1)^2+2\gamma^2\right]+2\gamma^4}{(l'+1)^2+\gamma^2}.
\end{aligned} \quad (3.20)$$

These expressions for circular states and the corresponding ones for ns-states are discussed and compared with the Schrödinger values as a function of the principal quantum number n for the pionic system with nuclear charge $Z = 68$ in Figure 3.3. We observe that both centroid and variance ratios increase very rapidly with n, being the rate of this behaviour much faster for circular than for $S$-states. This indicates that the charge compression provoked by relativity in a given system (i.e. for fixed $Z$) (i) decreases when $n$ ($l$) is increasing for fixed $l$ ($n$). This can also be noticed in Figure 3.4, where the two previous ratios have been plotted as a function of $l$ for different values of $n$.

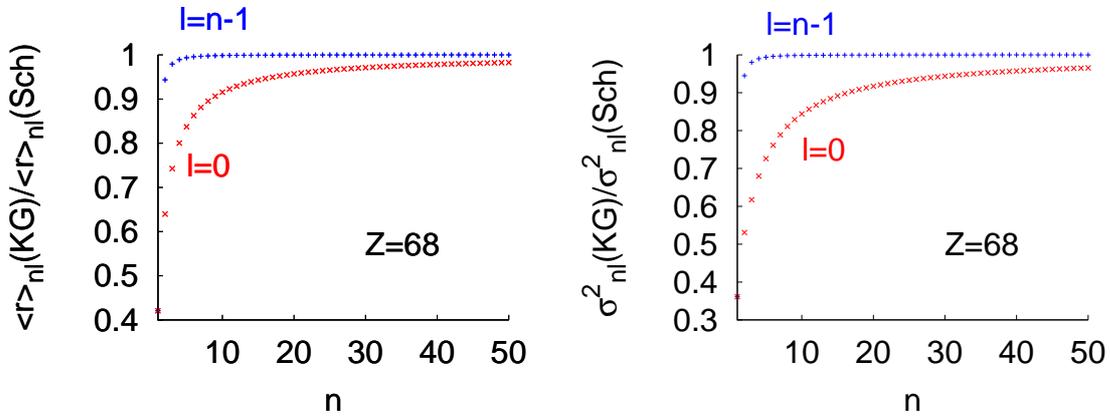

FIGURE 3.3: Comparison of the Klein-Gordon and Schrödinger values for the centroid (Left) and the variance (Right) of the pionic $1S$ and the circular states as a function of the quantum number $n$ with $Z = 68$.

Then, we have plotted these two ratios in terms of the nuclear charge $Z$ of the system in Figure 3.5 for the states $1S$, $2S$ and $2P$. We find that both the centroid and the variance ratios monotonically decrease as the nuclear charge $Z$ increases. Moreover, the decreasing rate is much faster for the states $1S$, than for the states $2S$ and $2P$. These two observations illustrate that the relativistic charge compression effect is bigger in heavier systems for a given ($nl$)-state. Moreover, we see here again that for a given system it increases both when $n$ decreases for fixed $l$ and when $l$ decreases for fixed $n$. The quantum number $m$ doesn't affect both ratios because the radial part of the density is not a function of it.

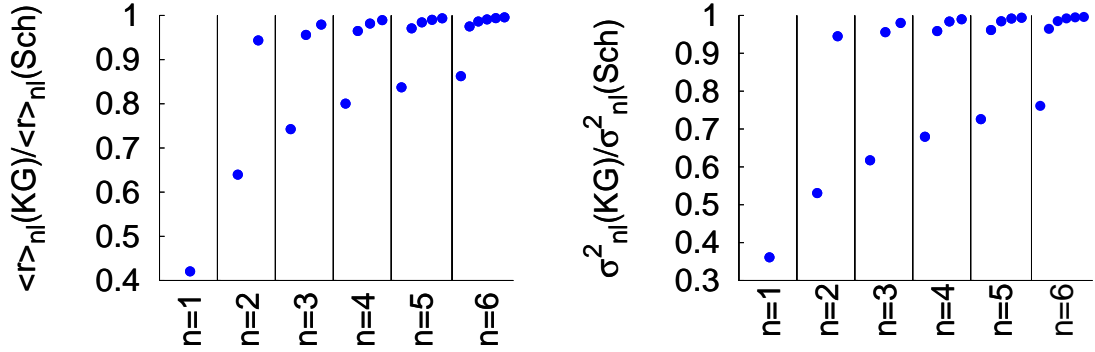

FIGURE 3.4: Comparison of the Klein-Gordon and Schrödinger values for the centroid (Left) and the variance (Right), as a function of the quantum number $l$ varying from 0 to $n-1$, for different values of $n$.

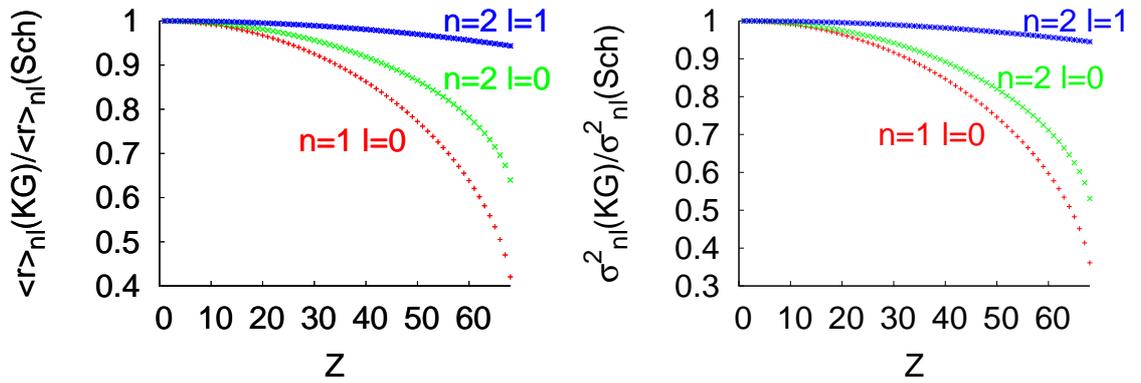

FIGURE 3.5: Comparison of the Klein-Gordon and Schrödinger values for the centroid (Left) and the variance (Right), as a function of the nuclear charge $Z$ for the pionic states $1S$, $2S$ and $2P$.

Finally, let highlight that in all figures the Klein-Gordon values tend towards the Schrödinger values in the non-relativistic limit of large $n$ or small $Z$.

### 3.2.2 Shannon and Fisher information measures

Here we study numerically the relativistic effects on the charge spreading of pionic systems of hydrogenic type by means of the following information-theoretic measures of the associated charge distribution $\rho_{nlm}(\vec{r})$ given by Eq (3.9): The Shannon entropy power and the Fisher information.

The Shannon entropic power of a negatively-charged Klein-Gordon particle characterized by the charge density $\rho_{nlm}(\vec{r})$ is defined by [15]

$$N_{nlm} \equiv N[\rho_{nlm}] = \frac{1}{2\pi e}\exp\left(\frac{2}{3}S_{nlm}\right), \qquad (3.21)$$

where $S_{nlm}$ is the Shannon entropy of $\rho_{nlm}(\vec{r})$ given by the expectation value of $-\log(\rho_{nlm}(\vec{r}))$, i. e.

$$S_{nlm} \equiv S[\rho_{nlm}] = -\int_{\mathbb{R}^3} \rho_{nlm}(\vec{r}) \log \rho_{nlm}(\vec{r}) d^3r, \qquad (3.22)$$

which quantifies the total extent of the charge spreading of the system. Taking into account the above-mentioned ansatz for $\psi(\vec{r})$ and Eqs. (3.7), (3.9) and (3.22), this expression can be separated out into radial and angular parts

$$S_{nlm} = S[R_{nl}] + S[Y_{lm}],$$

as it is explained in full detail in [21], being $R_{nl}$ and $Y_{lm}$ the radial and angular parts of the density. We should keep in mind that the angular part is the same for both Klein-Gordon and Schrödinger cases.

The Fisher information is defined by [10]

$$I_{nlm} \equiv I[\rho_{nlm}] = \int_{\mathbb{R}^3} \frac{|\nabla \rho(\vec{r})|^2}{\rho(\vec{r})} d^3r. \qquad (3.23)$$

Remark that we are not using here the parameter dependent Fisher information originally introduced (and so much used) by statisticians [81], but its translationally invariant form that does not depend on any parameter; see ref. [10, 82] for further details. It is worthy to point out that the Fisher information is a measure of the gradient content of the charge distribution: so, when $\rho(\vec{r})$ has a discontinuity at a certain point, the local slope value drastically changes and the Fisher information strongly varies. This indicates that it is a local quantity in contrast to the Heisenberg measure $\sigma_{nl}^2$ and the Shannon entropy $S[\rho]$ (and its associated power), which have a global character because they are powerlike and logarithmic functionals of the density, respectively.

Unlike the moment-based quantities discussed in the previous Section, these complementary measures do not depend on a special point, either the origin as the ordinary moments or the centroid as the Heisenberg measure. These quantities, first used by statisticians and electrical engineers and later by quantum physicists, have been shown to be measures of disorder or smoothness of the density $\rho_{nlm}(\vec{r})$ [10, 15]. Let us highlight that the Fisher information does not only measure the charge spreading of the system in a complementary and qualitatively different manner as the Heisenberg and Shannon measures but also it quantifies their oscillatory character, indicating the local charge concentration all over the space [10].

The relativistic (Klein-Gordon) and non-relativistic (Schrödinger) values of the Shannon entropic power are numerically discussed and compared in Figure 3.6 for the pionic system. Therein, on the left, we plot the ratio $N_{nl}(\text{KG})/N_{nl}(\text{Sch})$ between these two values as a function of the principal quantum number $n$ for the system with nuclear charge $Z = 68$. We notice that the Shannon ratio systematically increases when $n$ is increasing, approaching to unity for large n, for both circular and $S$-states. Moreover, we find that this approach is much faster for circular states, what indicates once more that the relativistic effects are much more important for states S. In addition, on the right of Figure 3.6, we show the dependence of the Shannon ratio with the nuclear charge $Z$ for the $1S$, $2S$ ant $2P$ states. We observe, here again, that the ratio is a decreasing function of $Z$ for any state, indicating that the relativistic effects are much more important for heavy systems. Moreover, for a given system (i.e. fixed $Z$) the relativistic effects increase when $n$ ($l$) decreases for fixed $l$ ($n$). The quantum number $m$ affects the absolute value of the Shannon entropic power but it doesn't affect the ratio.

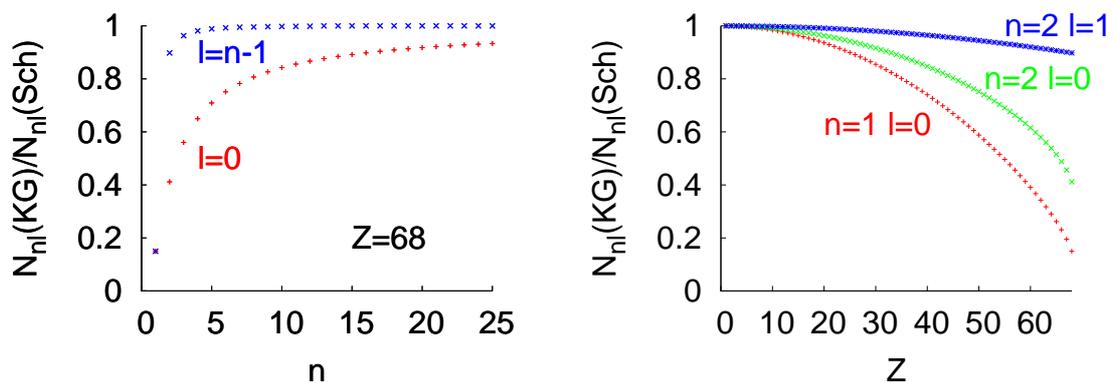

FIGURE 3.6: Comparison of the Klein-Gordon and Schrödinger values for the Shannon entropic power as a function of the principal quantum number $n$ (Left) and the nuclear charge $Z$ (Right).

Figure 3.7 shows the dependence of the ratio of the non-relativistic and relativistic values of the Fisher information for various states with $l \neq 0$ on their quantum numbers $(n, l, m)$ for the pionic system with $Z = 68$ (left graph) and on the nuclear charge $Z$ (right graph). The Fisher information for $S$-states is not defined because the involved integral diverges. First we should remark that here, contrary to the previous quantities considered in this work, the Schrödinger values are always less than the Klein-Gordon ones; this is strongly related to the local character of the Fisher information, indicating that the localized internodal charge concentration is always larger in the relativistic case. Second, we observe that for fixed $l$ the Fisher ratio $I_{nl}(\text{Sch})/I_{nl}(\text{KG})$ monotonically increases when $n$ is getting bigger, approaching to unity at a rate which grows as $l$ is increasing. Third, we find that the Fisher ratio decreases for all states in a systematic way as the nuclear charge increases. Moreover, for a given $Z$ value this ratio increases as either the quantum numbers $n$ and/or $l$ increase.

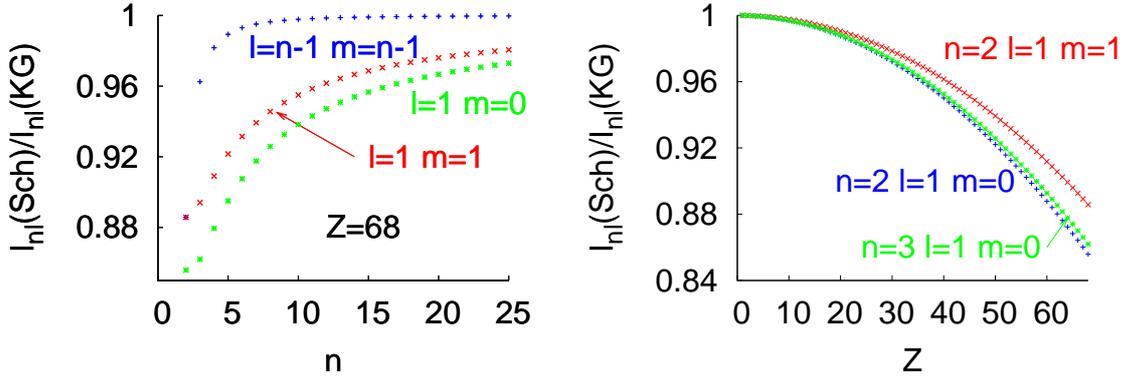

FIGURE 3.7: Comparison of the Klein-Gordon and Schrödinger values for the Fisher information as a function of the principal quantum number $n$ (Left) and the nuclear charge $Z$ (Right).

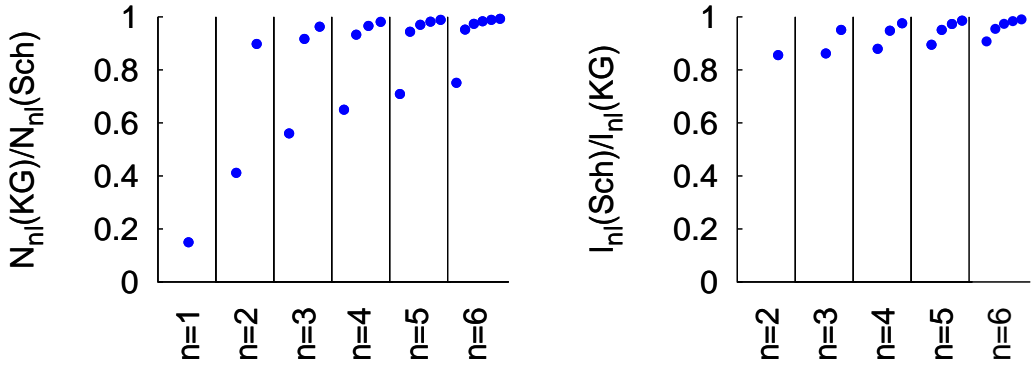

FIGURE 3.8: Comparison of the Klein-Gordon and Schrödinger values for the Shannon entropic power as a function of $l$ varying from 0 to $n-1$ (Left) and the Fisher information as a function of $l$ varying from 1 to $n-1$ (Right) for different values of $n$.

For completeness, the behaviour of the Shannon and Fisher ratios in terms of the orbital quantum number $l$ for a fixed $n$ is more explicitly shown of the left and right graphs, respectively, of Figure 3.8.

Finally, in Figure 3.9, the dependence of the Fisher ratio on the magnetic quantum number $m$ is studied. Notice that the ratio is bigger when $|m|$ is increasing, indicating that the lower $|m|$ is, the more concentrated is the charge density of the state and the more important are the relativistic effects.

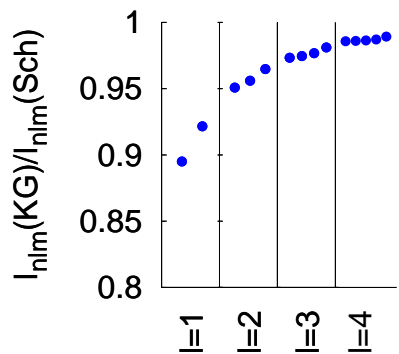

FIGURE 3.9: Comparison of the Klein-Gordon and Schrödinger values for the Fisher information as a function of $m$ varying from 0 to $l$, for different values of $l$.

## 3.3 Complexity analysis of Klein-Gordon single-particle systems

Numerous phenomena and properties of many-electron systems have been qualitatively characterized by information-theoretic means. In particular, various single and composite information-theoretic measures have been proposed to identify and analyse the multiple facets of the internal disorder of non-relativistic quantum systems; see e.g. Ref. [11, 13, 14, 46, 63, 83–89].

They are often expressed as products of two quantities of local (e. g. the Fisher information) and/or global (e. g. the variance or Heisenberg measure, the Shannon entropy, the Renyi and Tsallis entropies and the disequilibrium or linear entropy $\langle\rho\rangle$) character, which describe the charge spreading of the system in a complementary and more complete manner than their individual components. This is the case of the disequilibrium-Shannon or Lopez-Ruiz-Mancini-Calvet (LMC in short) [86], disequilibrium-Heisenberg [89], Fisher-Shannon [11, 13, 46, 63, 87, 88, 90] and the Cramer-Rao [11, 13, 15, 88] complexities, which have their minimal values at the extreme ordered and disordered limits.

Recently these studies have been extended to take into account the relativistic effects in atomic physics. Relativistic quantum mechanics [70] tells us that special relativity provokes (at times , severe) spatial modifications of the electron density of many-electron systems, what produces fundamental and measurable changes in their physical properties. The qualitative and quantitative evaluation of the relativistic modification of the spatial redistribution of the electron density of ground and excited states in atomic and molecular systems by information-theoretic means is a widely open field. In the last three years the relativistic effects of various single and composite information-theoretic quantities of the ground states of hydrogenic [55] and neutral atoms [52, 91] have been investigated in different relativistic settings.

First Borgoo et al [52] (see also [92]), in a Dirac-Fock setting, find that the LMC shape complexity of the ground-state atoms (i) has an increasing dependence on the nuclear charge (also observed by Katriel and Sen [55] in Dirac ground-state hydrogenic systems), (ii) manifest shell and relativistic effects, the latter being specially relevant in the disequilibrium ingredient (which indicates that they are dominated by the innermost orbital). Then, Sañudo and López-Ruiz [91] (see also [93]) show a similar trend for both LMC and Fisher-Shannon complexities in a different setting which uses the fractional occupation probabilities of electrons in atomic orbitals instead of the continuous electronic wavefunctions; so, they use discrete forms for the information-theoretic ingredients of the complexities. Moreover, their results allow to identify the shell structure of noble gases and the irregular shell filling of some specific elements; this phenomenon is specially striking in the Fisher-Shannon case as the authors explicitly point out.

The present Section contributes to this new field with the quantification of the relativistic compression of both ground and excited states of the Klein-Gordon single-particle wavefunctions in a Coulombian well by means of the Fisher-Gordon complexity. This quantity is defined by

$$C_{FS}[\rho] := I[\rho] \times J[\rho], \tag{3.24}$$

where

$$I[\rho] = \int \rho(\vec{r}) \left[\frac{d}{dx}\log\rho(\vec{r})\right]^2 d\vec{r}, \ J[\rho] = \frac{1}{2\pi e}\exp\left(2S[\rho]/3\right), \tag{3.25}$$

are the Fisher information and the Shannon entropic power of the density $\rho(\vec{r})$, respectively. The latter quantity, which is an exponential function of the Shannon entropy $S[\rho] = -\langle log\rho \rangle$, measures the total extent to which the single-particle distribution is in fact concentrated [15]. The Fisher information $I[\rho]$, which is closely related to the kinetic energy [82], is a local information-theoretic quantity because it is a gradient functional of the density, so being sensitive to the single-particle oscillations. Then, contrary to the remaining complexities published in the literature up until now, the Fisher-Shannon complexity has a property of locality and it takes simultaneously into account the spatial extent of the density and its (strong) oscillatory nature.

### 3.3.1 The Fisher-Shannon measure of pionic systems

Let us now numerically discuss the relativistic effects in the Fisher-Shannon complexity of a pionic system. First, we center our attention in the dependence on the nuclear charge of the system. As we can see in Figure 3.10, the Fisher-Shannon complexity of the Klein-Gordon case depend on the nuclear charge $Z$, contrary to the non-relativistic description. The Schrödinger or non-relativistic value of the Fisher-Shannon complexity has been recently shown to be independent of the nuclear charge $Z$ for any hydrogenic system, It is apparent that this quantity is a very good indicator of the relativistic effects as has been recently pointed out by Sañudo and López-Ruiz [91, 93] in other relativistic settings. These effects are bigger when the nuclear charge increases, so the relativistic Fisher-Shannon complexity enhances. This behaviour is easy to understand because when we take into account the relativistic effects, the charge probability density is more compressed towards the nucleus than in the non-relativistic case [70] (see Figure 3.2).

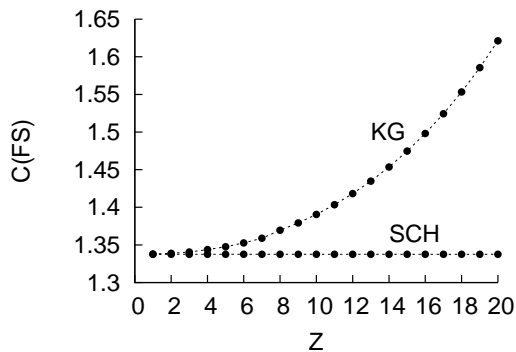

FIGURE 3.10: Fisher-Shannon complexity ratio for the ground state Klein-Gordon (KG) and Schrödinger (SCH) pionic atom in terms of the nuclear charge $Z$ (atomic units are used).

To measure the relativistic effects we define the quantity $\zeta_{\text{FS}} = 1 - \frac{C_{SCH}(FS)}{C_{KG}(FS)}$. This quantity varies from 0 to 1, so that $\zeta_{\text{FS}} \sim 0$ when the relativistic effects are negligible and $\zeta_{\text{FS}} \sim 1$ in the ultrarelativistic limit. In Figure 3.11 we can see the effects of the relativity model for different values of the nuclear charge $Z$ and various $S(n,0,0)$ states.

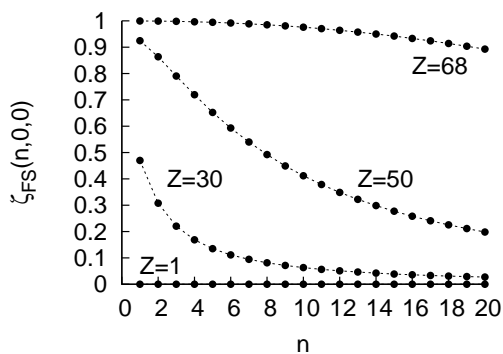

FIGURE 3.11: Fisher-Shannon complexity ratio for various $S$ states of the Klein-Gordon and Schrödinger pionic atom.

First, we observe that the relativistic effects increase when the nuclear charge is increasing not only for the ground state (as already pointed out) but also for all the excited states. Second, the relativistic effects decrease when the principal quantum number is increasing. Third, this decreasing behaviour with $n$ has a strong dependence with $Z$, being slower as bigger is the nuclear charge.

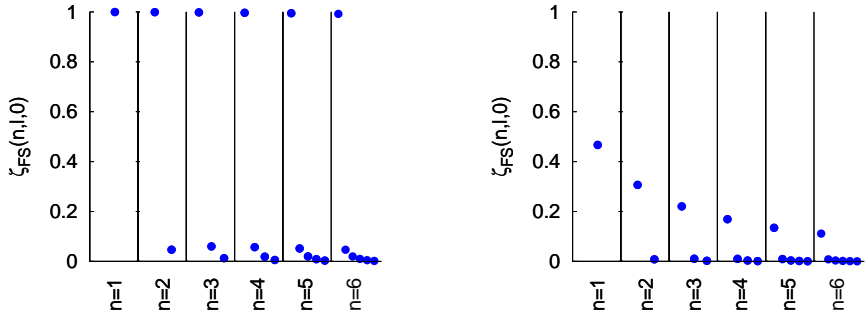

FIGURE 3.12: Fisher-Shannon complexity ratio for various states $(n, l, 0)$ of the Klein-Gordon and Schrödinger pionic atoms with $Z = 68$ (left) and $Z = 30$ (right).

In Figure 3.12 we can observe that the relativistic effects are practically negligible when the angular quantum number is different to zero even when the nuclear charge is big. This dependence in $l$ is more important than the dependence on the principal quantum number $n$. For completeness, let us point out that the relativistic effects are practically negligible when the magnetic quantum number $m$ varies for $(Z, n, l)$ fixed.

### 3.3.2    The LMC shape complexity of pionic systems

Here we study the LMC shape complexity [86] of the pionic systems defined by Equation 1.6 in a similar way as done previously with the Fisher-Shannon complexity. The principal results are given in Figures 3.13, 3.14, 3.15. In Figure 3.13 the values of the Klein-Gordon (KG) and Schrödinger (SCH) shape complexities for the ground-state pionic atom are shown in terms of the nuclear charge $Z$. It is apparent that the behaviour of this measure with the nuclear charge $Z$ is similar to that of the Fisher-Shannon complexity displayed in Figure 3.10. We first observe that, opposite to the Schrödinger setting (where the shape complexity is constant), the relativistic Klein-Gordon shape complexity varies indeed with $Z$. This indicates that this measure is also a good indicator of the relativistic effects. These effects clearly enhance when the nuclear charge is increasing. This enhancement can be quantified by means of the ratio $\zeta_{\text{SC}} = 1 - \frac{C_{SCH}(SC)}{C_{KG}(SC)}$, whose value for various states $S$ in terms of the principal quantum number $N$ is given in Figure 3.14 for the pionic systems with nuclear charge $Z = 1$, 30, 50 and 68. The relativistic effects manifest much more weakly in the shape complexity than in the Fisher-Shannon measure. This relative behaviour is mainly due to the fact that the shape complexity does not grasp the strong oscillatory character of the pionic wavefunctions, while the Fisher-Shannon complexity really does.

Finally in Figure 3.15 we show the relative behaviour if the shape complexity in the relativistic Klein-Gordon and non-relativistic cases in terms of the orbital quantum number $l$ for various states $(n, l, 0)$ with fixed $n$. This behaviour is similar to the Fisher-Shannon case shown in Figure 3.12. again, we observe that the relativistic effects are negligible for $l$ other than 0, even for atoms with a large nuclear charge.

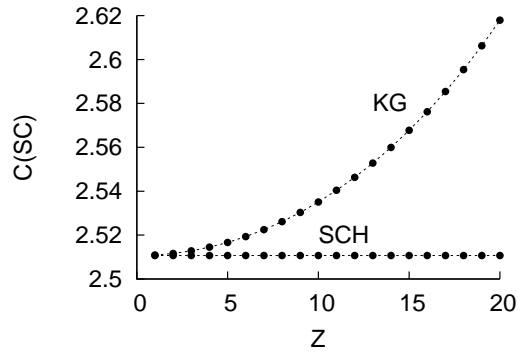

FIGURE 3.13: Shape complexity for the ground state Klein-Gordon (KG) and Schrödinger (SCH) pionic atom in terms of the nuclear charge $Z$.

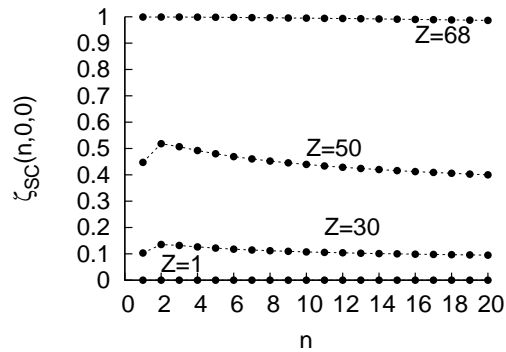

FIGURE 3.14: Shape complexity for $S$ states of the Klein-Gordon and Schrödinger pionic atom with nuclear charges $Z = 1$, 30, 50 and 68.

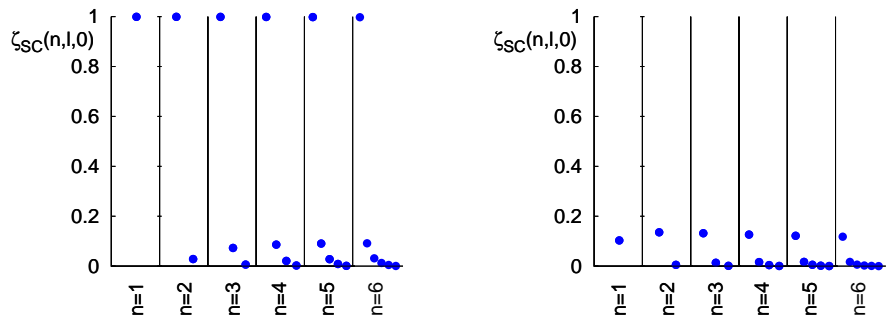

FIGURE 3.15: Shape complexity for various states $(n, l, 0)$ of the Klein-Gordon and Schrödinger pionic atoms with $Z = 68$ (left) and $Z = 30$ (right).

For completeness, let us also point out that the relativistic effects are practically negligible when the magnetic number $m$ varies for pionic states with given quantum numbers $(n, l)$ of atoms with a nuclear charge $Z$.

## 3.4  Conclusions

The relativistic charge compression of spinless Coulomb particles has been quantitatively investigated by means of the Heisenberg, Shannon and Fisher spreading measures. These three complementary quantities show that the relativity effects are larger (i. e. the charge compresses more towards the origin) for the lower energetic states and when the Coulomb strength (i. e. the nuclear charge $Z$) increases. Moreover, a detailed analysis of these quantities on the quantum numbers $(n, l, m)$ characterising the physical states of a given system (i. e. for a fixed $Z$) indicate that the relativistic effects increase when $n$ ($l$) decreases for fixed $l$ ($n$). Furthermore, the study of the Fisher information shows that the relativistic effects also increase when the magnetic quantum number $|m|$ is increasing for fixed $(n, l)$.

We have explored relativistic effects on the behaviour of the Fisher-Shannon complexity of pionic systems with nuclear charge $Z$ in the Klein-Gordon framework too. We have done it for both ground and excited states. First we found that the relativistic Fisher-Shannon complexity grows when the nuclear charge increases, in contrast with the non-relativistic case for both ground and excited states. A similar behaviour has been recently observed in the case of the ground state of systems governed by the Dirac equation [52, 91–93]. We found that this trend remains for excited states in a damped way, so that the relativistic effects enhance with $Z$ for a given $(n, l, m)$ state and, for a given $Z$, decrease when the principal and/or orbital quantum numbers are increasing. Let us also highlight that the non-relativistic limits at large principal quantum number $n$ for a given $Z$ (see Figs. 3.11 and 3.12) and at small values of $Z$ (see Fig. 3.10) are reached. On the other hand it is pertinent to underline that the finite nuclear volume effects are very tiny for any information-theoretic and complexity measure because of its macroscopic character.

We have also investigated the relativistic Klein-Gordon effects in pionic atoms by means of the LMC shape complexity [86] $C(\text{LMC}) = \langle \rho \rangle \exp S[\rho]$. We found that the relativistic effects are also identified by this quantity but in a much weaker way than the Fisher-Shannon complexity $C(\text{FS})$. Apparently this is because of the property of locality of $C(\text{FS})$ coming through its gradient-dependent Fisher ingredient, which grasps much better the (strong) oscillatory condition of the pionic densities.

# Chapter 4

# Information-theoretic lengths of orthogonal polynomials

Let $\{p_n(x)\}$ denote a sequence of real orthonormal polynomials with respect to the weight function $\omega(x)$ on the interval $\Delta \equiv (a,b) \subseteq \mathbb{R}$ (see e.g. [94, 95]), i.e.

$$\int_\Delta p_n(x)p_m(x)\omega(x)dx = \delta_{n,m}, \quad m,n \in \mathbb{N}. \tag{4.1}$$

The distribution of these polynomials along the orthogonality interval can be complementarity measured by means of the spreading properties of the normalized-to-unity density function

$$\rho_n(x) \equiv \rho[p_n] = p_n^2(x)\omega(x), \tag{4.2}$$

which is called Rakhmanov's density of the polynomial $p_n(x)$, to honour the pioneering work [96] of this mathematician who has shown that this density governs the asymptotic ($n \to \infty$) behaviour of the ratio $p_{n+1}/p_n$. Physically, this probability density characterizes the stationary states of a large class of quantum-mechanical potentials [94]. Beyond the variance, the spreading of the orthogonal polynomials is best analysed by the information-theoretic properties of their associated Rakhmanov probability densities; namely, the Fisher information [81], the Rényi entropy [97] and the Shannon entropy [2].

The information-theoretic knowledge of the orthogonal polynomials is reviewed in Ref [24] up to 2001, where the quantum-mechanical motivation and some physical applications are also given, and up today in [98], where emphasis is made on asymptotics. Therein, it is pointed out that the study of the information-theoretic measures of orthogonal polynomials was initiated in the nineties with the asymptotic computation of the Rényi and Shannon entropies of the classical orthogonal polynomials [99–101]. Up until now, however, the explicit expressions of these two spreading measures are not known save for the Shannon measure for some particular subclasses of the Jacobi polynomials; namely the Chebyshev polynomials of first and second type [100, 101] and the Gegenbauer polynomials $C_n^{(\lambda)}(x)$ with integer parameter $\lambda$ [102, 103]. On the other hand, the Fisher information for all classical orthogonal polynomials on a real interval (Hermite, Laguerre, Jacobi) has been calculated in a closed form in 2005 [104]. This has allowed us to find, more recently, the Cramer-Rao information plane (i.e. the plane defined by both the Fisher information and the variance) for these systems [105].

This Chapter has two main purposes. First, in Section 4.1 to introduce a more appropriate set of spreading measures (to be called information-theoretic-based spreading lengths of Rényi, Shannon and Fisher types) in the field of orthogonal polynomials. The spreading lengths are direct measures of position spread of the variable $x$, as the root-mean-square or standard deviation $\Delta x$. So, they have the following properties: same units as $x$, invariance under translations and reflections, linear scaling with $x$ and vanishing in the limit as $\rho_n(x)$ approaches a delta function. The second purpose is to compute all these lengths for the Hermite (see Section 4.2) and Laguerre polynomials 4.3. In each case we first compute their moments around the origin and, as well, the standard deviation of these objects. Then we compute the entropic or frequency moments [106–108] of these polynomials by means of the Bell polynomials, so useful in Combinatorics [109, 110], in the Hermite and Laguerre case and, alternatively, by the linearization technique of Srivastava-Niukkanen in the Laguerre case. This achievement allows us to obtain the Rényi lengths of Hermite and Laguerre polynomials by means of an efficient algorithm based on the known properties of Bell's polynomials. With respect to the Shannon length, since its explicit expression cannot be determined, we obtain its asymptotics and we derive a family of sharp upper bounds by use of an information-theoretic-based optimization procedure. Moreover, the Fisher information is explicitly calculated. Finally, these information-theoretic quantities are applied to the computation of the spreading lengths of the main quantum-mechanical prototype of physical systems: the harmonic oscillator.

## 4.1 Spreading lengths of classical orthogonal polynomials

Here we shall refer to the hypergeometric-type polynomials, i.e. the polynomials $\{p_n(x)\}$ defined by the orthogonality relation given by Eq. (4.1). It is well-known that they can be reduced to one of the three classical families of Hermite, Jacobi and Laguerre by appropriate linear changes of the variable [94]. The spreading of these polynomials along their orthogonality interval $\Delta \equiv (a,b)$ can be measured by means of (i) the moments around a particular point of the orthogonality interval; usually it is chosen the origin (moments around the origin: $\mu'_k = \langle x^k \rangle_n$) or the centroid $\langle x \rangle$ (central moments or moments around the centroid: $\mu_k = \langle (x - \langle x \rangle_n)^k \rangle_n$) [111] and (ii) the frequency moments $W_k[\rho_n] := \langle [\rho_n(x)]^k \rangle$ [107, 111, 112] of the associated Rakhmanov density $\rho_n(x)$ given by Eq. (4.2). The symbol $\langle f(x) \rangle \equiv \langle f(x) \rangle_n$ denotes the expectation value of $f(x)$ with respect to the density $\rho_n(x)$ as

$$\langle f(x) \rangle_n := \int_\Delta f(x) \rho_n(x) dx = \int_\Delta f(x) \omega(x) p_n^2(x) dx.$$

These two classes of ordinary and frequency moments are complementary spreading measures of global type. While the ordinary moments measures the distribution of the probability with respect to a particular point of the support interval $\Delta$, the frequency moments measure the extent to which the probability is in fact distributed. So it happens that the latter moments are, at times, much better probability estimators than the ordinary ones [107, 108]. Moreover, they are fairly efficient in the range where the ordinary moments are fairly inefficient [106]; see also the brief, recent summary about these quantities done in Ref. [113]. Therein we learn that the frequency moments are also called "entropic moments" because they are closely connected to the Rényi entropies [97]

$$R_q[\rho_n] := \frac{1}{1-q} \log \langle [\rho_n(x)]^{q-1} \rangle; \quad q > 0; \quad q \neq 1, \tag{4.3}$$

and the Tsallis entropies [114]

$$T_q[\rho_n] := \frac{1}{q-1} \left[ 1 - \langle [\rho_n(x)]^{q-1} \rangle \right]; \quad q > 0; \quad q \neq 1, \tag{4.4}$$

as well as to other information-theoretic quantities, such as the Brukner-Zeilinger entropy [115] and the linear entropy [116].

Among the first class of moments we highlight the familiar root-mean-square or standard deviation $(\Delta x)_n$ given by

$$(\Delta x)_n = \left( \langle x^2 \rangle_n - \langle x \rangle_n^2 \right)^{\frac{1}{2}},$$

because it is a direct measure of spreading [72] in the sense of having the same units as the variable, and has various interesting properties: invariance under translations and reflections, linear scaling (i.e., $\Delta y = \lambda \Delta x$ for $y = \lambda x$) and vanishing as the density approaches a Dirac delta density (i.e. in the limit that $x$ tends towards a given definite value).

From the second class of moments we fix our attention on the Rényi lengths [72] defined by

$$\mathcal{L}_q^R[\rho_n] \equiv \exp(R_q[\rho_n]) = \langle [\rho_n(x)]^{q-1} \rangle^{-\frac{1}{q-1}} = \left\{ \int_\Delta dx [\rho_n(x)]^q \right\}^{-\frac{1}{q-1}}, \quad q > 0, \quad q \neq 1, \tag{4.5}$$

which are not only direct spreading measures but also have the three above-mentioned properties of the standard deviation. Among them we highlight the second-order Rényi length (also called Onicescu information [8], Heller length [117], disequilibrium [4, 6] and inverse participation ratio in other contexts [118, 119])

$$\mathcal{L}[\rho_n] := \mathcal{L}_2^R[\rho_n] = \langle \rho_n(x) \rangle^{-1} = \left\{ \int_\Delta dx [\rho_n(x)]^2 \right\}^{-1},$$

and, above all, the Shannon length [2, 72]

$$N[\rho_n] := \lim_{q \to 1} \mathcal{L}_q^R[\rho_n] = \exp(S[\rho_n]) \equiv \exp\left[-\int_\Delta dx \rho_n(x) \log \rho_n(x)\right],$$

where $S[\rho_n]$ is the Shannon information entropy [2].

There exists a third class of spreading measures for the classical orthogonal polynomials, which are qualitatively different in the sense that, in contrast to the previous ones, they have a local character. Indeed they are functionals of the derivative of the associated Rakhmanov density $\rho_n(x)$, so that they are very sensitive to local rearrangements of the variable. The most distinctive measure of this class is the so-called Fisher length [81, 120, 121] defined by

$$(\delta x)_n := \frac{1}{\sqrt{F[\rho_n]}} \equiv \left\langle \left[\frac{d}{dx} \log \rho_n(x)\right]^2 \right\rangle^{-\frac{1}{2}} = \left\{ \int_\Delta dx \frac{[\rho_n'(x)]^2}{\rho_n(x)} \right\}^{-\frac{1}{2}}, \qquad (4.6)$$

where $F[\rho_n]$ denotes the Fisher information of the classical orthogonal polynomials [104]. This quantity measures the pointwise concentration of the probability along the orthogonality interval, and quantifies the gradient content of the Rakhmanov density providing (i) a quantitative estimation of the oscillatory character of the density and of the polynomials and (ii) the bias to particular points of the interval, so that it measures the degree of local disorder.

It is worthy to remark that the Fisher length, as the Heisenberg and Rényi lengths, is a direct spreading measure and has the three properties of translation and reflection invariance, linear scaling and vanishing when the density tends to a delta density. In addition, the Fisher length is finite for all distributions [120]. Moreover, the direct spreading measures just mentioned have an uncertainty/certainty property: see Refs. [72] for the Heisenberg, Shannon and Rényi cases, and [122, 123] for the Fisher case. Finally, they fulfil the inequalities [72]

$$(\delta x)_n \leq (\Delta x)_n, \quad \text{and} \quad N[\rho_n] \leq (2\pi e)^{\frac{1}{2}} (\Delta x)_n, \qquad (4.7)$$

where the equality is reached if and only if $\rho_n(x)$ is a Gaussian density.

## 4.2 Spreading lengths of Hermite polynomials

In this section we calculate the moments-with-respect-to-the-origin $\langle x^k \rangle$, $k \in \mathbb{Z}$, the entropic moments $W_q[\rho_n] \equiv \langle [\rho_n(x)]^q \rangle$, of integer order $q$, and the Rényi, Shannon and Fisher spreading lengths of the Rakhmanov density of the orthonormal Hermite polynomials $\tilde{H}_n(x)$ given by

$$\tilde{H}_n(x) = \sum_{l=0}^{n} c_l x^l, \tag{4.8}$$

with

$$c_l = \frac{(-1)^{\frac{3n-l}{2}} n!}{(2^n n! \sqrt{\pi})^{\frac{1}{2}} \left(\frac{n-l}{2}\right)! l!} \frac{2^l}{} \frac{(-1)^l + (-1)^n}{2}, \tag{4.9}$$

where the last factor vanishes when $l$ and $n$ have opposite parities. These polynomials are known to fulfil the orthonormality relation (4.1) with the weight function $\omega_H(x) = e^{-x^2}$ on the whole real line [94, 95]. Thus, according to Eq. (4.2), the expression

$$\rho_n(x) \equiv \rho[\tilde{H}_n(x)] = e^{-x^2} \tilde{H}_n^2(x), \tag{4.10}$$

gives the Rakhmanov density of the orthonormal Hermite polynomial $\tilde{H}_n(x)$.

### 4.2.1 Ordinary moments and the standard deviation

The moments of the Hermite polynomials $\tilde{H}_n(x)$ are defined by the moments of its associated Rakhmanov density (4.2); that is

$$\langle x^k \rangle_n := \int_{-\infty}^{\infty} x^k \tilde{H}_n^2(x) e^{-x^2} dx, \; k = 0, 1, 2, \ldots \tag{4.11}$$

The use of (4.8), (4.1) and (4.11) gives

$$\langle x^k \rangle_n = \begin{cases} \frac{k!}{2^k \Gamma(\frac{k}{2}+1)} \, {}_2F_1\left( \begin{array}{c} -n, -\frac{k}{2} \\ 1 \end{array} \middle| 2 \right), & \text{even } k \\ 0, & \text{odd } k \end{cases}, \tag{4.12}$$

which can be also obtained from some tabulated special integrals [124].

In the particular cases $k = 0, 1$ and $2$ we obtain the first few moments around the origin:

$$\langle x^0 \rangle_n = 1, \; \langle x \rangle_n = 0, \; \langle x^2 \rangle_n = n + \frac{1}{2},$$

so that the second central moment of the density

$$(\Delta x)_n = \sqrt{\langle x^2 \rangle_n - \langle x \rangle_n^2} = \sqrt{n + \frac{1}{2}}, \tag{4.13}$$

describes the standard deviation.

### 4.2.2  Entropic moments and Rényi lengths of integer order $q$

The $q$th-order frequency or entropic moment of the Hermite polynomials $\tilde{H}_n(x)$ is defined by the corresponding quantity of its associated Rakhmanov density given by Eq. (4.2); that is,

$$W_q[\rho_n] = \left\langle \{\rho_n\}^{q-1} \right\rangle = \int_{-\infty}^{\infty} \{\rho_n(x)\}^q \, dx = \int_{-\infty}^{\infty} e^{-qx^2} \{\tilde{H}_n(x)\}^{2q} dx; \; q \geq 1 \quad (4.14)$$

The evaluation of this quantity in a closed form is not at all a trivial task despite numerous efforts published in the literature [125, 126] save for some special cases. Here we will do it by means of the Bell polynomials which play a relevant role in Combinatorics [127, 128]. We start from the explicit expression (4.8) for the Hermite polynomial $\tilde{H}_n(x)$; then, taking into account the Appendix B we have that its $p$th-power can be written down as

$$[\tilde{H}_n(x)]^p = \left[\sum_{k=0}^{n} c_k x^k\right]^p = \sum_{k=0}^{np} \frac{p!}{(k+p)!} B_{k+p,p}(c_0, 2!c_1, \ldots, (k+1)!c_k) x^k, \quad (4.15)$$

with $c_i = 0$ for $i > n$, and the remaining coefficients are given by Eq. (4.9). Moreover, the Bell polynomials are given by

$$B_{m,l}(c_1, c_2, \ldots, c_{m-l+1}) = \sum_{\hat{\pi}(m,l)} \frac{m!}{j_1! j_2! \cdots j_{m-l+1}!} \left(\frac{c_1}{1!}\right)^{j_1} \left(\frac{c_2}{2!}\right)^{j_2} \cdots \left(\frac{c_{m-l+1}}{(m-l+1)!}\right)^{j_{m-l+1}}, \quad (4.16)$$

where the sum runs over all partitions $\hat{\pi}(m,l)$ such that

$$j_1 + j_2 + \cdots + j_{m-l+1} = l, \quad \text{and} \quad j_1 + 2j_2 + \cdots + (m-l+1)j_{m-l+1} = m. \quad (4.17)$$

The substitution of the expression (4.15) with $p = 2q$ into Eq. (4.14) yields the value

$$\begin{aligned} W_q[\rho_n] &= \sum_{k=0}^{2nq} \frac{(2q)!}{(k+2q)!} B_{k+2q,2q}(c_0, 2!c_1, \ldots, (k+1)!c_k) \int_{-\infty}^{\infty} e^{-qx^2} x^k dx \\ &= \sum_{j=0}^{nq} \frac{\Gamma\left(j + \frac{1}{2}\right)}{q^{j+\frac{1}{2}}} \frac{(2q)!}{(2j+2q)!} B_{2j+2q,2q}(c_0, 2!c_1, \ldots, (2j+1)!c_{2j}), \end{aligned} \quad (4.18)$$

where the parameters $c_i$ are given by Eq. (4.9), keeping in mind that $c_i = 0$ for every $i > n$, so that the only non-vanishing terms correspond to those with $j_{i+1} = 0$ so that

$c_i^{j_{i+1}} = 1$ for every $i > n$. In the particular cases $q = 1$ and 2, we obtain the value

$$W_1[\rho_n] = \int_{-\infty}^{\infty} \rho_n(x)dx = 1$$

for the normalization of $\rho_n(x)$, as one would expect, and

$$\begin{aligned} W_2[\rho_n] &= \int_{-\infty}^{\infty} \{\rho_n(x)\}^2 \, dx = \int_{-\infty}^{\infty} e^{-2x^2}(\tilde{H}_n(x))^4 dx \\ &= \sum_{j=0}^{2n} \frac{\Gamma\left(j+\frac{1}{2}\right)}{2^{j+\frac{1}{2}}} \frac{4!}{(2j+4)!} B_{2j+4,4}(c_0, 2!c_1, \ldots, (2j+1)!c_{2j}), \end{aligned} \quad (4.19)$$

for the second-order entropic moment.

As well, the entropic moments of arbitrary order $q$ for the Hermite polynomials of lowest degree (e.g., $n = 0, 1$ and 2) have the values

$$W_q[\rho_0] = \sqrt{\frac{\pi^{1-q}}{q}}, \quad W_q[\rho_1] = \frac{2^q}{\pi^{\frac{q}{2}} q^{q+\frac{1}{2}}} \Gamma\left(q+\frac{1}{2}\right), \quad W_q[\rho_2] = \frac{\pi^{\frac{1-q}{2}} 2^q (2q)!}{q^{2q+\frac{1}{2}}} L_{2q}^{(-2q-\frac{1}{2})}\left(-\frac{q}{2}\right),$$

where $L_n^{(\alpha)}(x)$ denotes a Laguerre polynomial.

Then, we can now calculate all the information-theoretic measures of the Hermite polynomials which are based on their entropic moments $W_q[\rho_n]$ just calculated, such as the Rényi and Tsallis entropies given by Eqs. (4.3) and (4.4), respectively, and the Brukner-Zeilinger and linear entropy which are closely related to the first-order entropic moment $W_2[\tilde{H}_n]$ given by Eq. (4.19). Particularly relevant are the values

$$\begin{aligned} \mathcal{L}_q^R[\rho_n] &= \{W_q[\rho_n]\}^{-\frac{1}{q-1}} = \\ &= \left( \sum_{j=0}^{nq} \frac{\Gamma\left(j+\frac{1}{2}\right)}{q^{j+\frac{1}{2}}} \frac{(2q)!}{(2j+2q)!} B_{2j+2q,2q}(c_0, 2!c_1, \ldots, (2j+1)!c_{2j}) \right)^{-\frac{1}{q-1}} \quad ; q = 2, 3, \ldots \end{aligned} \quad (4.20)$$

for the Rényi lengths (4.5) of Hermite polynomials, and

$$\mathcal{L}[\rho_n] = \{W_2[\rho_n]\}^{-1} = \left( \sum_{j=0}^{2n} \frac{\Gamma\left(j+\frac{1}{2}\right)}{2^{j+\frac{1}{2}}} \frac{4!}{(2j+4)!} B_{2j+4,4}(c_0, 2!c_1, \ldots, (2j+1)!c_{2j}) \right)^{-1},$$

for the Heller length of Hermite polynomials.

Let us just explicitly write down the values

$$\mathcal{L}_q^R[\rho_0] = \pi^{\frac{1}{2}} q^{\frac{1}{2(q-1)}}$$

$$\mathcal{L}_q^R[\rho_1] = \left(\frac{\pi^{\frac{1}{2}} q}{2}\right)^{\frac{q}{q-1}} q^{\frac{1}{2(q-1)}} \left(\Gamma\left(q+\frac{1}{2}\right)\right)^{-\frac{1}{q-1}}$$

$$\mathcal{L}_q^R[\rho_2] = \pi^{\frac{1}{2}} 2^{\frac{q}{1-q}} q^{\frac{4q+1}{2q-2}} \left( (2q)! L_{2q}^{(-2q-\frac{1}{2})}\left(-\frac{q}{2}\right) \right)^{-\frac{1}{q-1}},$$

for the Rényi lengths of the first few Hermite polynomials with lowest degrees, and the values

$$\mathcal{L}[\rho_0] = \sqrt{2\pi}, \qquad \mathcal{L}[\rho_1] = \frac{4}{3}\sqrt{2\pi}, \qquad \mathcal{L}[\rho_2] = \frac{64}{41}\sqrt{2\pi},$$

for the Onicescu-Heller length of these polynomials.

### 4.2.3  Shannon length: Asymptotics and sharp bounds

Here we determine the asymptotics of the Shannon length $N[\rho_n]$ of the Hermite polynomials $\tilde{H}_n(x)$ and its relation with its standard deviation. Moreover we use an information-theoretic-based optimization procedure to find sharp upper bounds to $N[\rho_n]$ and we discuss their behaviour in a numerical way.

The Shannon length or exponential entropy of the Hermite polynomials $\tilde{H}_n(x)$ is defined by

$$N[\rho_n] = \exp\{S[\rho_n]\}, \tag{4.21}$$

where

$$S[\rho_n] = -\int_{-\infty}^{\infty} e^{-x^2} \tilde{H}_n^2(x) \log\left[e^{-x^2} \tilde{H}_n^2(x)\right] dx, \tag{4.22}$$

is the Shannon entropy of the Hermite polynomial of degree $n$. Simple algebraic operations yield to

$$S[\rho_n] = n + \frac{1}{2} - \int_{-\infty}^{\infty} e^{-x^2} \tilde{H}_n^2(x) \log \tilde{H}_n^2(x) dx. \tag{4.23}$$

The logarithmic integral involved in this expression has not yet been determined in spite of serious attempts for various authors [99–101, 129]. Nevertheless these authors have found its value for large $n$ by use of the strong asymptotics of Hermite polynomials; it is given as

$$\int_{-\infty}^{\infty} e^{-x^2} \tilde{H}_n^2(x) \log \tilde{H}_n^2(x) dx = n + \frac{3}{2} - \log \pi - \log \sqrt{2n} + o(1). \tag{4.24}$$

Then, from (4.23) and (4.24) one has the asymptotical value

$$S[\rho_n] = \log \sqrt{2n} + \log \pi - 1 + o(1) = \log \frac{\pi\sqrt{2n}}{e} + o(1),$$

for the Shannon entropy, and

$$N[\rho_n] \simeq \frac{\pi\sqrt{2n}}{e}; \quad n \gg 1, \tag{4.25}$$

for the Shannon length of the Hermite polynomial $\tilde{H}_n(x)$. From Eqs. (4.13) and (4.25) one finds the following relation

$$N[\rho_n] \simeq \frac{\pi\sqrt{2}}{e}(\Delta x)_n; \ n \gg 1, \tag{4.26}$$

between the asymptotical values of the Shannon length and the standard deviation of the Hermite polynomial $\tilde{H}_n$ [130]. Remark that this relation fulfils the general inequality (4.7) which mutually relate the Shannon length and the standard deviation.

Since the evaluation of the Shannon $S(\tilde{H}_n)$ given by Eq. (4.23) is not yet possible for a generic degree, it seems natural to try to find sharp bounds to this quantity. Here we do that by use of the non-negativity of the Kullback-Leibler entropy of the two arbitrary probability densities $\rho(x)$ and $f(x)$ defined by

$$I_{\text{KL}}[\rho, f] = \int_{-\infty}^{\infty} \rho(x) \log \frac{\rho(x)}{f(x)} dx.$$

Indeed, for $\rho(x) = \rho_n(x)$, the Rakhmanov density (4.2) associated to the Hermite polynomials, the non-negativity of the corresponding Kullback-Leibler functional yields the following upper bound to $S(\tilde{H}_n)$:

$$S[\rho_n] \leq -\int_{-\infty}^{\infty} \rho_n(x) \log f(x) dx. \tag{4.27}$$

Then we make the choice

$$f(x) = \frac{ka^{\frac{1}{k}}}{2\Gamma\left(\frac{1}{k}\right)} e^{-x^k}, \ k = 2, 4, \ldots \tag{4.28}$$

duly normalized to unity, as prior density, The optimization of the upper bound (4.27) for the choice (4.28) with respect to the parameter $a$ yields the following optimal bound

$$S[\rho_n] \leq \log\left[A_k \langle x^k \rangle_n^{\frac{1}{k}}\right], \ k = 2, 4, \ldots$$

with the constant

$$A_k = \frac{2(ek)^{\frac{1}{k}}}{k}\Gamma\left(\frac{1}{k}\right).$$

Finally, taking into account the expectation values $\langle x^k \rangle_n$ of the Rakhmanov density $\rho_n$ of the Hermite polynomial $\tilde{H}_n(x)$ given by Eq. (4.12), we have that

$$S[\rho_n] \leq \log\left[\frac{(ek)^{\frac{1}{k}}}{k}\Gamma\left(\frac{1}{k}\right)\left[\frac{k!}{\Gamma\left(\frac{k}{2}+1\right)}\,{}_2F_1\left(\begin{array}{c}-n,-\frac{k}{2}\\1\end{array}\bigg|2\right)\right]^{\frac{1}{k}}\right],$$

| $n$ | 0 | 1 | 2 | 3 | 4 | 5 | 6 | 7 | 8 | 9 | 10 | 11 | 12 |
|---|---|---|---|---|---|---|---|---|---|---|---|---|---|
| $k_{\text{opt}}$ | 2 | 6 | 8 | 10 | 12 | 14 | 16 | 16 | 18 | 20 | 22 | 22 | 24 |
| $c_{k,n}$ | 2.92 | 4.54 | 5.57 | 6.40 | 7.11 | 7.75 | 8.33 | 8.86 | 9.36 | 9.83 | 10.30 | 10.70 | 11.10 |

TABLE 4.1: Values of $k$ which provides the best upper bound $c_{k,n}$ for the first few values of $n$.

and

$$N[\rho_n] \leq \left[\frac{(ek)^{\frac{1}{k}}}{k}\Gamma\left(\frac{1}{k}\right)\left[\frac{k!}{\Gamma\left(\frac{k}{2}+1\right)}\,{}_2F_1\left(\begin{array}{c}-n,-\frac{k}{2}\\1\end{array}\bigg|2\right)\right]^{\frac{1}{k}}\right] \equiv c_{k,n}, \qquad (4.29)$$

as optimal bounds for the Shannon entropy and Shannon length of $\tilde{H}_n(x)$, respectively. In Table 4.1 we give the value $k_{\text{opt}}$ of $k$ which provides the best upper bound $c_{k,n}$ for the first few values of $n$.

### 4.2.4 Fisher length

Finally, let us point out that the Fisher information of the Hermite polynomials $\tilde{H}_n(x)$ is given [104] by

$$F[\rho_n] = \int_{-\infty}^{+\infty}\left\{\frac{d}{dx}\rho_n(x)\right\}^2\frac{dx}{\rho_n(x)} = 4n+2. \qquad (4.30)$$

Thus, the Fisher length of these polynomials is

$$(\delta x)_n = \frac{1}{\sqrt{F[\rho_n]}} = \frac{1}{\sqrt{4n+2}}, \qquad (4.31)$$

whose values lie down within the interval $[0,\frac{1}{\sqrt{2}}]$. The comparison of this expression with Eq. (4.13) allows us to point out that not only the general relation (4.7) between the Fisher length and the standard deviation is fulfilled, but also that $(\delta x)_n(\Delta x)_n = \frac{1}{2}$.

### 4.2.5 Effective computation of the spreading lengths

In this Subsection we examine and discuss some computational issues relative to the spreading lengths of Hermite polynomials $H_n(x)$ and their mutual relationships. In contrast with the polynomials orthogonal on a segment of the real axis for which an efficient algorithm based on the three-term recurrence relation has been recently discovered for the computation of the Shannon entropy by V. Buyarov et al. [23], up until now there is no explicit formula or at least a stable numerical algorithm for the computation of the Rényi and Shannon lengths of unbounded orthogonal polynomials, such as Hermite polynomials, for any reasonable $n \in \mathbb{N}$. A naive numerical evaluation of these Hermite functionals by means of quadratures is not often convenient except for the lowest-order polynomials since the increasing number of integrable singularities spoils any attempt to achieve a reasonable accuracy for arbitrary $n$.

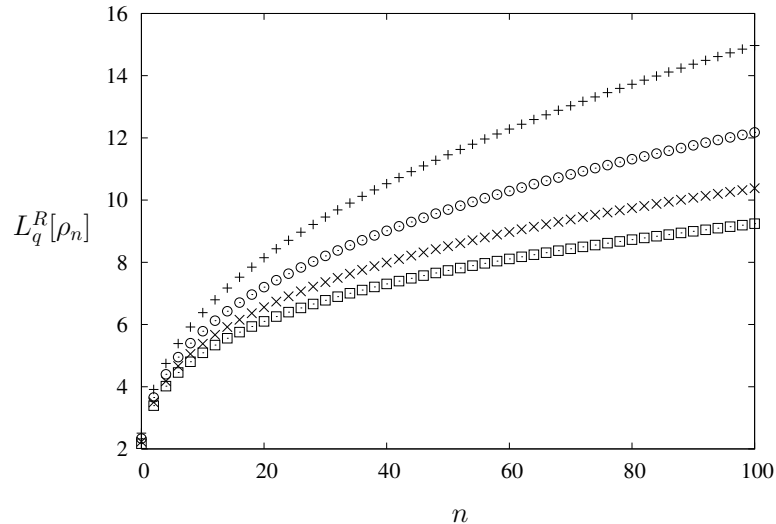

FIGURE 4.1: Rényi lengths $\mathcal{L}_q^R[\rho_n]$ with $q = 2$ (+), 3 (⊙), 4 (×), and 5 (⊡), in terms of the degree $n$.

We propose an analytical, error-free and easily programmable computing approach for the entropic moments $W_q[\rho_n]$ and the Rényi spreading lengths $\mathcal{L}_q^R[\rho_n]$ of the orthonormal polynomials $\tilde{H}_n(x)$, which are given by Eqs. (4.18) and (4.21), respectively, in terms of the combinatorial multivariable Bell polynomials defined by Eq. (4.16). This approach requires the knowledge of the expansion coefficients $c_l$ given by Eq. (4.9) and the determination of the partitions $\tilde{\pi}(m, l)$ given by Eq. (4.17).

In Figures 4.1 and 4.2 we have shown the Rényi lengths $\mathcal{L}_q^R[\rho_n]$ with $q = 2, 3, 4$ and 5, and the Shannon length $N[\rho_n]$ of the orthonormal Hermite polynomials in terms of the degree $n$ within the range $n \in [0, 100]$, respectively. We observe that both Shannon and Rényi lengths with fixed $q$ have an increasing parabolic dependence on the degree $n$ of the polynomials. Moreover, this behaviour is such that for fixed $n$ the Rényi length decreases when $q$ is increasing. As well, in Figure 4.2 we plot the optimal upper bound $c_{k_{\text{opt}},n}$ for the Shannon length $N[\rho_n]$ according to Eq. (4.29).

Finally, for completeness, we have numerically studied the relation of the Shannon length $N[\rho_n]$ and the Onicescu-Heller length $\mathcal{L}[\rho_n]$ of these polynomials with the standard deviation $(\Delta x)_n$ within the range $n \in [0, 100]$. The corresponding results are shown in Figures 4.3 and 4.4, respectively. We observe that when $n$ varies from 0 to 100 the two spreading lengths have a quasilinear behaviour in terms of the standard deviation. Moreover, we found the fit

$$N[\rho_n] = 1.723(\Delta x)_n + 2.00 \quad (4.32)$$

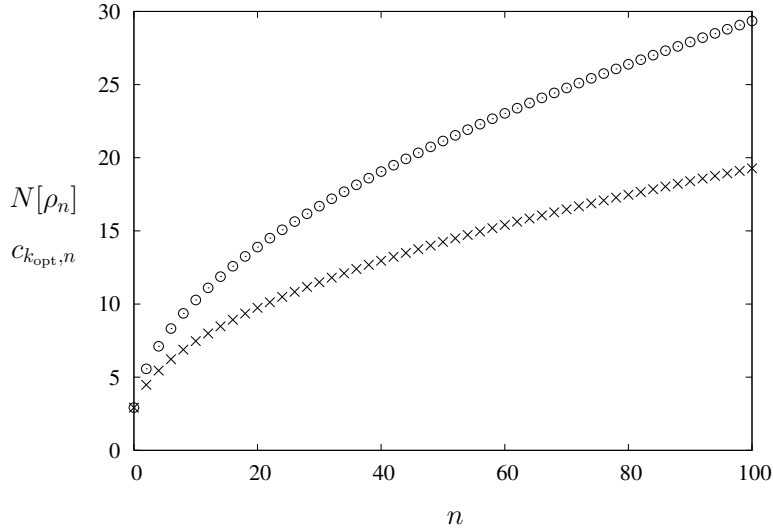

FIGURE 4.2: Shannon length $N[\rho_n]$ ($\times$) and its optimal upper bound $c_{k_{\text{opt}},n}$ ($\odot$) in terms of the degree $n$.

for the Shannon length, with a correlation coefficient $R = 0.999998$, and

$$\mathcal{L}[\rho_n] = 1.204(\Delta x)_n + 2.92,$$

for the Onicescu-Heller length or second-order Rényi length with a correlation coefficient $R = 0.99994$. In fact, the global dependence of $N[\rho_n]$ on $(\Delta x)_n$ is slightly concave, being only asymptotically linear according to the rigorous expression (4.26), i.e. $N[\rho_n] \simeq \frac{\pi\sqrt{2}}{e}(\Delta x)_n \simeq 1.63445(\Delta x)_n$ for $n \gg 1$, which is in accordance with (4.32). It is worthwhile remarking (i) the $n^\alpha$ behaviour with $\alpha > 0$ of the global spreading lengths (standard deviation, Rényi and Shannon lengths) and (ii) the $n^{-\frac{1}{2}}$-law which is followed by the (local) Fisher length. This difference may be associated to the gradient-functional form (4.6) of the Fisher length, indicating a locality property in the sense that it is very sensitive to the oscillatory character of the polynomials.

### 4.2.6   Application

The states of the one-dimensional harmonic oscillator, with a potential $V(x) = \frac{1}{2}\lambda^2 x^2$, are determined by the following probability density

$$\rho_n^{\text{HO}}(x) = \frac{\sqrt{\lambda}}{2^n n! \sqrt{\pi}} e^{-\lambda x^2} \left(H_n(\sqrt{\lambda}x)\right)^2,$$

where $H_n(x)$ are the orthogonal Hermite polynomials [94]. Since $\tilde{H}_n(x) = (2^n n! \sqrt{\pi})^{-\frac{1}{2}} H_n(x)$, this density can be expressed in terms of the Rakhmanov density of the orthonormal

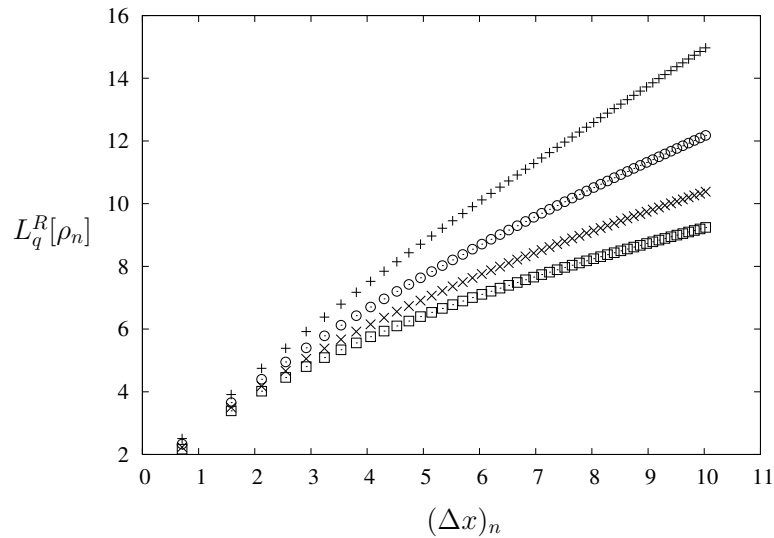

FIGURE 4.3: Rényi lengths $\mathcal{L}_q^R[\rho_n]$ with $q = 2$ (+), 3 (⊙), 4 (×), and 5 (⊡), in terms of the standard deviation $(\Delta x)_n$ for $n \in [0, 100]$.

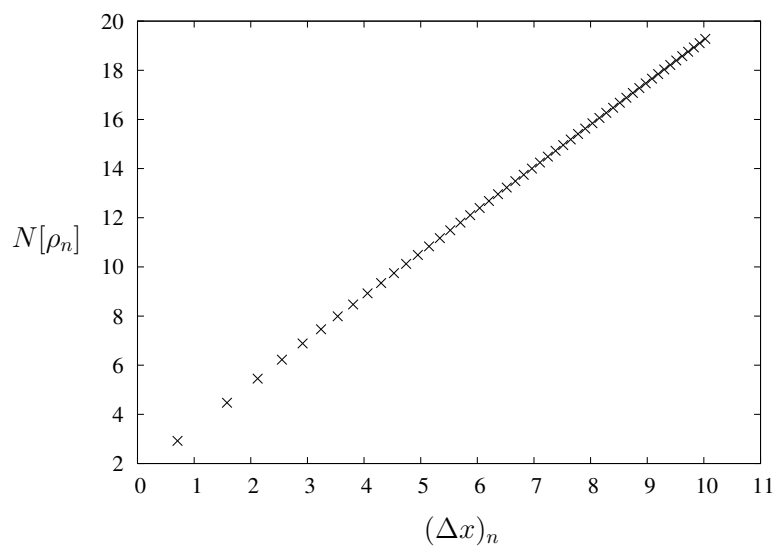

FIGURE 4.4: Shannon length $N[\rho_n]$ in terms of the standard deviation $(\Delta x)_n$ for $n \in [0, 100]$.

Hermite polynomials $\tilde{H}_n(x)$ defined by Eq. (4.10) as follows:

$$\rho_n^{\text{HO}}(x) = \sqrt{\lambda}\rho_n(\sqrt{\lambda}x).$$

Taking this into account, we obtain for the harmonic oscillator the following expressions

$$\langle x^k \rangle_n^{\text{HO}} = \lambda^{-\frac{k}{2}} \langle x^k \rangle_n;\ k = 0, 1, 2, \ldots$$

$$W_q[\rho_n^{\text{HO}}] = \lambda^{\frac{q-1}{2}} W_q[\rho_n];\ q = 0, 1, 2, \ldots,$$

for the ordinary and entropic moments, and

$$S[\rho_n^{\text{HO}}] = -\log\sqrt{\lambda} + S[\rho_n],$$

$$F[\rho_n^{\text{HO}}] = \lambda F[\rho_n],$$

for the Shannon and Fisher information measures in terms of the corresponding quantities of the Hermite polynomials, whose values are given by Eqs. (4.12), (4.18), (4.23) and (4.30), respectively. Then, it is straightforward to find that all the uncertainty measures of the harmonic oscillator (namely, standard deviation and Rényi, Shannon and Fisher lengths) are equal, save for a multiplication factor $\lambda^{-\frac{1}{2}}$, to the corresponding spreading lengths of the Hermite polynomials which control their wavefunctions. This is to say that

$$(\Delta x)_n^{\text{HO}} = \lambda^{-\frac{1}{2}}(\Delta x)_n, \quad \mathcal{L}_q^R[\rho_n^{\text{HO}}] = \lambda^{-\frac{1}{2}}\mathcal{L}_q^R[\rho_n]$$

$$N[\rho_n^{\text{HO}}] = \lambda^{-\frac{1}{2}}N[\rho_n], \quad (\delta x)_n^{\text{HO}} = \lambda^{-\frac{1}{2}}(\delta x)_n,$$

whose values can be obtained by keeping in mind Eqs. (4.13), (4.21)-(4.23), and (4.31) respectively.

### 4.2.7 Open problems

Here we want to pose the following unsolved problems: to find the asymptotics of the frequency or entropic moments $W_q[\rho_n]$ defined by

$$W_q[\rho_n] = \int_{-\infty}^{\infty}\left[\omega_H(x)\tilde{H}_n^2(x)\right]^q dx = \int_{-\infty}^{\infty} e^{-qx^2}\left[\tilde{H}_n(x)\right]^{2q} dx, \; q \geq 1 \quad (4.33)$$

in the two following cases:

$$\begin{aligned}&\bullet \; n \to \infty, \text{ and } q \in \mathbb{R} \text{ fixed.} \\ &\bullet \; q \to \infty, \text{ and } n \in \mathbb{N} \text{ fixed.}\end{aligned} \quad (4.34)$$

Up until now the only known result is due to A. Aptekarev et al [98, 99, 129] who have shown that these quantities, which are weighted $\mathcal{L}^{2q}$-norms of the orthonormal Hermite polynomials $\tilde{H}_n(x)$ as

$$W_q[\rho_n] = \parallel \tilde{H}_n(x) \parallel_{2q}^{2q},$$

behave asymptotically ($n \to \infty$) as

$$W_q[\rho_n] = \left(\frac{2}{\pi}\right)^q \frac{\Gamma\left(q+\frac{1}{2}\right)\Gamma\left(1-\frac{q}{2}\right)}{\Gamma(q+1)\Gamma\left(\frac{3}{2}-\frac{q}{2}\right)}(2n+1)^{\frac{1-q}{2}}(1+o(1)),$$

when $q \in [0, \frac{4}{3}]$. To extend and generalize this result in the sense mentioned above, it might be useful for the interested reader to keep in mind the two following related results:

(a) L. Larsson-Cohn [131] studied the $L^p$-norms of the monic Hermite polynomials $h_n(x)$, orthogonal with respect to the Gaussian weight $\omega_G(x) = (2\pi)^{-\frac{1}{2}} \exp\left(-\frac{x^2}{2}\right)$, defined by

$$\|h\|_p = \left(\frac{1}{2\pi}\right)^{\frac{p}{2}} \left\{\int_{-\infty}^{\infty} e^{-\frac{x^2}{2}} |h_n(x)|^p dx\right\}^{\frac{1}{p}}.$$

Keeping in mind that $\|h\|_n = \sqrt{n!}$, he found the following asymptotical $(n \to \infty)$ result

$$\|h\|_p = \begin{cases} c(p) n^{-\frac{1}{4}} \sqrt{n!} (1 + O(n^{-1})), & \text{if } 0 < p < 2 \\ c(p) n^{-\frac{1}{4}} \sqrt{n!} (p-1)^{\frac{n}{2}} (1 + O(n^{-1})), & \text{if } 2 < p < \infty, \end{cases}$$

with the values

$$c(p) = \begin{cases} \left(\frac{2}{\pi}\right)^{\frac{1}{4}} \left(\frac{2}{2-p}\right)^{\frac{1}{2p}} \left\{\frac{\Gamma\left(\frac{p+1}{2}\right)}{\sqrt{\pi}\Gamma\left(\frac{p+2}{2}\right)}\right\}^{\frac{1}{p}}, & \text{for } p < 2 \\ \left(\frac{2}{\pi}\right)^{\frac{1}{4}} \left(\frac{p-1}{2p-4}\right)^{\frac{p-1}{2p}}, & \text{for } p > 2 \end{cases}$$

However, the asymptotics of the entropic moments $W_q[h_n]$ of the Hermite polynomials $h_n(x)$ defined by

$$W_q[h_n] = \int_{-\infty}^{\infty} \left[\omega_G(x) h_n^2(x)\right]^q dx = \left(\frac{1}{2\pi}\right)^{\frac{q}{2}} \int_{-\infty}^{\infty} e^{-\frac{q}{2}x^2} [h_n(x)]^{2q} dx, \qquad (4.35)$$

has not yet been found.

(b) R. Azor et al [125] have used combinatorial techniques for the asymptotics $(n \to \infty)$ of the following functionals of the Hermite polynomials $H_n(x)$ orthogonal with respect to $e^{-x^2}$:

$$Z_k[H_n] := \int_{-\infty}^{\infty} e^{-x^2} [H_n(x)]^k dx, k \in \mathbb{N}.$$

They were only able to solve it in the particular non-trivial case $k = 4$, where

$$Z_4[H_n] = \frac{3}{4n} \sqrt{\frac{3}{\pi}} \frac{6^{2n}}{(n!)^2} \left\{1 - \frac{1}{4n} + \frac{3}{16n^2} + O(n^{-3})\right\}.$$

Besides, they have also considered the asymptotics $(k \to \infty)$ of $Z_k[H_n]$ as well as that of the quantities

$$D_k[H_n] := \sqrt{\frac{2}{\pi}} \int_{-\infty}^{\infty} e^{-2x^2} \left[\tilde{H}_n(x)\right]^k dx,$$

finding the following results

$$Z_k[H_n] \sim \left[1 + (-1)^{kn}\right] \left(2^{kn-1}\pi\right)^{\frac{1}{2}} \left(\frac{kn}{e}\right)^{\frac{kn}{2}} e^{-\frac{n-1}{2}}, \text{ for } k \gg 1$$

$$D_k[H_n] \sim \frac{1 + (-1)^{kn}}{\sqrt{2}} \left(\frac{kn}{e}\right)^{\frac{kn}{2}} e^{-(n-1)}, \text{ for } k \gg 1,$$

The solution of the problems (4.33)-(4.34) and/or (4.35), and the full determination of the asymptotics of $Z_k[H_n]$ and $D_k[H_n]$ for ($n \to \infty$, $k \in \mathbb{R}$), either by means of an approximation-theoretic methodology [98, 99, 129, 131] or by use of combinatorial techniques [125, 126], is of great relevance from both mathematical and applied points of view from obvious reasons, keeping in mind the physico-mathematical meaning of these quantities as previously mentioned. It is worthwhile remarking that, in particular, the quantity $W_q[\rho_n]$ is not only the ($2q$)th-power of the $L^{2q}$-norm of the Hermite polynomials but it also represents the entropic moment of order $q$ of the harmonic oscillator and their isospectral physical systems, which fully determines the Rényi and Tsallis entropies of these objects.

## 4.3   Direct spreading measures of Laguerre polynomials

The Laguerre polynomials $\left\{L_n^{(\alpha)}(x)\right\}$ are real hypergeometric polynomials orthonormal with respect to the weight function $\omega_\alpha(x) = x^\alpha e^{-x}$ on the interval $[-1, 1]$. They play a crucial role in numerous branches of applied mathematics [132–134], mathematical physics [94, 95], quantum physics [135]. This is mainly because their algebraic properties (orthogonality relation, three-term recurrence relation, second-order differential equation, ladder relation,...) are simple and widely known [94, 132, 133], which have allowed us to describe a great deal of scientific and technological phenomena. Let us just mention that the Laguerre polynomials appear in the wavefunctions which describe the quantum states of one and many-body systems with a great diversity of quantum-mechanical potentials in ordinary [135–140] and supersymmetric [141] quantum mechanics. The Coulomb and Morse potentials are only two particularly relevant cases in atomic and molecular physics (see e.g. [136, 138, 142–144] as well as in $D$-dimensional physics [145], where the radial wavefunctions are controlled by Laguerre polynomials.

In this work we study the spreading measures of Laguerre polynomials $L_n^{(\alpha)}$, which quantify the distribution of its Rakhmanov associated probability density

$$\rho_{n,\alpha}(x) = \frac{1}{d_n^2} \left[L_n^{(\alpha)}(x)\right]^2 \omega_\alpha(x), \tag{4.36}$$

where $d_n^2$ is a normalization constant. Physically, this probability density characterizes the stationary states of a large class of quantum-mechanical potentials [94, 136, 137,

139, 140, 146]. The most familiar spreading measure is the simple root-mean-square or standard deviation

$$(\Delta x)_{n,\alpha} = \left(\langle x^2\rangle_{n,\alpha} - \langle x\rangle_{n,\alpha}^2\right)^{\frac{1}{2}}, \qquad (4.37)$$

where the expectation value of a function $f(x)$ is defined by

$$\langle f(x)\rangle_{n,\alpha} := \int_0^\infty f(x)\rho_{n,\alpha}(x)dx. \qquad (4.38)$$

The information-theoretic-based spreading measures of the Laguerre polynomials are not so well known. We refer to the Fisher information

$$F\left[L_n^{(\alpha)}\right] := \left\langle\left[\frac{d}{dx}\log\rho_{n,\alpha}(x)\right]\right\rangle = \int \frac{[\rho'_{n,\alpha}(x)]^2}{\rho_{n,\alpha}(x)}dx, \qquad (4.39)$$

to the Rényi entropy of order $q$

$$R_q\left[L_n^{(\alpha)}\right] := \frac{1}{1-q}\log\left\langle[\rho_{n,\alpha}(x)]^{q-1}\right\rangle, \qquad (4.40)$$

and its $(q \to 1)$ limit, the Shannon entropy

$$S\left[L_n^{(\alpha)}\right] := -\int_0^\infty \rho_{n,\alpha}(x)\log\rho_{n,\alpha}(x)dx,$$

which measure the distribution of the Laguerre polynomial $L_n^{(\alpha)}$ all over the orthogonality interval without reference to any specific point of the interval, so providing alternative and complementary measures for the spreading of the Laguerre polynomials. The knowledge of these measures and some quantum-mechanical applications is reviewed in Ref. [24] up to 2001. Their behaviour for large $n$ and fixed $\alpha$ has been recently surveyed [98].

Since the Fisher, Rényi and Shannon measures of a given density $\rho(x)$ have particular units, which are different from that of the variable $x$, it is much more useful to use the related information-theoretic lengths [72, 120, 147–149]; namely, the Fisher length given by

$$(\delta x)_{n,\alpha} = \frac{1}{\sqrt{F\left[\rho_{n,\alpha}\right]}}, \qquad (4.41)$$

and the $q$th-order Rényi and Shannon lengths defined by

$$\mathcal{L}_q^R\left[\rho_{n,\alpha}\right] = \exp\left(R_q\left[\rho_{n,\alpha}\right]\right); \quad q > 0, q \neq 1, \qquad (4.42)$$

and

$$N\left[\rho_{n,\alpha}\right] = \lim_{q\to 1}\mathcal{L}_q^R\left[\rho_{n,\alpha}\right] = \exp\left(S\left[\rho_{n,\alpha}\right]\right), \qquad (4.43)$$

respectively. Following Hall [72, 120, 148], these three quantities together with the standard deviation will be referred as the direct spreading measures of the density $\rho_{n,\alpha}$ because they share the following properties: translation and reflection invariance and linear scaling under adequate boundary conditions, same units as the variable, and vanishing when the density tends to a delta density. Moreover, they have an associated uncertainty property [72, 120, 123, 148] and fulfil the inequalities

$$(\delta x)_{n,\alpha} \leq (\Delta x)_{n,\alpha} \qquad \text{and} \qquad N[\rho_{n,\alpha}] \leq (2\pi e)^{\frac{1}{2}} (\Delta x)_{n,\alpha}. \tag{4.44}$$

Here we will investigate the direct spreading measures of the Laguerre polynomials mentioned above. First, in Subsection 4.3.1, we give the known values of the ordinary moments, the standard deviation and the Fisher length of these polynomials. Second, in Subsection 4.3.2, the entropic moments $\left\langle [\rho_{n,\alpha}(x)]^k \right\rangle$ and the Rényi lengths are computed by use of two different approaches; one makes use of the Srivastava-Niukkanen linearization relation [150] of Laguerre polynomials, and another one which is based on the combinatorial multivariable Bell polynomials [128, 149]. Third, in Subsection 4.3.3, the asymptotics of the Shannon length is given and some sharp bounds to this measure are found. Then, in Subsection 4.3.4, all the four spreading measures are computationally discussed. Finally, some open problems and conclusions are given.

### 4.3.1 Ordinary moments, standard deviation and Fisher length

In this section the known values for the moments-around-the-origin $\left\langle x^k \right\rangle_{n,\alpha}$ ($k \in \mathbb{Z}$), the standard deviation $(\Delta x)_{n,\alpha}$ and the Fisher length $(\delta x)_{n,\alpha}$ of the Laguerre polynomials are given. Let us start writing the orthonormality relation

$$\int_0^\infty \widetilde{L}_n^{(\alpha)}(x) \widetilde{L}_m^{(\alpha)}(x) \omega_\alpha(x) dx = \delta_{nm},$$

for the orthonormal Laguerre polynomials

$$\tilde{L}_n^{(\alpha)}(x) = \left[ \frac{n!}{\Gamma(\alpha + n + 1)} \right]^{\frac{1}{2}} L_n^{(\alpha)}(x). \tag{4.45}$$

Then, the moment-around-the-origin of order $k \in \mathbb{Z}$ is defined, according to Eq. (4.38), by

$$\left\langle x^k \right\rangle_{n,\alpha} = \int_0^\infty x^k \rho_{n,\alpha}(x) dx = \int_0^\infty x^{k+\alpha} e^{-x} \left[ \widetilde{L}_n^{(\alpha)}(x) \right]^2 dx.$$

This integral can be calculated by different means; in particular, by use of the expression [37, 80, 151–153]

$$\int_0^\infty x^s e^{-x} L_n^{(\alpha)}(x) L_m^{(\beta)}(x) dx = \Gamma(s+1) \sum_{r=0}^{\min(n,m)} (-1)^{n+m} \binom{s-\alpha}{n-r} \binom{s-\beta}{m-r} \binom{s+r}{r},$$

one finds that

$$\left\langle x^k \right\rangle_{n,\alpha} = \frac{n!\Gamma(k+\alpha+1)}{\Gamma(n+\alpha+1)} \sum_{r=0}^{n} \binom{k}{n-r}^2 \binom{k+\alpha+r}{r},$$

where the binomial number is $\binom{a}{b} = \frac{\Gamma(a+1)}{\Gamma(b+1)\Gamma(a-b+1)}$. Then, taking into account Eq. (4.37) and the values $\left\langle x^k \right\rangle_{n,\alpha}$ for $k = 1, 2$, one has the following expression for the standard deviation of the Laguerre polynomials [105]

$$(\Delta x)_{n,\alpha} = \sqrt{2n^2 + 2(\alpha+1)n + \alpha + 1}. \tag{4.46}$$

The Fisher information of the Laguerre polynomials defined by Eq. (4.39) has been recently shown [104] to have the value

$$F\left(L_n^{(\alpha)}\right) = \begin{cases} 4n+1; & \alpha = 0, \\ \frac{(2n+1)\alpha+1}{\alpha^2-1}; & \alpha > 1, \\ \infty; & \alpha \in (-1,+1], \alpha \neq 0, \end{cases}$$

so that the Fisher length of these polynomials has, according to (4.41), the value

$$(\delta x)_{n,\alpha} = \begin{cases} \frac{1}{\sqrt{4n+1}}; & \alpha = 0, \\ \sqrt{\frac{\alpha^2-1}{(2n+1)\alpha+1}}; & \alpha > 1, \\ 0; & \alpha \in (-1,+1], \alpha \neq 0. \end{cases}$$

It is worth remarking that the inequality $(\delta x)_{n,\alpha} \leq (\Delta x)_{n,\alpha}$ is clearly satisfied.

### 4.3.2 Rényi lengths

In this section the Rényi lengths of the Laguerre polynomials $\mathcal{L}_q^R[\rho_{n,\alpha}]$ defined by Eq. (4.42) will be computed by two different approaches: an algebraic approach which is based on the Srivastava-Niukkanen linearization relation [150], and a combinatorial method which utilizes the multivariable Bell polynomials [128].

According to Eqs. (4.40) and (4.42), the Rényi length of order $q$ is given by

$$\mathcal{L}_q^R[\rho_{n,\alpha}] = \{W_q[\rho_{n,\alpha}]\}^{-\frac{1}{q-1}}, \quad q > 0, q \neq 1, \tag{4.47}$$

where

$$W_q[\rho_{n,\alpha}] := \left\langle [\rho_{n,\alpha}(x)]^{q-1} \right\rangle = \int_0^\infty [\rho_{n,\alpha}(x)]^q\, dx = \int_0^\infty \left[\widetilde{L}_n^\alpha(x)\right]^{2q} x^{q\alpha} e^{-qx} dx, \tag{4.48}$$

are the frequency or entropic moments of the Rakhmanov density (4.36) of the Laguerre polynomials. In spite of the efforts of numerous researchers [126, 132, 154–156], these quantities have not yet been calculated. Here we will compute them by use of two different approaches.

#### 4.3.2.1    Algebraic approach

To calculate the entropic moment $W_q[\rho_{n,\alpha}]$, we first use (4.45) and (4.48) to write

$$W_q[\rho_{n,\alpha}] := \left[\frac{n!}{\Gamma(\alpha+n+1)}\right]^q I_q\left[L_n^{(\alpha)}\right], \tag{4.49}$$

where

$$I_q\left[L_n^{(\alpha)}\right] := \int_0^\infty x^{\alpha q} e^{-qx} \left[L_n^{(\alpha)}(x)\right]^{2q} dx. \tag{4.50}$$

This functional of the orthogonal Laguerre polynomial can be calculated by use of the linearization formula of Srivastava-Niukkanen [150] for the products of various Laguerre polynomials given by

$$x^\mu L_{m_1}^{(\alpha_1)}(t_1 x) \cdots L_{m_r}^{(\alpha_r)}(t_r x) = \sum_{k=0}^\infty \Theta_k(\mu; t_1, \cdots, t_r) L_k^{(\beta)}(x),$$

where the coefficients $\Theta_k(\mu; x_1, \cdots, x_r)$ can be expressed as

$$\Theta_k(\mu; t_1, \cdots, t_r) = (\beta+1)_\mu \binom{m_1+\alpha_1}{m_1} \cdots \binom{m_r+\alpha_r}{m_r}$$
$$\times F_A^{(r+1)}[\beta+\mu+1, -m_1, \cdots, -m_r, -k; \alpha_1+1, \cdots, \alpha_r+1, \beta+1; t_1, \cdots, t_r, 1],$$

in terms of the Lauricella's hypergeometric functions of $(r+1)$ variables [157]. The Pochhammer symbol is $(a)_n = \frac{\Gamma(a+n)}{\Gamma(a)}$. This general relation with the values ($\beta = 0$, $\alpha_1 = \cdots = \alpha_r = \alpha$, $m_1 = \cdots = m_r = n$, $x = qt$, $t_1 = \cdots = t_r = \frac{1}{q}$, $\mu = \alpha q, r = 2q$) readily yields the following linearization result for the powers of Laguerre polynomials:

$$(qt)^{2q}\left[L_n^{(\alpha)}(t)\right]^{2q} = \sum_{k=0}^\infty \Theta_k\left(\alpha q; \frac{1}{q}, \cdots, \frac{1}{q}\right) L_k^{(0)}(qt), \tag{4.51}$$

where

$$\Theta_k\left(\alpha q; \frac{1}{q}, \cdots, \frac{1}{q}\right) = \Gamma(\alpha q+1) \binom{n+\alpha}{n}^{2q}$$
$$\times F_A^{(2q+1)}\left(\alpha q+1; -n, \cdots, -n; -k; \alpha+1 \cdots, \alpha+1, 1; \frac{1}{q}, \cdots, \frac{1}{q}, 1\right).$$

Taking into account (4.50), (4.51) and the orthogonality relation of the polynomials $L_n^{(\alpha)}(x)$, one finally has that the term with $k = 0$ is the only non-vanishing contribution

*Chapter 4 Information-theoretic lengths of orthogonal polynomials*        79to $I_q\left[L_n^{(\alpha)}\right]$, so that

$$I_q\left[L_n^{(\alpha)}\right] = \frac{1}{q^{\alpha q+1}}\Theta_0\left(\alpha q; \frac{1}{q}, \cdots, \frac{1}{q}\right), \qquad (4.52)$$

with

$$\Theta_0 = \Gamma(\alpha q+1)\binom{n+\alpha}{n}^{2q}$$
$$\times F_A^{(2q+1)}\left(\alpha q+1; -n, \cdots, -n; 0; \alpha+1, \cdots, \alpha+1, 1; \tfrac{1}{q}, \cdots, \tfrac{1}{q}, 1\right). \qquad (4.53)$$

Then, the entropic moments of the Laguerre polynomials have, according to Eqs. (4.49) and (4.52), the following expresion

$$W_q\left[\rho_{n,\alpha}\right] = \left[\frac{n!}{\Gamma(\alpha+n+1)}\right]^q \frac{1}{q^{\alpha q+1}}\Theta_0\left(\alpha q; \frac{1}{q}, \cdots, \frac{1}{q}\right). \qquad (4.54)$$

Finally, from Eqs. (4.47), (4.53) and (4.54) one has that the Rényi entropy of order $q$ of the Laguerre polynomials is given by

$$\begin{aligned}\mathcal{L}_q^R[\rho_{n,\alpha}] &= \left[\left(\frac{n!}{\Gamma(\alpha+n+1)}\right)^q \frac{1}{q^{\alpha q+1}}\Theta_0\left(\alpha q; \frac{1}{q}, \cdots, \frac{1}{q}\right)\right]^{-\frac{1}{q-1}}\\ &= \left[\left(\frac{n!}{\Gamma(\alpha+n+1)}\right)^q \frac{1}{q^{\alpha q+1}}\Gamma(\alpha q+1)\binom{n+\alpha}{n}^{2q}\right.\\ &\quad\left.\times F_A^{(2q+1)}\left(\alpha q+1, -n, \cdots, -n, 0; \alpha+1, ..., \alpha+1, 1; \frac{1}{q}, \cdots, \frac{1}{q}, 1\right)\right]^{-\frac{1}{q-1}},\end{aligned}$$

for every $q > 0$, $q \neq 1$. Some examples follow:

- For $n = 0$

$$\mathcal{L}_q^R[\rho_{0,\alpha}] = \left[\frac{1}{\Gamma(\alpha+1)^q}\frac{\Gamma(\alpha q+1)}{q^{\alpha q+1}}\right]^{-\frac{1}{q-1}}.$$

- For $n = 1$

$$\mathcal{L}_q^R[\rho_{1,\alpha}] = \left[\frac{\Gamma(\alpha q+1)(\alpha+1)^{2q}}{\Gamma(\alpha+2)^q q^{\alpha q+1}}(-(1+\alpha)q)^{-2q}U(-2q, -(2+\alpha)q, -(1+\alpha)q)\right]^{-\frac{1}{q-1}},$$

where $U(a,b,z)$ is a Tricomi confluent hypergeometric function [158].

#### 4.3.2.2   Combinatorial approach

In this approach we begin with the explicit expression of the Laguerre polynomials given by

$$\widetilde{L}_n^{(\alpha)}(x) = \sum_{k=0}^{n} c_k x^k,$$

with

$$c_k = \sqrt{\frac{\Gamma(n+\alpha+1)}{n!}} \frac{(-1)^k}{\Gamma(\alpha+k+1)} \binom{n}{k}, \tag{4.55}$$

Recently ([149]; see appendix B) it has been found that an integer power of a polynomial can be expressed by use of the multivariable Bell polynomials of Combinatorics [128]. This result applied to the Laguerre polynomials gives

$$\left[\widetilde{L}_n^{(\alpha)}(x)\right]^p = \sum_{k=0}^{np} \frac{p!}{(k+p)!} B_{k+p,p}(c_0, 2!c_1, ..., (k+1)!c_k) x^k, \tag{4.56}$$

with $c_i = 0$ for $i > n$, and the remaining coefficients are given by Eq. (4.55). Moreover, the Bell polynomials are given by

$$B_{m,l}(c_1, c_2, \ldots, c_{m-l+1}) = \sum_{\hat{\pi}(m,l)} \frac{m!}{j_1! j_2! \cdots j_{m-l+1}!} \left(\frac{c_1}{1!}\right)^{j_1} \left(\frac{c_2}{2!}\right)^{j_2} \cdots \left(\frac{c_{m-l+1}}{(m-l+1)!}\right)^{j_{m-l+1}},$$

where the sum runs over all partitions $\hat{\pi}(m,l)$ such that

$$j_1 + j_2 + \cdots + j_{m-l+1} = l, \quad \text{and} \quad j_1 + 2j_2 + \cdots + (m-l+1)j_{m-l+1} = m.$$

The replacement of expression (4.56) with $p = 2q$ into Eq. (4.48) yields the value

$$\begin{aligned}W_q[\rho_{n,\alpha}] &= \sum_{k=0}^{2nq} \frac{(2q)!}{(k+2q)!} B_{k+2q,2q}(c_0, 2!c_1, \ldots, (k+1)!c_k) \int_0^\infty x^{q\alpha} e^{-qx} x^k dx = \\ &= \sum_{j=0}^{2nq} \frac{\Gamma(\alpha q + k + 1)}{q^{\alpha q + k + 1}} \frac{(2q)!}{(k+2q)!} B_{k+2q,2q}(c_0, 2!c_1, ..., (k+1)!c_k),\end{aligned} \tag{4.57}$$

where the parameters $c_i$ are given by Eq. (4.55), keeping in mind that $c_i = 0$ for every $i > n$, so that the only non-vanishing terms correspond to those with $j_{i+1} = 0$ so that $c_i^{j_{i+1}} = 1$ for every $i > n$.

It is worthwhile to check that for $q = 1$ one has that

$$W_1[\rho_{n,\alpha}] = \int_0^\infty \rho_{n,\alpha}(x) dx = 1,$$

and that for $q = 2$ we have

$$W_2[\rho_{n,\alpha}] = \sum_{k=0}^{4n} \frac{\Gamma(2\alpha + k + 1)}{2^{2\alpha + k + 1}} \frac{24}{(k+4)!} B_{k+4,4}(c_0, 2!c_1, ..., (k+1)!c_k),$$

for the Onicescu information [8] of the Laguerre polynomials. Finally, from Eqs. (4.47) and (4.57) one has the following alternative expression for the $q$th-order Rényi length of the Laguerre polynomial for $q = 2, 3, \ldots$,

$$\mathcal{L}_q^R [\rho_{n,\alpha}] = \left( \sum_{j=0}^{2nq} \frac{\Gamma(\alpha q + k + 1)}{q^{\alpha q + k + 1}} \frac{(2q)!}{(k+2q)!} B_{k+2q,2q} \left( c_0, 2!c_1, \ldots, (k+1)!c_k \right) \right)^{-\frac{1}{q-1}},$$

which for $q = 2$ yields the value

$$\mathcal{L}_2^R [\rho_{n,\alpha}] = \left[ \sum_{n=0}^{4n} \frac{\Gamma(2\alpha + k + 1)}{2^{2\alpha + k + 1}} \frac{24}{(k+4)!} B_{k+4,4} \left( c_0, 2!c_1, \ldots, (k+1)!c_k \right) \right]^{-1},$$

for the Onicescu or second-order Rényi length [8] of the Laguerre polynomials.

With these expressions we obtain the same values of $\mathcal{L}_q^R[\rho_{0,\alpha}]$ and $\mathcal{L}_q^R[\rho_{1,\alpha}]$ as in the previous subsection.

### 4.3.3 Shannon length: Asymptotics and sharp bounds

The goal of this Subsection is twofold. First, to study the asymptotics of the Shannon spreading length $N[\rho_{n,\alpha}]$ of the orthonormal Laguerre polynomials $\widetilde{L}_n^{(\alpha)}$ and its relation to the standard deviation $(\Delta x)_{n,\alpha}$. Second, to find sharp upper bounds to $N[\rho_{n,\alpha}]$ by use of an information-theoretic optimization procedure.

Although many results have been recently published in the literature (see e.g. [159, 160]) about the asymptotics of the Laguerre polynomials themselves, they have not yet been successfully used to obtain the asymptotics of functionals of these mathematical functions beyond the $L_p$-norm method of Aptekarev et al [44, 99]. Here we use the results provided by this method to fix the asymptotics of the Shannon length of these polynomials and its relation to the standard deviation. From Eq. (4.43) we have that

$$N[\rho_{n,\alpha}] = \exp(S[\rho_{n,\alpha}]), \tag{4.58}$$

where

$$S[\rho_{n,\alpha}] := -\int_0^\infty \omega_\alpha(x) \left[ \widetilde{L}_n^{(\alpha)}(x) \right]^2 \log \left\{ \omega_\alpha(x) \left[ \widetilde{L}_n^{(\alpha)}(x) \right]^2 \right\} dx = E_n \left[ \widetilde{L}_n^{(\alpha)} \right] + J_n \left[ \widetilde{L}_n^{(\alpha)} \right],$$

with the following entropic functionals [24, 161]

$$E_n \left[ \widetilde{L}_n^{(\alpha)} \right] = -\int_0^\infty \omega_\alpha(x) \left[ \widetilde{L}_n^{(\alpha)}(x) \right]^2 \log \left[ \widetilde{L}_n^{(\alpha)}(x) \right]^2 dx, \tag{4.59}$$

and

$$J_n \left[ \widetilde{L}_n^{(\alpha)} \right] = -\int_0^\infty \omega_\alpha(x) \left[ \widetilde{L}_n^{(\alpha)}(x) \right]^2 \log \omega_\alpha(x) dx = 2n + \alpha + 1 - \alpha \psi(\alpha + n + 1). \tag{4.60}$$

Moreover, the use to the $L_p$-norm method of Aptekarev et al [99] has permitted to find [24] the following values for the asymptotics of $E_n\left[\widetilde{L}_n^{(\alpha)}\right]$

$$E_n\left[\widetilde{L}_n^{(\alpha)}\right] = -2n + (\alpha+1)\log(n) - \alpha - 2 + \log(2\pi) + o(1). \tag{4.61}$$

Then, according to Eqs. (4.59), (4.60) and (4.61), one has that the asymptotical behaviour

$$S[\rho_{n,\alpha}] = (\alpha+1)\log(n) - \alpha\psi(\alpha+n+1) - 1 + \log(2\pi) + o(1),$$

for the Shannon entropy, and

$$N[\rho_{n,\alpha}] \simeq 2\pi n^{\alpha+1} e^{-\alpha\psi(\alpha+n+1)-1}, \tag{4.62}$$

for the Shannon length of the orthonormal Laguerre polynomials. Moreover, from Eqs. (4.46) and (4.62) one finds that

$$N[\rho_{n,\alpha}] \simeq \frac{\pi\sqrt{2}}{e}(\Delta x)_{n,\alpha}; \qquad n \gg 1, \tag{4.63}$$

between the asymptotical values of the Shannon length and the standard deviation of the polynomial $\widetilde{L}_n^{(\alpha)}$. It is worth noting that this relation fulfils the general inequality (4.44) which mutually relates the Shannon length and the standard deviation for general densities. Moreover, the relation (4.63) for the Rakhmanov densities of Laguerre polynomials is also satisfied by the Rakhmanov densities of Hermite (See Eq. 4.26) [130, 149] and Jacobi [130, 147] polynomials.

Let us now find sharp upper bounds to the Shannon length $N[\rho_{n,\alpha}]$ by taking into account the non-negativity of the relative Shannon entropy (also called Kullback-Leibler entropy) of two arbitrary probability densities $\rho(x)$ and $f(x)$. This yields that the Shannon entropy of $\rho(x)$ is bounded from above by means of

$$S[\rho_{n,\alpha}] \leq -\int_0^\infty \rho(x)\log f(x)dx.$$

For $\rho(x) = \rho_{n,\alpha}(x)$ this expression produces an infinite set of upper bounds to the Shannon entropy of Laguerre polynomials. The choice of $f(x)$ in the form

$$f(x) = \frac{ba^{\frac{1+m}{b}}}{\Gamma\left(\frac{1+m}{b}\right)}x^m e^{-ax^b}; \; m > -1, a > 0, b \in \mathbb{N}^+, 0 \leq x < \infty, \tag{4.64}$$

(which is normalized to unity) followed by the optimization of the upper bound with respect to the parameter $a$ gives rise to

$$S[\rho_{n,\alpha}] \leq \log\frac{\Gamma(\beta)e^\beta}{b\beta^\beta}\langle x^b\rangle^\beta - m\langle\log x\rangle; \; b > 0, \; m > -1, \; \beta = \frac{1+m}{b}, \tag{4.65}$$

following the lines of Refs [28, 162]. Then, according to Eqs. (4.58) and (4.65), we have the following set of infinite sharp bounds

$$N\left[\rho_{n,\alpha}\right] \leq \frac{\Gamma\left(\beta\right)e^{\beta}}{b\beta^{\beta}}\left\langle x^{b}\right\rangle^{\beta}e^{-m\langle\log x\rangle};\ m>-1,\ b>0, \qquad (4.66)$$

for the Shannon length of the Laguerre polynomial $L_n^{(\alpha)}(x)$. For $m = 0$ we have the upper bound

$$N\left[\rho_{n,\alpha}\right] \leq \frac{\Gamma\left(\frac{1}{b}\right)(be)^{\frac{1}{b}}}{b}\left\langle x^{b}\right\rangle^{\frac{1}{b}},\ b>0. \qquad (4.67)$$

This bound is particularly interesting because it only depends on the expectation value $\left\langle x^{b}\right\rangle$; the expectation value $\langle \log x \rangle$ is, at times, unavailable or difficult to evaluate.

### 4.3.4 Some computational issues

In this section we study various computational issues of the direct spreading measures of Laguerre polynomials. It is worth pointing out that there is no stable numerical algorithm for the computation of the Rényi and Shannon lengths of these polynomials in contrast to the case of orthogonal polynomials on a finite interval for which an efficient algorithm based on the three-term recurrence relation has been recently found by Buyarov et al [23]. Moreover, a naive numerical evaluation of these Laguerre functionals by means of quadratures is not often convenient except for the lowest-order polynomials since the increasing number of integrable singularities spoils any attempt to achieve a reasonable accuracy for arbitrary $n$. Here we carry out the following numerical study. First, we examine the numerical accuracy of the optimal bounds (4.66) and (4.67) to the Shannon length of the Laguerre polynomials $L_n^{(\alpha)}(x)$, $\alpha$ fixed, for various degrees $n$ by taking into account the optimal values of the parameter $b$ in (4.67) and the optimal values of $(b, m)$ in (4.66). Second, we study the mutual comparison of Fisher, Shannon and Onicescu lengths and the standard deviation of $L_n^{(\alpha)}(x)$ for fixed $\alpha$ and various degrees $n$. Finally, we discuss the correlation of the Shannon length $N(L_n^{(\alpha)})$ and the standard deviation $(\Delta x)_n$ for various pairs $(n, \alpha)$, which allow us to find, at times, linear relations between their components.

In Figure 4.5 it is numerically studied the accuracy of the bounds (4.66) and (4.67) to the Shannon length of the Laguerre polynomial $L_n^{(\alpha=0)}(x)$ given by the optimal values $(b_{\text{opt}}, 0)$ in (4.67). This is done by comparing the corresponding optimal bounds with the "numerically exact" value of the lengths $N\left[\rho_{n,0}\right]$ for the polynomials with degree $n$ from 0 to 10. The graph on the right of the figure gives the relative ratio of the bound given by Eq. (4.67) with $b = b_{\text{opt}}$, and the ratio of the bound given by Eq. (4.66) with $b = b_{\text{opt}}$ and $m = m_{\text{opt}}$. Notice that the latter bound is always better than the former, as we have the parameter $m$ to adjust. The values of optimal pairs $(b_{\text{opt}}, 0)$ and $(b_{\text{opt}}, m_{\text{opt}})$ are shown in Tables 4.2 and 4.3, respectively. Note that the optimum value $b_{\text{opt}}$ is different when considering the bound (4.66) or (4.67). Also notice that the optimum values for $n = 0$ are $(b_{\text{opt}}, m_{\text{opt}}) = (1, 0)$, where the density $f(x)$ defined in (4.64) equals the Rakhmanov density for $n = 0$. Remark that the best bounds are obtained for expectation values $\langle x^b \rangle$ where $b = b_{\text{opt}}$ is an increasing function of the degree $n$ of the polynomial in both cases; this is directly connected with the larger spreading of the polynomial when its degree has higher values.

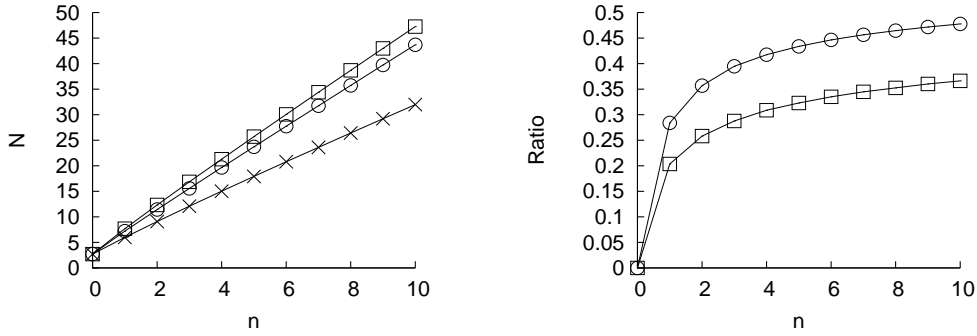

FIGURE 4.5: Left: Shannon length ($\times$), upper bound with $m = 0$ and $b = b_{\text{opt}}$ ($\odot$), and upper bound with $m = m_{\text{opt}}$ and $b = b_{\text{opt}}$ ($\boxdot$) of the Laguerre polynomials $L_n^{(0)}(x)$, as a function of the degree $n$. Right: Relative ratios of the bounds with $m = 0$ and $b = b_{\text{opt}}$ ($\odot$), and with $m = m_{\text{opt}}$ and $b = b_{\text{opt}}$ ($\boxdot$), as a function of $n$.

| $n$ | 0 | 1 | 2 | 3 | 4 | 5 | 6 | 7 | 8 | 9 | 10 |
|---|---|---|---|---|---|---|---|---|---|---|---|
| $b_{\text{opt}}$ | 1 | 3 | 4 | 6 | 7 | 8 | 9 | 10 | 11 | 12 | 13 |

TABLE 4.2: Values $b_{\text{opt}}$ of the parameter $b$ which yield the best (i.e. lowest) upper bounds (4.67) to the Laguerre polynomial $L_n^{(0)}(x)$ for various degrees $n$.

| $n$ | 0 | 1 | 2 | 3 | 4 | 5 | 6 | 7 | 8 | 9 | 10 |
|---|---|---|---|---|---|---|---|---|---|---|---|
| $b_{\text{opt}}$ | 1 | 4 | 6 | 7 | 9 | 10 | 11 | 12 | 14 | 15 | 16 |
| $m_{\text{opt}}$ | 0 | -0.332 | -0.338 | -0.322 | -0.332 | -0.327 | -0.324 | -0.321 | -0.322 | -0.320 | -0.319 |

TABLE 4.3: Values $(b_{\text{opt}}, m_{\text{opt}})$ of the parameters $(b, m)$ which yield the best (i.e. lowest) upper bound (4.66) to the Laguerre polynomial $L_n^{(0)}(x)$ for various degrees $n$.

To study the behaviour of the accuracy of the two previous bounds with respect to $\alpha$, we have done in Figure 4.6 a study of the Shannon lengths $N[\rho_{n,5}]$ similar to that done in Figure 4.5 for $L_n^{(0)}(x)$. The corresponding values $(b_{\text{opt}}, 0)$ and $(b_{\text{opt}}, m_{\text{opt}})$ are given in Tables 4.4 and 4.5 respectively. The two graphs of the figure show qualitatively similar and quantitatively better results than those found in Figure 4.5.

In Figures 4.7 and 4.8 we study the mutual comparison of various direct spreading measures (namely, the standard deviation $\Delta x$ and the Fisher, Shannon and the Onicescu or second-order Rényi lengths) of the Laguerre polynomials $L_n^{(0)}(x)$ and $L_n^{(5)}(x)$, respectively, when the degree $n$ varies from 0 to 10. Several observations are in order.

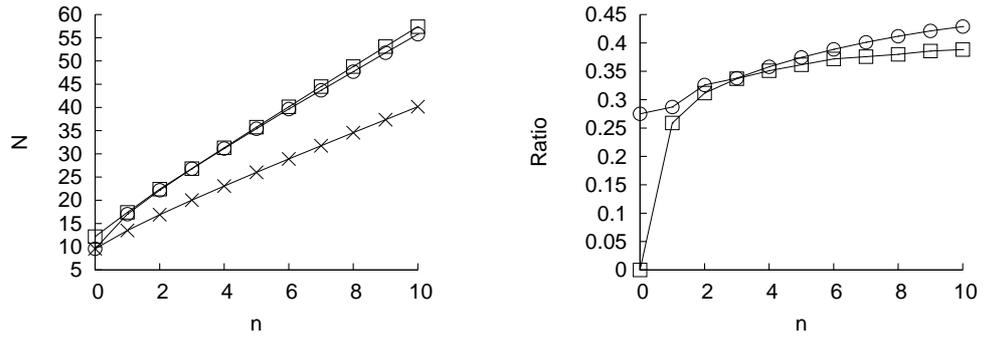

FIGURE 4.6: Left: Shannon length ($\times$), upper bound with $m = 0$ and $b = b_{\mathrm{opt}}$ ($\odot$), and upper bound with $m = m_{\mathrm{opt}}$ and $b = b_{\mathrm{opt}}$ ($\boxdot$) of the Laguerre polynomials $L_n^{(5)}(x)$, as a function of the degree $n$. Right: Relative ratios of the bounds with $m = 0$ and $b = b_{\mathrm{opt}}$ ($\odot$), and with $m = m_{\mathrm{opt}}$ and $b = b_{\mathrm{opt}}$ ($\boxdot$), as a function of $n$.

| $n$ | 0 | 1 | 2 | 3 | 4 | 5 | 6 | 7 | 8 | 9 | 10 |
|---|---|---|---|---|---|---|---|---|---|---|---|
| $b_{\mathrm{opt}}$ | 5 | 6 | 7 | 8 | 10 | 11 | 12 | 13 | 14 | 15 | 16 |

TABLE 4.4: Values $b_{\mathrm{opt}}$ of the parameter $b$ which yield the best (i.e. lowest) upper bounds (4.67) to the Laguerre polynomial $L_n^{(5)}(x)$ for various degrees $n$.

| $n$ | 0 | 1 | 2 | 3 | 4 | 5 | 6 | 7 | 8 | 9 | 10 |
|---|---|---|---|---|---|---|---|---|---|---|---|
| $b_{\mathrm{opt}}$ | 1 | 5 | 7 | 9 | 10 | 11 | 13 | 14 | 15 | 16 | 17 |
| $m_{\mathrm{opt}}$ | 5 | 0.288 | 0.053 | -0.049 | -0.098 | -0.131 | -0.160 | -0.177 | -0.190 | -0.201 | -0.210 |

TABLE 4.5: Values $(b_{\mathrm{opt}}, m_{\mathrm{opt}})$ of the parameters $(b, m)$ which yield the best (i.e. lowest) upper bound (4.66) to the Laguerre polynomial $L_n^{(5)}(x)$ for various degrees $n$.

First, all the measures with global character (standard deviation, Shannon and Rényi lengths) grow linearly or quasilinearly when the degree of the polynomial is increasing; essentially because the polynomial spreads more and more. Moreover, they behave so that $\Delta x < \mathcal{L}_2 < N$. Second, the (local) Fisher length decreases when the degree $n$ is increasing; essentially, because the polynomial becomes more and more oscillatory, so growing its gradient content. Third, the Fisher length has always a value smaller than all the global spreading measures.

Finally, in Figure 4.9 we have numerically studied the connection of the Shannon length $N[\rho_{n,\alpha}]$ and the standard deviation $(\Delta x)_{n,\alpha}$ of the Laguerre polynomials $L_n^{(\alpha)}(x)$, with $\alpha = 0$ and $5$, when the degree $n$ varies from 0 to 20. We obtain the fits

$$N[\rho_{n,0}] = 1.9144(\Delta x)_n + 3.611, \tag{4.68}$$

and

$$N[\rho_{n,5}] = 1.8951(\Delta x)_n + 5.966, \tag{4.69}$$

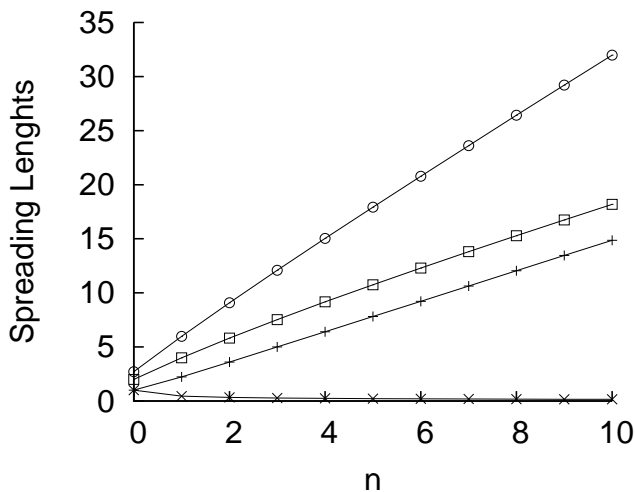

FIGURE 4.7: Standard deviation $\Delta x$ (+), Fisher length $\delta x$ (×), Onicescu length $\mathcal{L}_2$ (□), and Shannon length $N$ (⊙) of the Laguerre polynomial $L_n^{(0)}(x)$ as a function of $n$.

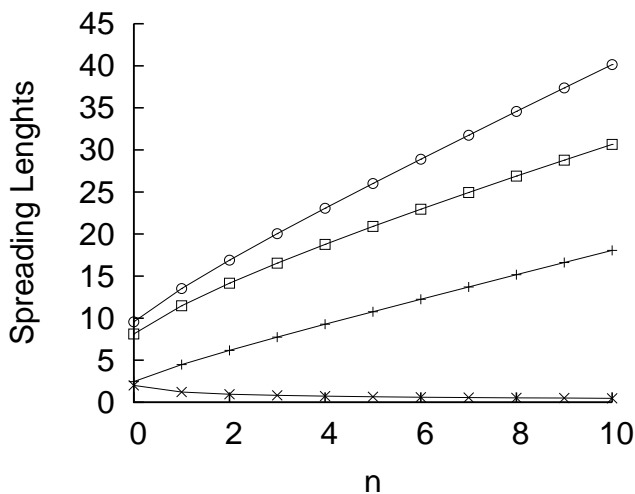

FIGURE 4.8: Standard deviation $\Delta x$ (+), Fisher length $\delta x$ (×), Onicescu length $\mathcal{L}_2$ (□), and Shannon length $N$ (⊙) of the Laguerre polynomial $L_n^{(5)}(x)$ as a function of $n$.

for data with $n \geq 10$, and correlation coefficient $R = 0.9999$ in both cases. This apparent quasilinear behaviour of the Shannon length with respect to the standard deviation is, in fact, slightly concave, being linear only asymptotically in accordance to the rigorous expression (4.63), i.e. $N\left[\widetilde{L}_n^{(\alpha)}\right] \simeq 1.63445 \, (\Delta x)_{n,\alpha}$ for $n >> 1$, which is very close to (4.68) and (4.69).

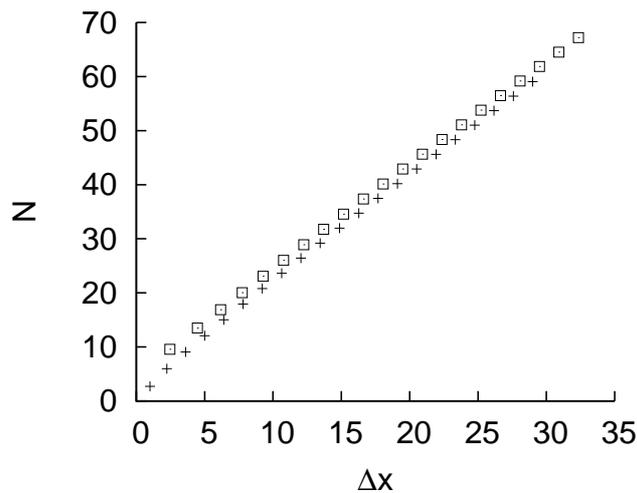

FIGURE 4.9: Shannon length $N$ as a function of the standard deviation $\Delta x$ for the Laguerre polynomials $L_n^{(0)}(x)$ (+) and $L_n^{(5)}(x)$ ($\square$), when the degree $n$ varies from 0 to 20.

## 4.4 Conclusions

In this Chapter we have introduced new direct measures of orthogonal-polynomials spreading other than the root-mean-square or standard deviation $\Delta x$ which have an information-theoretic origin; namely, the Rényi, Shannon and Fisher length. They share with $\Delta x$ various interesting properties: same units as the variable, invariance under translations and reflections, linear scaling, vanishing in the limit that the variable tends towards a given definite value, and global character. In contrast with $\Delta x$, they do not depend on any particular point of the orthogonality interval what allow them to be considered as proper spreading lengths. The Rényi and Shannon lengths are powerlike and logarithmic functionals of the polynomial $p_n(x)$ while Fisher length is a gradient functional of it, what allows one to state that the former lengths are global measures and the latter one has a locality property.

In Section 4.2 we have developed a computational methodology for the Rényi length of the orthonormal Hermite polynomials $\tilde{H}_n(x)$ which is based on the combinatorial multi-variable Bell polynomials whose arguments are controlled by the expansion coefficients of $\tilde{H}_n(x)$. For the Shannon length of these polynomials, since it cannot be calculated explicitly, we give (i) its asymptotics and the relation with the standard deviation, and (ii) sharp upper bounds by means of an information-theoretic-based optimization procedure. Moreover, it is computationally found that the second-order Rényi length (also called Onicescu-Heller length) and the Shannon length linearly depend on the standard deviation of the polynomials. On the other hand, the Fisher length is analytically shown to have a reciprocal behaviour with $\Delta x$, mainly because of its local character.

The previous results have been applied to the quantum-mechanical harmonic oscillator and some open problems related to the asymptotics of the frequency or entropic moments of Hermite polynomials are posed. Their solution would allow not only to complete this work but also to extend considerably the related findings of various authors in connection to the $L^p$-norms of Hermite polynomials [98, 99, 125, 129, 131].

In Section 4.3 the direct spreading measures of the Laguerre polynomials $L_n^{(\alpha)}(x)$ are analytically and numerically studied. Beyond the ordinary moments $\langle x^k \rangle$, $k \in \mathbb{Z}$, and the standard deviation $(\Delta x)_n$, which have been explicitly given in terms of $(n, \alpha)$, we have developed two theoretical approaches of algebraic and combinatorial types to obtain two equivalent analytical expressions for the Rényi lengths of arbitrary order. For the Shannon length, whose explicit value is not yet known (in fact, its calculation is a formidable task!), we have found sharp bounds in terms of the expectation value $\langle x^k \rangle$ and/or the logarithmic expectation value $\langle \log x \rangle$ by means of an information-theoretic-based optimization procedure.

Moreover, the linear correlation of the Shannon length and the standard deviation for the Laguerre polynomials $L_n^{(\alpha)}(x)$ with large degree $n$ is underlined. In fact, the correlation factor is not only independent on the parameter $\alpha$ but, most importantly, it is the same as for the remaining hypergeometrical families on a finite interval (Jacobi polynomials) [130, 147] or on the whole real line (Hermite polynomials) [130] (see section below).

Then we carried out a numerical study of the four direct spreading measures of Laguerre polynomials. Let us remark, among other results, that the Fisher length has the smallest value, and the Shannon length depend quasilinearly on the standard deviation.

Finally, let us highlight a number of open information-theoretic problems related to Laguerre polynomials: (i) to find the asymptotics of the entropic moments and, subsequently, the Rényi lengths in the spirit of [44, 99], (ii) to identify the most general class of polynomials for which the asymptotical relation (4.63) of the Shannon length and the standard deviation is fulfilled, and (iii) to characterize the most general class of polynomials for which the ratio between these two direct spreading measures is a constant (i.e., it does not depend on the degree nor the parameters of the polynomials) as already pointed out in [130].

# Chapter 5

# Quantum learning

Learning can be defined as the changes in a system that result in an improved performance over time on tasks that are similar to those performed in the system's previous history. Although learning is often thought of as a property associated with living things, machines or computers are also able to modify their own algorithms as a result of training experiences. This is the main subject of the broad field of "machine learning". Recent progress in quantum communication and quantum computation [163] – development of novel and efficient ways to process information on the basis of laws of quantum theory – provides motivations to generalize the theory of machine learning into the quantum domain [164]. For example, quantum learning algorithms have been developed for extracting information from a "black-box" oracle for an unknown Boolean function [165, 166].

The main ingredient of the quantum machine is a feed-back system that is capable of modifying its initial quantum algorithm in response to interaction with a "teacher" such that it yields better approximations to the intended quantum algorithm. In the literature there have been intensive and extensive studies by employing feed-back systems. They include quantum neural networks [167], estimation of quantum states [168], and automatic engineering of quantum states of molecules or light with a genetic algorithm [169–171]. Quantum neural networks deal with many-body quantum systems and refer to the class of neural network models which explicitly use concepts from quantum computing to simulate biological neural networks [172]. Standard state-engineering schemes optimize unitary transformations to produce a given target quantum state. The present approach of quantum automatic control contrasts with these methods. Instead of quantum state it optimizes *quantum operations (e.g. unitary transformations) to perform a given quantum information task*. It is also different than the problems studied in Ref. [165, 166], where one does not learn a task but rather a specific property of a black-box oracle.

An interesting question arises in this context: (1) Can a quantum machine learn to perform a given quantum algorithm? This question has been answered affirmative for special tasks, such as quantum pattern recognition [173], matching of unknown quantum states [174], and for learning quantum computational algorithms such as the Deutch algorithm [175], the Grover search algorithm and the discrete Fourier transform [176].

Another interesting question is: (2) Can one have quantum improvements in the speed of learning in a sense that a quantum machine requires *fewer* steps than the best classical machine to learn some *classical* task? By "classical task" we mean an operation or a function which has classical input and classical output. Quantum machines such as quantum state discriminator, universal quantum cloner or programmable quantum processor [177] do not fall into this category. Quantum computational algorithms do perform classical tasks, but no investigation has been undertaken to compare speed of learning of these algorithms with that of their classical counterparts. To our knowledge the question (2) is still open thus far.

In this Chapter we will give evidence for the first explicit classical computational task that quantum machines can learn *faster* than their classical counterparts. In both cases certain set of independent parameters must be optimized to learn the task. We will show that the fraction of the space of parameters, which correspond to (approximate) successful completion of the task, is exponentially smaller for the classical machine than for the quantum one. This analytical results supports our numerical simulation showing that quantum machine learns faster than the classical one.

## 5.1 The speed of quantum and classical learning

We first define a family of problems of our interest: let $m$-th member ($m \in \mathbb{N}$) of this family be the $k$-th root of NOT with $k = 2^m$, where the roots of NOT are defined as follows:

**Definition 1.** The operation is *$k$-th root of NOT* if, when applied subsequently $nk$ times on the Boolean input of 0 or 1, it returns the input for even $n$'s and its negation for odd $n$'s. We denote this operation with $\sqrt[k]{\text{NOT}}$. (Remark: With this definition we want to discard the cases for which, for example, the operation returns $k$-th root of NOT when performed once, but does not return identity when performed twice.)

The machine that performs this operation takes one input bit and returns one output bit. This bit will be called "target bit". In general, however, the machine could use many more auxiliary bits that might help the performance. Specifically, in the classical case the input $\vec{i}$ and output $\vec{j}$ are vectors with binary components. Any operation is defined by a probability distribution $p(\vec{i}, \vec{j})$ which gives the probability that the machine will generate the output $\vec{j}$ from the input $\vec{i}$. Thus, one has $\sum_{\vec{j}} p(\vec{i},\vec{j}) = 1$. The readout of the target bit is a map: $\vec{j} \to \{0,1\}$. Without loss of generality we assume that the target bit is the first component of the input and the output vector. The remaining components are auxiliary bits which play the role of the machine's memory.

In quantum case no auxiliary (qu)bits are necessary as only one qubit is enough to implement any $\sqrt[k]{\text{NOT}}$. The input of the machine is a single qubit and the machine itself is a unitary transformation. The input state will be either $|0\rangle$ or $|1\rangle$ corresponding to the Boolean values of classical bits "0" and "1", respectively. The readout procedure is the measurement in the computational basis $\{|0\rangle, |1\rangle\}$ and we consider the state that the qubit is projected to as the output of the machine.

In both cases the term learning is used for the process of approximating the function $\sqrt[k]{\text{NOT}}$ to which we will refer as the target function. We will consider that learning has been accomplished when the learning machine returns with high probability correct outputs for both inputs. Then a learning process is reduced to approximating the target function in a sequence of taking the inputs, performing transformations on the inputs, returning the outputs, estimating the fidelity between the actual outputs and the ones that the target function would have produced and correspondingly of making adjustments to the transformations. The schematic diagrams depicting both types of machines are shown in the Fig. 5.1. Now we will describe the learning in both cases in more detail.

*Quantum learning*: In every learning trial the following steps are performed:

1. Select a new unitary operator $U$ using a Gaussian random walk (The first $U$ is initialized randomly using the Haar measure).

2. Run the unitary $U^k$ on an input qubit state chosen to be $|0\rangle$ or $|1\rangle$ with equal probability. Measure the output qubit in the computation basis. Repeat this on $M$ input states and store the results (classical bits). The number $M$ defines the size of teachers (classical) memory of the quantum machine.

3. Estimate how close is the actual operation to the target one. To achieve this count the number of times the operation is successful in approximating the target function (i.e. it produces $|1\rangle$, when the input was in $|0\rangle$, and it produces $|0\rangle$, when it was in $|1\rangle$). The number of successes is denoted by $new_s$ and $old_s$ in the executed and the previous trial, respectively.

4. If $new_s \geq old_s$, go to 1 with the current unitary operator as the center of the Gaussian; Otherwise, go to 1 with the unitary operator chosen in the previous trail as the center of the Gaussian.

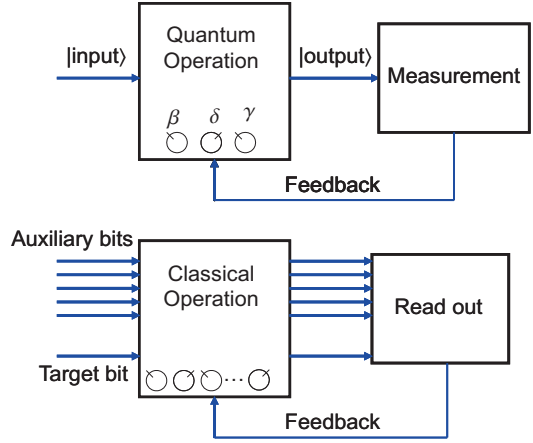

FIGURE 5.1: Diagram of classical and quantum learning machines. The learning procedure consists in a sequence of taking the inputs, performing transformations on them, returning the outputs, estimating the figure of merit between the outputs obtained and the expected ones and correspondingly making adjustments to the transformations. For the task of extracting the $k$-th root of NOT (see text for definition) the dimension of the space of parameters for a classical machine is log 2k larger than that for a quantum machine.

Any single qubit rotation can be parametrized by Euler's angles as follows:

$$U = e^{i\alpha} \begin{pmatrix} e^{-i\left(\frac{\beta}{2}+\frac{\delta}{2}\right)} \cos\left(\frac{\gamma}{2}\right) & -e^{i\left(-\frac{\beta}{2}+\frac{\delta}{2}\right)} \sin\left(\frac{\gamma}{2}\right) \\ e^{i\left(\frac{\beta}{2}-\frac{\delta}{2}\right)} \sin\left(\frac{\gamma}{2}\right) & e^{i\left(\frac{\beta}{2}+\frac{\delta}{2}\right)} \cos\left(\frac{\gamma}{2}\right) \end{pmatrix}. \quad (5.1)$$

Since the global phase $\alpha$ is irrelevant for the present application, we are left with the parameters $\delta \in [0, 2\pi]$, $\beta \in [0, 2\pi]$, $\gamma \in [0, \pi]$. In every new learning trial these parameters will be selected independently with a normal probability distribution entered around the values from the previous run and the widths of the Gaussians are taken as free parameters of the simulation. There are two free parameters of the learning procedure: $\sigma_\gamma$ and $\sigma_\beta$ ($\sigma_\delta = \sigma_\beta$). In all simulations these parameters are optimized to minimize the number of learning steps.

Note that if quantum machine performs the task for $n = 1$ perfectly, then it will also perform the task perfectly for all $n$. This is why our quantum machine is trained only to learn the task for $n = 1$. Nevertheless, after the learning has been completed one should compare how close the performance of the learning machine is to this of the target operation for all $n$. We define a set of figures of merit $\{P^n\}_{n=1}^{\infty}$ as follows:

$$\begin{aligned}
P^1 &= \frac{1}{2}(|\langle 0|U^k|1\rangle|^2 + |\langle 1|U^k|0\rangle|^2) \\
P^2 &= \frac{1}{4}(|\langle 0|U^k|1\rangle|^2 + |\langle 1|U^k|0\rangle|^2 + |\langle 0|U^{2k}|0\rangle|^2 + \langle 1|U^{2k}|1\rangle|^2) \quad (5.2) \\
P^n &= \frac{1}{2n}(|\langle 0|U^k|1\rangle|^2 + |\langle 1|U^k|0\rangle|^2 + \cdots + |\langle 0|U^{nk}|b\rangle|^2 + |\langle 1|U^{nk}|b\oplus 1\rangle|^2),
\end{aligned}$$

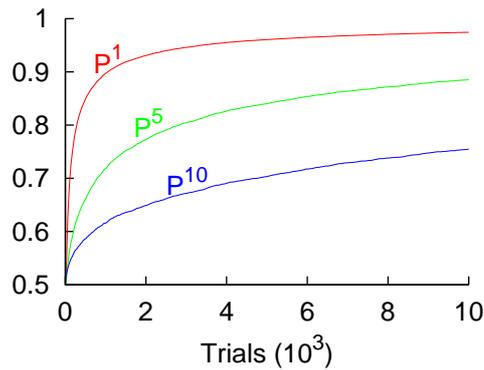

FIGURE 5.2: Quantum learning for performing the 4-th root of NOT. Different figures of merit $P^s$ ($s = 1, 5, 10$) as a function of the number of learning trials (x $10^3$). The size of teachers memory $M$ is varied to achieve the maximal value of the figures of merits for a given number of trials. The free parameters have the values $\sigma_\gamma = \frac{\pi}{4}$ and $\sigma_\alpha = \sigma_\beta = \frac{\pi}{8}$.

where $b = 0$ if $n$ is even, and $b = 1$ if $n$ is odd, and $\oplus$ denotes sum modulo 2. Note that each subsequent $P^n$ is more demanding in the sense that more constraints from the definition of the $\sqrt[k]{\text{NOT}}$ are being taken into account. This is reflected by the results, which are presented in Fig. 5.2.

The memory size of the teacher $M$ is another free parameter of the quantum machine. The learning ability has a very strong dependence on $M$ as can be seen from Fig. 5.3. For lower values of $M$ the learning is faster at the beginning (up to about 4x$10^4$ trials), before it slows down and saturates. At the saturation the size of the memory does not allow distinguishing between sufficiently "good" operations all for which $new_s = M$. For higher $M$ values the learning is slower, but it reaches higher fidelities. To combine the high speed with the high fidelity of learning we apply the learning procedure with variable $M$: The machine starts with $M = 1$ and whenever it obtains the number of successes $new_s = M$ it increments $M$ by one. With this kind of algorithm the learning has one less free parameter. All our simulations were done for variable $M$, unless stated otherwise.

Next we describe the classical learning procedure.

*Classical learning*: The classical learning is an iterative process of finding the optimal probability distribution $p(\vec{i}, \vec{j})$ for the classical machine to extract the $\sqrt[k]{\text{NOT}}$. The speed of learning depends on the number $N^2 - N$ of independent parameters (independent probabilities $p(\vec{i}, \vec{j})$), where $N = 2^{dim(i)}$ and $dim(i)$ is the dimension of the input $\vec{i}$ and the output vector $\vec{j}$. We will refer to $N$ as the memory size of the classical learning machines because it is equal to the total number of distinguishable internal states of the machine. To minimize the number of learning trials required to complete the learning and thus to maximize the speed of learning, we are interested in the minimal number of internal states $N$ for which it is possible to construct a classical machine that is able to extract $\sqrt[k]{\text{NOT}}$.

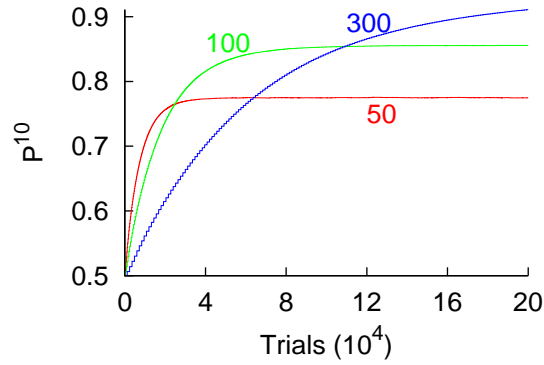

FIGURE 5.3: Quantum learning for performing the 4-th root of NOT. Figure of merit $P^{10}$ as a function of the number of learning trials (x $10^4$) for different sizes of teachers memory $M$ (blue = 300, green =100, red=50). The free parameters have the values $\sigma_\gamma = \frac{\pi}{4}$ and $\sigma_\alpha = \sigma_\beta = \frac{\pi}{8}$

**Lemma**: Any classical machine that performs $k$-th root of NOT perfectly must have at least $2k$ internal states if $k = 2^m$ and $m \in \mathbb{N}$.

**Proof.** Each probabilistic classical machine can be considered as a convex combination of deterministic ones. If it performs some task perfectly, then there must also be deterministic machine that does the same. This means that we can restrict ourselves in this proof only to deterministic machines without any loss of generality. Any (deterministic, classical) machine can be represented as an oriented graph, with vertices corresponding to the internal states. Edge pointing from vertex $\vec{i}$ to $\vec{j}$ will mean that the operation on input $\vec{i}$ generates the output $\vec{j}$. Any (finite) machine must have at least one loop and, if the machine is run subsequently a large number of times, it will eventually end up in that loop. Since the definition of the task involves arbitrary large $n$'s we may start our analysis from $n$ large enough such that the machine is already in the loop. Since we will prove the lemma by giving constraint from below on the size of the loop, we may assume that the whole graph is a one loop and each vertex is a part of it.

Let the length of the loop be $N$. Let $g$ be the greatest common divisor $g = GCD(k, N)$. Then there exist numbers $x$ and $y$ such that

$$k = gx \quad N = gy \quad GCD(x, y) = 1 \tag{5.3}$$

If the machine is initially in a vertex that corresponds to input "1" of the target bit and we apply the operation $Nk$ times we will always end up in the same vertex "1", since $Nk = 0 \mod N$. Since, however, the task is defined such that for $N$ odd the ending vertex should correspond to "0" value for the target bit, one concludes that $N$ must be even. Therefore, we can write $N$ as $N = 2^K c$, where $c$ is odd and $K \geq 1$. We also have $Nx = 0 \mod N$, but since $Nx = gyx = ky$, then $ky = 0 \mod N$. According to our definition of $\sqrt[k]{\text{NOT}}$ this implies that $y$ is even and, since $GCD(y, x) = 1$, $x$ is odd. Also

$$y = \frac{N}{g} = \frac{Nx}{k} = 2^{K-m} xc \tag{5.4}$$

Since $y$ is even and both $x$ and $c$ odd, then $K \geq m + 1$ must hold. We conclude with $N = 2^K c \geq 2^K \geq 2^{m+1} = 2k$.

The Lemma implies that if the machine is to perform $\sqrt[k]{\text{NOT}}$ perfectly it needs to have $\log k = m$ auxiliary bits in addition to the target bit. It is easy to check that this is not only necessary but also a sufficient condition. One just needs to design machine that is a loop of length $2k$ where the vertices corresponding to initial target input bits 0 and 1 are at a distance $k$ from each other. The number of functions with this property divided by the total number of functions $f : \{0,1\}^{2k} \to \{0,1\}^{2k}$ gives the fraction of the target functions:

$$R = \frac{(2k-4)!(2k-2)(k^2-2)}{(2k)^{2k}} \simeq O\left(\frac{1}{k4^k}\right). \tag{5.5}$$

The target functions thus constitute an exponential small fraction of all functions. Next, we will consider probabilistic classical machines which in order to approximate the target functions with high probability need to be sufficiently "close" (e.g. in the sense of Kullback-Leibler divergence) in the probability space. In such a way both the quantum and classical machines "search" in a continuous space of parameters, however, the relative fraction of this space that is close to the target functions is obviously much larger for the quantum case.

In the case of quantum machine any root of NOT can be performed with only one qubit. The operation that performs $\sqrt[k]{\text{NOT}}$ is $\sqrt[k]{\sigma_x}$, where $\sigma_x$ is spin matrix along direction $x$. Therefore, the memory requirements for our family of problems grows as $\log k$ in the classical case, while remaining constant in the quantum one.

Next, we introduce the classical learning procedure. We assume that the classical machine is initially in a "random" state for which $p(\vec{i}, \vec{j}) = \frac{1}{2k}$. The learning process consist of the following steps:

1. Set initially the internal state of the machine such that its first bit (target bit) is in 0 or 1 with equal probability. All auxiliary bits are in 0.

2. Apply the operation $k$ times and after each of them read out the output: $\vec{j}_r$, with $r \in \{1, ..., k\}$. We observe a sequence $\vec{i} \equiv \vec{j}_0 \to \vec{j}_1 \to \vec{j}_2 \to ... \to \vec{j}_k$ of machine' states. If the target bit of the final state $\vec{j}_k$ is inverse of the target bit of initial state $\vec{i}$, move to step 3. Otherwise move to 4.

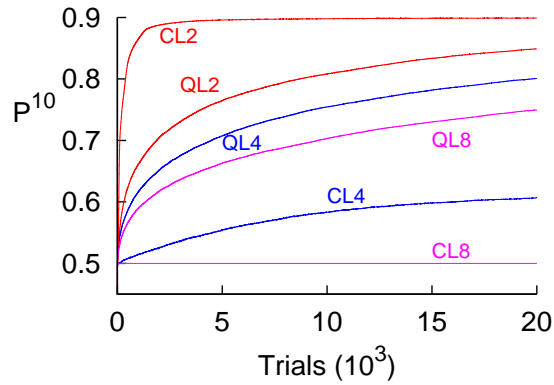

FIGURE 5.4: The figure of merit ($P^{10}$) of classical (CL) and quantum (QL) learning for performing different $k$-th ($k = 2, 4, 8$) roots of NOT as a function of the number of learning trials (x$10^3$). The values of free parameters are chosen to maximize the figure of merit. Already for $k = 4$ quantum learning is faster than the classical one. For the 8-th root of NOT, the figure of merit of classical learning is as for a random choice (= 0.5) at the given time scale. The free parameters have the values $\sigma_\gamma = \frac{\pi}{4}$ and $\sigma_\alpha = \sigma_\beta = \frac{\pi}{8}$ (for all roots) for the quantum case and the values $K_s = K_f = 0.25$ (2nd root), $K_s = K_f = 0.75$ (4th root) and $K_s = 0.75$ $K_f = 0.25$ (8th root) for the classical one.

3. Increase every probability $p(\vec{j}_{r-1}, \vec{j}_r)$ that led to success by adding a factor $1 \geq K_s \geq 0$. Renormalize the probability distribution such that $\sum_{\vec{j}} p(\vec{i}, \vec{j}) = 1$ and go back to step 2.

4. Decrease every probability $p(\vec{j}_{r-1}, \vec{j}_r)$ that led to a failure by subtracting a factor $1 \geq K_f \geq 0$ (if then the probability is negative, put it to be 0). Renormalize the probability distribution and go back to step 1.

Note that repeating the steps 2. and 3. the classical machine gradually learns to perform the task for all $n$. The learning has two free parameters $K_s$ and $K_f$, exactly like the quantum learning (with a variable teachers memory size $M$). To estimate how close is machine's functioning to the one of the target machine we use the set of figures of merit for all $n$: $\{P^n\}_{n=1}^\infty$, which are similar to those of Eq. (5.2). For example, $P^2 \equiv P_k(0,1) + P_k(1,0) + P_{2k}(0,0) + P_{2k}(1,1)$, where $P_k(1,0)$ is the probability that the target bit has been changed from 1 to 0 after applying the transformation $k$ times and other probabilities are similarly defined.

## 5.2 Conclusions

Recent progress in quantum communication and quantum computation (development of novel and efficient ways to process information on the basis of laws of quantum theory) provides motivation to generalize the theory of machine learning into the quantum domains. We tackle the following question: Can one have quantum improvements in the speed of learning in a sense that quantum machine requires fewer steps than the best classical machine to learn some classical task?

In this Chapter we have performed computer simulations of the both quantum and classical learning process. The results are presented in Fig. 5.4. We see that the learning in the quantum case is much faster for $k > 2$. This speed-up can be understood if one realizes that for the present problem the process of learning is an optimization of a square matrix: unitary transformation $U$ in the quantum case, and a matrix with entries $p(\vec{i}, \vec{j})$ in the classical one. While the size of $U$ remains 2 (with complex entries), the size of the matrix with entries $p(\vec{i}, \vec{j})$ grows linearly with $k$. It is clear that optimization of significantly larger matrices requires more iterative steps and thus leads to slower learning.

The classical learning algorithm given is not the most general and might not be optimal. The general framework for finding optimal learning procedures is still not fully understood. We have chosen the quantum and classical learning algorithms such that the comparison between them is most evident. The two tasks, i.e. finding a unitary operator for the $k$-th root of NOT, and finding a classical probability distribution that generates the $k$-th root of NOT, though are different from the physical point of view, both require optimization of matrix elements. Since for a given task, the classical machines require a significantly larger number of independent parameters (of which only a small fraction leads to the desired matrix) to be optimized, it is natural to assume that they also require a larger number of learning steps to accomplish learning, regardless of the explicit learning procedure employed. This is exactly what our numerical simulations show.

Briefly, quantum information processing has been recently shown to allow a speed-up over the best possible classical algorithms in computation and has advantages over its classical counterpart in communication tasks, such as secure transmission of information or communication complexity. Here, we extend the list with a novel task from the field of machine learning: learning to perform the $k$-th root of NOT.

# Chapter 6

# Quantum Entanglement

In this Chapter we briefly review some aspects of the phenomenon of quantum entanglement.

Quantum entanglement is one of the most essential features[1] of quantum mechanics [163, 180–182]. Entanglement constitutes a fundamental resource for the implementation of quantum information processes of technological relevance, such as quantum teleportation, superdense coding, and quantum computation [163, 183]. Recent developments related to the study of quantum entanglement are also leading to a deeper understanding of various basic aspects of quantum physics, such as, for example, the foundations of quantum statistical mechanics [184] and the quantum-to-classical transition via the decoherence process [185].

Some of the counter-intuitive properties of entangled quantum states were first pointed out in 1935 by Einstein, Podolski and Rosen in their celebrated "EPR" article [186]. These authors argued that the peculiar features exhibited by entangled states suggest that quantum mechanics does not provide a "complete" description of reality. Schrödinger [187] (who introduced the term "entanglement" in its german version: *Verschränkung*) hailed quantum entanglement as *"the characteristic trait of Quantum Mechanics, the one that enforces its entire departure from classical lines of thought"*.

Quantum entanglement is closely related to the tensor-product structure of the Hilbert space that is used in quantum mechanics to describe composite systems. This tensor product structure allows for the existence of pure states of composite systems that cannot be factorized as the product of states associated with each subsystem. The simplest quantum system admitting entangled states is one composed by two particles each described by a Hilbert space of dimension two (qubits). The composite system is then described by a four-dimensional Hilbert space. Let $|0\rangle_1$ ($|0\rangle_2$) and $|1\rangle_1$ ($|1\rangle_2$) denote

---
[1]Another basic, non-classical aspect of quantum systems, that is nowadays attracting increasing attention is contextuality [178, 179]

the members of two orthonormal basis for particles 1 and 2. A simple example of a non entangled -or separable- state is

$$|\psi\rangle_{\text{sep}} = |00\rangle = |0\rangle_1 |0\rangle_2. \tag{6.1}$$

On the other hand, an example of an entangled state is provided by the Bell state

$$|\psi\rangle_{\text{ent}} = \frac{1}{\sqrt{2}} \left( |01\rangle + |10\rangle \right). \tag{6.2}$$

In 1964 Bell [188] formalized the ideas of EPR of a local hidden variable model (LHVM). It was based in the following assumptions:

1. Realism: The outputs of the measurements are determined by the properties of the system and not by the measurement process.

2. Locality: The output of a local measurement is independent of any other action or event with a space-like separation from the measurement.

3. Free will: The setting of local apparatus is independent of the properties of the system which determine the local result.

With these assumptions Bell derived an inequality for the statistical correlations of measurements performed on a bipartite system. Bell proved that this inequality is violated by some states of two qubits systems. The Bell original inequality was the inspiration for the formulation by Clauser, Horne, Shimony and Holt (CHSH) [189] of an inequality that can be tested experimentally. In a bipartite system with the dichotomic variables $(A_1, A_2)$ for the subsystem $A$ and $(B_1, B_2)$ for the subsystem $B$ (the possible values are $\pm 1$) the assumptions of the LHVM gives the following statistical constraint

$$|\langle A_1 B_1 \rangle + \langle A_1 B_2 \rangle + \langle A_2 B_1 \rangle - \langle A_2 B_2 \rangle| \leq 2 \tag{6.3}$$

where $\langle A_i B_j \rangle$ is the mean value of the product $(A_i B_j)$.

It can be shown that if a two-qubit system is in the Bell state (6.2), it is possible to find appropriate quantum observables $\hat{A}_i$, $\hat{B}_i$ such that the value of the left part of Eq. (6.3) is $2\sqrt{2}$. Therefore, the CHSH inequality is violated for this kind of quantum states. Bell inequalities can be considered as the first procedure to distinguish entangled from non-entangled states.

## 6.1 Composite systems with distinguishable subsystems

Consider a pure state $|\psi\rangle_{A_1,...,A_m} \in \mathscr{H}_{A_1,...,A_m} = \mathscr{H}_{A_1} \otimes \cdots \otimes \mathscr{H}_{A_m}$, where the Hilbert space of the complete system $\mathscr{H}_{A_1,...,A_m}$ is the tensor product of the Hilbert spaces $\mathscr{H}_{A_i}$ of all the subsystems. This state is fully separable if and only if it can be factorized as:

$$|\psi\rangle_{A_1,...,A_m} = |\psi\rangle_{A_1} \otimes \cdots \otimes |\psi\rangle_{A_m}. \tag{6.4}$$

Pure states that do not admit this factorization are entangled states [2]. Examples of pure entangled states for the Hilbert space $\mathscr{H} = \mathscr{H}_1 \otimes \mathscr{H}_2$ with $\dim(\mathscr{H}_1) = \dim(\mathscr{H}_2) = 2$ are the four Bell states

$$\begin{aligned}|\psi^{\pm}\rangle &= \frac{1}{\sqrt{2}}\left(|01\rangle \pm |10\rangle\right) \\ |\phi^{\pm}\rangle &= \frac{1}{\sqrt{2}}\left(|00\rangle \pm |11\rangle\right).\end{aligned} \tag{6.5}$$

A mixed state $\rho_{A_1,...,A_m}$ is fully separable if it can be written as

$$\rho = \sum_{i=1}^{N} \lambda_i \rho_i^{(A_1)} \otimes \cdots \otimes \rho_i^{(A_m)}, \tag{6.6}$$

where $\lambda_i$ are positive weights satisfying $\sum_{i=1}^{N} \lambda_i = 1$. Alternatively a equivalent definition is that the state $\rho_{A_1,...,A_m}$ is fully separable if it can be written as

$$\rho = \sum_{i=1}^{N} p_i |\phi_i^{(A_1)}\rangle\langle\phi_i^{(A_1)}| \otimes \cdots \otimes |\phi_i^{(A_m)}\rangle\langle\phi_i^{(A_m)}|, \tag{6.7}$$

where the $p_i$ are, again, positive weights adding up to one. Mixed states, that cannot be written as in Eq. (6.6) and (6.7) are entangled.

Separable states can be characterized physically as those that can be prepared by distant agents, each operating locally on his/her subsystem, that can only communicate classically.

Let us consider a bipartite quantum system consisting of two subsystems $A$ and $B$. If a pure state $|\psi\rangle$ describing the composite system is entangled, the marginal density

---
[2]Note that when we have three o more subsystems, and the global state cannot be written as in (6.4), this does not imply that every subsystem is entangled with the rest of the system. Some subsystems may be entangled among themselves and disentangled from the rest.

matrices

$$\rho_A = Tr_B \left(|\psi\rangle\langle\psi|\right)$$
$$\rho_B = Tr_A \left(|\psi\rangle\langle\psi|\right), \tag{6.8}$$

describing the subsystems correspond to <u>mixed states</u>. The degree of mixedness of these marginal density matrices can be regarded as a measure of the amount of entanglement of the pure state $|\psi\rangle$. The most fundamental of these measures is the entropy of entanglement

$$E\left(|\psi\rangle\right) = S_{VN}(\rho_A) = S_{VN}(\rho_B), \tag{6.9}$$

where $S_{VN}(\rho) = -Tr\rho\log\rho$ denotes the von Neumann entropy of the statistical operator. Another usefull indicator of the amount of entanglement of $|\psi\rangle$ is provided by the linear entropy of the marginal density matrices

$$S_L(\rho_A) = S_L(\rho_B), \tag{6.10}$$

where $S_L(\rho) = 1 - Tr\left(\rho^2\right)$.

One of the most fundamental entanglement measure for mixed states $\rho$ of bipartite systems is the entanglement of formation $E(\rho)$, defined as,

$$E\left(\rho\right) = \min_{\{\lambda_i, |\psi_i\rangle\}} \sum_i \lambda_i E\left[|\psi_i\rangle\right], \tag{6.11}$$

where the minimum is taken over all the statistical mixtures $\{\lambda_i, |\psi_i\rangle\}$ that lead to the same state $\rho$,

$$\rho = \sum_i \lambda_i |\psi_i\rangle\langle\psi_i|, \tag{6.12}$$

with $0 \leq \lambda_i \leq 1$ and $\sum_i \lambda_i = 1$.

The evaluation of $E\left[\rho\right]$ is in general quite difficult and has to be done numerically. A closed analytical expression for $E\left[\rho\right]$ (Wootters' formula [190]) is known only for systems of two qubits.

The existence of non-separable states leads naturally to the following problem: given a mixed state $\rho$ of a bipartite system determine if $\rho$ represents a separable state or not. A simple example of a separability test is to check if the state violate a Bell inequality like (6.3). However, in 1989 Werner proved that some non-separable states admit a LHVM, so they cannot violate any Bell inequality [191]. Important examples of this kind of states are provided by some members of the Werner family of states. The Werner states of two-qubits are of the form,

$$\rho = p|\psi\rangle\langle\psi| + \frac{1-p}{4}\mathbb{I}, \tag{6.13}$$

where $p \in [0,1]$, $|\psi\rangle = \frac{1}{\sqrt{2}}(|01\rangle - |10\rangle)$ is a Bell state (see Eq. 7.110) and $\mathbb{I}$ is the identity. When $p < \frac{1}{\sqrt{2}}$ these states accept a LHVM and, consequently, don't violate any Bell inequality.

In 1996 Peres [192] derive a separability criterion for density matrices that can detect entanglement in the system of Eq. (6.13) better than Bell inequalities. A density matrix corresponding to a bipartite system is regarded as separable if

$$\rho_{m\mu,n\nu} = \sum_i \lambda_i (\rho_i^{A_1})_{mn} (\rho_i^{A_2})_{\mu\nu}, \tag{6.14}$$

where Latin indices refer to the first subsystem and Greek indices to the second one.

Let me now consider the new matrix

$$\sigma_{m\mu,n\nu} \equiv \rho_{n\mu,m\nu}. \tag{6.15}$$

So the Latin indices have been transposed, but not the Greek ones. Note that the new matrix is hermitian. If the system is separable $\sigma$ can be expressed as

$$\sigma = \sum_i \lambda_i (\rho_i^{A_1})^T \otimes \rho_i^{A_2}. \tag{6.16}$$

Since the the transpose matrices $\left(\rho_i^{A_1}\right)^T \equiv \left(\rho_i^{A_1}\right)^*$ are non-negative, normalized matrices, they are also density matrices. Then none of the eigenvalues of $\sigma$ is negative if Eq. (6.14) is fulfilled, so this gives us a necessary condition of entangled states. It is proved that this necessary condition becomes also a sufficient one for composite systems having dimensions $2 \times 2$ and $2 \times 3$.

Peres criterion applied to systems defined by Eq. (6.13) detects entanglement for $p > \frac{1}{3}$, that is all the range of $p$ with entanglement.

## 6.2  Systems of identical fermions

In a system of identical fermions the definition of separable system given by Eq. (6.4) is useless because the Hilbert space of the system can not be expressed as the tensor product of the single particle Hilbert spaces. Therefore, a new definition of entanglement is required for these systems. Following the discussion of reference [193] let us illustrate the problem of entanglement in fermion systems by recourse to the system of Fig. 6.1. This system consists of two identical particles of spin $\frac{1}{2}$ separated by a potential barrier.

When the barrier is very high the overlap of the particles' wave function is negligible and they can be regarded as non-identical, distinguishable, particles. The state of the system can then be represented as

$$|\Psi_{\text{init}}\rangle = [|L\rangle |\uparrow\rangle_A \otimes |R\rangle |\downarrow\rangle_B], \qquad (6.17)$$

where $|L\rangle(|R\rangle)$ denote spatial single-particle wave functions localized in the left (right) part of the barrier and $(|\uparrow\rangle, |\downarrow\rangle)$ are the eigenstates of $\hat{S}_z$. In this example we regard the system as described by an effective four-dimensional single-particle Hilbert space $\{|L\uparrow\rangle, |L\downarrow\rangle, |R\uparrow\rangle, |R\downarrow\rangle\}$.

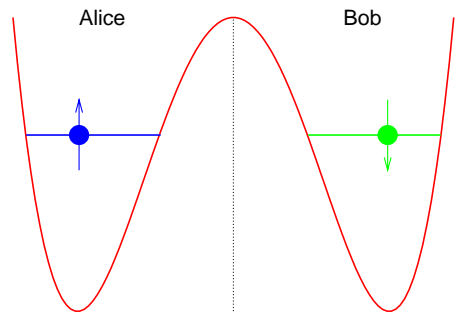

FIGURE 6.1: Two identical particles separated by a high potential barrier, so they can be considered distinguishable.

The state (6.17) is clearly separable. If after a time $t_1$ the wall between the particles goes down, the antisymmetry of the wave function must be taken explicitly into account; then the state becomes

$$|\psi(t_1)\rangle = \frac{1}{\sqrt{2}} \left[ |L\rangle |\uparrow\rangle_1 \otimes |R\rangle |\downarrow\rangle_2 - |R\rangle |\downarrow\rangle_1 \otimes |L\rangle |\uparrow\rangle_2 \right], \qquad (6.18)$$

so the particles are not independently accessible anymore. The correlations of this system are non-accessible, so the system must me considered "non-entangled". Note that the "non-entangled" state can be written as a single Slater determinant.

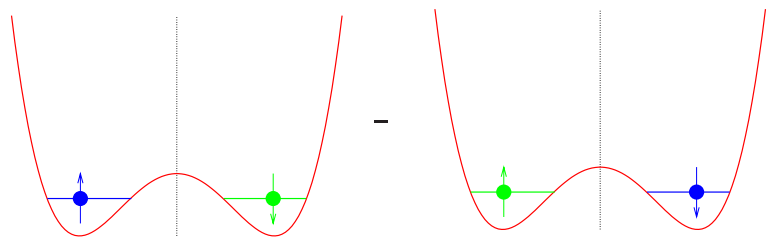

FIGURE 6.2: State resulting if the wall goes down. The particles become indistinguishable.

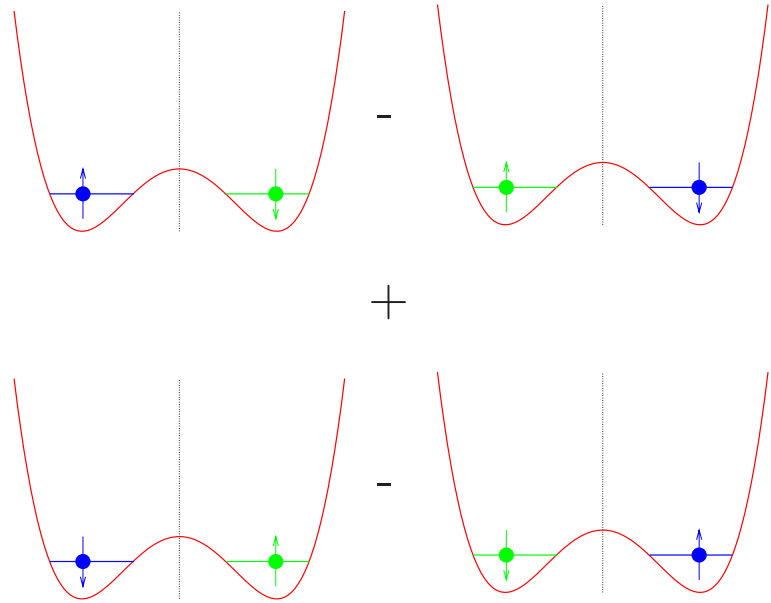

FIGURE 6.3: Entangled system of identical particles.

Now take the system of identical fermions described by

$$|\psi_{\text{init2}}\rangle = \frac{1}{2}\left[|L\rangle|\uparrow\rangle \otimes |R\rangle|\downarrow\rangle - |R\rangle|\downarrow\rangle \otimes |L\rangle|\uparrow\rangle + |L\rangle|\downarrow\rangle \otimes |R\rangle|\uparrow\rangle - |R\rangle|\uparrow\rangle \otimes |L\rangle|\downarrow\rangle\right], \tag{6.19}$$

that is represented in Fig. 6.3. This state contains some usefull correlations so if, after a time $t_2$, we localize the particles by raising the potential barrier. The new state of the system is

$$|\psi(t_2)\rangle = \frac{1}{\sqrt{2}}\left[|L\rangle|\uparrow\rangle \otimes |R\rangle|\downarrow\rangle + |L\rangle|\downarrow\rangle \otimes |R\rangle|\uparrow\rangle\right], \tag{6.20}$$

that is a maximally entangled state as is represented in Fig. 6.4.

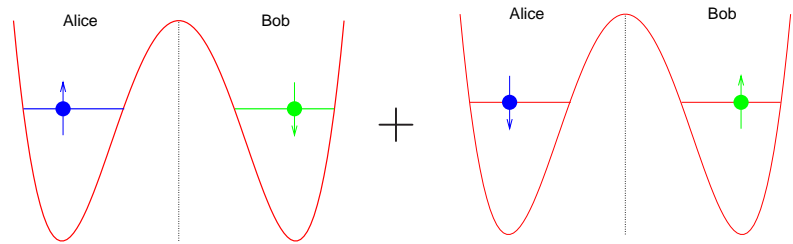

FIGURE 6.4: Final entangled state.

The above discussion indicates, then, that a system composed of identical particles should be considered separable if and only if it can be described as a single Slater determinant [193], also denominated as having Slater rank 1. Entangled states, on the other hand, are those that cannot be expressed as a single Slater determinant. There is nowadays growing consensus on this conception of entanglement between particles in many-fermion systems, as is clearly expressed by Eckert et al in Ref. [193]: *"Quantum correlations in systems of indistinguishable fermions arise if more than one Slater determinant is involved, i.e., if there is no single-particle basis such that a given state of N indistinguishable fermions can be represented as an elementary Slater determinant (i.e. a fully antisymmetric combination of N orthogonal single-particle states). These correlations are the analogue of quantum entanglement in separated systems and are essential for quantum information processing in non-separated systems"*. Also in the excellent review on entanglement in many-body systems by Amico et al [180] we read: *"A pure fermion state is entangled if and only if its Slater rank is larger than 1"*.

In the particular case of a system composed for two identical fermions a separable state can be defined as
$$\frac{1}{\sqrt{2}}\left\{|\phi_1\rangle|\phi_2\rangle - |\phi_2\rangle|\phi_1\rangle\right\}, \tag{6.21}$$
where $|\phi_1\rangle$, $|\phi_2\rangle$ are two orthogonal and normalized single-particle states.

When studying entanglement-related properties of pure states of bipartite quantum systems it is sometimes convenient to use the Schmidt decomposition of the joint state. For any pure state $|\psi\rangle$ of a quantum system consisting of two distinguishable subsystems $A_1$ and $A_2$ it is possible to find two orthonormal basis $\left\{|\phi_i^{(A_1)}\rangle\right\}$ and $\left\{|\phi_i^{(A_2)}\rangle\right\}$ such that
$$|\psi\rangle = \sum_i \sqrt{\lambda_i}|\phi_i^{(A_1)}\rangle \otimes |\phi_i^{(A_2)}\rangle, \tag{6.22}$$
where
$$0 \geq \lambda_i \geq 1 \quad \text{and} \quad \sum_i \lambda_i = 1. \tag{6.23}$$

The numbers $\lambda_1$'s are called the Schmidt coefficients of the state $|\psi\rangle$. The Schmidt decomposition admits a natural generalization in the case of a system of two identical fermions. In this case, given a pure state $|\psi\rangle$ it is possible to find an orthonormal basis $\{|i\rangle,\ i = 0, 1, ...\}$ of the single-particle Hilbert space, such that the two-fermions pure state $|\psi\rangle$ can be written as

$$|\psi\rangle = \sum_i \sqrt{\frac{\lambda_i}{2}}\left(|2i\rangle|2i+1\rangle - |2i+1\rangle|2i\rangle\right), \tag{6.24}$$

where the Schmidt coefficients $\lambda_i$'s verify (6.23)

The simplest fermionic system admitting entanglement is composed by two fermions with a single-particle Hilbert space of dimension four, as the system I have used in the example before. This system can be regarded as the *"fermionic version"* of the standard two-qubit system (that is, a system of two distinguishable qubits).

# Chapter 7

# Identical fermions, entanglement and two-electron systems

Entanglement constitutes an essential ingredient in the quantum mechanical description of the physical world [181, 185]. It is also a physical resource with important technological implications [163]. A fundamental first step in the study of the entanglement properties of a given class of quantum systems is the establishment of appropriate separability criteria. That is, to establish criteria that enables us to tell if a given quantum state is separable or entangled. A good separability criterion, besides its obvious importance as a tool for determining the presence or absence of entanglement, is also relevant as the possible basis of quantitative measures of entanglement. An appropriate measure of the deviation of the actual properties of a given quantum state from those required by the separability criterion may provide a valuable estimation of the amount of entanglement exhibited by that state. Here we are going to consider practical separability criteria and entanglement measures for pure states of $N$ identical fermions.

The study of the entanglement features of atomic systems can be regarded as part of a broader field that has attracted considerable attention in recent years: The application of information theory to atomic physics provides an interesting new point of view in the study of atomic structure that has been explored in various recent research works [63, 88, 122, 194–201]. This line of enquiry has several points of contact with the field of quantum information theory, particularly in connection with the study of the entanglement-related properties exhibited by atomic systems. Besides its intrinsic theoretical interest, this area of research is also of practical relevance, because some of the systems studied by contemporary atomic physics, such as ion traps, constitute important candidates for the experimental implementation of quantum information technology. Moreover, the investigation of information-theoretical aspects of atomic structure proved to be related to other areas of physics, such as the theory of critical phenomena [199].

The aims of this Chapter are: i) To investigate some aspects of separability criteria and entanglement measures for pure states of $N$ identical fermions, (see Section 7.1), ii) to explore the entanglement-relates properties of the eigenstates of two electron systems, (see Section 7.2) and iii) to analyse a new criterion proposed by Walborn et al [1] in 2009 for the detection of entanglement in quantum systems with continuous variables (see Section 7.3).

## 7.1　Separability criteria and entanglement measures for pure states of $N$ identical fermions

The study of the entanglement features of systems consisting of $N$ identical fermions has attracted the attention of many researchers in recent years [180, 193, 202–209]. Entanglement between fermionic particles has been studied in connection with various physical scenarios. To mention just a few examples, researchers have recently investigated entanglement in two-electrons atomic states [194], entanglement between pairs of electrons in a conducting band [208], entanglement dynamics in two-electrons scattering processes [203], and the role of entanglement in time-optimal evolutions of fermionic systems [202, 209], among many others. Appropriate separability criteria (and entanglement measures) for pure states of two identical fermions have been recently derived (using the Schmidt decomposition) and applied to the study of various physical systems and processes [203, 204, 208]. Alas, the aforementioned derivations of separability criteria cannot be extended to situations involving more than two fermions because in such cases the Schmidt decomposition doesn't exist.

Some separability criteria for more than two fermions have been proposed in the recent quantum information literature, but they are difficult to implement in practice and exhibit a growing degree of complexity when one increases the number of particles of the system or the dimensionality of the single-particle Hilbert space. The necessary and sufficient criterion introduced by Eckert, Schliemann, Bruss, and Lewenstein [193] (from now on ESBL) is based on a projection operator acting upon an $N$-fermion state and resulting in an $(N-1)$-fermion state. This operator depends on an arbitrary single-particle state $|a\rangle$. The ESBL criterion says that a pure $N$-fermion state $|\Psi\rangle$ has Slater rank one (that is, it is a separable state) if and only if the result of applying the projector operator on $|\Psi\rangle$ is, *for any single-particle state $|a\rangle$*, either equal to an $(N-1)$-fermion state of Slater rank 1 or equal to zero. The ESBL separability criterion has been recently hailed [180] as the main result known so far on necessary and sufficient separability criteria for $N$-fermion pure states. The ESBL criterion certainly is of considerable relevance from the fundamental and conceptual points of view, but it is of little practical use. To check if a given state $|\Psi\rangle$ fulfils the ESBL criterion is, in general, basically as difficult as the original problem of finding out if $|\Psi\rangle$ has Slater rank equal to 1 or not. The ESBL criterion can be iterated $N-2$ times, leading to a chain of separability tests eventually ending with a separability test to be performed on a two-fermion state. However, this procedure does not reduce the difficulty of the criterion, since each link in the aforementioned chain involves a relation that has to be checked for an arbitrary single particle state $|a\rangle$ [193]. A different approach employing sophisticated techniques from algebraic geometry has been advanced in [206]. According to this proposal, however, to be identified as separable a quantum $N$-fermion state has to comply with several relations (that is, not just with one identity as in the criterion proposed by us), their number increasing with the number of fermions in the system.

### 7.1.1 Separability criteria

The aim of the present Section is to derive two inequalities verified, respectively, by the purity $Tr\left(\rho_r^2\right)$ and the von Neuman entropy $-Tr\left(\rho_r \log \rho_r\right)$ of the single particle reduced density matrix $\rho_r$ of an $N$-fermions pure state. These inequalities lead to simple separability criteria and suggest practical entanglement measures. These separability criteria turn out to be closely related to some previous results from the theory of Hartree-Fock wave functions that, even though themselves constituting useful necessary and sufficient separability criteria, doesn't seem to have been recognized as such in the recent literature. Our derivations are different from (and simpler than) the ones followed in the aforementioned works on the Hartree-Fock wave functions. Moreover, our developments clarify why those previous results have not been believed to provide sufficient separability criteria for $N$-fermions states.

Let us consider a system consisting of a *constant* number $N$ of identical fermions with a single particle Hilbert space of dimension $D$, with $N \leq D$ (if $N > D$ it is not possible to construct an antisymmetric $N$-fermion state). A pure state of such a system is separable (that is, non-entangled) if it has Slater rank equal to one [180]. That is to say, the state is non entangled if it can be expressed as a single Slater determinant,

$$a_{i_1}^\dagger ... a_{i_N}^\dagger |0\rangle, \tag{7.1}$$

where $a_i^\dagger$ are fermionic creation operators acting upon the vacuum state $|0\rangle$ and leading to an orthonormal basis $\{|i\rangle = a_i^\dagger |0\rangle\}$ of the single-particle Hilbert space. A pure state of the $N$-fermion system that cannot be written in the above way has a finite amount of entanglement. Correlations between the $N$ fermions that are due solely to guarantee the antisymmetric character of the fermionic states do not contribute to the state's amount of entanglement [193, 204, 205]. There are profound physical reasons for this. On the one hand, these correlations (exhibited by states with Slater rank 1) can't be used as a resource to implement non-classical information transmission or information processing tasks [193]. On the other hand, the non-entangled character of states represented by one Slater determinant is consistent with the possibility of associating complete sets of properties to both parts of the composite system (see [204, 205] for an interesting, detailed discussion of this approach).

When discussing the entanglement properties of systems of $N$ identical fermions the relevant group of "local transformations" is isomorphic to the group $SU(D)$ of (special) unitary transformations acting on the $D$-dimensional single-particle Hilbert space [193]. Given a transformation $U \in SU(D)$ the corresponding "local transformation" acts on a general $N$-fermions state according to $\sum w_{i_1,\ldots,i_N} a^\dagger_{i_1} \ldots a^\dagger_{i_N} |0\rangle \to \sum w_{i_1,\ldots,i_N} \tilde{a}^\dagger_{i_1} \ldots \tilde{a}^\dagger_{i_N} |0\rangle$, where $\tilde{a}^\dagger_i |0\rangle = |\tilde{i}\rangle$ and $U|i\rangle = |\tilde{i}\rangle$, $(i = 1, \ldots, D)$. The set of non-entangled fermionic states is closed under the action of these "local transformations". Furthermore, the entanglement measures that I am going to consider in this work are invariant under those transformations.

A simple illustration of the fact that the correlations associated with a fermionic state of Slater rank 1 cannot be used as a resource for quantum information tasks is provided by a two-electrons system with a four dimensional relevant single-particle Hilbert space [193]. Let us assume that the relevant single-particle Hilbert space admits a basis of the form $\{|\phi_1\rangle|+\rangle, |\phi_1\rangle|-\rangle, |\phi_2\rangle|+\rangle, |\phi_2\rangle|-\rangle\}$, where $|\phi_{1,2}\rangle$ are two spatial wave functions and $|\pm\rangle$ corresponds to the spin degree of freedom. The two electrons can be treated as effectively distinguishable entities if they are spatially localized. This can occur if the moduli of $\langle \vec{r}|\phi_1\rangle$ and $\langle \vec{r}|\phi_2\rangle$ are non-overlapping. The single particle basis can then be partitioned between two agents (Alice and Bob), $\{|\phi_1\rangle|+\rangle, |\phi_1\rangle|-\rangle\}$ being the basis of Alice's space and $\{|\phi_2\rangle|+\rangle, |\phi_2\rangle|-\rangle\}$ the basis of Bob's space. Under these circumstances, a state of the form $\frac{1}{\sqrt{2}}\Big(|\phi_1\rangle|+\rangle \otimes |\phi_2\rangle|-\rangle - |\phi_2\rangle|-\rangle \otimes |\phi_1\rangle|+\rangle\Big)$ given by a single Slater determinant (and describing two particles localized in different spatial regions) effectively behaves as the non-entangled (in the usual sense) state $|\phi_1\rangle|+\rangle_A \otimes |\phi_2\rangle|-\rangle_B$ describing two distinguishable objects ($A$ and $B$). On the other hand, a state describing two localized electrons that cannot be cast as one single Slater determinant effectively behaves as an entangled state (in the standard sense corresponding to distinguishable subsystems) that is useful for performing non-classical information related tasks (see [193] for a more detailed discussion).

The amount of entanglement associated with an $N$-fermion state corresponds, basically, to the quantum correlations exhibited by the state on top of the minimum correlations needed to comply with the antisymmetric constraint on the fermionic wave function. Note that here we are considering entanglement between *particles*, and not entanglement between *modes* (see [210] for a comprehensive discussion of entanglement between modes).

Given a single particle orthonormal basis $\{|i\rangle = a^\dagger_i |0\rangle, \ i = 1, \ldots, D\}$, any pure state of the $N$-fermion system can be expanded as,

$$|\Psi\rangle = \sum_{i_1,\ldots,i_N=1}^{D} w_{i_1,\ldots,i_N}\, a_{i_1}^{\dagger} \ldots a_{i_N}^{\dagger} |0\rangle, \tag{7.2}$$

where the complex coefficients $w_{i_1,\ldots,i_N}$ are antisymmetric in all indices and comply with the normalization condition

$$\sum_{i_1,\ldots,i_N=1}^{D} |w_{i_1,\ldots,i_N}|^2 = \frac{1}{N!}. \tag{7.3}$$

The single-particle reduced density matrix $\rho_r$ associated with the $N$-fermion pure state (7.2) has matrix elements,

$$\langle i|\rho_r|j\rangle = \frac{1}{N} \langle \Psi|a_j^{\dagger} a_i|\Psi\rangle, \tag{7.4}$$

where the factor $1/N$ guaranties that $\rho_r$ is normalized to unity,

$$Tr\rho_r = 1. \tag{7.5}$$

Let $F_i \equiv \langle i|\rho_r|i\rangle$ denote the diagonal elements of $\rho_r$. After some algebra it is possible to verify that,

$$F_i = \sum_{\substack{(i_1,\ldots,i_n) \\ i_1<i_2<\ldots<i_n}} (N!)^2 |w_{i_1,\ldots,i_N}|^2\, f_i^{(i_1,\ldots,i_N)}, \quad i=1,\ldots,D, \tag{7.6}$$

where

$$f_i^{(i_1,\ldots,i_N)} = \begin{cases} \frac{1}{N}, & \text{if } i \in (i_1,\ldots,i_N), \\ 0 & \text{otherwise.} \end{cases} \tag{7.7}$$

Note that the sum in (7.6) has only $\binom{D}{N} = \frac{D!}{N!(D-N)!}$ terms because it doesn't run over all the $D^N$ possible $N$-uples $(i_1,\ldots,i_N)$; it runs only over the $\binom{D}{N}$ $N$-uples whose indices are all different and listed in increasing order. Thus, the vector $\vec{F}$ (with components $\{F_i,\ i=1,\ldots,D\}$) can be expressed as a linear combination of the $\binom{D}{N}$ vectors $\vec{f}^{(i_1,\ldots,i_N)}$ (with components $\{f_i^{(i_1,\ldots,i_N)},\ i=1,\ldots,D\}$). Each one of these vectors has $D$ components, $N$ of them being equal to $1/N$ and the rest equal to zero. To simplify notation it is convenient to introduce a single global label $k$, $1 \leq k \leq \binom{D}{N}$, to characterize the coefficients $(N!)^2 |w_{i_1,\ldots,i_N}|^2$ and the vectors $\vec{f}^{(i_1,\ldots,i_N)}$. Equation (7.6) can then be recast in a more compact way as,

$$F_i = \sum_{k=1}^{M} d_k f_{ik}, \tag{7.8}$$

where $M = \binom{D}{N}$ and the identifications

$$\begin{aligned} (N!)^2 |w_{i_1,\ldots,i_N}|^2 &\to d_k \\ f_i^{(i_1,\ldots,i_N)} &\to f_{ik} \end{aligned} \tag{7.9}$$

have been made. We have $0 \leq d_k \leq 1$, $(1 \leq k \leq M)$, $0 \leq f_{ik} \leq 1$, $(1 \leq k \leq M; 1 \leq i \leq D)$, and,

$$\sum_{k=1}^{M} d_k = 1; \quad \sum_{i=1}^{D} f_{ik} = 1; \quad \sum_{i=1}^{D} f_{ik}^2 = \frac{1}{N}. \tag{7.10}$$

The vector $\vec{F}$ and each of the vectors $\vec{f_k}$ can be regarded as properly normalized probability distributions, and the vector $\vec{F}$ is a *convex* linear combination of the vectors $\vec{f_k}$.

Let us now consider the sum of the squares of the components of the vector $\vec{F}$,

$$\begin{aligned}
\sum_{i=1}^{D} F_i^2 &= \sum_{i=1}^{D} \left\{ \left( \sum_{k=1}^{M} d_k^2 f_{ik}^2 \right) + 2 \left( \sum_{k<k'} d_k d_{k'} f_{ik} f_{ik'} \right) \right\} = \\
&= \sum_{i=1}^{D} \left\{ \left( \sum_{k=1}^{M} d_k \left( 1 - \sum_{k' \neq k} d_{k'} \right) f_{ik}^2 \right) + 2 \left( \sum_{k<k'} d_k d_{k'} f_{ik} f_{ik'} \right) \right\} = \\
&= \sum_{i=1}^{D} \left\{ \left( \sum_{k=1}^{M} d_k f_{ik}^2 \right) - \left( \sum_{k \neq k'} d_k d_{k'} f_{ik}^2 \right) + 2 \left( \sum_{k<k'} d_k d_{k'} f_{ik} f_{ik'} \right) \right\} = \\
&= \sum_{i=1}^{D} \left\{ \left( \sum_{k=1}^{M} d_k f_{ik}^2 \right) - \sum_{k<k'} d_k d_{k'} \left( f_{ik}^2 + f_{ik'}^2 - 2 f_{ik} f_{ik'} \right) \right\} = \\
&= \left\{ \sum_{k=1}^{M} d_k \left( \sum_{i=1}^{D} f_{ik}^2 \right) \right\} - \left\{ \sum_{k<k'} d_k d_{k'} \sum_{i=1}^{D} (f_{ik} - f_{ik'})^2 \right\}.
\end{aligned} \tag{7.11}$$

Since $\sum_{i=1}^{D} f_{ik}^2 = \frac{1}{N}$ for all $k$, it follows from (7.11) that,

$$\sum_{i=1}^{D} \langle i |\rho_r| i \rangle^2 = \sum_{i=1}^{D} F_i^2 = \frac{1}{N} - \underbrace{\sum_{k<k'} d_k d_{k'} \sum_{i=1}^{D} (f_{ik} - f_{ik'})^2}_{\geq 0} \leq \frac{1}{N}. \quad (7.12)$$

The inequality in (7.12) can also be obtained applying Jensen inequality to the square of the right hand side of (7.8) and taking into account the first and the third equations in (7.10).

The only way for the equality sign to hold in (7.12) is to have one of the $d_k$ equal to 1 and the rest equal to 0, meaning that there is only one term in the original expansion for $|\Psi\rangle$. This implies that $|\Psi\rangle$ has Slater rank one, and can thus be expressed as one single Slater determinant. Since I didn't impose any restriction on the single-particle basis $\{|i\rangle\}$, equation (7.12) holds for any such a basis. In particular, it holds for the eigenbasis of the single-particle reduced statistical operator $\rho_r$, implying that

$$Tr\left(\rho_r^2\right) \leq \frac{1}{N}. \quad (7.13)$$

It is easy to see that when the Slater rank of the $N$-fermions state $|\Psi\rangle$ is one we have $Tr\left(\rho_r^2\right) = \frac{1}{N}$. On the other hand, $Tr\left(\rho_r^2\right) = \frac{1}{N}$ implies that there exists a single-particle basis for which the equal sign holds in (7.12), implying in turn that the state under consideration has Slater rank 1 and it is then separable.

Summing up, the following double implication obtains,

$$|\Psi\rangle \text{ has Slater rank one } \iff Tr\left(\rho_r^2\right) = \frac{1}{N}. \quad (7.14)$$

In other words, *a pure state of $N$ identical fermions is separable if and only if the purity of the reduced single-particle density matrix is equal to $1/N$.*

It is possible to formulate a separability criterion equivalent to (7.14) in terms of the von Neumann entropy of the single particle density matrix $\rho_r$. Let us consider the Shannon entropies of $\vec{F}$ and $\vec{f_k}$ (regarded as probability distributions),

$$S[\vec{F}] = -\sum_{i=1}^{D} F_i \log F_i; \quad S[\vec{f_k}] = -\sum_{i=1}^{D} f_{ik} \log f_{ik} \quad (7.15)$$

Using the concavity property of the Shannon entropy [15], it follows from (7.8) that,

$$S[\vec{F}] \geq \sum_{k=1}^{M} d_k S[\vec{f_k}] = \log N, \quad (7.16)$$

where the inequality reduces to an equality if and only if all the probability vectors $\vec{f}_k$ appearing in the sum in the middle term in (7.16) are equal to each other. This can only happen if one of the $d_k$'s is equal to 1 and the rest are equal to zero. That is, it can happen only if the $N$-fermion state can be written as a single Slater determinant. Equation (7.16) holds for any single-particle basis $\{|i\rangle\}$. In particular, it holds for the eigenbasis of $\rho_r$, which leads to

$$S[\rho_r] \geq \log N. \tag{7.17}$$

It is plain that an $N$-fermion pure state with Slater rank one leads to a single-particle reduced density matrix verifying $S[\rho_r] = \log N$. Conversely, the relation $S[\rho_r] = \log N$ implies that there exists a single-particle basis such that $-\sum \langle i|\rho_r|i\rangle \log\langle i|\rho_r|i\rangle = S[\vec{F}] = \log N$ which, as we have already seen, implies that the $N$-fermion pure state can be written as a single Slater determinant and, consequently, describes a separable state. Summarizing,

$$|\Psi\rangle \text{ has Slater rank one} \iff -Tr(\rho_r \log \rho_r) = \log N. \tag{7.18}$$

A particular instance of the separability criterion (7.18), corresponding to systems of two identical fermions, has already been discussed by Ghirardi and Marinatto in [204]. The derivation of the $N=2$ case of (7.18) given by Ghirardi and Marinatto is based upon the Schmidt decomposition for systems of two fermions. Unfortunately, the Schmidt decomposition does not exist when $N \geq 3$ and, consequently, the developments presented in [204] cannot be extended to situations involving systems of three or more identical fermions. Our present treatment, besides providing a necessary and sufficient separability criterion valid for arbitrary values of the number $N$ of particles, is also of interest as yielding an alternative way of obtaining the $N=2$ criterion without recourse to the Schmidt decomposition.

The necessary and sufficient condition for separability $Tr[\rho_r^2] = 1/N$ is closely related to the condition

$$\rho_r^2 = \frac{1}{N}\rho_r \tag{7.19}$$

that the single particle reduced density matrix has to verify if the global wave function can be expressed as a Slater determinant. Condition (7.19) has been discussed in the past in the context of atomic physics [211, 212] and actually constitutes a classicall result from the theory of the Hartree-Fock approximation. However, the relevance of condition (7.19) as a useful separability criterion for $N$-fermions pure states has not been properly appreciated within the field of quantum entanglement theory. In fact, condition (7.19) has been in the recent literature regarded as not providing a necessary separability criterion. In fact, in connection with $N$-fermions states leading to a reduced density matrix verifying (7.19) it has been recently stated that "*... a wave function of this kind can in general not be written as a single Slater determinant constructed from orthogonal states*" [193]. As we are going to show next, our present results show in a direct and manifest way that the alluded wave functions can indeed be written as a single Slater determinant constructed from orthogonal states (that is, they have Slater rank 1).

Note that condition (7.19) is not, by itself, equivalent to either the relation (7.14) or to the entropic relation (7.18). It is plain that a density matrix $\rho_r$ complying with (7.19) must necessarily verify relations (7.14) and (7.18). However, the reciprocal implication doesn't hold. A density matrix verifying (7.14) (or verifying (7.18)) does not necessarily fulfil (7.19). For instance, if $\rho_r$ has eigenvalues $(\frac{1}{2}, \frac{1}{2\sqrt{2}}, \frac{1}{2\sqrt{2}}, 0)$ we have that $Tr[\rho_r^2] = \frac{1}{2}$ but $\rho_r^2 \neq \frac{1}{2}\rho_r$. However, it follows from our proof of the separability conditions (7.14) and (7.18) that either of the relations $Tr[\rho_r^2] = \frac{1}{N}$ or $S[\rho_r] = \log N$, *together with the additional information that the single particle statistical operator $\rho_r$ comes from an $N$-fermion pure state*, guarantee that equation (7.19) is verified (since in that case we have an equality in equation (7.12) and the global state must have Slater rank 1, implying that the only possible values for the eigenvalues of $\rho_r$ are $1/N$ and 0). In other words, *in the special case of statistical operators $\rho_r$ that are reduced single particle matrices arising from an $N$-fermion state* we have the double implication

$$Tr\left(\rho_r^2\right) = \frac{1}{N} \iff \rho_r^2 = \frac{1}{N}\rho_r. \tag{7.20}$$

Consequently, and contrary to some current beliefs, equation (7.19) does provide a necessary and sufficient criterion for separability of $N$-fermion states.

### 7.1.2 Entanglement measures

On the light of the separability criteria (7.14) and (7.18) it is reasonable to regard the differences

$$\begin{aligned} \mathcal{E}_L &= \frac{1}{N} - Tr\left(\rho_r^2\right) \\ \mathcal{E}_{VN} &= S[\rho_r] - \log N, \end{aligned} \tag{7.21}$$

as measures of the amount of entanglement exhibited by a pure state of a system of $N$ identical fermions. The quantities (7.21) have already been proposed as measures of entanglement for fermions (particularly for two-fermion systems, see the excellent review [180] on entanglement in many-particle systems) but our present results lend considerable further support to that proposal, because we now know with certainty that the measures (7.21) are non-negative quantities that vanish if and only if the fermionic pure state under consideration is separable. In the particular case of systems of two fermions with $D = 4$ the quantity $4\mathcal{E}_L$ reduces to the entanglement measure (usually referred to as squared concurrence) studied in [193] (see also [202]).

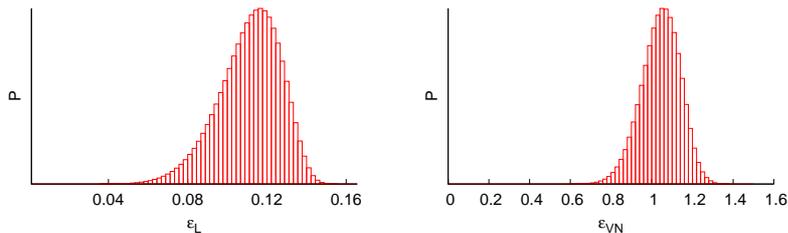

FIGURE 7.1: Probability distribution of the entanglement values for the linear entropy (left) and the Von Neumann entropy (right) measures for three particles in a six dimensional single particle space.

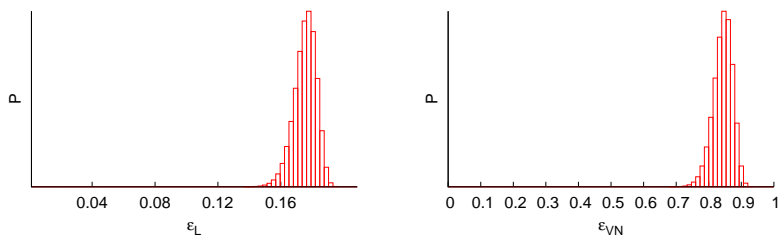

FIGURE 7.2: Probability distribution of the entanglement values for the linear entropy (left) and the Von Neumann entropy (right) measures for three particles in a eight dimensional single particle space.

In order to explore the typical values adopted by these measures I have generated random quantum states and made histograms with the probability of appearance for a entanglement value. The results corresponding to a system with three particles are shown in Figures 7.1 and 7.2.

As can be seen in in Figures 7.1 and 7.2, the probability distributions of both entanglement measures are qualitatively very similar. Both figures indicate that the probability of finding separable states decreases with the dimension of the single- particle Hilbert space. Consistently with this trend, it is also observed it is also observed that the average entanglement exhibited by random quantum states also increases with the single particle Hilbert space dimension.

As an analytical example I have selected the following states [206] of a three fermions state with single-particle dimension equal to six.

$$\begin{aligned}|\Psi\rangle &= \frac{1}{\sqrt{3}}\left(\sqrt{2}|135\rangle + |246\rangle\right) \\ |\Phi\rangle &= \frac{1}{\sqrt{3}}\left(|123\rangle + |345\rangle + |156\rangle\right),\end{aligned} \qquad (7.22)$$

where the ket $|ijk\rangle$ means the Slater determinant

$$|ijk\rangle = \frac{1}{\sqrt{6}}\begin{vmatrix} |i\rangle & |i\rangle & |i\rangle \\ |j\rangle & |j\rangle & |j\rangle \\ |k\rangle & |k\rangle & |k\rangle \end{vmatrix}, \qquad (7.23)$$

where $|i\rangle$, $|j\rangle$, $|k\rangle$ are normalized and orthogonal single particle states.

The reduced density matrix corresponding with these states is

$$\rho_r(\psi) = \rho_r(\phi) = \frac{1}{9}\begin{pmatrix} 2 & 0 & 0 & 0 & 0 & 0 \\ 0 & 1 & 0 & 0 & 0 & 0 \\ 0 & 0 & 2 & 0 & 0 & 0 \\ 0 & 0 & 0 & 1 & 0 & 0 \\ 0 & 0 & 0 & 0 & 2 & 0 \\ 0 & 0 & 0 & 0 & 0 & 1 \end{pmatrix}. \qquad (7.24)$$

So the entanglement measures (7.21) have the values

$$\mathcal{E}_L = \frac{1}{3} - Tr\left(\rho_r^2\right) = 0.148 \qquad (7.25)$$
$$\mathcal{E}_{VN} = S[\rho_r] - \log 3 = 0.636 \qquad (7.26)$$

and if these measures are normalized to unity (it means making the maximum value equal to 1) the results are

$$\mathcal{E}_L^{\text{norm}} = 0.889 \qquad (7.27)$$
$$\mathcal{E}_{VN}^{\text{norm}} = 0.918, \qquad (7.28)$$

so both measures shows that these states have a high amount of entanglement.

## 7.2 Quantum entanglement in two-electron atomic models

Quantum entanglement in two-electron systems has attracted the attention of several researchers [194, 195, 200, 201, 213–215]. Interesting results concerning the entanglement-related features of the eigenstates of a one-dimensional atomic model with Coulomb-like interactions [195], the ground state of the Hooke system [213], the spherically averaged helium-like model near the ionization threshold [215], and quantum dots with a step-like confining potential well [214] have been recently reported. However, the systematic characterization of the entanglement properties of the eigenstates of two-electron systems, particularly in the case of excited states, remains largely an open problem. The aim of the present Section is to investigate the entanglement-related properties of the energy eigenstates of two exactly soluble systems composed by two charged fermions: the Crandall [216] and the Hooke [213, 217] atomic models. These are three-dimensional atomic models consisting of two identical, spin-$\frac{1}{2}$ fermions ("electrons") in an external harmonic confining potential, with an electron-electron interaction potential having the $r^{-2}$ form (in the case of the Crandall atom) and the standard Coulomb form in the case of the Hooke atom. Exactly soluble atomic models provide valuable foil systems where some fundamental aspects of atomic physics, such as the basic entanglement features exhibited by atomic states, can be studied in detail. The information gained about the eigenstates of this kind of toy models can be used as a valuable guide to interpret the properties exhibited by more realistic systems, as well as to develop useful approximation techniques to treat them [213, 218]. Some results related to the entanglement properties of a soluble two-electron system have been already obtained for the Moshinsky model, both for the ground state [194, 200] and for excited states [201]. However, the Moshinsky system is a very special system, because the interaction between the two constituent particles (the "electrons") is harmonic. Here we consider the entanglement properties of the ground and the first few excited states of the two aforementioned models. Furthermore, we compute numerically the entanglement of the ground and first excited states of Helium-like atoms (using high-quality eigenfunctions of the Kinoshita type [219]) and investigate its dependence on both the states' energy and the nuclear charge.

The Section is organized as follows. In Subsection 7.2.1 we provide a brief discussion of quantum entanglement in systems of two identical fermions. The Crandall and Hooke atomic models are reviewed in Subsection 7.2.2. Then, in Subsections 7.2.3 and 7.2.4, the entanglement properties of the eigenstates of these model atoms are investigated. The entanglement features of the ground and first excited states of Helium-like atoms are considered in Subsection 7.2.5.

### 7.2.1 Quantum entanglement in systems of two identical fermions

The Schmidt decomposition of pure states of systems constituted by two identical fermions leads to a natural and physically sensible measure of the amount of entanglement exhibited by these states [208]. Given a pure state $|\Phi\rangle$ of two identical fermions there always exists an orthonormal basis $\{|i\rangle, \ i = 0, 1, \ldots\}$ of the single-particle Hilbert space such that $|\Phi\rangle$ can be cast as

$$|\Phi\rangle \,=\, \sum_i \sqrt{\frac{\lambda_i}{2}} \, \Big(|2i\rangle|2i+1\rangle - |2i+1\rangle|2i\rangle\Big), \tag{7.29}$$

where the Schmidt coefficients $\lambda_i$ verify $0 \leq \lambda_i \leq 1$ and $\sum_i \lambda_i = 1$ (if the single-particle Hilbert space has a finite dimension $N$, we assume that $N$ is even and that the sums on the index $i$ go from $i = 0$ to $i = N/2$). A useful measure of the amount of entanglement exhibited by the pure state $|\Phi\rangle$ is [69, 208]

$$\mathcal{E}(|\Phi\rangle) \,=\, 1 - \sum_i \lambda_i^2 \,=\, 1 - 2Tr(\rho_1^2), \tag{7.30}$$

where $\rho_1 = Tr_2(|\Phi\rangle\langle\Phi|)$ is the single-particle reduced density matrix obtained after tracing the two-particle density matrix $\rho = |\Phi\rangle\langle\Phi|$ over one of the particles. The entanglement measure (7.30) has been recently applied to the analysis of electron-electron scattering processes [220] and to the study of entanglement-related aspects of quantum brachistochrone evolutions [202].

According to the entanglement measure (7.30), correlations between the two fermions that are due solely to the antisymmetric character of the global two-particle state do not contribute to the entanglement of the state. Indeed, the amount of entanglement of a two-fermion state is associated with the quantum correlations exhibited by the state on top of the minimum correlations required by the antisymmetry of the global wave function [69, 180, 193, 202, 204, 205, 208, 220]. For example, in the case of a two-fermion state whose wave function can be expressed as a single Slater determinant one of the Schmidt coefficients is equal to 1 and the rest are equal to zero. It is clear from (7.30) that such a state has no entanglement. In fact, there are profound physical arguments indicating that two-fermion states represented by a single Slater determinant must be regarded as non-entangled [69, 180, 193, 202, 204, 205, 208, 209, 220]. First, the correlations exhibited by such states are not useful as a resource to perform non-classical information transmission or information processing tasks [193]. Second, the non-entangled character of states that can be represented as one Slater determinant is consistent with the possibility of assigning complete sets of properties to both parts of the composite system [204, 205].

Let us now consider the application of the above measure to a pure state of a two-electron system. For our present purposes it is sufficient to consider states described by wave functions of the type

$$\Phi = \Psi(\vec{r}_1, \vec{r}_2) \chi(\sigma_1, \sigma_2), \tag{7.31}$$

where the global wave function $\Phi$ can be factorized as the product of a coordinate wave function $\Psi(\vec{r}_1, \vec{r}_2)$ and a spin wave function $\chi(\sigma_1, \sigma_2)$, $\vec{r}_1$ and $\vec{r}_2$ being the vector positions of the two electrons. The density matrix corresponding to a wave function of the type (7.31) has the form

$$\rho = \rho^{(\text{coord.})} \otimes \rho^{(\text{spin})}, \tag{7.32}$$

where the matrix elements of $\rho^{(\text{coord.})}$ are

$$\langle \vec{r}_1{}', \vec{r}_2{}' | \rho^{(\text{coord.})} | \vec{r}_1, \vec{r}_2 \rangle = \Psi(\vec{r}_1{}', \vec{r}_2{}') \Psi^*(\vec{r}_1, \vec{r}_2). \tag{7.33}$$

To evaluate the entanglement measure (7.30) on a state with the wave function (7.31) we have to consider separately the cases of a spin wave function describing parallel spins or antiparallel spins. If we have parallel spins (that is, if the coordinate wave function is antisymmetric and the spin wave function is either $\chi_{++}$ or $\chi_{--}$), the entanglement measure (7.30) corresponding to a two-electron state of the form (7.31) reduces to

$$\mathcal{E}(|\Phi\rangle) = 1 - 2 \int |\langle \vec{r}_1{}' | \rho_r | \vec{r}_1 \rangle|^2 \, d\vec{r}_1{}' d\vec{r}_1, \tag{7.34}$$

On the other hand, if we have anti-parallel spins (that is, if the coordinate wave function is symmetric and the spin wave function is $\frac{1}{\sqrt{2}}(\chi_{+-} - \chi_{-+})$ or, alternatively, if the coordinate wave function is antisymmetric and the spin wave function is $\frac{1}{\sqrt{2}}(\chi_{+-} + \chi_{-+})$) the amount of entanglement is given by

$$\mathcal{E}(|\Phi\rangle) = 1 - \int |\langle \vec{r}_1{}' | \rho_r | \vec{r}_1 \rangle|^2 \, d\vec{r}_1{}' d\vec{r}_1. \tag{7.35}$$

In equations (7.34) and (7.35) we have

$$\langle \vec{r}_1{}' | \rho_r | \vec{r}_1 \rangle = \int \Psi(\vec{r}_1{}', \vec{r}_2) \Psi^*(\vec{r}_1, \vec{r}_2) \, d\vec{r}_2. \tag{7.36}$$

Notice that a two-electron state with a wave function of the form

$$\frac{1}{\sqrt{2}} \left[ \psi_1(\vec{r}_1)\psi_2(\vec{r}_2) - \psi_2(\vec{r}_1)\psi_1(\vec{r}_2) \right] \chi_{kk}, \quad k = \pm, \tag{7.37}$$

with $\psi_1(\vec{r})$ and $\psi_2(\vec{r})$ orthogonal, normalized single-particle (coordinate) wave functions, has zero entanglement. This example illustrates an important point already mentioned. The wave function (7.37) is a Slater determinant. The associated correlations between the two electrons, due entirely to the anti-symmetry requirement on the fermionic state, do not contribute to the entanglement of the state.

### 7.2.2 The Crandall and Hooke atoms

**The Crandall atom**

The Crandall atom is a two "electron" model with an harmonic confining potential and an inverse cubic electron-electron repulsion force [216]. The total Hamiltonian of the system is

$$H = -\frac{1}{2}\left(\nabla_1^2 + \nabla_2^2\right) + \frac{1}{2}\omega^2\left(r_1^2 + r_2^2\right) + \frac{\lambda}{r_{12}^2} \tag{7.38}$$

where $\vec{r}_1$ and $\vec{r}_2$ are the vector positions of the two particles, $r_{12} = |\vec{r}_1 - \vec{r}_2|$, $\omega$ is the natural frequency of the external harmonic field, and $\lambda$ is the interaction parameter. We have used atomic units ($m = \hbar = 1$) throughout the paper. Introducing the new variables $\vec{u}$ and $\vec{v}$ [216],

$$\vec{u} = \frac{1}{\sqrt{2}}(\vec{r}_1 + \vec{r}_2), \qquad \vec{v} = \frac{1}{\sqrt{2}}(\vec{r}_1 - \vec{r}_2), \tag{7.39}$$

the Hamiltonian separates as,

$$H = H_{\vec{u}} + H_{\vec{v}} = -\frac{1}{2}\nabla_{\vec{u}}^2 + \frac{1}{2}\omega^2 u^2 - \frac{1}{2}\nabla_{\vec{v}}^2 + \frac{1}{2}\omega^2 v^2 + \frac{\lambda}{2v^2}, \tag{7.40}$$

admitting the factorized eigenfunctions

$$\Psi(\vec{r}_1, \vec{r}_2) = \Psi(\vec{u}, \vec{v}) = U_{n_2 l_2 m_2}(\vec{u}) V_{n_1 l_1 m_1}(\vec{v}), \tag{7.41}$$

with

$$U_{n_2 l_2 m_2}(\vec{u}) = e^{-\frac{\omega u^2}{2}} u^{l_2} L_{n_2}^{(l_2 + \frac{1}{2})}(\omega u^2) Y_{l_2 m_2}(\theta_u, \phi_u), \tag{7.42}$$

and

$$V_{n_1 l_1 m_1}(\vec{v}) = e^{-\frac{\omega v^2}{2}} v^a L_{n_1}^{(a + \frac{1}{2})}(\omega v^2) Y_{l_1 m_1}(\theta_v, \phi_v), \tag{7.43}$$

where $L_n^{(\alpha)}(x)$ denote the Laguerre polynomials and $a = \frac{1}{2}\left[\sqrt{1 + 4\lambda + 4l(l+1)} - 1\right]$. The variables $u, \theta_u, \phi_u, v, \theta_v, \phi_v$ are the spherical coordinates associated with the vectors $(\vec{u}, \vec{v})$. We will denote by $|n_1 l_1 m_1 n_2 l_2 m_2\rangle_{\vec{u},\vec{v}}$ the (spatial) eigenfunctions of the Hamiltonian (7.38), which are characterized by the quantum numbers $n_1$, $l_1$, $m_1$, $n_2$, $l_2$ and $m_2$ (to fully define the eigenstates of the two-electron system we have to specify also the spin wave function $\xi(\sigma_1, \sigma_2)$). The above quantum numbers adopt the values

$$n_1, l_1 = 0, 1, 2, 3, ... \quad m_1 = -l_1, ..., l_1,$$
$$n_2, l_2 = 0, 1, 2, 3, ... \quad m_2 = -l_2, ..., l_2, \tag{7.44}$$

and the corresponding eigenenergies are [216]

$$E = \frac{\omega}{2} \left\{ 5 + 4n_2 + 4n_1 + 2l_1 + [1 + 4\lambda + 4l_1(l_1 + 1)]^{\frac{1}{2}} \right\}. \tag{7.45}$$

All the (coordinate) wave functions $|n_1 l_1 m_1 n_2 l_2 m_2\rangle_{\vec{u},\vec{v}}$ have definite parity, which is determined by the quantum number $l_2$: even values of $l_2$ correspond to symmetric coordinate eigenfunctions and odd values of $l_2$ to antisymmetric ones. A final remark concerning our notation is in order. A cursory glance at the ket $|n_1 l_1 m_1 n_2 l_2 m_2\rangle_{\vec{u},\vec{v}}$ may suggest that it represents a separable state. However, in general, it represents an entangled state of the two electron system.

**The Hooke atom**

The Hooke atom is a two-electron atomic model with harmonic confining potential and Coulombic electron-electron repulsion force. The total Hamiltonian of the system is

$$H = -\frac{1}{2}\left(\nabla_1^2 + \nabla_2^2\right) + \frac{1}{2}\omega^2\left(r_1^2 + r_2^2\right) + \frac{1}{r_{12}}$$

where $\vec{r}_1$ and $\vec{r}_2$ are the coordinates of the two particles, $r_{12} = |\vec{r}_1 - \vec{r}_2|$ and $\omega$ is the natural frequency of the external harmonic field. Introducing the centre of mass and the relative position vectors [217]

$$\vec{R} = \frac{1}{2}(\vec{r}_1 + \vec{r}_2) \qquad \vec{r} = \vec{r}_1 - \vec{r}_2, \tag{7.46}$$

the Hamiltonian separates as follows

$$H = H_{\vec{R}} + H_{\vec{r}} = -\frac{1}{4}\nabla_{\vec{R}}^2 + \omega^2 \vec{R}^2 - \nabla_{\vec{r}}^2 + \frac{1}{4}\omega^2 r^2 + \frac{1}{r}. \tag{7.47}$$

The eigenfunctions of (7.47) can be factorized as

$$\Psi(\vec{r}_1, \vec{r}_2) = \Psi(\vec{R}, \vec{r}) = \psi_{n_1 l_1 m_1}(\vec{R}) \phi_{n_2 l_2 m_2}(\vec{r}), \tag{7.48}$$

leading to the pair of eigenvalue equations,

$$\left[-\frac{1}{2}\nabla_R^2 + \frac{1}{2}\omega_R^2 R^2\right]\psi(\vec{R}) = \eta'\psi(\vec{R}), \tag{7.49}$$

and

$$\left[-\frac{1}{2}\nabla_r^2 + \frac{1}{2}\omega_r^2 r^2 + \frac{1}{2r}\right]\phi(\vec{r}) = \epsilon'\phi(\vec{r}), \tag{7.50}$$

with $\omega_R = 2\omega$ and $\omega_r = \frac{1}{2}\omega$. The total energy of the eigenstate (7.48) is then $E = \eta + \epsilon$, where $\eta = \frac{1}{2}\eta'$ and $\epsilon = 2\epsilon'$.

Equation (7.49) is the eigenvalue equation corresponding to a three-dimensional, isotropic quantum harmonic oscillator, with well-known solutions of the form

$$\psi(\vec{R}) = N_{n_1 l_1} R^l e^{-\frac{\omega_R^2 R^2}{2}} L_{n_1}^{(l_1+1/2)}(\omega_R R^2) Y_{l_1 m_1}(\theta_R, \phi_R), \tag{7.51}$$

with

$$N_{n_1,l_1} = \left(\left(\frac{\omega_R^3}{4\pi}\right)^{\frac{1}{2}} \frac{2^{n_1+2l_1+3} n_1! \left(\frac{\omega_R}{2}\right)^{l_1}}{(2n_1 + 2l_1 + 1)!!}\right)^{\frac{1}{2}} \tag{7.52}$$

and

$$\eta' = \omega_R \left(2n_1 + l_1 + \frac{3}{2}\right). \tag{7.53}$$

On the other hand, the eigenvalue equation (7.50) admits a closed analytical solution only for certain particular states, each of them requiring a separate treatment. These analytical solutions can be determined by recourse to a power-series expansion with a three step recurrence in the quantum number $n_2$. The lowest energy state that can be calculated by this method corresponds to $n_2 = 2$. For $n_2 = 2$ and arbitrary $l_2, m_2$ the wavefunction is given by

$$\phi(\vec{r}) = k_2 r^{l_2} e^{-\frac{r^2}{8(l_2+1)}} \left(1 + \frac{r}{2(l_2+1)}\right) Y_{l_2 m_2}(\theta_r, \phi_r) \tag{7.54}$$

where

$$k_2 = \left[2^{1+2l_2}(1+l_2)^{l_2}\left(\sqrt{l_2+1}(5+4l_2)\Gamma\left(\frac{3}{2}+l_2\right) + 4(l_2+1)\Gamma(2+l_2)\right)\right]^{-\frac{1}{2}} \tag{7.55}$$

The concomitant eigenvalue is

$$\epsilon' = \frac{2l_2 + 5}{8(l_2 + 1)}. \tag{7.56}$$

By recourse to the solution (7.54) we can build the full wavefunctions for the states $(n_1, l_1, m_1, n_2 = 2, l_2, m_2)$ with $\omega = \frac{1}{2(l_2+1)}$. For $n_2 = 3$ and arbitrary $l_2, m_2$, one has

that

$$\phi(\vec{r}) = k_3 r^{l_2} e^{-\frac{r^2}{8(4l_2+5)}} \left(1 + \frac{r}{2(l_2+1)} + \frac{r^2}{4(l_2+1)(4l_2+5)}\right) Y_{l_2 m_2}(\theta_r, \phi_r) \quad (7.57)$$

with

$$k_3 = \frac{1}{16} \left[ \frac{2^{7+2l_2}(3+2l_2)(5+4l_2)^{2+l_2}\Gamma(1+l_2)}{1+l_2} + \frac{4^{2+l_2}(5+4l_2)^{3/2+l_2}(61+88l_2+32l_2^2)\Gamma(3/2+l_2)}{(1+l_2)^2} \right] \quad (7.58)$$

The associated eigenvalue is

$$\epsilon' = \frac{2l_2+7}{8(4l_2+5)}. \quad (7.59)$$

Using (7.57) we can build the complete wavefunctions for the states $(n_1, l_1, m_1, n_2 = 3, l_2, m_2)$ with $\omega = \frac{1}{2(4l_2+5)}$. Heretoforth we will denote by $|n_1 l_1 m_1 n_2 l_2 m_2\rangle_{\vec{r},\vec{R}}$ the eigenfunctions of the Hamiltonian (7.46), which are characterized by the quantum numbers $n_1$, $l_1$, $m_1$, $n_2$, $l_2$ and $m_2$. To fully define the eigenstates of the two-electron system we have to specify, of course, also the spin wave function $\chi(\sigma_1, \sigma_2)$.

### 7.2.3 Entanglement in the Crandall atom

The integrals appearing in equations (7.34-7.36), that have to be computed in order to evaluate the amount of entanglement of the eigenstates, cannot be computed analytically for general eigenstates of the Crandall model. We have evaluated these integrals by recourse to the MonteCarlo method. The main results obtained are summarized in Figures 7.3, 7.4, 7.5 and 7.6.

We encounter two general trends. First, the entanglement increases monotonically with the parameter $\lambda$ and, consequently, with the strength of the interaction between the particles. For high enough values of $\lambda$ the entanglement approaches its maximum value $\mathcal{E} = 1$. Second, the amount of entanglement also tends to increase when we consider higher excited states (that is, it increases with the energy).

Another interesting feature observed in Figures 7.3 and 7.4 is that the entanglement exhibited by excited eigenstates does not necessarily go to zero in the limit $\lambda \to 0$. In other words, *for an arbitrarily weak (but finite) interaction there already are excited eigenstates exhibiting a considerable amount of entanglement*. In the non-interacting case corresponding to $\lambda = 0$ these states have degenerate eigenenergies and the degeneracy enables one to construct an alternative set of non-entangled eigenstates sharing the same energy. However, when $\lambda > 0$ the interaction lifts the degeneracy and the aforementioned eigenstates become necessarily entangled. It is worth stressing that the finite amount of entanglement corresponding to the limit $\lambda \to 0$ is not due to the correlations arising exclusively from the antisymmetric nature of the (global) fermionic states. As already mentioned, these correlations do not contribute to the entanglement of the state. As can be appreciated in Figure 7.6, in the limit $\lambda \to 0$ the amount of entanglement tends to increase with the energy of the eigenstates.

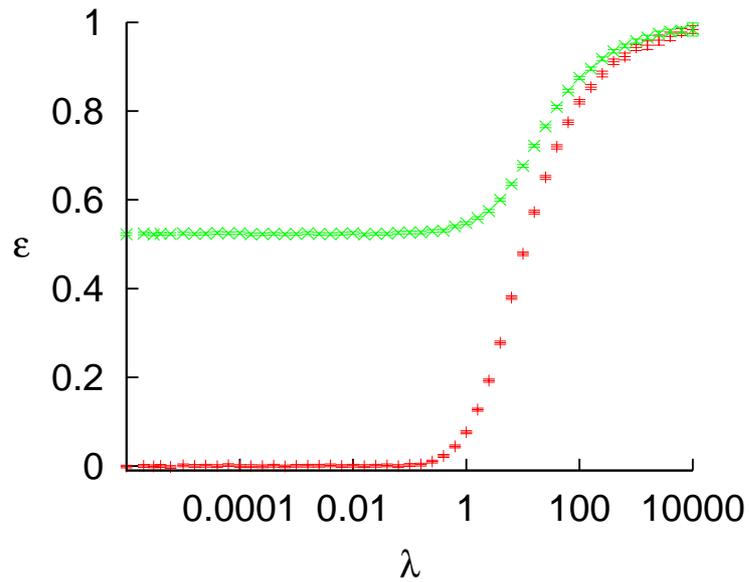

FIGURE 7.3: Entanglement of the ground (+) and first excited state (×), with antiparallel spins of the Crandall atom as a function of the parameter $\lambda$. Atomic units are used.

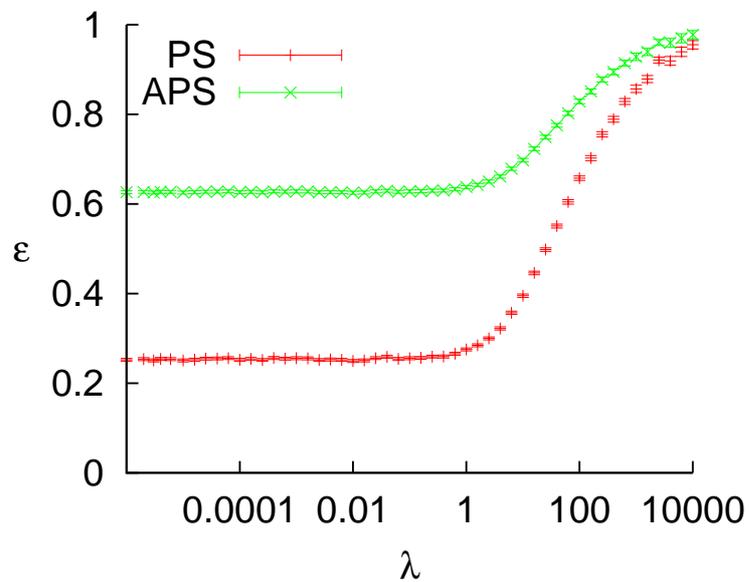

FIGURE 7.4: Entanglement of the ($n = l = 1$ and all the other quantum numbers equal to 0)-eigenstates, with parallel (PS) and anti-parallel spins (APS) of the Crandall atom as a function of the parameter $\lambda$. Atomic units are used.

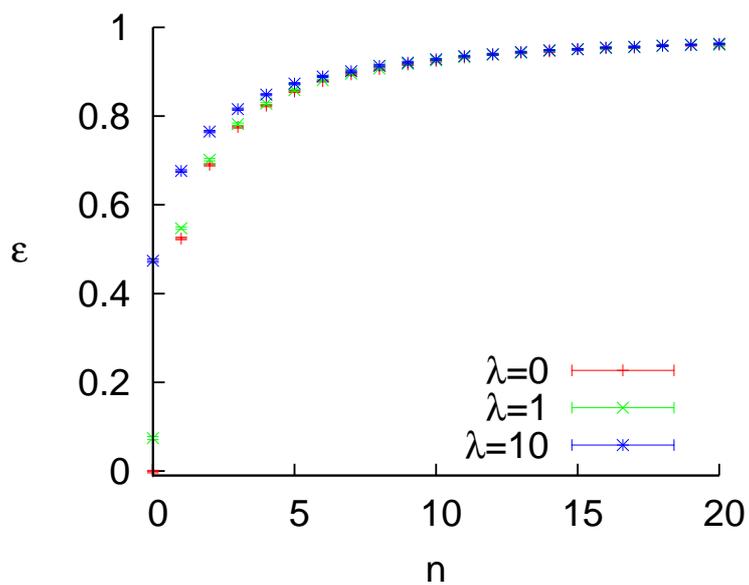

FIGURE 7.5: Entanglement as a function of the quantum number $n$ for the states with all the other quantum numbers equal to zero, at different values of the parameter $\lambda$. Atomic units are used.

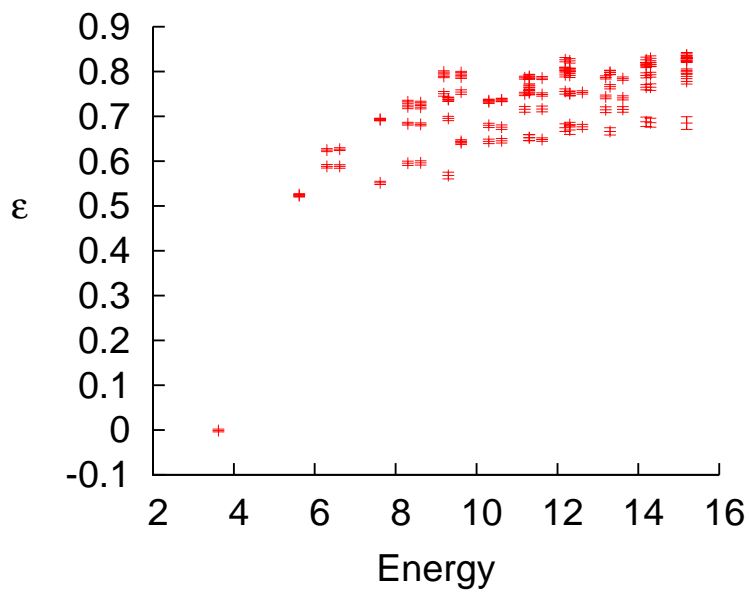

FIGURE 7.6: Entanglement for the ground state and the several excited states as a function of the energy of the system with the parameter $\lambda$ arbitrarily small. Atomic units are used.

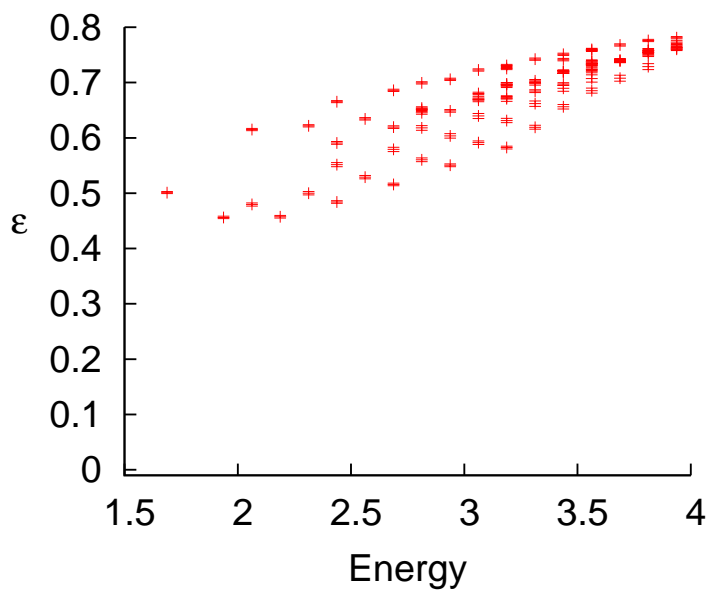

FIGURE 7.7: Entanglement for states with $n_2 = 2$, $l2 = 0$, $m_2 = 0$, $\omega = 0.5$ and arbitrary $n_1$, $l_1$, $m_1$ as a function of the energy of the system. Atomic units are used.

### 7.2.4 Entanglement in the Hooke atom

As in the case of the Crandall model, the integrals (7.34)-(7.36) needed to determine the amount of entanglement of the system's eigenstates do not admit analytical treatment and we evaluated them using a MonteCarlo approach similar to the one used in the calculations for the Crandall model. The entanglement properties exhibited by the eigenstates of the Hooke atom are similar to those characterizing the eigenstates of the Crandall model.

The entanglement of the eigenstates of the Hooke atom tends to increase with the eigenstates' energy, similarly as the Crandall model. That can be appreciated in Figure 7.7. In this Figure we depict the entanglement versus the total energy associated with eigenstates of the Hooke atom with $\omega = \frac{1}{2}$. The quantum numbers characterizing the states represented are $n_2 = 2$, $l_2 = 0$, $m_2 = 0$, $n_1 \in [0, 7]$, $l_1 \in [0, n_1 - 1]$, $m_1 \in [0, n_1 - 1]$.

### 7.2.5 Entanglement in helium-like atoms

It is interesting to explore to what extent the main entanglement features characterizing the exactly soluble models of Crandall and Hooke are also observed in systems whose confining potential is not harmonic. As a first step in this direction we are now going to compute the entanglement corresponding to the ground and first excited states of Helium-like atoms by means of the high-quality eigenfunctions of the Kinoshita type obtained by Koga [219].

| Z | state | energy |
|---|---|---|
| 1 | GS | -0.5277510165226 |
| 2 | GS | -2.903724377032 |
| 2 | 1s2s, $2\,^3S$ | -2.175229378225 |
| 2 | 1s2s, $2\,^1S$ | -2.145974045970 |
| 3 | GS | -7.279913412667 |
| 4 | GS | -13.65556623841 |
| 5 | GS | -22.03097158023 |

TABLE 7.1: Energies for the wavefunction of Helium-like atoms. GS means the $1s^2\,^1S$ ground state. Atomic units are used

The Hamiltonian of a Helium-like atom (in atomic units) reads

$$H = -\frac{1}{2}\nabla_1^2 - \frac{1}{2}\nabla_2^2 - \frac{Z}{r_1} - \frac{Z}{r_2} + \frac{1}{r_{12}}, \tag{7.60}$$

where $Z$ denotes the nuclear charge. The aforementioned eigenfunctions for Helium-like systems are represented by the following Kinoshita-type ansatz with half-integer powers [219],

$$\Psi_N = e^{-\xi s} \sum_{i=1}^{N} c_i s^{\frac{l_i}{2}} \left(\frac{t}{u}\right)^{m_i} \left(\frac{u}{s}\right)^{\frac{n_i}{2}}, \tag{7.61}$$

where $s$, $t$, and $u$ stand for the Hylleraas coordinates given by

$$s = |\vec{r_1}| + |\vec{r_2}| \qquad t = |\vec{r_1}| - |\vec{r_2}| \qquad u = |\vec{r_1} - \vec{r_2}|$$
$$s \in [0, \infty], \ u \in [0, s], \ t \in [-u, u]. \tag{7.62}$$

The optimization of the exponent $\xi$, the coefficients $c_i$ and the powers $\{l_i, m_i, n_i\}$ in the eigenfunctions given by equation (7.61) with $N = 100$ terms lead to the energies listed in Table 7.1.

The main results obtained here concerning the entanglement related features of Helium-like atoms are summarized in Figures 7.8 and 7.9.

The ground state and the excited state $2\,^1S$ have symmetric spatial wave function. Consequently, the spin part of the wave function is antisymmetric (singlet) and the entanglement of the state is given by equation (7.35). On the other hand, the excited state $2\,^3S$ has an anti-symmetric spatial wavefunction. In this case, the entanglement depicted in Figure 7.8 correspond to the member of the spin triplet having anti-parallel spins (that is, the spin wavefunction is $\frac{1}{\sqrt{2}}\left(\chi_{+-}+\chi_{-+}\right)$) and the state's entanglement is again given by (7.35).

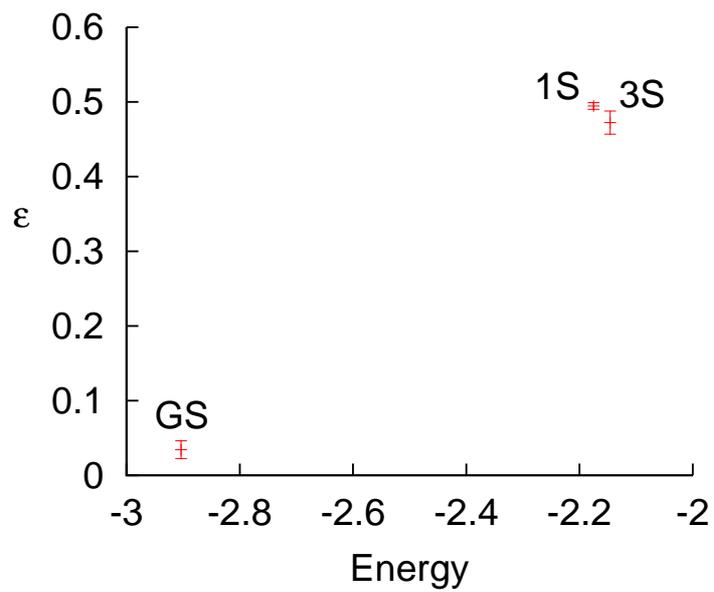

FIGURE 7.8: Entanglement of the ground and two excited states of the Helium atom as a function of the energy. Atomic units are used.

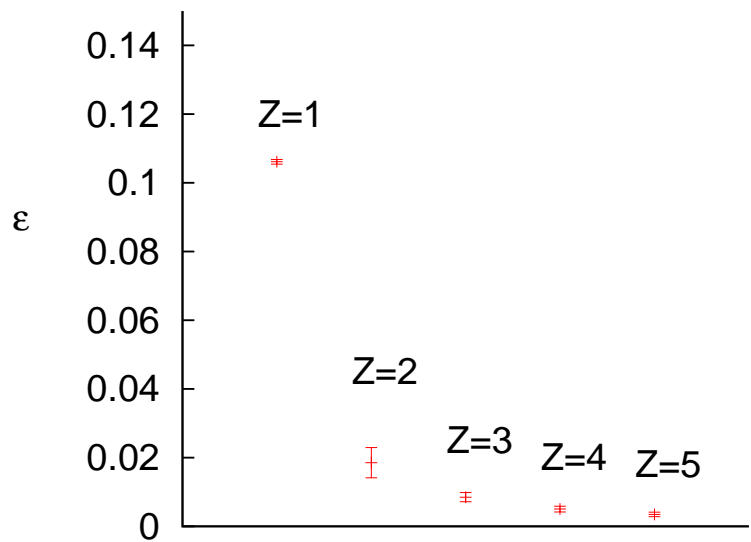

FIGURE 7.9: Entanglement of the ground states of the Helium atom for different values of the nuclear charge $Z$. Atomic units are used.

Our findings for the Helium-like atoms are fully consistent with the entanglement properties of the Crandall and Hooke models that were discussed in the previous Subsections. Indeed, the data depicted in Figure 7.8 suggest that the entanglement of the Helium eigenstates tends to increase with energy. On the other hand, Figure 7.9 clearly shows that the entanglement of the ground state of Helium-like systems decreases monotonically with the nuclear charge $Z$. This last parameter determines the strength of the nuclear Coulomb field, while the strength of the electron-electron interaction is constant. Consequently, the behaviour observed in Figure 7.9 can be construed as indicating that the system becomes more entangled when the relative strength of the electron-electron interaction (as compared with the nuclear-electron interaction) increases. This behaviour is similar to the ones exhibited by both the Crandall and Hooke atomic models.

## 7.3 Separability criteria for continuous quantum systems

In this section I am going to investigate some features of two separability criteria for continuous systems recently advanced by Walborn et al [1]. These criteria can be implemented experimentally. From the experimental point of view, they offers some advantages over alternative approaches to the problem of entanglement detection in continuous systems.

Consider a bipartite quantum system with a configuration space characterized by two continuous variables $x_1$ and $x_2$ (which correspond to the two subsystems). A pure state of this system can be described by a wave function $\Psi(x_1, x_2)$, while a mixed state is described by a density matrix with elements $\rho(x_1, x_2; x'_1 x'_2)$. The separability criteria advanced by Walborn et al are based upon entropic inequalities of the form

$$H[R_\pm] + H[S_\mp] \geq c, \tag{7.63}$$

verified by separable states of the system. Consequently, states violating (7.63) must be entangled. In Eq. (7.63), $H[R_\pm]$ and $H[S_\pm]$ denote the Shannon entropies of the probability densities $R_\pm$ and $S\pm$ respectively associated with the measurement of the variables

$$\begin{aligned} r_\pm &= r_1 \pm r_2, \quad r_j = x_j \cos\theta_j + p_j \sin\theta_j \\ s_\pm &= s_1 \pm s_2, \quad s_j = p_j \cos\theta_j - x_j \sin\theta_j, \end{aligned} \tag{7.64}$$

where $p_{1,2}$ are the momenta canonically conjugate to the coordinates $x_{1,2}$ (that is, $[x_i, p_j] = i\delta_{ij}$) and $\theta_{1,2}$ are appropriate angles.

Walborn et al derived two inequalities like (7.63). One of these inequalities, the so called "strong" one, is satisfied by separable pure states, while the other inequality (the "weak" one) is verified by all separable states (pure or mixed). The criteria obtained by Walborn et al are potentially relevant for experimentalists, because they are less expensive to implement experimentally than alternative criteria involving general moments of the variables $r_i$ and $s_i$. Walborn et al only tested their criteria for some particular states. However, to really asses the experimental value of these criteria it is necessary to explore it more systematically. Of particular interest from the experimental point of view is the weak criterion, since in practical situations one has imperfect preparations of quantum states, that lead to some degree of mixedness.

### 7.3.1 Description of the criteria

**Preliminaries**

To implement the separability criteria that we consider in this section it is necessary to express the wave function (or density matrix) of the system in terms of the variables $r_\pm$ and $s_\pm$. To do that, we need to determine the eigenfunctions of operators of the form

$$\Lambda = \cos\theta\, x - i\sin\theta\frac{\partial}{\partial x}. \tag{7.65}$$

In the $x$-representation, the observables $r_{1,2}$ and $s_{1,2}$ are represented by operator of the form 7.65.

The associated eigenvalue equation is

$$\Lambda\Phi_\lambda = \lambda\Phi_\lambda \rightarrow \left[\cos\theta\, x - i\sin\theta\frac{\partial}{\partial x}\right]\Phi_\lambda(x) = \lambda\Phi_\lambda(x) \tag{7.66}$$

whose normalized solution reads

$$\Phi_\lambda(x) = \frac{1}{\sqrt{2\pi\sin\theta}}\exp\left[\frac{i}{\sin\theta}\left\{\lambda x - \left(\frac{x^2+\lambda^2}{2}\right)\cos\theta\right\}\right] \tag{7.67}$$

where $\lambda$ stands for the eigenvalue of the operator $\Lambda$. The continuous eigenvalue $\lambda$ corresponds to the variables $r_{1,2}$ or $s_{1,2}$.

Given a quantum state $|\Psi\rangle$, one can obtain its wavefunction $\Psi(\lambda) = \langle\Phi_\lambda|\Psi\rangle$, expressed in the "$\lambda$-representation", from its wave function by recourse to the Fourier-like transform. So,

$$\langle\Phi_\lambda|\Psi\rangle = \int_{-\infty}^{\infty}\langle\Phi_\lambda|x\rangle\langle x|\Psi\rangle dx = \int_{-\infty}^{\infty}\Phi_\lambda^*(x)\Psi(x)dx \tag{7.68}$$

In order to implement the transformation $\Psi(x) \to \Psi(\lambda)$ we first express $\Psi(x)$ as a linear combination of the eigenfunctions of the harmonic oscillator,

$$\langle x|\Psi_n\rangle = k_n \, e^{-\frac{x^2}{2}} H_n(x), \tag{7.69}$$

where $(\hbar = \omega = m = 1)$ and $k$ is the normalization constant

$$k_n = \sqrt{\frac{1}{2^n n! \sqrt{\pi}}}. \tag{7.70}$$

Then we perfom the transformation $\Psi_n(x) \to \Psi_n(\lambda)$ for each member of the harmonic oscillator basis,

$$\begin{aligned}\Psi_n(\lambda) = \langle \Phi_\lambda | \Psi_n \rangle &= k \int_{-\infty}^{\infty} e^{-\frac{x^2}{2}} H_n(x) \exp\left[\frac{-i}{\sin\theta}\left(\lambda x - \cos\theta\left(\frac{x^2+\lambda^2}{2}\right)\right)\right] dx = \\ &= \frac{k}{\sqrt{2\pi \sin\theta}} e^{-i\frac{\lambda^2}{2}\cot g\theta} \int_{\infty}^{\infty} H_n(x) e^{-i\frac{\lambda x}{\sin\theta}} e^{-\frac{x^2}{2}(1-i\cot g\theta)}.\end{aligned} \tag{7.71}$$

For different $n$'s the solutions are

$$\begin{aligned}\Psi_0(\lambda) &= k e^{-\frac{\lambda^2}{2}} e^{i\left(\frac{\theta}{2}-\frac{\pi}{4}\right)} \\ \Psi_1(\lambda) &= k e^{-\frac{\lambda^2}{2}} H_1(\lambda) e^{-i\theta} e^{i\left(\frac{\theta}{2}-\frac{\pi}{4}\right)} \\ \Psi_2(\lambda) &= k e^{-\frac{\lambda^2}{2}} H_2(\lambda) e^{-2i\theta} e^{i\left(\frac{\theta}{2}-\frac{\pi}{4}\right)}\end{aligned} \tag{7.72}$$

The general solution is:

$$\Psi_n(\lambda) = k e^{-\frac{\lambda^2}{2}} H_n(\lambda) e^{-in\theta} e^{i\left(\frac{\theta}{2}-\frac{\pi}{4}\right)} \tag{7.73}$$

The global phase is irrelevant, so we can take

$$\Psi_n(\lambda) = k e^{-\frac{\lambda^2}{2}} H_n(\lambda) e^{-in\theta} \tag{7.74}$$

**Algorithm for pure states**

This is the algorithm for applying the criteria for pure states

1. Expand the wavefunction in the harmonic oscillator basis.

$$\psi(x_1, x_2) = \sum_{n_1 n_2=0}^{D} k_{n1,n2} c_{n1,n2} e^{-\frac{x_1^2}{2}} e^{-\frac{x_2^2}{2}} H_{n_1}(x_1) H_{n_2}(x_2). \tag{7.75}$$

2. Determine the wavefunction in the representation defined by the variables

$$r_1 = \cos\theta_1 x_1 + \sin\theta_1 p_1$$
$$r_2 = \cos\theta_2 x_2 + \sin\theta_2 p_2 \qquad (7.76)$$

The result is

$$\psi(r_1, r_2) = \sum_{n_1 n_2 = 0}^{D} k_{n1,n2} c_{n1,n2} e^{-\frac{r_1^2}{2}} e^{-\frac{r_2^2}{2}} H_{n_1}(r_1) H_{n_2}(r_2) e^{-i(n_1\theta_1 + n_2\theta_2)} \qquad (7.77)$$

3. Change to the variables $(s_1, s_2)$ that are canonical conjugates of $(r_1, r_2)$. Obtain

$$\psi(s_1, s_2) = \sum_{n_1 n_2 = 0}^{D} k_{n1,n2} c_{n1,n2} e^{-\frac{s_1^2}{2}} e^{-\frac{s_2^2}{2}} H_{n_1}(s_1) H_{n_2}(s_2) e^{-i(n_1(\theta_1 + \frac{\pi}{2}) + n_2(\theta_2 + \frac{\pi}{2}))}.$$
$$(7.78)$$

4. Implement the change of variables

$$r_+ = r_1 + r_2$$
$$r_- = r_1 - r_2 \qquad (7.79)$$

where the Jacobian of the transformation is $J = 2$. The result is

$$\psi(r_+, r_-) = \sum_{n_1 n_2 = 0}^{D} k_{n1,n2} c_{n1,n2} e^{-\frac{r_+^2}{4}} e^{-\frac{r_-^2}{4}} H_n\left(\tfrac{1}{2}(r_+ + r_-)\right) H_n\left(\tfrac{1}{2}(r_+ - r_-)\right) \times$$
$$e^{-i(n_1\theta_1 + n_2\theta_2)}. \qquad (7.80)$$

5. Implement the change of variables

$$s_+ = s_1 + s_2$$
$$s_- = s_1 - s_2 \qquad (7.81)$$

$$\psi(s_+, s_-) = \sum_{n_1 n_2 = 0}^{D} k_{n1,n2} c_{n1,n2} e^{-\frac{s_+^2}{4}} e^{-\frac{s_-^2}{4}} H_n\left(\tfrac{1}{2}(s_+ + s_-)\right) H_n\left(\tfrac{1}{2}(s_+ - s_-)\right) \times$$
$$e^{-i(n_1(\theta_1 + \frac{\pi}{2}) + n_2(\theta_2 + \frac{\pi}{2}))} \qquad (7.82)$$

6. Calculate the probability density functions

$$R(r_+, r_-) = \tfrac{1}{2}|\Psi(r_+, r_-)|^2 \qquad R(r_1, r_2) = |\Psi(r_1, r_2)|^2$$
$$S(s_+, s_-) = \tfrac{1}{2}|\Psi(s_+, s_-)|^2 \qquad S(s_1, s_2) = |\Psi(s_1, s_2)|^2 \qquad (7.83)$$

7. Determine the associated marginal probability densities

$$R_+(r_+) = \int dr_- R(r_+, r_-)$$
$$R_-(r_-) = \int dr_+ R(r_+, r_-)$$
$$R_1(r_1) = \int dr_2 R(r_1, r_2)$$
$$R_2(r_2) = \int dr_1 R(r_1, r_2)$$
$$S_+(s_+) = \int ds_- R(s_+, s_-)$$
$$S_-(s_-) = \int ds_+ R(s_+, s_-)$$
$$S_1(s_1) = \int ds_2 R(s_1, s_2)$$
$$S_2(s_2) = \int ds_1 R(s_1, s_2) \tag{7.84}$$

8. Calculate the 8 entropies

$$H[R_+] = -\int dr_+ R_+(r_+) \log R_+(r_+) \qquad H[R_-] = -\int dr_- R_-(r_-) \log R_-(r_-)$$
$$H[S_+] = -\int ds_+ S_+(s_+) \log S_+(s_+) \qquad H[S_-] = -\int ds_- S_-(s_-) \log S_-(s_-)$$
$$H[R_1] = -\int dr_1 R_1(r_1) \log R_1(r_1) \qquad H[R_2] = -\int dr_2 R_2(r_2) \log R_2(r_2)$$
$$H[S_1] = -\int ds_1 S_1(s_1) \log S_1(s_1) \qquad H[S_2] = -\int ds_2 S_2(s_2) \log S_2(s_2)$$
$$\tag{7.85}$$

9. Check the 2 strong inequalities

$$H[R_+] + H[S_-] \geq \tfrac{1}{2} \log \left\{ e^{2H[R_1]+2H[S_1]} + e^{2H[R_2]+2H[S_2]} + e^{2H[R_1]+2H[S_2]} + e^{2H[R_2]+2H[S_1]} \right\}$$
$$H[R_-] + H[S_+] \geq \tfrac{1}{2} \log \left\{ e^{2H[R_1]+2H[S_1]} + e^{2H[R_2]+2H[S_2]} + e^{2H[R_1]+2H[S_2]} + e^{2H[R_2]+2H[S_1]} \right\}$$
$$\tag{7.86}$$

If either of these inequalities is not satisfied, the state under consideration is entangled.

10. Check the 2 weak inequalities

$$H[R_+] + H[S_-] \geq \log(2\pi e)$$
$$H[R_-] + H[S_+] \geq \log(2\pi e) \tag{7.87}$$

If either of these inequalities is not satisfied, the state under consideration is entangled.

**Algorithm for mixed states**

In the case of mixed states the procedure is basically the same, this time formulated in terms of the density matrix describing the state of the system.

We consider mixed states described by a statistical operator whose matrix elements $\langle n_1 n_2 | \rho | n_3 n_4 \rangle = \rho_{n_1 n_2 n_3 n_4}$ (expressed in the harmonic oscillator basis) are non-zero only if $n_i \leq D$.

Therefore, the density matrix of the system is of the form,

$$\hat{\rho} = \sum_{n_1 n_2 n_3 n_4 = 0}^{D} \rho_{n_1 n_2 n_3 n_4} |n_1 n_2\rangle \langle n_3 n_4|. \tag{7.88}$$

In the $x$-representation the statistical operator is

$$\rho(x_1, x_2; x_1', x_2') = \langle x_1 x_2 | \hat{\rho} | x_1' x_2' \rangle = \sum_{n_1 n_2 n_3 n_4=0}^{D} \rho_{n_1 n_2 n_3 n_4} \langle x_1 x_2 | n_1 n_2 \rangle \langle n_3 n_4 | x_1' x_2' \rangle =$$
$$= \sum_{n_1 n_2 n_3 n_4=0}^{D} \rho_{n_1 n_2 n_3 n_4} \Psi_{n_1}(x_1) \Psi_{n_2}(x_2) \Psi_{n_3}^*(x_1') \Psi_{n_4}^*(x_2'). \tag{7.89}$$

The probability density function in the $(x_1, x_2)$-space is given by the diagonal of the density matrix

$$p(x_1, x_2) = \rho(x_1, x_2; x_1, x_2) = \sum_{n_1 n_2 n_3 n_4=0}^{D} \rho_{n_1 n_2 n_3 n_4} \Psi_{n_1}(x_1) \Psi_{n_2}(x_2) \Psi_{n_3}^*(x_1) \Psi_{n_4}^*(x_2). \tag{7.90}$$

The probability densities corresponding to the variables $(r_1, r_2)$, $(r_+, r_-)$, $(s_1, s_2)$ and $(s_+, s_-)$ are,

$$p(r_1, r_2) = \sum_{n_1 n_2 n_3 n_4=0}^{D} \rho_{n_1 n_2 n_3 n_4} k_{n_1 n_2} k_{n_3 n_4} e^{r_1^2 + r_2^2} H_{n_1}(r_1) H_{n_2}(r_2) H_{n_3}(r_3) H_{n_4}(r_4) \times$$
$$e^{-i[(n_1 - n_3)\theta_1 + (n_2 - n_4)\theta_2]} \tag{7.91}$$

$$p(r_+, r_-) = \sum_{n_1 n_2 n_3 n_4=0}^{D} \rho_{n_1 n_2 n_3 n_4} k_{n_1 n_2} k_{n_3 n_4} e^{\frac{r_+^2 + r_-^2}{2}} H_{n_1}\left(\tfrac{1}{2}(r_+ + r_-)\right) H_{n_2}\left(\tfrac{1}{2}(r_+ - r_-)\right) \times$$
$$H_{n_3}\left(\tfrac{1}{2}(r_+ + r_-)\right) H_{n_4}\left(\tfrac{1}{2}(r_+ - r_-)\right) e^{-i[(n_1 - n_3)\theta_1 + (n_2 - n_4)\theta_2]} \tag{7.92}$$

$$p(s_1, s_2) = \sum_{n_1 n_2 n_3 n_4=0}^{D} \rho_{n_1 n_2 n_3 n_4} k_{n_1 n_2} k_{n_3 n_4} e^{s_1^2 + s_2^2} H_{n_1}(s_1) H_{n_2}(s_2) H_{n_3}(s_3) H_{n_4}(s_4) \times$$
$$e^{-i\left[(n_1 - n_3)\left(\theta_1 + \frac{\pi}{2}\right) + (n_2 - n_4)\left(\theta_2 + \frac{\pi}{2}\right)\right]} \tag{7.93}$$

$$p(s_+, s_-) = \sum_{n_1 n_2 n_3 n_4=0}^{D} \rho_{n_1 n_2 n_3 n_4} k_{n_1 n_2} k_{n_3 n_4} e^{\frac{s_+^2 + s_-^2}{2}} H_{n_1}\left(\frac{1}{2}(s_+ + s_-)\right) H_{n_2}\left(\frac{1}{2}(s_+ - s_-)\right) \times$$
$$H_{n_3}\left(\frac{1}{2}(s_+ + s_-)\right) H_{n_4}\left(\frac{1}{2}(s_+ - s_-)\right) e^{-i\left[(n_1-n_3)\left(\theta_1 + \frac{\pi}{2}\right) + (n_2-n_4)\left(\theta_2 + \frac{\pi}{2}\right)\right]} \quad (7.94)$$

Then the entropies can be evaluated and the inequalities checked.

### 7.3.2 Analysis for pure states

To test the efficiency of the criteria for entanglement detection I have generated (for different values of $D$) random pure states of the form

$$|\psi\rangle = \sum_{n_1, n_2=0}^{D} c_{n_1 n_2} |n_1 n_2\rangle. \quad (7.95)$$

Here we denote by $\{|n\rangle, n = 0, 1, 2, ...\}$ the eigenstates of a one-dimensional harmonic oscillator. Therefore

$$\langle x|n\rangle = \Psi_n(x) = e^{-\frac{x^2}{2}} H_n(x), \qquad n = 0, 1, 2, ... \quad (7.96)$$

In these random states the coefficients $c_{n_1 n_2}$ have been generated uniformly distributed according to the Haar measure. In Figure 7.10 the percentage of detected states as a function of $D$ for the strong and weak inequality is plotted. The angles $\theta_1$ and $\theta_2$ have been scanned in intervals by $\pi/4$. These results are in agreement with the original ones [1]. However, the results summarized in Fig. 7.10 are considerably more reliable than those reported in [1], because I generated many more states and, consequently I have a better statistics. We see that the criteria sensibility is smaller for higher dimensions, principally for the weak inequality.

Also the efficiency of the criteria has been investigated as a function of the number of different values of $\theta_1$ and $\theta_2$ that have been scanned. In Figure 7.11 I plotted the percentage of states detected in function of the numbers of angles used for both $\theta_1$ and $\theta_2$. As can be seen, detection efficiency saturates when the number of angles growths.

The criteria has been also tested statistically for maximally entangled states within the subspace spaned by $|00\rangle$, $|01\rangle$, $|10\rangle$ $|11\rangle$ (corresponding to $D = 1$). For that I take the state

$$|\psi\rangle = \frac{1}{\sqrt{2}}\left(|00\rangle + |11\rangle\right), \quad (7.98)$$

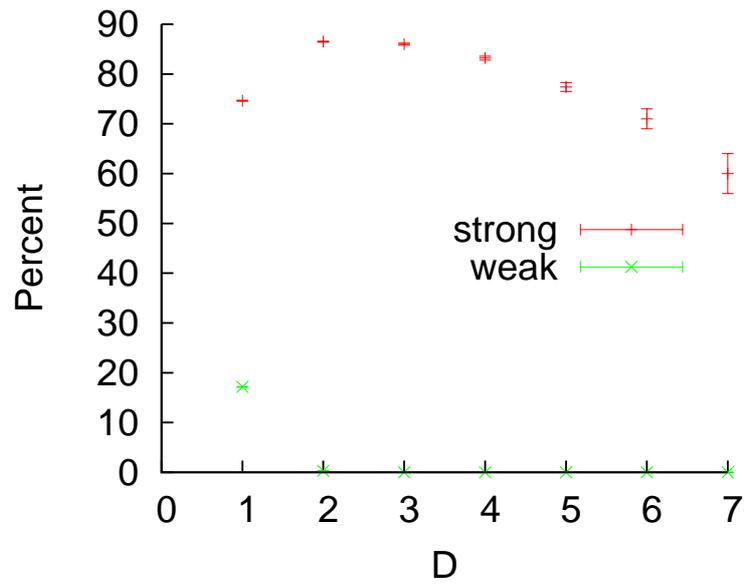

FIGURE 7.10: Percentage of detected states in function of the dimension, for both the strong and weak criteria. The angles $\theta_1$ and $\theta_2$ have been scanned in intervals by $\pi/4$.

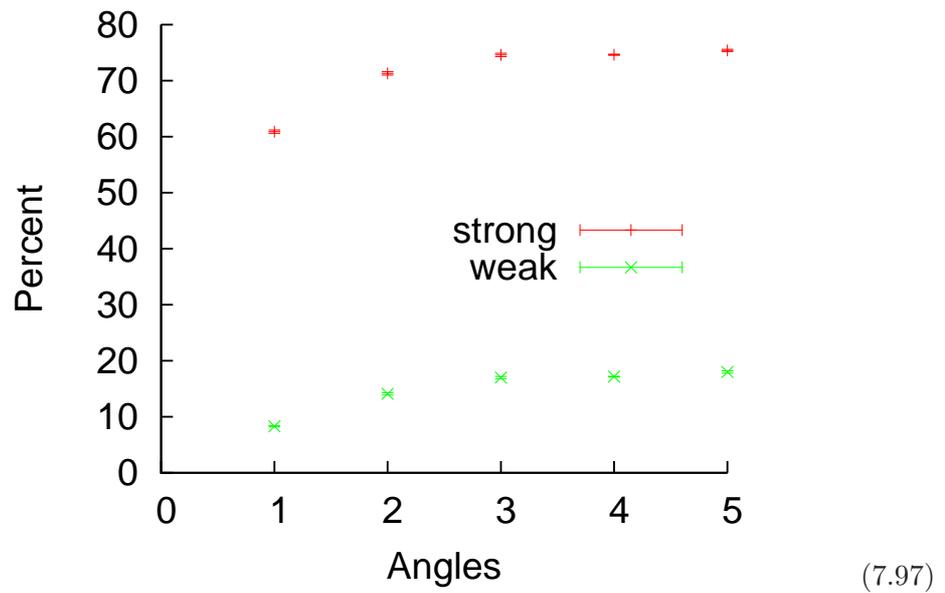

(7.97)

FIGURE 7.11: Percentage of detected states in function of the angles scanned for $\theta_1$ and $\theta_2$ ($D=1$).

and I apply a local random unitary operation to the first qubit. This operation is represented by the unitary operator,

$$U = U_1 \otimes \mathbb{1}_2, \tag{7.99}$$

where the subindex indicates the qubit upon which the operator acts. The unitary operator $U_1$ can be parametrized in terms of the Euler angles as

$$U = e^{i\alpha} \begin{pmatrix} e^{-i\left(\frac{\beta}{2}+\frac{\delta}{2}\right)} \cos\left(\frac{\gamma}{2}\right) & -e^{i\left(-\frac{\beta}{2}+\frac{\delta}{2}\right)} \sin\left(\frac{\gamma}{2}\right) \\ e^{i\left(\frac{\beta}{2}-\frac{\delta}{2}\right)} \sin\left(\frac{\gamma}{2}\right) & e^{i\left(\frac{\beta}{2}+\frac{\delta}{2}\right)} \cos\left(\frac{\gamma}{2}\right) \end{pmatrix}. \tag{7.100}$$

where $\delta \in [0, 2\pi]$, $\beta \in [0, 2\pi]$, $\gamma \in [0, \pi]$ and $\alpha \in [0, 2\pi]$ is the global phase.

The results can be seen in Figure 7.12, for different numbers of angles scanned for $\theta_1$ and $\theta_2$.

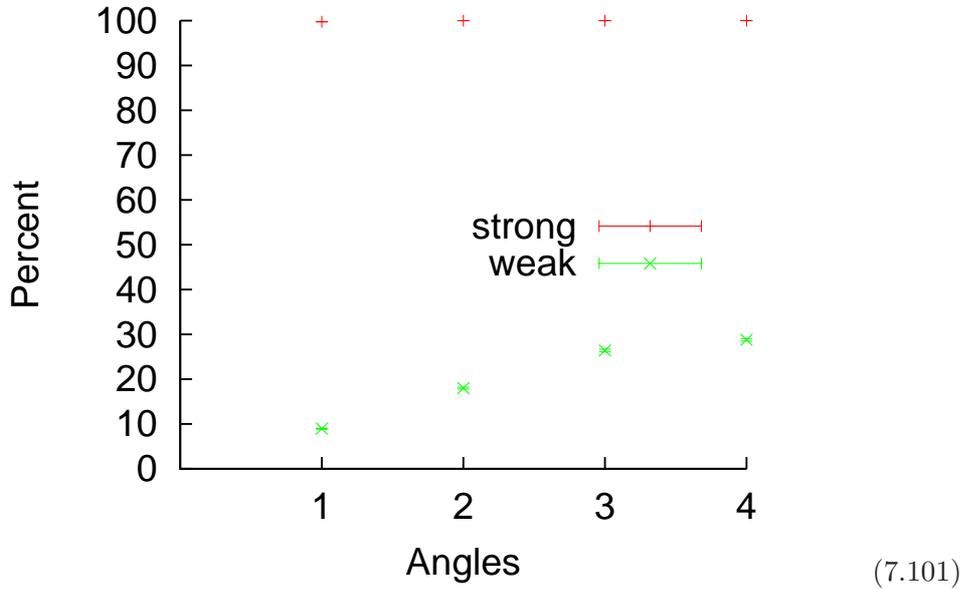

(7.101)

FIGURE 7.12: Percentage of detected states in function of the angles scanned for $\theta_1$ and $\theta_2$ for maximally entangled states.

I have also tested the detection capacity of the criterion as a function of the entanglement of the system. For that I have selected the state

$$|\psi\rangle = \cos\beta|00\rangle + \sin\beta|11\rangle, \tag{7.102}$$

that depends on one parameter $\beta$. The amount of entanglement of it system can be characterized as the von Neumann entropy $\epsilon = \cos^2\beta \log(\cos^2\beta) - \sin^2\beta \log(\sin^2\beta)$, so it depends only in $\beta$. For generating random states with the same amount of entanglement I applied a random operator on the state of Eq. 7.102, it has the form

$$U = U_1 \otimes U_2, \tag{7.103}$$

where $U_1$ and $U_2$ are unitary random operators.

As can be seen in Figure 7.13, the detection efficiency of the criterion increases with the entanglement of the system.

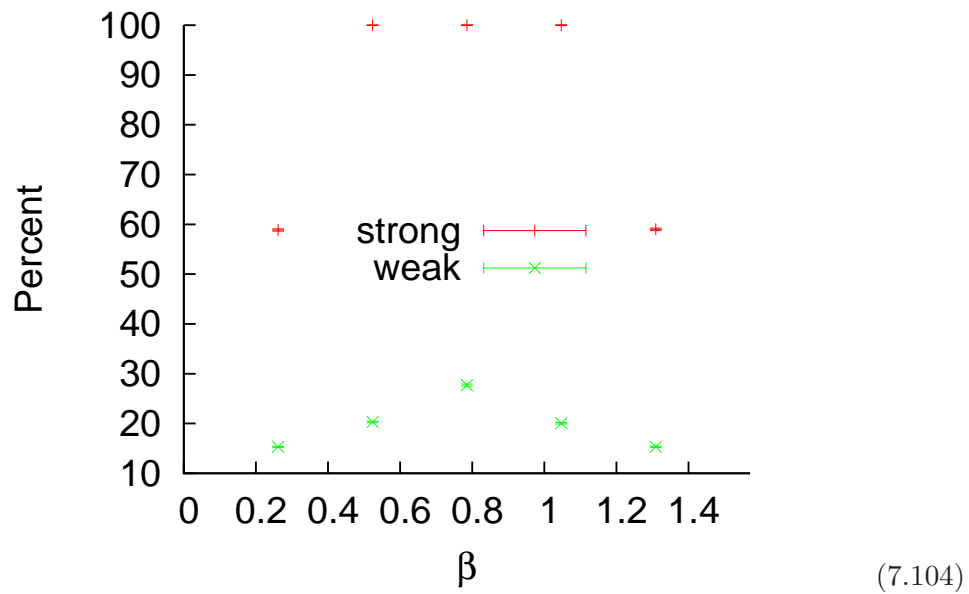

$$\tag{7.104}$$

FIGURE 7.13: Percentage of detected states for the random states generated by $\cos\beta|00\rangle + \sin\beta|11\rangle$ as a function of $\beta$. ($\theta_1$ and $\theta_2$ have been scanned in intervals by $\pi/4$).

The criteria can also detect entanglement for states "$D00D$" of the form $(|D0\rangle + |0D\rangle)/\sqrt{2}$ [1]. The strong criteria detects entanglement up to $D = 5$ with $\theta_1 = \theta_2 = 0$, except for $D = 2$ where it is detected for $\theta_1 = 0$ and $\theta_2 = \pi/2$. We have tested this inequality also for a state of the form

$$|\psi\rangle_{1001} = e^{i\alpha}\cos\beta|01\rangle + \sin\beta|10\rangle. \tag{7.105}$$

In Figure 7.14 can be seen the amount of the violation of strong inequalities as a function of $\beta$ and $\alpha$ for $\theta_1 = \theta_2 = 0$. When the two strong inequalities are violated the amount of the violation is added. A similar plot is shown in Figure 7.15 but, in this case the angles $\theta_1$ and $\theta_2$ are scanned in 4 values each. The weak inequalities can not detect any entanglement for this state.

Other state that have been tested is the "Bell-type" state. It has the form

$$|\psi\rangle_{\text{Bell}} = e^{i\alpha}\cos\beta|00\rangle + \sin\beta|11\rangle. \tag{7.106}$$

The results are shown in Figures 7.16 and 7.17.

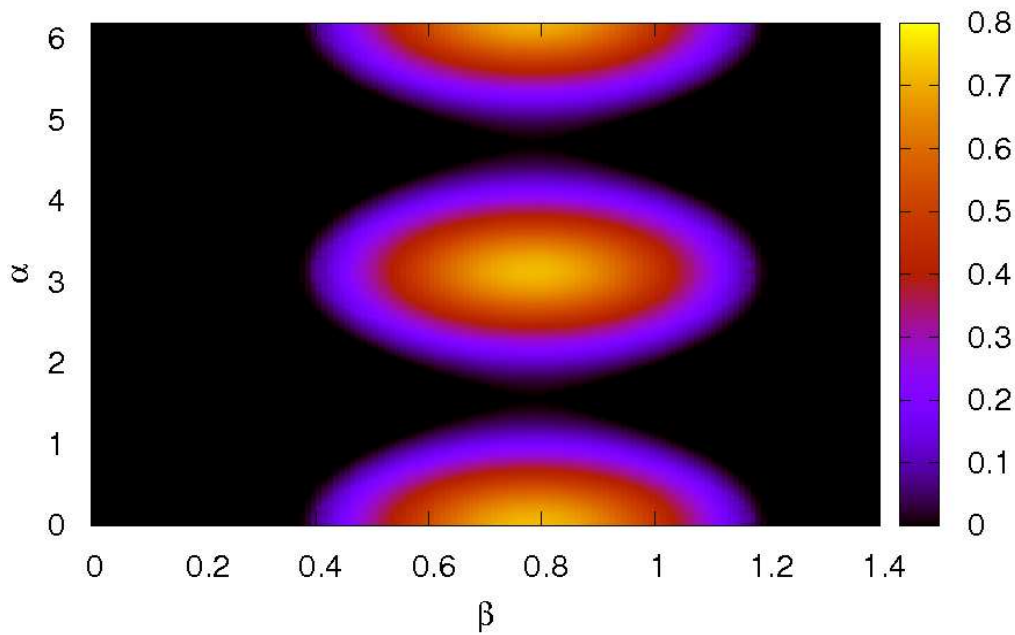

FIGURE 7.14: Amount of violation for the strong inequalities for "1001" states $(\theta_1 = \theta_2 = 0)$.

Finally I have studied also maximum entangled pure states (MEPS) with the form

$$|\psi\rangle_{\mathrm{MEPS}} = \frac{1}{\sqrt{2}} \left( \cos\beta |00\rangle + e^{i\phi} \sin\beta |01\rangle - e^{-i\phi} |10\rangle - \cos\beta |11\rangle \right). \tag{7.107}$$

The amount of violation of the inequalities for this state is shown in Figures 7.18 and 7.19 as a function of $\beta$ and $\phi$.

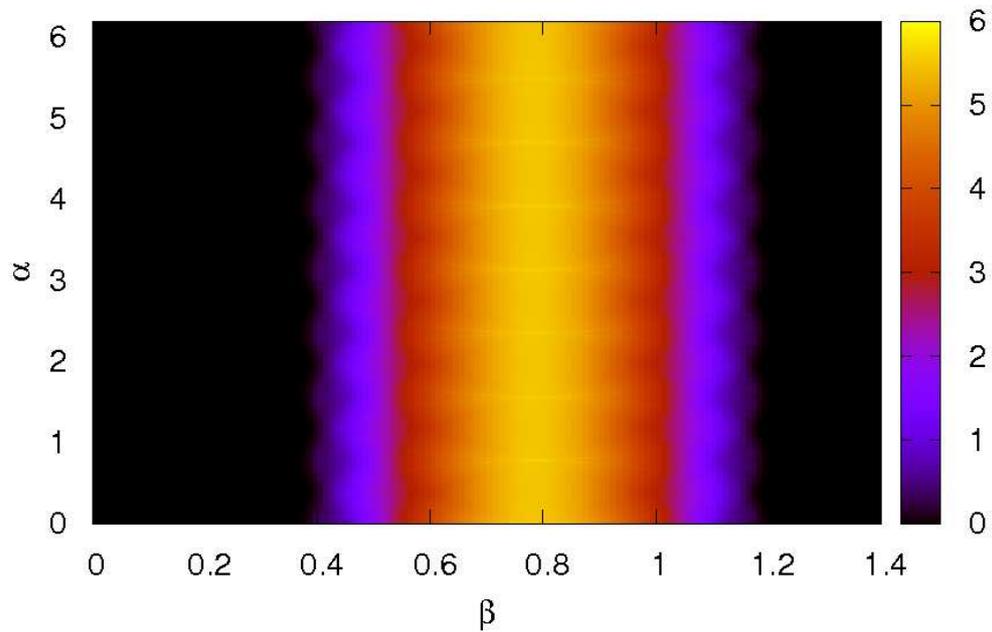

FIGURE 7.15: Amount of violation for the strong inequalities for "1001" states ($\theta_1$ and $\theta_2$ scanned in intervals of $\frac{\pi}{4}$).

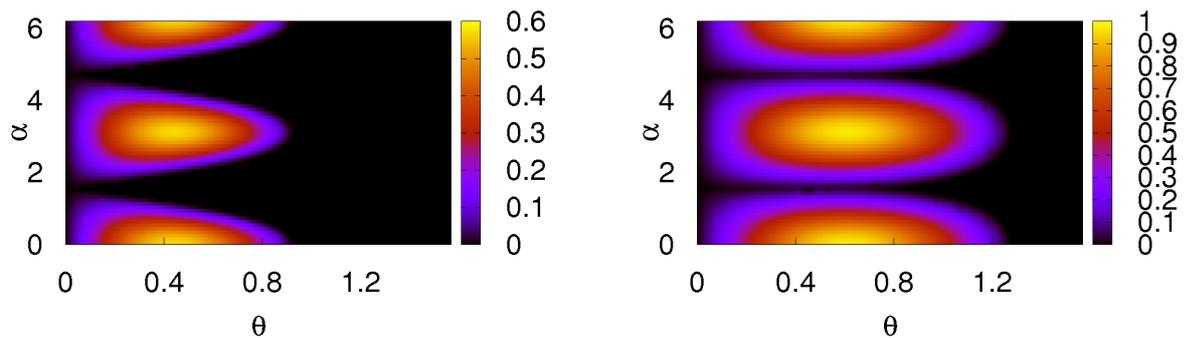

FIGURE 7.16: Amount of violation for the weak (left) and strong (right) inequalities for the "Bell-type" states ($\theta_1 = \theta_2 = 0$).

I have also explored the dependence of the separability criteria on the values of the angles $\theta_1$ and $\theta_2$. The amount of violation of the inequality (7.86) corresponding to the strong criterion and of inequality (7.87) (as a function of $\theta_1$ and $\theta_2$) is depicted in Figure 7.20 for the state $|\psi_+\rangle = \frac{1}{\sqrt{2}}\left(|00\rangle + |11\rangle\right)$ and in Figure 7.21 the amount of violation for the weak criteria for the state $|\phi_+\rangle = \frac{1}{\sqrt{2}}(|01\rangle + |10\rangle)$ is shown. The weak criterion never detects entanglement in the case of state $|\phi_+\rangle$.

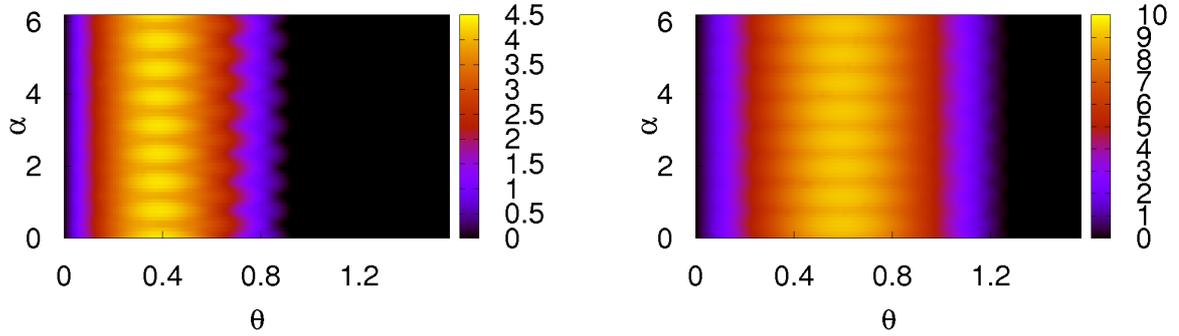

FIGURE 7.17: Amount of violation for the weak (left) and strong (right) inequalities for the "Bell-type" states ($\theta_1$ and $\theta_2$ scanned in intervals of $\frac{\pi}{4}$).

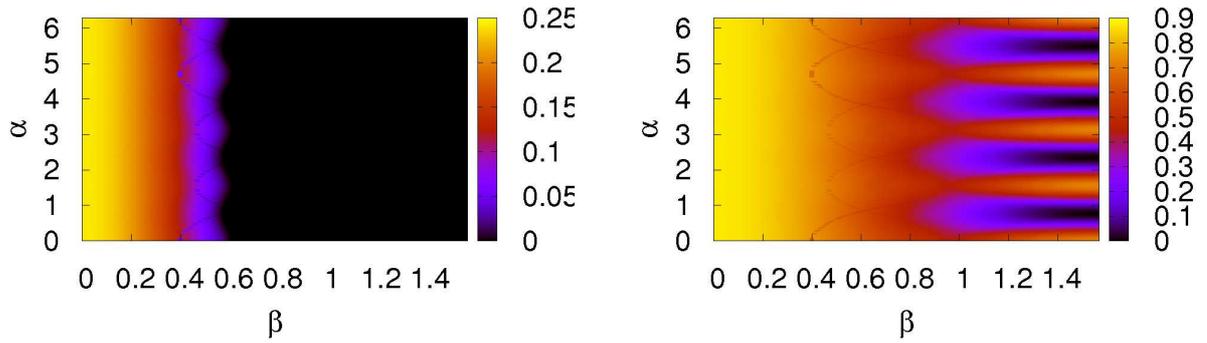

FIGURE 7.18: Amount of violation for the weak (left) and strong (right) inequalities for the MEPS states ($\theta_1 = \theta_2 = 0$).

### 7.3.3 Analysis for mixed states

First we have tested the detection capacity of the criterion for states of the form.

$$\rho = p|\psi_\text{B}\rangle\langle\psi_\text{B}| + \frac{1-p}{4}\mathbb{I} \qquad (7.108)$$

where $\mathbb{I}$ is the identity matrix, and $|\psi_\text{B}\rangle$ can be any of the Bell's states

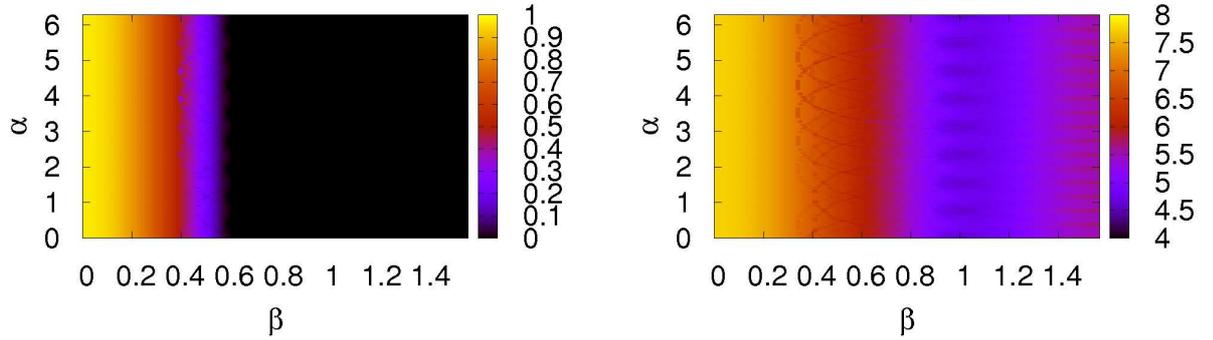

FIGURE 7.19: Amount of violation for the weak (left) and strong (right) inequalities for the MEPS states with $\theta_1$ and $\theta_2$ scanned in intervals of $\frac{\pi}{4}$.

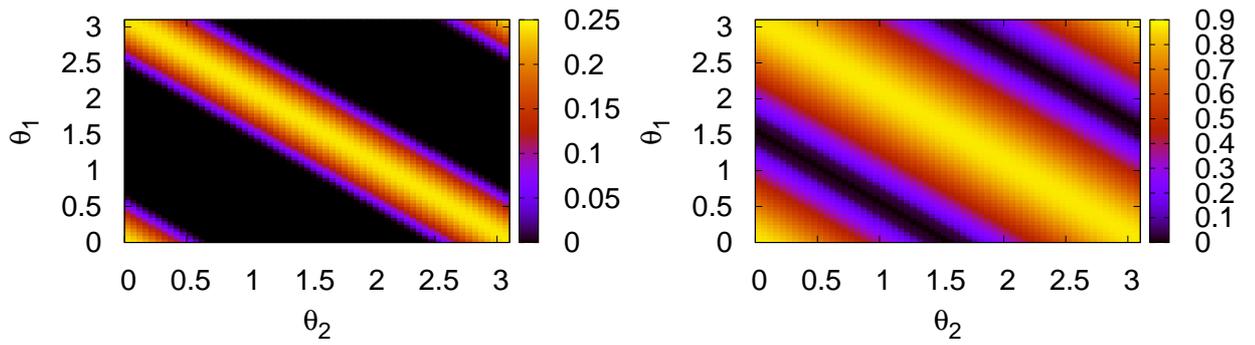

FIGURE 7.20: Amount of violation for the weak (left) and strong (right) inequalities for the state $\psi_+ = \frac{1}{\sqrt{2}}(|00\rangle + |11\rangle)$ as a function of $\theta_1$ and $\theta_2$.

$$|\Psi^+\rangle = \frac{1}{\sqrt{2}}(|00\rangle + |11\rangle)$$
$$|\Psi^-\rangle = \frac{1}{\sqrt{2}}(|00\rangle - |11\rangle) \quad (7.109)$$
$$|\Phi^+\rangle = \frac{1}{\sqrt{2}}(|01\rangle + |10\rangle)$$
$$|\Psi^-\rangle = \frac{1}{\sqrt{2}}(|01\rangle - |10\rangle)$$

$$(7.110)$$

Setting $\theta_1 = \theta_2 = 0$ I found that for the states $\rho_{\Psi^+}$ and $\rho_{\Psi^-}$ the weak criterion detects entanglement when $p \geq 0.828 \pm 0.001$. For the states $\rho_{\Phi^+}$ and $\rho_{\Phi^-}$ entanglement is not

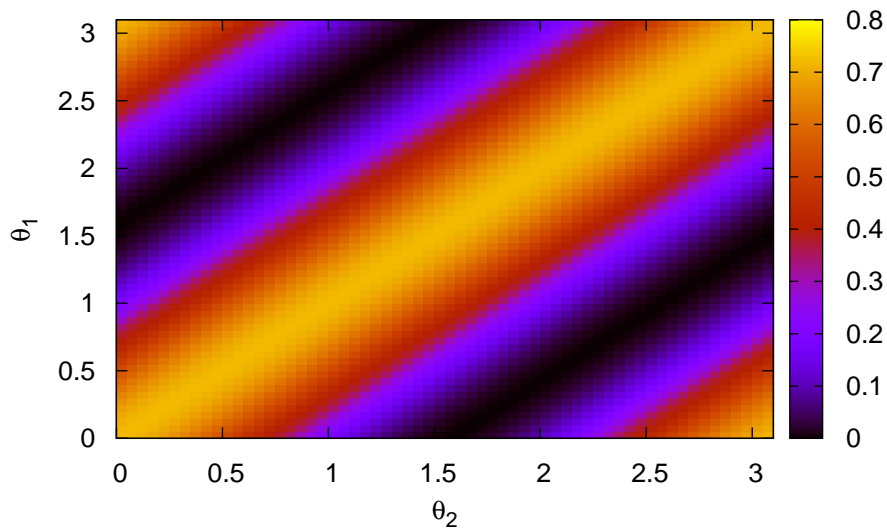

FIGURE 7.21: Amount of violation for the strong inequality for the state $|\phi_+\rangle = \frac{1}{\sqrt{2}}(|01\rangle + |10\rangle)$ as a function of $\theta_1$ and $\theta_2$.

detected. For understanding it we have plot in Figure 7.22 the amount of the violation of the inequality as a function of $p$

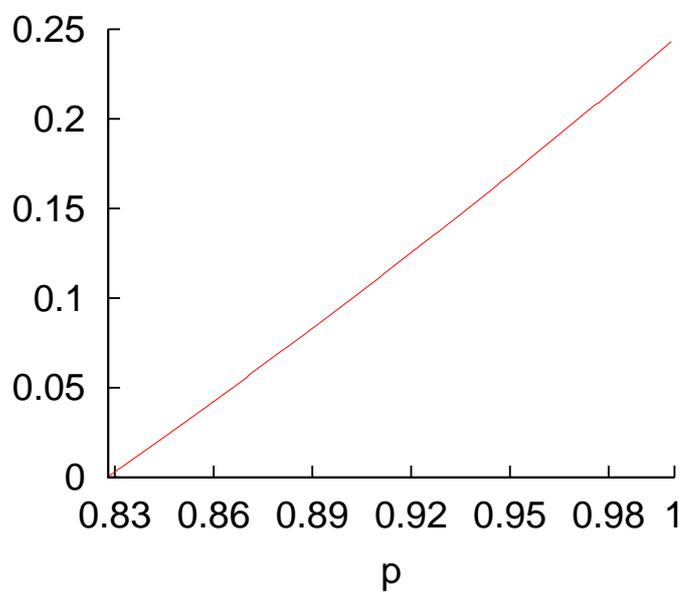

(7.111)

FIGURE 7.22: Magnitude $\log(2\pi e) - H[r_+] + H[s_-]$ for the state $\rho_{\Phi^+}$, that is the same as the magnitude $\log(2\pi e) - H[r_-] + H[s_+]$ for $\rho_{\Phi^-}$ ($\theta_1 = \theta_2 = 0$).

I have also accomplished an statistical study of states having the form

$$\rho = p|\psi'\rangle\langle\psi'| + \frac{1-p}{4}\mathbb{I}, \tag{7.112}$$

where $|\psi'\rangle$ is the result of applying a random unitary operation to the first qubit (like Eq. 7.99) on the state $|\Psi^+\rangle$. The results are shown in Figure 7.23 for different values of $p$. As can be seen the percentage of states detect increases linearly with $p$.

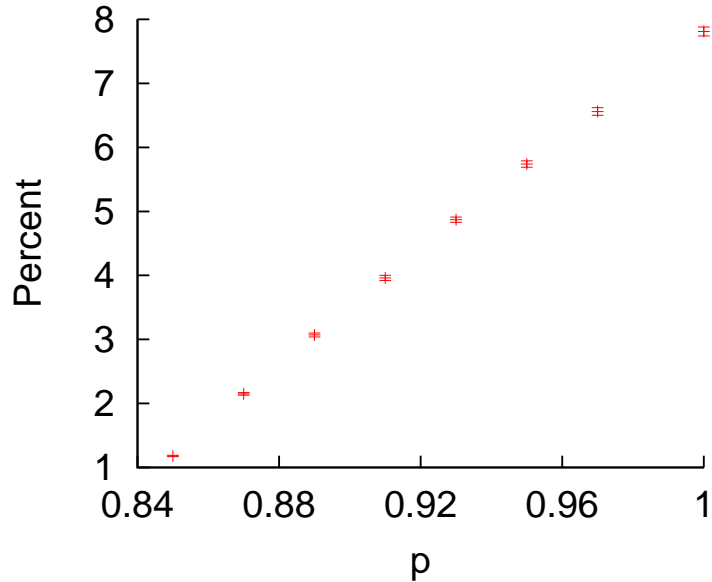

FIGURE 7.23: Percentage of states detected by the criteria for states (7.112) with $\theta_1 = \theta_2 = 0$.

I have also tested states similar to (7.112), but using

$$|\psi\rangle = \cos\beta|00\rangle + \sin\beta|11\rangle \tag{7.113}$$

The violation of the inequalities are shown in Figure 7.24 as a function of $\beta$ and $p$.

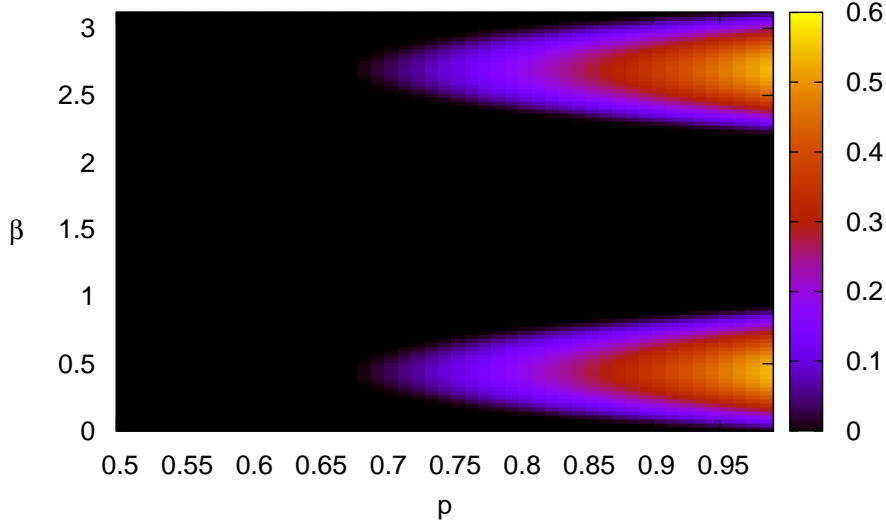

FIGURE 7.24: Amount of the violation of the inequality for states (7.113) as a function of $\beta$ and $p$.

Other interesting states are the MEMS states. These are two-qubit states of maximum entanglement for a given degree of mixedness [191]. They depend only on one parameter $\gamma$ and their density matrix is

$$\rho_{\text{MEM}} = g(\gamma)|00\rangle\langle 00| + \frac{\gamma}{2}|00\rangle\langle 11| + +(1 - 2g(\gamma))|01\rangle\langle 01| + \frac{\gamma}{2}|11\rangle\langle 00| + g(\gamma)|11\rangle\langle 11| \tag{7.114}$$

where

$$g(\gamma) = \begin{cases} \frac{\gamma}{2} & \gamma \geq \frac{2}{3} \\ \frac{1}{3} & \gamma < \frac{2}{3} \end{cases}$$

On the case of MEMS states I found that the criteria ($\theta_1 = \theta_2 = 0$) detects entanglement if $\gamma \geq 0.849 \pm 0.001$. In Figure 7.25 the amount of the violation of the inequality can be seen for $\theta_1 = \theta_2 = 0$

Finally we have tested diagonal mixed states in Bell basis (DMSBB). Its are mixed states that depend on 4 parameters $\lambda_i$. The explicit expression is

$$\rho_{\text{DMSBB}} = \lambda_1|\Psi^+\rangle\langle\Psi^+| + \lambda_2|\Psi^-\rangle\langle\Psi^-| + \lambda_3|\Phi^+\rangle\langle\Phi^+| + \lambda_4|\Phi^-\rangle\langle\Phi^-| \tag{7.115}$$

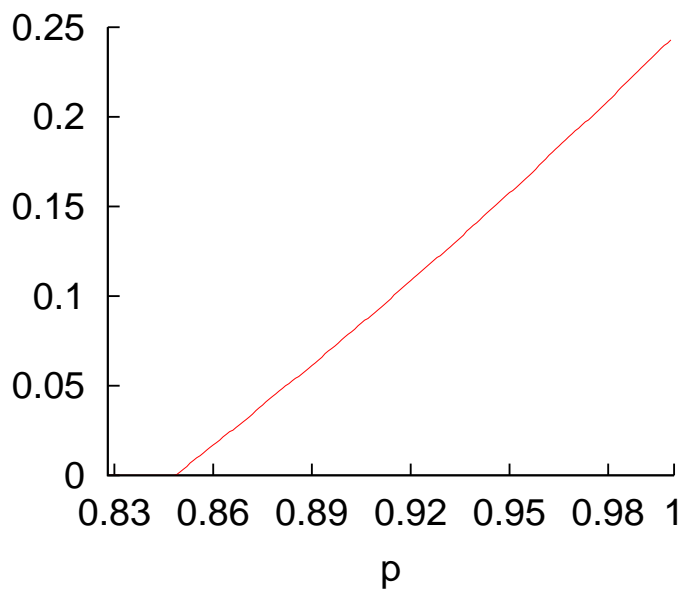

FIGURE 7.25: $H[r_-] + H[s_+] - \log(2\pi e)$ for MEMS state ($\theta_1 = \theta_2 = 0$).

where $\lambda_1 + \lambda_2 + \lambda_3 + \lambda_4 = 1$ and the pure states are the Bell states given by Eq. 7.110 There are 3 free parameters.

The result is that the criterion ($\theta_1 = \theta_2 = 0$) needs $\lambda_1 > 0.85$, or $\lambda_2 > 0.85$ for detecting entanglement, it is blind to the values of $\lambda_3$ and $\lambda_4$.

If we do $\lambda_3 = 0$ and let $\lambda_4$ as a normalization parameter we have 2 free parameters, $\lambda_1$ and $\lambda_2$. Te amount of violation for this case can be seen in Figure 7.26 for $\theta_1 = \theta_2 = 0$.

## 7.4 Conclusions

We have derived a couple of inequalities involving respectively the purity and the Von Neumann entropy of the single particle, reduced density matrix $\rho_r$ of an $N$-fermion pure state. These inequalities lead directly to simple and practical (necessary and sufficient) separability criteria based on the verification of a single identity. These criteria are drastically simpler than others that have been considered (for $N > 2$) in the recent literature. Moreover, the aforementioned inequalities also suggest two practical measures of entanglement for fermionic pure states. In the particular case of $N = 2$ the separability criteria discussed by us reduce to the criteria derived in [204] (see also [208]) by recourse to the fermionic Schmidt decomposition.

Also we have explored the entanglement-related properties of two models for systems composed by two charged fermions, the Crandall and Hooke atoms, in terms of the strength of the confining potential. We have considered particular values of the Hamiltonian parameters and eigenstates' quantum numbers that allow for exact solutions of the corresponding Schrödinger eigenvalue equation. Even though we have analytical expressions for the eigenfunctions of the models, the associated amounts of entanglement have to be computed numerically.

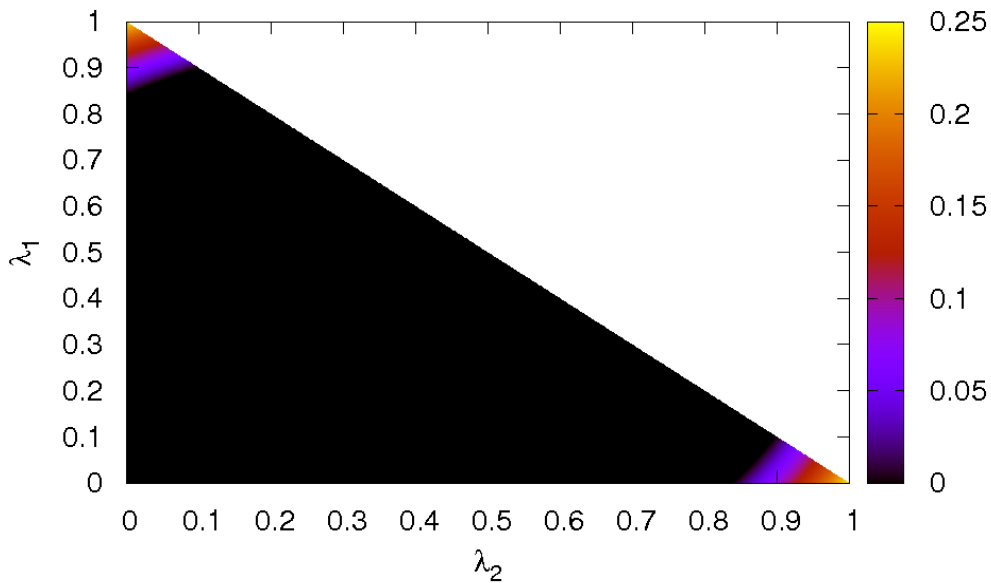

FIGURE 7.26: Amount of the violation for the DMSBB with $\lambda_3 = 0$ ($\theta_1 = \theta_2 = 0$).

The main entanglement features exhibited by the eigenstates of the Crandall and Hooke atoms are similar. In both cases the behaviour of the entanglement associated with the eigenstates of the system obeys the same general patterns. The amount of entanglement of the eigenstates tends to increase with the corresponding eigenenergy. The entanglement also tends to increase with the relative strength of the electron-electron interaction (as compared with the strength of the confining harmonic potential), approaching its maximum when the interaction becomes large enough. On the other hand, when the interaction tends to zero (but is still finite) the entanglement of the eigenstates does not necessarily go to zero. There are eigenstates endowed with a finite amount of entanglement for arbitrarily weak (but non vanishing) interaction.

It would be interesting to explore systematically the entanglement properties of other models, not confined within an harmonic well, in order to determine which (if any) of the above trends are shared by general two-particle systems, and which are special properties characterizing models with an external harmonic confining potential. As a first step towards this goal we have studied the entanglement of the ground and first excited states of Helium-like atoms. We found that the entanglement exhibited by the eigenfunctions of Helium-like atoms tends to increase with energy and decrease with the nuclear charge $Z$.

Finally in section 7.3 the behaviour of the separability criterion for quantum systems with continuous variable recently proposed by Walborn et al [1] has been studied. Some features of it are:

- The efficiency of the system decreases as the dimension of the system increases.

- Increasing the number of angles scanned for $\theta_1$ and $\theta_2$ increases the effectivity of the criterion but this growth saturates for more than 4 angles.

- For maximally entangled pure states within the $D = 1$ subspace the strong criterion detects practically all the entanglement.

- The detecting capacity of the criterion grows statistically with the amount of entanglement, but for given values of $\theta_1$ and $\theta_2$ there is not a perfect correlation between detection and entanglement.

- For the mixed families of mixed states (7.108) and the states that are diagonal in the Bell basis the criterion detects entanglement only in the case of very entangled states. It is sensitive only to the "parallel" states (like $(|00\rangle + |11\rangle)/\sqrt{2}$), and not for the "antiparallel" ones $((|01\rangle + |10\rangle)/\sqrt{2})$.

- The amount of violation of the inequalities for a determinate mixed state increases with the amount of entanglement of it.

# Appendix A

# Radial and angular contributions to the hydrogenic disequilibrium

Here we briefly describe the calculation of the radial and angular integrals involved in the determinatio of the disequilibrium of the hydrogenic orbital $(n, l, m)$ studied in Chapter 1.

To calculate the radial integral $K(n, l, m)$ given by equation (1.23) we have use the orthonormal Laguerre polynomial $L_k^{(\alpha)}(x)$, which has the relation

$$L_k^{(\alpha)}(x) = \left[\frac{\Gamma(k+\alpha+1)}{k!}\right]^{1/2} \tilde{L}_k^{(\alpha)}(x) \qquad (A.1)$$

with the orthonormal Laguerre polynomial $\tilde{L}_k^{\alpha}(x)$. Now, taking into account the linearization formula

$$\left[L_{n_r}^{(2l+1)}(x)\right]^2 = \frac{\Gamma(2l+n_r+2)}{2^{2n_r} n_r!} \sum_{k=0}^{n_r} \binom{2n_r - 2k}{n_r - k} \frac{(2k)!}{k!} \frac{1}{\Gamma(2l+2+k)} L_{2k}^{4l+2}(2x) \qquad (A.2)$$

and the orthogonality relation

$$\int_0^\infty \omega_\alpha(x) L_k^{(\alpha)}(x) L_{k'}^{(\alpha)}(x) = \frac{\Gamma(k+\alpha+1)}{k!} \delta_{kk'}, \qquad (A.3)$$

one has

$$\langle \rho \rangle_R = \frac{Z^3 2^{2-4n}}{n^5} \sum_{k=0}^{n_r} \binom{2n_r - 2k}{n_r - k}^2 \frac{(k+1)_k}{k!} \frac{\Gamma(4l+2k+3)}{\Gamma^2(2l+k+2)}$$

To calculate the angular integral $\langle \rho \rangle_Y$, we use the following linearization relation of the spherical harmonics

$$|Y_{lm}(\Omega)|^2 = \sum_{l'=0}^{2l} \frac{\hat{l}^2 \hat{l}'}{\sqrt{4\pi}} \begin{pmatrix} l & l & l' \\ 0 & 0 & 0 \end{pmatrix} \begin{pmatrix} l & l & l' \\ m & m & -2m \end{pmatrix} Y_{l',2m}^*(\Omega), \qquad (A.4)$$

where $\hat{a} = \sqrt{2a+1}$ and the $3j$-symbols [221] have been used. Then, taking into account Eqs. (1.21) and (A.4) one has that

$$\langle \rho \rangle_Y = \sum_{l'=0}^{2l} \frac{\hat{l}^2 \hat{l}'}{\sqrt{4\pi}} \begin{pmatrix} l & l & l' \\ 0 & 0 & 0 \end{pmatrix} \begin{pmatrix} l & l & l' \\ m & m & -2m \end{pmatrix} W(l,m), \tag{A.5}$$

where $W(l,m)$ denotes the following integral of three spherical harmonics

$$W(l,m) = \int_\Omega Y_{l,-m}(\Omega) Y_{l,-m}(\Omega) Y^*_{l,2m}(\Omega) \, d\Omega \tag{A.6}$$

Moreover, taking into account the known general integral

$$\int_\Omega Y_{l_1,m_1}(\Omega) Y_{l_2,m_2}(\Omega) Y^*_{l_3,m_3}(\Omega) \, d\Omega = \frac{\hat{l}_1 \hat{l}_2 \hat{l}_3}{\sqrt{4\pi}} \begin{pmatrix} l_1 & l_2 & l_3 \\ 0 & 0 & 0 \end{pmatrix} \begin{pmatrix} l_1 & l_2 & l_3 \\ m_1 & m_2 & m_3 \end{pmatrix},$$

one finally has

$$\langle \rho \rangle_Y = \sum_{l'=0}^{2l} \left(\frac{\hat{l}^2 \hat{l}'}{\sqrt{4\pi}}\right)^2 \begin{pmatrix} l & l & l' \\ 0 & 0 & 0 \end{pmatrix}^2 \begin{pmatrix} l & l & l' \\ m & m & -2m \end{pmatrix}^2.$$

# Appendix B

# $p$th-power of a polynomial of degree $n$ and Bell polynomials

Here we calculate the $p$th-power of a polynomial of degree $n$ in terms of the expansion coefficients $c_k (k = 1, ..., n)$, needed in Chapter 4

Consider the polynomial $y_n(x)$ with degree $n$,

$$y_n(x) = \sum_{k=0}^{n} c_k x^k.$$

Its $p$th-power is

$$[y_n(x)]^p = \left(\sum_{k=0}^{n} c_k x^k\right)^p = \sum_{\pi(p)} \frac{p!}{j_0! j_1! \cdots j_n!} (c_0 x^0)^{j_0} (c_1 x^1)^{j_1} \cdots (c_n x^n)^{j_n},$$

where the sum is defined over all the partitions $\pi(p)$ such that $j_0 + j_1 + \cdots + j_n = p$.

The previous expression of $[y_n(x)]^p$ can be written as

$$[y_n(x)]^p = \sum_{\pi(p)} \frac{p!}{j_0! j_1! \cdots j_n!} c_0^{j_0} c_1^{j_1} \cdots c_n^{j_n} x^{j_1 + 2j_2 + \cdots + nj_n} = \sum_{k=0}^{np} A_{k,p}(c_0, \ldots, c_n) x^k,$$

where

$$A_{k,p}(c_0, \ldots, c_n) = \sum_{\tilde{\pi}(k,p)} \frac{p!}{j_0! j_1! \cdots j_n!} c_0^{j_0} c_1^{j_1} \cdots c_n^{j_n}$$

where the sum is defined over all the partitions $\tilde{\pi}(k,p)$ such that

$$j_0 + j_1 + \cdots + j_n = p, \quad \text{and} \quad j_1 + 2j_2 + \cdots + nj_n = k.$$

The Bell polynomials are defined as

$$B_{m,l}(c_1, c_2, \ldots, c_{m-l+1}) = \sum_{\hat{\pi}(m,l)} \frac{m!}{j_1! j_2! \cdots j_{m-l+1}!} \left(\frac{c_1}{1!}\right)^{j_1} \left(\frac{c_2}{2!}\right)^{j_2} \cdots \left(\frac{c_{m-l+1}}{(m-l+1)!}\right)^{j_{m-l+1}}$$

where the sum is defined over all the partitions $\hat{\pi}(m,l)$ such that

$$j_1 + j_2 + \cdots + j_{m-l+1} = l, \quad \text{and} \quad j_1 + 2j_2 + \cdots + (m-l+1)j_{m-l+1} = m.$$

The relation between the coefficients $A_{k,p}(c_0, \ldots, c_{k-p+1})$ and the Bell polynomials can be established as

$$A_{k,p}(c_0, \ldots, c_n) = \frac{p!}{k!} \sum_{j_0=0}^{p} \frac{c_0^{j_0}}{j_0!} B_{k,p-j_0}(c_1, 2!c_2, \ldots, (k-p+j_0+1)!c_{k-p+j_0+1}),$$

where $c_i = 0$ for $i > n$.

From Equation (3l) from [109] we obtain that

$$\frac{1}{k!} \sum_{j_0=0}^{p} \frac{c_0^{j_0}}{j_0!} B_{k,p-j_0}(c_1, 2!c_2, \ldots, (k-p+j_0+1)!c_{k-p+j_0+1}) = \frac{1}{(k+p)!} B_{k+p,p}(c_0, 2!c_1, \ldots, (k+1)!c_k).$$

Then,

$$A_{k,p}(c_0, \ldots, c_n) = \frac{p!}{(k+p)!} B_{k+p,p}(c_0, 2!c_1, \ldots, (k+1)!c_k),$$

where, again, $c_i = 0$ for $i > n$.

# Appendix C

# Information and complexity measures

Here we list various information-theoretic measures of one (Shannon, Rényi and Tsallis entropies, Fisher Information) and two (Cramer-Rao, Fisher-Shannon and LMC complexities) components used in this work, which describe the spread of a probability density in a more appropriate (although complementary) way than the familiar variance.

### Variance

Although the variance of a random variable is not an information measure according to the usual definition, I have include it in this Appendix because it is the most familiar measure of the spreading of a probability distribution.

Let us consider a random variable $A$ with $N$ possible outcomes $\{A_i\}_{i=1}^{N}$ with the probability distribution $\{p_i\}_{i=1}^{N}$. An important quantity characterizing this kind of situation is given by uncertainty associated with $A$. The most common measure for this uncertainty is the *variance*

$$V[A] = \langle A^2 \rangle - \langle A \rangle^2, \tag{C.1}$$

where the mean value ($\langle \cdot \rangle$) is

$$\langle A \rangle = \sum_{i=1}^{N} p_i A_i. \tag{C.2}$$

This concept can be easily extended to continuous variables; for a probability distribution $\rho(x)$ of the variable $x \in [a, b]$ the *mean value* of the function $f(x)$ is

$$\langle f \rangle = \int_a^b f(x)\rho(x)dx \quad \text{(C.3)}$$

and then the expression of the variance $V[\rho]$ becomes

$$V[\rho] = \langle x^2 \rangle - \langle x \rangle^2 \quad \text{(C.4)}$$

For continuous variables the variance gives a measure of the spreading of the probability distribution function around it's centroid. This measure can be naturally extended to the case of a $D$ dimensional probability density defined in a region $\Omega^D \subset \mathbb{R}^D$

$$V[\rho] = \langle \vec{r}^2 \rangle - \langle \vec{r} \rangle^2 \quad \text{(C.5)}$$

where $\vec{r}$ is a vector such that $\vec{r} \in \Omega^D$ and the mean values are redefined as

$$\langle f \rangle = \int_{\Omega^D} f(\vec{r}) d^D r. \quad \text{(C.6)}$$

## Shannon entropy

The concept of *Shannon entropy* was introduced in 1948 by C.E. Shannon [71]. Shannon's entropic measure admits various possible axiomatic characterizations. One of the most appealing characterizations, from an intuitive point of view, is that given by the set of postulates advanced by Khinchin in 1957 [222].

Let us start with a random event $A$ with $N$ possible outcomes $\{A_i\}_{i=0}^{N}$ with probabilities $\{p_i\}_{i=0}^{N}$. The properties that we want for the uncertainty measure $H(p_i)$ are:

1. For a given set $\{p_i\}_{i=1}^{N}$ the function $H(p_i)$ takes the maximum value for the uniform distribution $p_i = \frac{1}{N}$.

2. $H(AB) = H(A) + H_A(B)$, where the symbol $H_A(B)$ stands for the "relative Shannon entropy" of the event $B$ with respect to the event $A$ (see discussion below).

3. The measure $H(p_1, ..., p_N) = H(p_1, ..., p_N, 0)$. The measure doesn't vary if you add an arbitrary set of measures with 0 probability.

The only function that fulfils all these properties is

$$H(A) = -\lambda \sum_{i=1}^{N} p_i \log p_i \quad \text{(C.7)}$$

where $\lambda$ can be any positive real constant; for simplicity we take $\lambda = 1$. The postulates 1 and 3 are easy to understand, but the postulate 2 needs a justification. Let us start with two random experiments $\{A_i, p_i\}_{i=1}^N$ and $\{B_i, q_i\}_{i=1}^N$ that are mutually dependent. The probability of having the outcome $B_i$ in the experiment $B$, given the occurrence of the outcome $A_j$ in the experiment $A$, is called *conditional probability* and is denoted as $p(B_i|A_j)$. These conditional probabilities are normalized, $\sum_i p(B_i|A_j) = 1$, and allow us to define the *conditional entropy* $H_j(B)$ of the experiment $B$, under the assumption that the event $A_j$ occurs in experiment $A$. The conditional entropy is,

$$H_j(B) = -\sum_{i=1}^N p(B_i|A_j) \log p(B_i|A_j). \tag{C.8}$$

The magnitude $H_j(B)$ depends on the particular outcome $A_j$ of the experiment $A$. Now we can define the conditional entropy between the experiment $B$ and $A$ as the average,

$$H_A(B) = \sum_{j=1}^N p_j H_j(B). \tag{C.9}$$

With these definitions for the relative entropies it is easy to see that the Shannon entropy given by Eq. C.7 fulfils all the assumptions proposed by Khinchin [222].

The Shannon entropy can also be used for continuous variables. If we have a probability density function $\rho(x)$ with $x \in [a, b]$ the Shannon entropy of the density is

$$S[\rho] = -\int_a^b \rho(x) \log \rho(x) dx \tag{C.10}$$

and to the case of a $D$ dimensional probability density in the space $\Omega^D$

$$S[\rho] = -\int_{\Omega^D} \rho(\vec{r}) \log \rho(\vec{r}) \, d^D r. \tag{C.11}$$

Note that the Shannon entropy for discrete variables is always positive but for continuous variables it can have negative values. To avoid this difficulty it is usual to use exponential functions like the *Shannon entropic power*

$$H[\rho] = \exp(S[\rho]) \tag{C.12}$$

Let us underline that: i) The Shannon entropy doesn't measure the spreading with respect to a specific point and ii) Shannon entropy is finite for some distributions where the variance diverges.

## Rényi entropies

Let us now reformulate the axioms leading to the Shannon entropy changing the second one for a less restrictive property.

2' If $A$ and $B$ are independent events then $H(AB) = H(A) + H(B)$.

There is a set of measures that fulfils conditions 1, 2' an 3 and is called Rényi entropies [97, 223]:

$$R_q[A] = \frac{1}{1-q} \log \left( \sum_{i=1}^{N} p_i^q \right), \tag{C.13}$$

with the continuous $D$-dimensional version:

$$R_q[\rho] = \frac{1}{1-q} \log \left( \int_{\Omega^D} [\rho(\vec{r})]^q \, \vec{r} \, d^D r \right). \tag{C.14}$$

The Shannon entropy can be considered as the limit of Rényi entropies when $q \to 1$.

## Tsallis entropies

Another important family of information measures closely related to the Rényi one, is given by the Tsallis entropies [114, 224]. These entropies are defined by

$$T_q[\rho] := \frac{1}{q-1} \left[ 1 - \left( \sum_{i=1}^{N} p_i^q \right) \right] \tag{C.15}$$

for the discrete case and

$$T_q[\rho] = \frac{1}{q-1} \left[ 1 - \int_{\Omega^D} [\rho(\vec{r})]^q \, \vec{r} \, d^D r \right], \tag{C.16}$$

for the continuous $D$ dimensional case.

The case $q = 2$ of Tsallis entropy is of special relevance in quantum theory. On the one hand, the measure $T_2[\hat{\rho}] = 1 - Tr\hat{\rho}^2$ (evaluated on a density matrix $\hat{\rho}$) constitutes a basic ingredient of usefull quantitative indicators of the amount of entanglement exhibited by pure states of composite systems [69, 225] (in these applications, the measure $T_2[\hat{\rho}]$, also known as linear entropy, is evaluated on the marginal density matrices associated with the subsystems of the system under consideration). On the other hand, intriguing recent development suggest that the quadratic measure $T_2$ may play an important role at the very foundations of quantum physics (see [226] and references therein).

## Fisher Information

The concept of Fisher information was introduced by R.A. Fisher in 1925 [81] in the context of statistical estimation theory [227]. Imagine that we have an experiment and we want to estimate the parameter $\theta$ making $N$ measurements in the system. The output of any measure will be

$$y_i = \theta + x_i, \tag{C.17}$$

where the $x_i$'s are random variables (noise). The purpose now is to calculate a value $\hat{\theta}$ as close as possible to the real value $\theta$.

The system is defined by a conditional probability given by the family of the probability densities $\rho_\theta(y_1, ..., y_N) \equiv \rho(y_1, ..., y_N, \theta)$. Then the Fisher information of the measurement is defined as

$$I(\theta) = \int_{\Omega^N} \left[\frac{\partial \log \rho_\theta(y_1, ..., y_N)}{\partial \theta}\right]^2 \rho_\theta(y_1, ..., y_N) dy_1...dy_N = \int_{\Omega^N} \frac{\left[\frac{\partial \rho_\theta(y_1,...,y_N)}{\partial \theta}\right]^2}{\rho_\theta(y_1, ..., y_N)} dy_1...dy_N \tag{C.18}$$

where $\Omega^N$ is the space of the variables $\{y_i\}_{i=0}^{N}$.

If we use the *mean-square error* of our estimation $\hat{\theta}(y_1,...y_N)$

$$\sigma^2 = \int_{\Omega^N} \left[\hat{\theta}(y_1,...y_N) - \theta\right]^2 \rho_\theta(y_1, ..., y_N) dy_1...dy_N, \tag{C.19}$$

it fulfils the Cramer-Rao inequality [15].

$$\sigma^2(\theta) I(\theta) \geq 1. \tag{C.20}$$

The Fisher information can be seen as a measure of the ability for determining the parameter $\theta$. It gives the minimum possible error in estimating $\theta$ for a given probability density $\rho_\theta(y_1, ..., y_N)$.

If we have only one measure $y$ in the range $[a, b]$, and the noise fluctuations are independent of the value of $\theta$ (shift invariance) the probability density fulfils

$$\rho_\theta(y) = \rho(y - \theta) = \rho(x) \tag{C.21}$$

so the expression C.18 becomes

$$I[\rho] = \int_a^b \frac{\left[\frac{\partial \rho(x)}{\partial x}\right]^2}{\rho(x)} dx \tag{C.22}$$

that is the *translationally-invariant Fisher information*; it measures the amount of gradient, so being sensitive to local changes of the probability distribution and oscillations. This definition can be also extended to a $D$ dimensional density defined in the space $\Omega^D \in \mathbb{R}^D$ as

$$I[\rho] = \int_{\Omega^D} \frac{[\nabla_D \rho(\vec{r})]^2}{\rho(\vec{r})} d^D r = \int_{\Omega^D} \rho(\vec{r}) \left|\nabla_D \log \rho(\vec{r})\right|^2 d^D r \tag{C.23}$$

## Complexity measures

Complexity is a hard and elusive concept. While everyone has an intuitive idea of complexity there is no consensus yet on the proper mathematical formulation of this concept. Considerable effort has been dedicated to the exploration of various possible ways to determine quantitatively the amount of "complexity" exhibited by a physical system or process [4, 11, 16, 86]. Here we are going to consider some mathematical measures of complexity for continuous probability distribution functions.

In spite of the aforementioned difficulties, there is general agreement on some basic properties that every complexity measure must fulfil. These requirements are:

1. Minimal value for the simplest probability densities (for a continuous variable in one dimension these are the uniform and Dirac-Delta distributions).

2. Invariance under replication, translation and rescaling transformation.

3. Easy mathematical formulation.

Three different measures of complexity that fulfil these conditions have been used in this work: the LMC shape complexity, the Cramer-Rao and the Fisher-Shannon measures.

The LMC shape complexity was first proposed by López-Ruiz, Mancini and Calbet [86] as a measure of complexity for discrete probability distributions. Anteneodo and Plastino investigated the behaviour of the LMC measure and pointed out some of its deficiencies [4]. This, in turn, motivates LMC to advance a new, improved version of their shape complexity measure given by

$$C_{SC}[\rho] = \langle \rho \rangle \times H[\rho], \tag{C.24}$$

where $\langle \rho \rangle$ is called disequilibrium and is defined (in $D$ dimensions) as

$$\langle \rho \rangle = \int_{\Omega^D} [\rho(\vec{r})]^2 \, d^D r \tag{C.25}$$

and $H[\rho]$ is the Shannon entropic power given by Eq. (C.12).

The Shannon entropic power is a measure of the spreading of the probability distribution, while the disequilibrium measures it's average height. For the Dirac-Delta distribution the Shannon entropy is zero and so is the disequilibrium for the uniform distribution.

The Cramer-Rao complexity has a deep relation with the inequality (C.20). It is defined by the product [11, 16]:

$$C_{CR}[\rho] = I[\rho] \times V[\rho], \tag{C.26}$$

being $I[\rho]$ the Fisher information, that measure the amount of gradient of the distribution and $V[\rho]$ is the variance, that is an spreading measure.

The Fisher-Shannon measure was proposed as a complexity measure by Angulo et al [13] but the Fisher-Shannon information plane has been considered before for studying other problems [228]. The Fisher-Shannon complexity measure is defined as

$$C_{FS}[\rho] = I[\rho] \times J[\rho], \tag{C.27}$$

with

$$J[\rho] = \frac{1}{2\pi e} \exp\left(2S[\rho]/3\right). \tag{C.28}$$

This measure is similar to the Cramer-Rao measure because it is composed by the Fisher information and a measure of spreading. In this case the variance is changed by the Shannon entropic power, that is a global spreading measure that doesn't depend in any specific point. As the Shannon entropy exists for some distribution where the variance doesn't, this measure of complexity can be applied to some distributions where the Cramer-Rao cannot.